\documentclass[final,3p,12pt]{elsarticle}

\usepackage{amssymb}
\usepackage{graphicx}
\usepackage{latexsym}
\usepackage{amsmath}
\usepackage{bbm}
\usepackage{color}
\usepackage{mathrsfs}
\usepackage{hyperref}

\newcommand{\SU}{\mathrm{SU}}
\newcommand{\U}{\mathrm{U}}
\newcommand{\SO}{\mathrm{SO}}

\newcommand{\SL}{\mathrm{SL}}
\newcommand{\Gtwo}{\mathrm{G}_2}
\newcommand{\LambdaQCD}{\Lambda_{\tiny{\mbox{QCD}}}}
\newcommand{\LambdaLat}{\Lambda_{\tiny{\mbox{lat}}}}
\newcommand{\MSbar}{$\overline{\textrm{MS}}$}
\newcommand{\Leff}{\mathcal{L}_{\tiny{\mbox{eff}}}}
\newcommand{\N}{\mathbb{N}}
\newcommand{\Z}{\mathbb{Z}}

\newcommand{\ide}{\mathbbm{1}}
\newcommand{\real}{{\rm{Re}}}

\newcommand{\diag}{{\rm{diag}}}

\newcommand{\dd}{{\rm{d}}}
\newcommand{\Tr}{{\rm Tr}}
\newcommand{\tr}{{\rm Tr}}
\newcommand{\beq}{\begin{eqnarray}}
\newcommand{\eeq}{\end{eqnarray}}
\newcommand{\topsusc}{\chi_{\mbox{\tiny{topol}}}}

\newcommand{\subs}[1]{\ensuremath{_{\textrm{#1}}}}

\journal{Physics Reports}

\begin{document}

\begin{frontmatter}
\begin{flushright} HIP-2012-24/TH\\ 
NSF-KITP-12-190
\end{flushright}



\title{$\SU(N)$ gauge theories at large $N$}


\author{Biagio Lucini}
\ead{b.lucini@swansea.ac.uk}
\address{College of Science, Swansea University, Singleton Park,
Swansea SA2 8PP, UK}

\author{Marco Panero}
\ead{marco.panero@helsinki.fi}
\address{Department of Physics and Helsinki Institute of Physics,
University of Helsinki, FIN-00014 Helsinki, Finland}
\address{Kavli Institute for Theoretical Physics, University of California, Santa Barbara, CA 93106, USA}

\begin{abstract}
We review the theoretical developments and conceptual advances that stemmed from the generalization of QCD to the limit of a large number of color charges, originally proposed by 't~Hooft. Then, after introducing the gauge-invariant non-perturbative formulation of non-Abelian gauge theories on a spacetime lattice, we present a selection of results from recent lattice studies of theories with a different number of colors, and the findings obtained from their extrapolation to the 't~Hooft limit. We conclude with a brief discussion and a summary.
\end{abstract}


\end{frontmatter}

\newpage
\thispagestyle{empty}
\tableofcontents
\newpage

\setcounter{page}{1}
\pagenumbering{arabic}

% \linenumbers

\section{Introduction}
\label{sec:intro}

In 1974, two seminal papers were published in the theoretical physics literature: on April 18, the article titled ``A planar diagram theory for strong interactions'', by Gerard~'t~Hooft, appeared in Nuclear Physics B~\cite{'tHooft:1973jz}, while, on October 15, Physical Review D published ``Confinement of quarks'', by Kenneth Geddes Wilson~\cite{Wilson:1974sk}. The former paper studied a non-Abelian gauge theory in the double limit of an infinite number of color charges $N$ and vanishing gauge coupling $g$, at fixed $g^2 N$, showing that, in this limit, Feynman diagrams could be arranged in an expansion in powers of $1/N$, in which the dominant contributions come from a limited class of diagrams with well-defined topological properties, and revealing a close analogy with the string model for strong interactions. The latter paper, on the other hand, introduced a regularization for non-Abelian gauge theories, by discretizing the continuum Euclidean spacetime on a grid, and proved the confinement of color charges in the strong-coupling limit of this model; this formulation laid the foundations of modern lattice field theory,\footnote{While the origin of lattice QCD can be identified with Wilson's article~\cite{Wilson:1974sk}, ideas related to the regularization of field theories on a spacetime grid were also discussed in earlier works by Wentzel~\cite{Wentzel:1940aa}, by Schiff~\cite{Schiff:1953zza}, by Wegner~\cite{Wegner:1984qt}, and in unpublished works by Smit and by Polyakov (see ref.~\cite{Kronfeld:2012ym}).} which provides a mathematically rigorous, gauge-invariant, non-perturbative definition of non-Abelian gauge theory,\footnote{An alternative computational framework for defining non-Abelian gauge theory in the strong coupling regime is provided by the formulation in light-cone quantization---see ref.~\cite{Brodsky:1997de} and references therein.} and makes it amenable to investigation by statistical field theory methods, including Monte Carlo simulations.

Both papers have been highly influential during the course of almost four decades, but the developments they inspired followed somewhat different historical paths. In particular, the elegant mathematical simplifications of the large-$N$ limit, the appeal of its qualitative phenomenological predictions, a record of successful applications of large-$N$ techniques in deriving analytical results for statistical spin systems, and the (semi-)analytical solution of the spectrum of large-$N$ QCD in two spacetime dimensions~\cite{'tHooft:1974hx}, led to early expectations that this approach could soon disclose the key to understand the low-energy dynamics of strong interactions. These expectations, however, turned out to be delusive, and, although this approach led to a number of fruitful theoretical developments (including, for example, the discovery of a closed set of equations for physical gauge-invariant operators~\cite{Makeenko:1979pb}, and, later, the formulation of a systematic $1/N$-expansion for baryons~\cite{Dashen:1993jt}), it did not provide an exact solution for QCD in four spacetime dimensions.

On the contrary, the early results (both analytical and numerical) in lattice field theory were mostly limited to a domain of parameters (gauge couplings, quark masses, lattice sizes, as well as simulation statistics) characterized by large discretization effects and systematic and/or statistical uncertainties, and provided valuable, but only qualitative, information about the continuum theory. Although QCD (with its full continuum symmetries) is expected to emerge as a low-energy effective description for the lattice model when the lattice spacing $a$ tends to zero, the suppression of artifacts of the lattice discretization in numerical simulations relies on the separation between the  typical energy scale $\mu$ relevant for a given physical quantity, and the intrinsic lattice cut-off $\pi/a$, so Monte Carlo computations have to be performed in a parameter range for which $\mu \ll a^{-1}$; in addition, at the same time the lattice sizes should be sufficiently large, to suppress finite-volume effects. The combination of these requirements poses a non-trivial computation challenge, which is made even tougher by the technical complications that arise when fermionic fields are included. For these reasons, until the end of the last century, many lattice QCD simulations were still performed on relatively coarse lattices, and/or within the so-called \emph{quenched approximation} (i.e., the effects of dynamical quark loops were neglected); the difficulties in approaching the continuum and large-volume limits in simulations including the dynamical effects of sufficiently light quarks were summarized by Ukawa at the international Lattice conference in 2001~\cite{Ukawa:2002pc}, in a presentation including a famous plot of the computational costs, which looked like an unsurmountable barrier, and was henceforth dubbed ``the Berlin wall'' of lattice QCD. 

Fortunately, dramatic algorithmic and machine-power improvements during the last decade have radically changed this scenario. Today, large-scale dynamical lattice simulations, at physically realistic values of the parameters, have become the routine, and can provide theoretical predictions for quantities such as hadron masses, from the first principles of QCD~\cite{Kronfeld:2012uk, Fodor:2012gf}. At the same time, there has been similar progress also for other challenging lattice field theory computations, including, in particular, simulations of large-$N$ gauge theories. This is particularly timely, given that during the last fifteen years, the interest in large-$N$ gauge theories has received a further, tremendous boost, with the conjecture of a duality between four-dimensional gauge theories and string theories defined in a higher-dimensional, curved spacetime~\cite{Maldacena:1997re, Gubser:1998bc, Witten:1998qj}, which is often called AdS/CFT, or holographic, correspondence. According to the ``dictionary'' relating quantities on the two sides of this correspondence, the strong coupling limit of the gauge theory corresponds to the  limit in which the spacetime where the dual string theory is defined is weakly curved; moreover, when the number of color charges in the gauge theory becomes large, the string coupling in the dual string theory tends to zero---so that the string theory reduces to its classical limit. In fact, it is only in this limit that string theory can be studied with analytical methods: hence all holographic computations rely on the approximation of an infinite number of color charges for the gauge theory. For this reason, understanding the \emph{quantitative} relevance of the large-$N$ limit for real-world QCD becomes a particularly important issue---one which can be reliably addressed in a first-principle approach via numerical computations on the lattice (at least for a large class of observables in non-supersymmetric theories). In addition, the level of control over systematic and statistical uncertainties in present lattice computations is also sufficient for reliable tests of some of the phenomenological predictions that have been derived using $1/N$ expansions and other (non-holographic) large-$N$ methods.

This motivated us to write the present review article, whose purpose is twofold: on the one hand, we would like to summarize the results of recent lattice computations in large-$N$ gauge theories, and to communicate them to the broader high-energy physics community, presenting them in a concise format, readily accessible also to people who are not lattice practitioners. In doing this, we point out some important fundamental aspects of lattice field theory, and its mathematical robustness, but also some of the non-trivial technical and computational challenges, that one has to cope with, in lattice studies of certain physical problems. On the other hand, we would also like to present an overview of the theoretical aspects of large-$N$ gauge theories, and to draw the attention of our colleagues working in lattice QCD to the many phenomenologically interesting implications of the large-$N$ limit---with an encouragement to compare the results of their computations (not only those for theories with more than three colors, but also, and especially, those for ``real-world QCD'' with $N=3$ colors) with the quantitative predictions for many physical observables, that have been worked out using $1/N$ expansions. 

In order to make this review as self-contained as possible, we decided to include an introduction to the large-$N$ limit of QCD (and to its main phenomenological implications) in section~\ref{sec:large_N_limit}, as well as an introductory overview of the lattice regularization in sec.~\ref{sec:lattice}. These sections are very pedagogical, and (each of them separately) should be suitable as a general introduction to the topics therein covered. Then, section~\ref{sec:from_factorization_to_orbifold} discusses some aspects related to the property of factorization in large-$N$ gauge theories: these include volume independence, loop equations in the continuum, the lattice Eguchi-Kawai model, and the interpretation of these properties in terms of large-$N$ ``orbifold'' equivalences; the results of various numerical investigations of reduced large-$N$ models are also critically reviewed. Finally, the main lattice results for physical quantities extrapolated to the large-$N$ limit are presented in section~\ref{sec:results}, while the conclusive section~\ref{sec:conclusions} summarizes the present status of this field, and outlines those that, in our view, are potentially interesting research directions for the future.

The ideal target of this article encompasses a broad audience, including people working in various fields of high-energy physics, and with a varying degree of expertise on the topics that are presented. Throughout the various sections, we tried to keep the discussion at a pedagogical level, which should also be suitable and easily understandable for graduate or undergraduate students. In particular, as we said, we hope that some sections, like sec.~\ref{sec:large_N_limit} and sec.~\ref{sec:lattice}, could be useful on their own, and could provide a concise overview for the readers interested in large-$N$ QCD or in lattice field theory, respectively.

We conclude this introduction with some disclaimers. First of all, the implications of the large-$N$ limit for QCD and for QCD-like theories have been studied in literally \emph{thousands} of scientific papers, for almost four decades: while it would be impossible to cover all of the relevant literature in the present article, we encourage the readers to integrate the discussion presented here with some of the many excellent reviews that are already available. The most significant early works on this subject are collected in ref.~\cite{Brezin_Wadia}, and introductory lectures on the topics discussed therein can be found, e.g., in refs.~\cite{Witten:1979pi, Coleman:1980nk, Lebed:1998st, Manohar:1998xv, Makeenko:1999hq, Makeenko:2004bz, Teper:2009uf}. In addition, we would like to mention the proceedings of the ``Phenomenology of Large N\subs{c} QCD'' conference held at the Arizona State University in Tempe, Arizona, US in 2002~\cite{Lebed:2002tj}, and the slides of the ``Large N {@} Swansea'' workshop held at the University of Swansea, UK in 2009, which are available online~\cite{largeNatSwansea}. Besides, there exist various review articles focused on specific topics, like, for example, the interpretation of the large-$N$ limit in terms of coherent states~\cite{Yaffe:1981vf}, baryon phenomenology~\cite{Jenkins:1998wy}, loop equations~\cite{Migdal:1984gj}, Eguchi-Kawai models~\cite{Das:1984nb}, early lattice results from simulations of Yang-Mills theories with two, three and four colors~\cite{Teper:1998kw} and aspects related to the dependence on a topological $\theta$-term~\cite{Vicari:2008jw}. Most of the lattice results in large-$N$ lattice gauge theories, that we discuss here, have also been reported in plenary presentations at various recent editions of the International Symposium on Lattice Field Theory~\cite{Narayanan:2007fb, Teper:2008yi, Panero:2012qx}.

Finally, although we tried to summarize the contents of the articles cited herein as clearly and as accurately as possible (compatibly with the tight constraints of a review article), we apologize with the authors of those works, which we may have presented in an unsatisfactory way. %If they feel that we misinterpreted their results, or presented them in a misleading fashion, we encourage them to contact us to provide constructive comments and suggest corrections: we shall be willing to amend the preprint version of this review accordingly, before proceeding to journal submission.

\section{The large-$N$ limit}
\label{sec:large_N_limit}
 
Expansions around the large-$N$ limit are a mathematical tool to study statistical models and quantum field theories characterized by invariance under a certain (local or global) group $G$ of transformations of their internal degrees of freedom, whose number is related to a parameter $N$. In contrast to the na\"{\i}ve expectation, that the dynamics might get more and more complicated when $N$ becomes arbitrarily large, in many cases it turns out that the opposite is true: often, the theory becomes analytically more tractable---or even solvable---in the limit for $N \to \infty$, and corrections accounting for the finiteness of $N$ can be arranged in an expansion around such limit, in powers of the ``small'' parameter $1/N$. 

In statistical spin systems, early applications of this idea date back to the seminal works by Stanley~\cite{Stanley:1968gx} and by various other authors~\cite{Ma:1973zu, Brezin:1972se, Brezin:1976qa, Bardeen:1976zh, Okabe:1978nn}: in particular, this approach showed the deep connections between the nature of critical phenomena and the spacetime dimensionality, and laid a theoretical basis for mean-field computations. In addition, when applied to matrix models, large-$N$ techniques can be used to derive exact analytical solutions~\cite{Brezin:1977sv}, and reveal the connection of these models to discretized random surfaces and to quantum gravity in two spacetime dimensions~\cite{David:1984tx, Ambjorn:1985az, Kazakov:1985ds, Kazakov:1985ea, Brezin:1990rb, Douglas:1989ve, Gross:1989vs}. Generalizations to higher dimensions can be formulated in terms of tensor models, or group field theory~\cite{Ambjorn:1990ge, Sasakura:1990fs, Godfrey:1990dt, Freidel:2005qe, Oriti:2009wn}. Although explicit computations in these models have been hindered by the difficulties in formulating a viable, systematic generalization of the $1/N$ expansion, significant progress has been recently achieved, with the introduction of colored tensor models~\cite{Gurau:2009tw}---see the very recent ref.~\cite{Gurau:2012vk} for a discussion.

For non-Abelian gauge theories---in particular, quantum chromodynamics (QCD)---the large-$N$ expansion was first studied in the 1970's by 't~Hooft~\cite{'tHooft:1973jz}, who suggested a generalization of the theory in which the number of color charges is the parameter $N$, which is sent to infinity (in an appropriate way, as discussed below). The large-$N$ limit of QCD is particularly interesting, because in this limit the theory undergoes a number of simplifications---which, in certain cases, allow one to derive exact solutions with analytical or semi-analytical methods. Well-known examples can be found in two spacetime dimensions, with the computation of the meson spectrum~\cite{'tHooft:1974hx}, the identification of a third-order phase transition separating the strong- and weak-coupling regimes in a lattice regularization of the theory~\cite{Gross:1980he, Wadia:1979vk, Goldschmidt:1979hq}, and the determination of the properties of the spectral density associated to Wilson loops in the continuum theory~\cite{Durhuus:1980nb}.

In the physical case of four spacetime dimensions, the large-$N$ limit does not make QCD analytically solvable; nevertheless, it discloses a considerable amount of information not captured by conventional perturbative computations around the weak-coupling limit. In addition, it also offers a simple interpretation---often based on elementary combinatorics arguments---for some empirical facts observed in hadronic interactions. Finally, a persistent \emph{Leitmotiv} in the scientific discourse on the subject, has been the idea that the large-$N$ limit of QCD may correspond to some kind of string theory.

In this section, we first introduce the 't~Hooft limit of QCD in subsection~\ref{subsec:planar_limit}, and discuss its main implications for the meson and glueball spectrum (subsec.~\ref{subsec:mesons_and_glueballs}) and for baryons (subsec.~\ref{subsec:baryons}). Then, after presenting some implications of the large-$N$ limit for the topological properties of the theory in subsec.~\ref{subsec:topology}, we review the expectations for the phase diagram at finite temperature and/or finite density in subsec.~\ref{subsec:phase_diagram}. Finally, in subsection~\ref{subsec:gauge_string} we briefly discuss the r\^ole of the large-$N$ limit in the conjectured correspondence between gauge and string theories~\cite{Maldacena:1997re, Gubser:1998bc, Witten:1998qj}, which has been widely regarded as a powerful analytical tool to study the dynamics of strongly coupled gauge theories for the last fifteen years.

Although, due to space limits, we cannot cover these topics in the present review, we would like to mention that there also exist interesting studies of the implications of the large-$N$ limit for hadron scattering amplitudes in the limit of large invariant energy $s$ at fixed transferred momentum $t$~\cite{Lipatov:1994xy, Faddeev:1994zg} and physics at the partonic level~\cite{Mueller:1993rr, Mueller:1994jq, Diakonov:1996sr, Pobylitsa:1996rs, Dressler:1999zg, Pobylitsa:2000tt, Ji:1998zb, Chen:2001et}, for QCD evolution equations~\cite{Ali:1991em, Balitsky:1996uh, Braun:1998id, Ball:1998sk, Braun:1999te, Belitsky:1999qh, Balitsky:2001gj, Belitsky:1999ru, Belitsky:1999bf, Derkachov:1999ze, Braun:2000av, Braun:2001qx, Ferreiro:2001qy, Korchemsky:2001nx}, and for small-$x$ physics~\cite{JalilianMarian:1997jx, JalilianMarian:1997gr, Kovchegov:1999ua, Kovchegov:1999yj, Iancu:2000hn} (see also the discussion in the reviews~\cite{Braun:2003rp, Belitsky:2004cz}), for the hadronic contribution to the muon anomalous magnetic moment~\cite{Bijnens:1995cc, Bijnens:1995xf, Knecht:2001qf}, for the physics of electroweak processes like, e.g., those relevant for kaon physics~\cite{Bardeen:1986uz, Bardeen:1986vz, Bardeen:1987vg, Bijnens:1995br, Bijnens:1998ee, Pallante:2001he}, for studies of the ``large-$N$ Standard Model'', in which large-$N$ QCD is combined with the usual electro-weak sector of the Standard Model and implications for Grand Unification are derived~\cite{Abbas:1990kd, Chow:1995by, Shrock:1995bp, Shrock:2002kp, Shrock:2007ai}. Ref.~\cite{Bai:2012zn} proposed to use mathematical tools relevant for the large-$N$ limit---specifically: random matrix theory---to study neutrino masses and mixing angles, considering the number of generations in the Standard Model\footnote{Note that this is different from the possibility of a large number of right-handed neutrinos, which was considered in refs.~\cite{Eisele:2007ws, Ellis:2007wz, Feldstein:2011ck, Heeck:2012fw}.} as the ``large'' parameter $N$. Finally, we would like to mention a very recent article discussing large-$N$ gauge theories from a philosophy of science point of view~\cite{Bouatta:2012st}.

\subsection{The planar limit of QCD}
\label{subsec:planar_limit}

In the Standard Model of elementary particle physics, strong interactions are described by QCD: a non-Abelian vector gauge theory based on local invariance under the $\SU(3)$ color gauge group. The QCD Lagrangian density in Minkowski spacetime reads:
\begin{equation}
\label{QCD_Lagrangian}
\mathcal{L} = -\frac{1}{2} \Tr \left( F_{\alpha\beta}F^{\alpha\beta} \right) + \sum_{f=1}^{n_f} \overline{\psi}_f \left( i \gamma^\alpha D_\alpha - m_f \right) \psi_f,
\end{equation}
where $g$ is the (bare) gauge coupling, $D_\mu=\partial_\mu-igA_\mu^a(x) T^a$ is the gauge covariant derivative (with the $T^a$'s denoting the eight generators of the Lie algebra of $\SU(3)$, in their representation as traceless Hermitian matrices of size $3 \times 3$, with the normalization: $\Tr (T^a T^b) = \delta^{ab}/2$), $F_{\alpha\beta}=(i/g)[D_\alpha,D_\beta]$ is the non-Abelian field strength, the $\gamma^\alpha$'s are the Dirac matrices, while $\psi(x)$ and $\overline{\psi}(x)=\psi^\dagger \gamma^0$ respectively denote the spinor associated with the quark fields and its conjugate. Quark fields are in the fundamental representation of the gauge group, and occur in $n_f$ different species (``flavors'', labelled by the $f$ subscript in the equation above), with generically different masses $m_f$. 

The 't~Hooft limit of QCD~\cite{'tHooft:1973jz} is a generalization of the theory, in which the gauge group is taken to be $\SU(N)$, and the number of color charges $N$ is assumed to be arbitrarily large. Standard weak-coupling computations show that, in order for this limit to make sense at least perturbatively, it is necessary that at the same time the coupling $g$ be taken to zero, holding the 't~Hooft coupling $\lambda = g^2 N$ fixed. As for the way the number~of quark flavors $n_f$ should be scaled in the large-$N$ limit, 't~Hooft's original proposal was to hold it fixed. Then, for example, it is easy to see that for $N\to \infty$ the two-loop perturbative QCD $\beta$-function:
\begin{equation}
\label{betafunction}
\mu \frac{d g}{d \mu} = - \frac{1}{(4\pi)^2} \left( \frac{11N -2 n_f}{3} \right)g^3  
- \frac{1}{(4\pi)^4} \left( \frac{34N^3 - 13 N^2 n_f + 3n_f}{3N} \right)g^5 + O(g^7)
\end{equation}
turns into the a renormalization group equation for $\lambda$ with finite coefficients:
\begin{equation}
\label{betafunction_for_lambda}
\mu \frac{d \lambda}{d \mu} = - \frac{11}{24\pi^2} \lambda^2  
- \frac{17}{192\pi^4} \lambda^3 + O(\lambda^4).
\end{equation}
Note that, since the coefficient of the first term appearing on the right-hand side of eq.~(\ref{betafunction_for_lambda}) is negative, perturbation theory predicts that the 't~Hooft limit of QCD is an asymptotically free theory. While higher-order coefficients of the $\beta$-function are scheme-dependent, inspection of the results obtained, e.g., in the minimal-subtraction scheme~\cite{vanRitbergen:1997va, Czakon:2004bu} shows that it remains true that the coefficients appearing on the r.h.s. of eq.~(\ref{betafunction_for_lambda}) tend---rather quickly~\cite{Ryttov:2010iz}---to finite values in the large-$N$ limit. It is then natural to assume that the $\LambdaQCD$ scale parameter of strong interactions is held fixed for $N \to \infty$.
 
Another interesting feature of eq.~(\ref{betafunction_for_lambda}) is that it does not depend on $n_f$: this is simply due to the fact that the number of quark degrees of freedom is $O(n_fN)$, i.e. $O(N)$ in the 't~Hooft limit, and hence subleading with respect to the number of gluon degrees of freedom, which is $O(N^2)$. In fact, a different large-$N$ limit of QCD (Veneziano limit) can be obtained, if $n_f$ is also sent to infinity, holding the $n_f/N$ ratio fixed~\cite{Veneziano:1976wm}; however it turns out that this choice leads to generally more complicated computations, and, hence, has received less attention in the literature. Another inequivalent large-$N$ limit of QCD (Corrigan-Ramond limit) is obtained, by assuming that $n_f$ is fixed, but (some of) the fermions (``larks'') are in the two-index antisymmetric representation of the color gauge group~\cite{Corrigan:1979xf, Kiritsis:1989ge} (which, for $N=3$, is nothing but the antifundamental representation), whose number of components scales like $O(N^2)$ at large $N$. Finally, yet another type of large-$N$ limit, in which $\lambda$ grows like $N^c$ (with $c>0$) for $N \to \infty$, has been recently proposed in refs.~\cite{Fujita:2012cf, Azeyanagi_12103601}. In the following, unless where otherwise stated, we shall focus on the 't~Hooft limit. 

The properties of QCD in the 't~Hooft limit can be studied in terms of so-called large-$N$ counting rules, and are determined by the combined effects that arise from the number of colors becoming large, and the coupling becoming small. Since this double limit is taken at fixed $\lambda$, it is convenient to write all Feynman rules by replacing $g$ with $\sqrt{\lambda/N}$. Furthermore, an easy way to keep track of the number of independent (fundamental) color indices appearing in Feynman diagrams is based on the so-called double-line, or ribbon graph, notation: since quarks are fields in the fundamental representation of the $\SU(N)$ gauge group, a generic quark propagator of the form $\langle \psi^i (x) \overline{\psi}^j(y) \rangle$, (where $i$ and $j$ are fundamental color indices, and we assume a suitable gauge-fixing, like, e.g., Feynman gauge) is proportional to a Kronecker delta $\delta^{ij}$, and can be associated to a single oriented line. By contrast, the properties of the $\SU(N)$~algebra generators imply that a propagator for gluon field components $\langle A^i_{\mu \; j} (x) A^k_{\nu \; l}(y) \rangle$ is proportional to $\left( \delta^i_l \delta^k_j -\delta^i_j \delta^k_l/N\right)$, and, hence, can be denoted by a pair of oppositely-oriented lines. Here and in the following, we neglect the $\delta^i_j \delta^k_l/N$ term appearing in the expression of the gluon propagator; strictly speaking, this amounts to replacing the $\SU(N)$ gauge group with $\U(N)$, and induces a subleading, $O(N^{-2})$ relative correction to the results derived in the large-$N$ limit.\footnote{The subleading corrections due to this difference between the $\U(N)$ and the $\SU(N)$ gauge theories can be accommodated, by introducing an unphysical $\U(1)$ ``phantom'' field, which cancels the extra $\U(1)$ degree of freedom of $\U(N)$~\cite{Canning:1975zr}.}

\begin{figure}[-t]
\centerline{\includegraphics[width=0.31\textwidth]{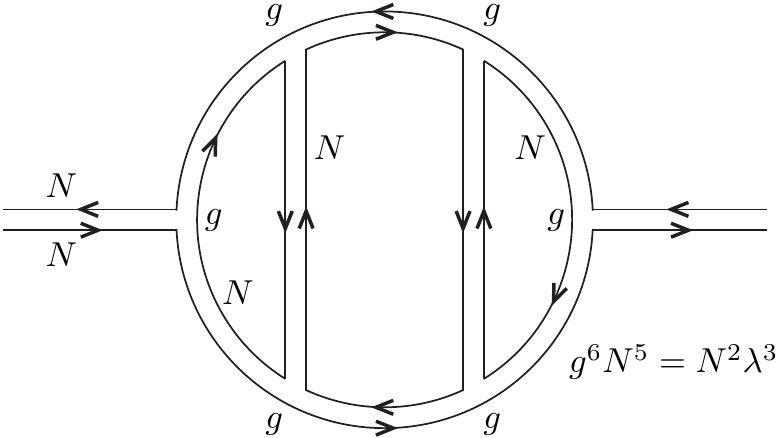}\hfill \includegraphics[width=0.31\textwidth]{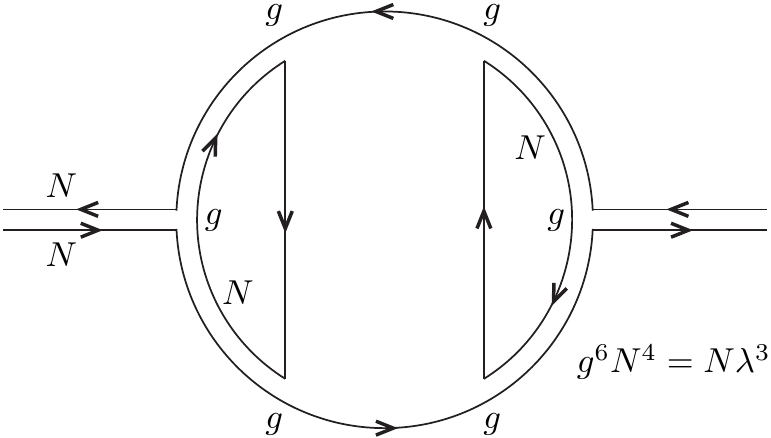}\hfill\includegraphics[width=0.31\textwidth]{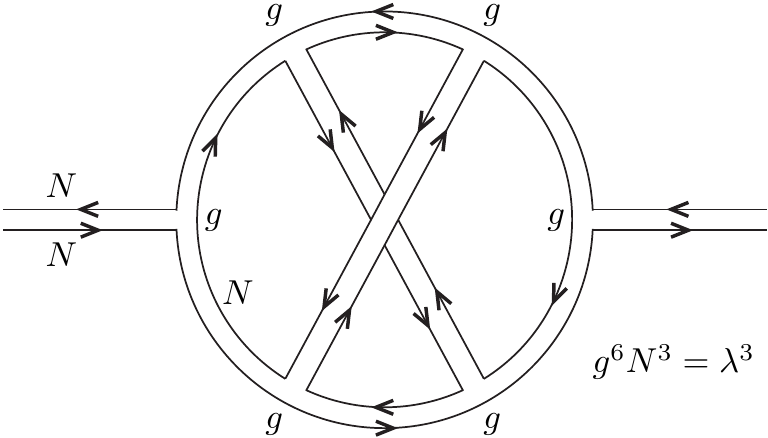}}
\caption{Different types of three-loop Feynman diagrams contributing to the gluon propagator; in the 't~Hooft limit, only the diagram on the left panel yields a non-negligible contribution.\label{fig:planar_quarkloop_nonplanar}}
\end{figure}

With this notation, it is easy to see that, in the 't~Hooft limit, the amplitudes for physical processes are dominated by diagrams which are planar in index space, and which do not contain dynamical quark loops. This can be clarified by the example depicted in fig.~\ref{fig:planar_quarkloop_nonplanar}, which shows three different types of diagrams contributing to the gluon propagator at three-loop order for generic $N$: the planar diagram on the left panel, which contains virtual gluons only, is the one with the largest number of independent color lines, and (if the color indices of the external gluon are not fixed) is proportional to $g^6 N^5 = N^2 \lambda^3$. Note that this diagram can be drawn on the plane (or, equivalently, on the surface of a two-sphere) without crossing lines, and is thus called ``planar''. By contrast, replacing an internal gluon loop with a quark loop (central panel) removes a color line, reducing the overall power of $N$ by one, down to $N \lambda^3$. Finally, the diagram on the right panel of fig.~\ref{fig:planar_quarkloop_nonplanar}, which includes a line crossing and is proportional to $g^6 N^3=\lambda^3$, is suppressed by two powers of $N$.

Part of the mathematical simplification of QCD in the 't~Hooft limit is due to the fact that in this limit the dynamics is only sensitive to planar diagrams (whose number grows only exponentially with the order of the diagram, in contrast to the total number of diagrams, which grows factorially---see ref.~\cite{Marino:2012zq} and references therein). In spite of this simplification, the summation of all planar Feynman diagrams of QCD in four spacetime dimensions is a daunting task, which has not been possible to complete---see refs.~\cite{Brezin:1977sv, Koplik:1977pf, Thorn:1977kx, Brower:1978fp, 'tHooft:1982kt, 'tHooft:1982cx, Rivasseau:1983jj} for a discussion. 

Another interesting simplification taking place in the 't~Hooft limit is that virtual quark loops can be neglected. Note that this feature, which corresponds to the so-called ``quenched approximation'' in lattice QCD, or to the ``probe approximation'' in the context of string theory, arises dynamically at infinite $N$, and, in particular, does not imply any of the fundamental problems that occur in quenched lattice QCD at finite $N$ (which include, e.g., the lack of a well-defined Hilbert space, unitarity violations and negative-norm states, et c.). In relation to the suppression of sea quark effects, we note two observations, suggesting that the 't~Hooft limit may be a ``good'' (i.e. ``quantitatively accurate'') approximation of real-world QCD. The first is the phenomenological success of quark model spectroscopy: the masses of experimentally observed hadronic states can be described surprisingly well, in terms of just the flavor symmetry patterns for \emph{valence} quarks, neglecting sea-quark effects. The second is the remarkable success of quenched lattice QCD calculations~\cite{Aoki:1999yr}, which are performed neglecting the dynamical effects (virtual fermion loops) induced by the determinant of the Dirac operator---see section~\ref{sec:lattice} for a discussion.

Note that the diagrams appearing in the central and right panel of fig.~\ref{fig:planar_quarkloop_nonplanar} can be drawn without crossing lines on the surface of a punctured sphere and of a two-torus, respectively. This is a manifestation of a general property of the classification of Feynman diagrams in the 't~Hooft limit: a generic amplitude $ \mathcal{A} $ for a physical process can be expressed in a double series, in powers of the coupling $ \lambda $ and in powers of $1/N$, where the latter expansion is of topological nature: the power of $ 1/N $ associated to a given diagram is related to the number of ``handles'' ($h$) and ``boundaries'' ($b$) of the simplest Riemann surface on which the diagram can be drawn without crossing lines (and with quark lines along the boundaries), and equals its Euler characteristic:
\begin{equation}
\label{large_N_amplitude_expansion}
\mathcal{A} = \sum_{h,\,b=0}^{\infty} N^{2-2h-b}\sum_{n=0}^{\infty} c_{h,\,b;\,n} \lambda^{n}.
\end{equation}
Since at large $N$ the leading contributions in eq.~(\ref{large_N_amplitude_expansion}) are $ O(N^2) $~and correspond to planar ($h=0$) diagrams with no quark loops ($b=0$), the 't~Hooft limit is also called ``planar limit''. Obviously, in the study of quantities involving gauge-invariant fermionic bilinears, there exists at least one closed quark line (which is conventionally drawn as the exterior boundary of the ribbon graphs), hence in this case the leading contributions come from terms corresponding to $b=1$, $h=0$, which are $O(N)$.

We conclude this subsection with an observation about generating functionals for connected and non-connected planar graphs: in general, due to some subtleties related to cyclic ordering of operators, exponentiating the generating functional for connected planar graphs \emph{does not} yield the generating functional for planar graphs only. The correct definition of the generating functionals for planar diagrams can be formulated in terms of appropriate non-commuting sources~\cite{Cvitanovic:1980jz}.

\subsection{Phenomenological implications of the large-$N$ limit for mesons and glueballs}
\label{subsec:mesons_and_glueballs}

Under the assumption that the 't~Hooft limit of QCD is a confining theory (see the discussion in section~\ref{sec:lattice}), the large-$N$ counting rules defined in subsection~\ref{subsec:planar_limit} lead to a number of interesting phenomenological implications, in particular for the lightest physical states: glueballs and mesons.

To see this, following the discussion in, e.g., ref.~\cite{Manohar:1998xv}, it is convenient to rescale the fields appearing in the Lagrangian density defined by eq.~(\ref{QCD_Lagrangian}), according to:
\begin{equation}
\label{field_rescaling}
A_\mu^a(x) \rightarrow \frac{1}{g} A_\mu^a(x),\qquad \psi_f(x) \rightarrow \sqrt{N} \psi_f(x),
\end{equation}
so that $\mathcal{L}$ can be written as: $\mathcal{L}=N \tilde{\mathcal{L}}$, with:
\begin{equation}
\label{rescaled_Lagrangian}
\tilde{\mathcal{L}}=-\frac{1}{4\lambda} \left( F^a_{\mu\nu} F^{a\,\mu\nu} \right) + \sum_{f=1}^{n_f} \overline{\psi}_f \left( i \gamma^\mu D_\mu - m_f \right) \psi_f.
\end{equation}
Na\"{\i}vely, one could then imagine that the quantum theory defined by the functional path integral:
\begin{equation}
\label{partition_function}
\mathcal{Z}_0 = \int \mathcal{D} A \mathcal{D} \overline{\psi} \mathcal{D} \psi \exp \left\{ i N \int \dd t~ \dd^3 x~\tilde{\mathcal{L}}[A,\psi,\overline{\psi}] \right\}
\end{equation}
reduces to its classical limit~\cite{Witten:1979pi}, obtained as the stationary point of $\tilde{\mathcal{L}}$. This, however, is not the case, because an ``entropic'' term~\cite{Haan:1981ks}, of the same order in $N$, also arises from the field measure, and, as a consequence, the large-$N$ limit \emph{is not} equivalent to the classical limit (see also the discussion in subsec.~\ref{subsec:classical_limit}).

Connected correlation functions of gauge-invariant (local or non-local) single-trace operators $O_a$, built from gauge fields and possibly fermionic bilinears, can be studied by adding corresponding source terms of the form $(N J_a O_a)$ to $\mathcal{L}$:
\begin{equation}
\label{partition_function_with_sources}
\mathcal{Z}_J = \int \mathcal{D} A \mathcal{D} \overline{\psi} \mathcal{D} \psi \exp \left\{ i N \int \dd t~\dd^3 x~ \left( \tilde{\mathcal{L}} [A,\psi,\overline{\psi}] + J_a O_a \right) \right\}
\end{equation}
and taking appropriate functional derivatives:
\begin{equation}
\label{vevs}
\langle O_1(x_1) \dots O_n(x_n) \rangle_c = (iN)^{-n} \frac{\delta}{\delta J_1(x_1)} \dots \frac{\delta}{\delta J_n(x_n)} \left. \ln \mathcal{Z}_J\right|_{J=0}.
\end{equation}
As discussed in subsection~\ref{subsec:planar_limit}, the leading contribution to the sum of connected vacuum graphs in the 't~Hooft limit is $O(N^2)$ (or $O(N)$, if fermionic bilinears are considered): as a consequence, $n$-point connected correlation functions like the one appearing in eq.~(\ref{vevs}) are dominated by diagrams of planar gluon loops, and scale like $O(N^{2-n})$, in the pure-glue sector, or like $O(N^{1-n})$, for the case of correlation functions involving quark bilinears.

Let $\mathcal{G}_i$ be a purely gluonic, gauge-invariant, Hermitian operator, with the appropriate quantum numbers to describe a given glueball; as a two-point connected correlation function of the form $\langle \mathcal{G}_i \mathcal{G}_i \rangle_c$ is $O(1)$ in the large-$N$ limit, the operator $\mathcal{G}_i$ creates a glueball state with an amplitude of order one. Glueball-glueball interactions (including decays, scatterings, et c.) are then described by higher-order connected correlation functions $\langle \mathcal{G}_1 \dots \mathcal{G}_n \rangle_c$, with $n \ge 3$: as they generically scale like $O(N^{2-n})$, they are suppressed by at least one power of $1/N$ relative to the free case. 

Similarly, meson states can be described by gauge-invariant, Hermitian operators $\mathcal{M}_i$, involving quark bilinears, with the appropriate quantum numbers. In this case, $\langle \mathcal{M}_i \mathcal{M}_i \rangle_c$ scales proportionally to $1/N$ in the large-$N$ limit, so that $\sqrt{N}\mathcal{M}_i$ is an operator creating a meson state with amplitude $O(1)$. Note that, as a consequence of this, the pion decay constant $f_\pi$, defined in terms of the matrix element describing the overlap between the isovector axial current and a pion state:
\begin{equation}
\label{fpi}
\langle 0 | N \overline{\psi}(x) \gamma_\mu \gamma_5 T_j \psi(x) | \pi_l(k) \rangle = -i f_\pi k_\mu \delta_{jl} \exp( -i k x)
\end{equation}
(in which the $N$ factor in the definition of the axial current is necessary, in order to express this quantity in terms of the rescaled quark fields) is proportional to $\sqrt{N}$ in the large-$N$ limit. 

Meson-meson interactions are described by connected correlators of the form: 
\begin{equation}
\label{meson_interactions}
N^{n/2} \langle \mathcal{M}_1 \dots \mathcal{M}_n \rangle_c,
\end{equation}
with $n \ge 3$: they scale like $O(N^{1-n/2})$, hence they are suppressed at large $N$. Finally, meson-glueball interactions and mixing processes are described by correlators of the form $\langle \mathcal{G}_1 \sqrt{N} \mathcal{M}_1 \rangle_c$ (or of higher order), which scale (at most) like $N^{-1/2}$, and thus are also suppressed.

To summarize, if the 't~Hooft limit of QCD is a confining theory, then its low-lying spectrum consists of stable, non-interacting glueballs and mesons. Exotic states like tetraquarks, molecules, et c. are absent in the 't~Hooft limit, because the leading-order contribution to their propagators comes from terms which correspond to the propagation of mesons or glueballs. At finite but ``large'' values of $N$ (or, better: at finite but ``small'' values of $1/N$), hadron interactions are suppressed by powers of $1/\sqrt{N}$, so that, in the 't~Hooft limit, QCD is a theory of weakly-coupled hadrons, with interactions described by a ``coupling'' which vanishes for $N \to \infty$. It is worth noting that, although the dependence of this ``hadronic coupling'' on $N$ is of the form $1/\sqrt{N}$, the parameter determining the actual numerical accuracy of large-$N$ expansions may be proportional to a power of it, and possibly be further suppressed by numerical factors.

Combining the two pieces of information, that at low energies the large-$N$ spectrum consists of non-interacting, infinitely narrow hadrons, and that at high energies the theory is asymptotically free (see eq.~(\ref{betafunction_for_lambda})), one can prove that the number of stable glueball and meson states is infinite by \emph{reductione ad absurdum}: since two-point correlation functions in momentum space can be expressed as a linear combination of hadron propagators with ``sharp'' poles, proportional to $(p^2-m^2)^{-1}$ (denoting the four-momentum as $p^\mu$), if the number of hadrons were finite, then it would be impossible to reproduce the functional form expected for meson correlators, which involves a dependence on the logarithm of $p^2$.

As discussed, for example, in ref.~\cite{Shifman:1998rb}, the features of the large-$N$ spectrum also have interesting relations with the QCD sum rules due to Shifman, Vainshtein and Zakharov~\cite{Shifman:1978bx, Novikov:1981xi}, which are based on the idea of expressing a generic correlation function of quark currents in terms of an operator product expansion~\cite{Wilson:1969zs}, with a matching to a sum over hadron states.

\begin{figure}[-t]
\centerline{\includegraphics[width=0.4\textwidth]{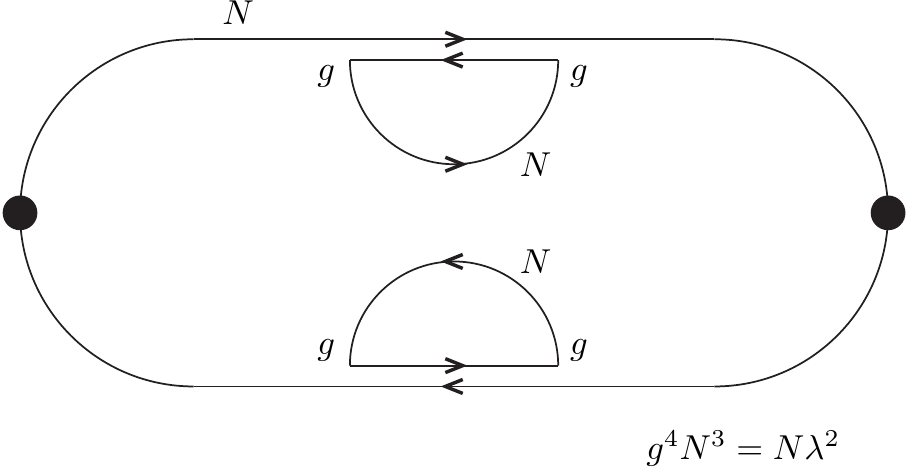}\hspace{16mm} \includegraphics[width=0.4\textwidth]{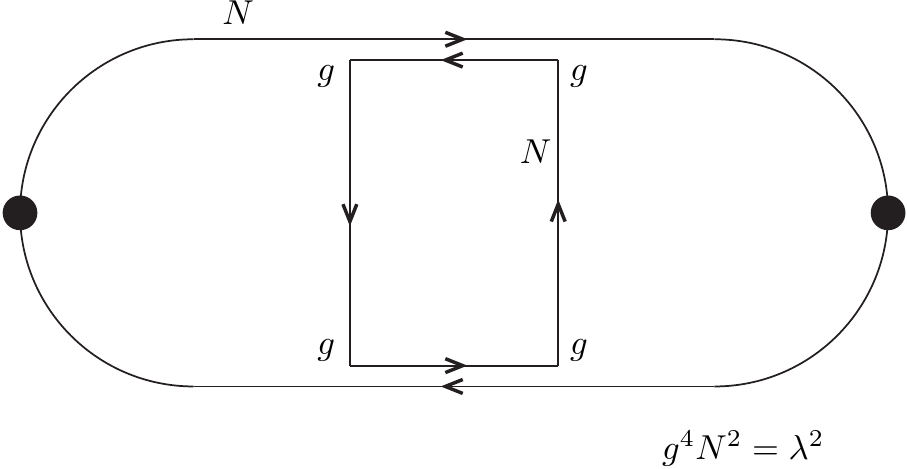}}
\caption{Large-$N$ counting rules offer a simple interpretation for the Okubo-Zweig-Iizuka rule: the figure shows an example of Feynman diagrams relevant for the propagation of a meson: the diagram on the right, in which the process goes through an intermediate stage where only virtual gluons appear, is suppressed by a power of $1/N$ with respect to the one on the left.\label{fig:ozirule}}
\end{figure}

Another interesting phenomenological implication of the large-$N$ counting rules relevant in the 't~Hooft limit of QCD, is that they provide an explanation for the empirical rule due to Okubo, Zweig and Iizuka~\cite{Okubo:1963fa, Zweig:1981pd, Iizuka:1966fk}, i.e. the suppression of strong-interaction processes described by Feynman diagrams that can be split in two by cutting only internal gluon lines. Fig.~\ref{fig:ozirule} shows an example of this, for two different Feynman diagrams describing the propagation of a meson (denoted by a blob): the diagram on the l.h.s., in which the quark/antiquark lines originating from the initial state propagate all the way to the final state (getting dressed by two virtual gluons on the way), is proportional to $\lambda N$, whereas the one on the r.h.s., in which the same process goes through the annihilation of the initial quark and antiquark, the intermediate emission of two virtual gluons, and then the creation of a quark/antiquark pair that ends up in the final meson, is proportional to $\lambda$, and therefore suppressed by one power of $N$ relative to the former diagram.

Large-$N$ counting rules also provide phenomenological information relevant for the meson sector---which, in particular, allows one to discuss a low-energy effective theory for the light mesons~\cite{Rosenzweig:1979ay, Nath:1979ik, Witten:1980sp, DiVecchia:1980ve, Kawarabayashi:1980dp, Kawarabayashi:1980uh} (see also ref.~\cite{Pich:2002xy} and references therein). In particular, the very existence of light, pseudoscalar mesons is related, by the Nambu-Goldstone mechanism, to the spontaneous breakdown of chiral symmetry. In the 't~Hooft limit, this phenomenon was first studied by Coleman and Witten in ref.~\cite{Coleman:1980mx}.

As it is well-known, in QCD with a generic number $N$ of color charges and $n_f$ massless quark flavors,\footnote{Here and in the following, we restrict our attention to $n_f \ge 2 $. Note that the $n_f=1$ case, albeit of limited phenomenological interest (because in nature both the up and down quarks are very light), is somewhat special~\cite{Creutz:2006ts}.} the Dirac operator $i \gamma^\mu D_\mu$ anticommutes with the chirality operator $\gamma_5$, hence for $n_f \ge 2$ the Lagrangian in eq.~(\ref{QCD_Lagrangian}) is classically invariant under a global, non-Abelian chiral symmetry $\U_R(n_f) \times \U_L(n_f)$. This symmetry describes the independent rotations of the right- and left-handed components $\psi_{R,L} = P_{R,L} \psi$ of the quark fields (and the corresponding components of the conjugate spinor $\overline{\psi}$), where the right- and left-projectors are defined by: 
\begin{equation}
\label{chiral_projectors}
P_{R,L}=\frac{\ide \pm \gamma_5}{2}.
\end{equation}
The global $\U_R(n_f) \times \U_L(n_f)$~symmetry of the classical Lagrangian for massless quarks can be re-expressed in terms of vector ($V$) and axial ($A$) transformations as: $\SU_V(n_f) \times \U_V(1) \times \SU_A(n_f) \times \U_A(1)$. 

At the quantum level, the $\U_V(1)$ symmetry remains exact, and corresponds to the conservation of the baryon number $\mathcal{B}$ in QCD. Also the $\SU_V(n_f)$ symmetry remains exact, and, in the real world, is only mildly explicitly broken by the small mass differences between the light quarks; this symmetry manifests itself in the approximate isospin and strangeness degeneracy patterns observed in the light hadron spectrum.

By contrast, the $\SU_A(n_f)$ symmetry is spontaneously broken: the QCD vacuum is characterized by a non-vanishing chiral condensate $\langle \overline{\psi} \psi \rangle$. Correspondingly, $(n_f^2-1)$ light (pseudo-)Nambu-Goldstone bosons appear in the spectrum: the three pions (if one assumes $n_f=2$ light quark flavors), or the three pions, together with the four kaons and the $\eta$ (if one also considers the strange quark as approximately massless).

Finally, the fate of the $\U_A(1)$ symmetry is particularly interesting, as it is related to the so-called $\U_A(1)$ problem: were the QCD vacuum invariant under this symmetry, then the spectrum of light hadrons should consist of (nearly) mass-degenerate partners of opposite parity, which are not seen experimentally. On the other hand, if the $\U_A(1)$ symmetry was spontaneously broken, then there should exist a light non-strange, isoscalar, pseudoscalar meson---but such particle is not observed either: the lightest meson with these quantum numbers is the $\eta^\prime$ meson, with a mass of $957.78(6)$~MeV~\cite{Nakamura:2010zzi}, much heavier than the pions, kaons and the $\eta$. The solution of this problem is the following: the $\U_A(1)$ is explicitly broken at the quantum level, due to the non-invariance of the path integral measure for quark fields~\cite{'tHooft:1976up, 'tHooft:1976fv, Fujikawa:1979ay}. The corresponding anomaly, however, is proportional to $g^2$ (for details, see the discussion in subsec.~\ref{subsec:topology}), i.e. it is suppressed like $1/N$ in the 't~Hooft limit, so that, for vanishing quark masses, the $\eta^\prime$ also becomes massless in the $N \to \infty$ limit. As a consequence, the spectrum of large-$N$ QCD with $n_f$ massless (or light) quarks features $n_f^2$ light pseudoscalar mesons, as the (pseudo-)Nambu-Goldstone bosons associated with the spontaneously broken axial symmetries.

Since, in general, the masses of other hadrons will be separated from these by a finite gap, the low-energy dynamics of QCD can be modelled by chiral perturbation theory ($\chi$PT)~\cite{Gasser:1983yg, Gasser:1984gg}---see also refs.~\cite{Leutwyler:1993iq, Ecker:1994gg, Pich:1995bw} for reviews of the topic---, which is an effective theory for the fields describing the light mesons. Assuming that the theory has $n_f=3$ light quark flavors, the latter can be conveniently packaged into the components of a $\U(3)$-valued matrix field $\mathcal{U}(x)$ defined by:
\begin{equation}
\label{chiral_effective_field}
\mathcal{U}= \exp \left( i \sqrt{2} \Phi / F \right),
\end{equation}
where the entries of $\Phi$ are linear combinations of the meson fields (see, e.g., ref.~\cite{Pich:2002xy} for an explicit expression). According to the usual construction of effective field theories, the Lagrangian $\Leff$ describing the dynamics of $\mathcal{U}$ must be compatible with the global symmetries of QCD, but, a priori, it is otherwise unspecified; in addition, it can also include couplings to external Hermitian matrix fields $L_\alpha$, $R_\alpha$, $S$ and $P$, to account for finite quark mass, electromagnetic- and weak-interaction effects.\footnote{In particular, the quark masses are encoded in the diagonal entries of $S$.} Defining $\tilde{\partial}_\alpha \mathcal{U} = \partial \mathcal{U} -i R_\alpha \mathcal{U} + i \mathcal{U} L_\alpha$ and $\tilde{\chi}=2B_0(S+iP)$, it is possible to organize the most general expression for $\Leff$ according to the number of derivatives it contains (i.e., in powers of momentum). At the lowest order in derivatives and quark masses, the expression for $\Leff$ reads:
\begin{equation}
\label{chiral_Lagrangian}
\Leff = \frac{F^2}{4} \tr \left( \tilde{\partial}^\alpha \mathcal{U}^\dagger \, \tilde{\partial}_\alpha \mathcal{U} + \tilde{\chi}^\dagger \mathcal{U} + \mathcal{U}^\dagger \tilde{\chi} \right),
\end{equation}
and involves two low-energy constants: $F$ can be interpreted as the meson decay constant,\footnote{Note that, with the conventions of eq.~(\ref{chiral_effective_field}) and eq.~(\ref{chiral_Lagrangian}), the phenomenological value of $F$ (or, more precisely, of $f_\pi$) is about $92.4$~MeV.} while $B_0$ is proportional to the quark condensate~\cite{GellMann:1968rz}. Note that $F^2 \sim f_\pi^2 = O(N)$, so $\Leff$ is $O(N)$, hence for $N \to \infty$---see also refs.~\cite{Rosenzweig:1979ay, Nath:1979ik, Witten:1980sp, DiVecchia:1980ve}---the tree-level approximation for the effective theory becomes exact.\footnote{Note that---in contrast to our discussion of the functional path integral in eq.~(\ref{partition_function})---in the present case there are no subtleties related to the large-$N$ scaling of the functional measure, because the degrees of freedom of the effective theory are hadrons, whose number \emph{does not} grow with $N$ in the 't~Hooft limit.} Moreover, since the definition of $\mathcal{U}$ in terms of the meson fields in eq.~(\ref{chiral_effective_field}) also contains a $1/F$ factor, the expansion of $\mathcal{U}$ generates meson interactions which are suppressed by powers of $N^{-1/2}$ in the 't~Hooft limit. Hence, large-$N$ $\chi$PT can be interpreted as a consistent combined expansion in powers of $1/N$, of the (squared) momentum, and of the quark masses~\cite{Kaiser:2000gs, HerreraSiklody:1996pm, HerreraSiklody:1997kd}.

The expression of $\Leff$ at the next-to-leading order involves ten new low-energy constants:\footnote{In principle, at this order one could also include a contribution reproducing the effect of QCD fermion anomalies: the effect of non-Abelian anomalies can be accounted for by a Wess-Zumino-Witten term~\cite{Wess:1971yu, Witten:1983tw}, with no free parameters, whereas the axial $\U_A(1)$ anomaly can be mimicked by a $\theta$-term~\cite{Kaiser:2000gs}.} their numerical values, fixed using experimental information on various decays~\cite{Ecker:1994gg, Pich:1995bw}, agree with the expectations from large-$N$ counting rules---which predict some of them to be $O(N)$, and some others to be $O(1)$. Quantitative large-$N$ predictions, however, can also be obtained, by working in the so-called single-resonance approximation~\cite{Ecker:1989yg}, i.e. by neglecting the contributions of resonance states above the nonet of lowest-lying mesons with a given set of quantum numbers, and are in very good agreement with the values determined from experimental input (see, e.g., table 1 in ref.~\cite{Pich:2002xy}). Combining (unitarized) chiral perturbation theory with large-$N$ arguments, it is also possible to investigate the nature of different physical mesons~\cite{Pelaez:2003dy, Pelaez:2004xp, Pelaez:2006nj, Geng:2008ag}. In addition, the large-$N$ limit also allows one to get a better matching between perturbative QCD predictions at high energies, and $\chi$PT predictions at low energies~\cite{Peris:1998nj}.\footnote{A large-$N$ effective model for quarks, gluons and pions, which could also provide a quantitatively accurate description of hadronic phenomena at moderate and intermediate energies, has been recently proposed in ref.~\cite{Weinberg:2010bq}.} Further details about the applications of the large-$N$ limit in $\chi$PT can be found in ref.~\cite{Kaiser:2000gs}. Other implications of the large-$N$ limit, which are relevant for mesons, have been derived for electroweak processes at low energies~\cite{Knecht:1998nn, Knecht:1998sp, Knecht:1999gb, Peris:2000sw, Peris:2000tw, Hambye:2003cy}. There are also a number of works (e.g.~\cite{Masjuan:2007ay, Rosell:2004mn, Rosell:2005ai} and the works reviewed in ref.~\cite{Pich:2002xy}) discussing the sense in which large-$N$ QCD provides a theoretical justification for computations of various physical quantities in the context of the vector meson dominance model~\cite{O'Connell:1995wf}. A very recent example is given by the study of generalized form factors carried out in ref.~\cite{Masjuan:2012sk} (to which we refer, for further references on this topic).

Finally, in the literature there exists a very large number of articles studying the properties of mesons in QCD-like models in the large-$N$ limit, using holographic methods. Since it would not be possible to discuss all of these works here, we refer the interested reader to the review~\cite{Erdmenger:2007cm}. However, we shall present a brief discussion of the r\^ole of the large-$N$ limit in holographic computations in subsection~\ref{subsec:gauge_string}.

\subsection{Phenomenological implications of the large-$N$ limit for baryons}
\label{subsec:baryons}

The basic features of baryons in the 't~Hooft limit of QCD have been known for more than three decades~\cite{Witten:1979kh}. Baryons are defined as completely antisymmetric, color-singlet states made of $N$ quarks, thus, in contrast to mesons, the Feynman diagrams that represent them (not just the associated combinatorial factors) depend on $N$. Another difference with respect to mesons is that baryon masses scale as $O(N)$: the mass of a baryon receives contributions from the constituent quark masses, from their kinetic energies (both of which are proportional to $N$) and from the total potential energy associated to quark-quark interactions (for each pair of quarks, the interaction strength is proportional to $g^2$, i.e. to $N^{-1}$, while the number of distinct quark pairs scales proportionally to $N^2$, so that their product is, again, $O(N)$). 

To get a quantitative picture of the ground state baryon energy in the 't~Hooft limit, one can study the large-$N$ baryon in a Hamiltonian approach, and resort to an approximate Hartree-Fock-like solution (or to a relativistic generalization thereof): this is based on the idea that the total potential felt by each quark can be interpreted as a background, mean-field-like potential resulting from the interaction with all other quarks in the baryon. Although this approach does not lead to an exact analytical solution for the large-$N$ baryon spectrum, it reveals that, while the mass of a baryon is $O(N)$, its ``form factor'' (i.e., its charge distribution) is independent of the number of colors in the 't~Hooft limit.

Arguments based on large-$N$ counting rules also show that the amplitude for baryon-baryon (or for baryon-antibaryon) scattering scales as $O(N)$, i.e. is of the same order as the baryon mass. Baryon-meson scattering, on the contrary, turns out to be $O(1)$, so this physical process does not affect the motion of the baryon, but only of the meson (whose mass is also $O(1)$).

Finally, there exist physical processes involving baryons, which are missed \emph{at all orders} in the large-$N$ expansion. One example is baryon-antibaryon pair production from high-energy collisions of lepton-antilepton pairs. This process proceeds via an intermediate virtual photon, which then creates a quark-antiquark pair. In order to form the baryon and the antibaryon in the final state, however, this process must take place $N$ times, and the probability for this to happen is exponentially suppressed as $e^{-N}$. This dependence on $N$ is never captured at any order in the expansion in powers of $N^{-1}$, hence the corresponding physical process is absent at all order in large-$N$ QCD. Similarly, the processes of baryon-antibaryon annihilation into a pair of leptons, and lepton-baryon scattering (which are related to the previous example by crossing symmetry) are also exponentially suppressed with $N$. 

These features reveal an intriguing analogy, suggested in ref.~\cite{Witten:1979kh}, between the $1/N$ expansion of QCD, and the usual weak-coupling expansion of quantum electrodynamics (QED), with mesons and baryons playing a r\^ole analogous to electrons and 't~Hooft-Polyakov monopoles, respectively. Indeed, electrons are the physical states in QED, their interactions can be successfully studied in a perturbative expansion in powers of $\alpha_{\mbox{\tiny{QED}}}=e^2/(4\pi)$, and their mass is independent of $e$, i.e. scales like $O(e^0)$. 't~Hooft-Polyakov monopoles, on the other hand, do not appear as elementary objects or as simple bound states in the theory (neither do they appear in any Feynman diagram), but rather as particular topological solutions; their masses are inversely proportional to $\alpha_{\mbox{\tiny{QED}}}$, and, for small $\alpha_{\mbox{\tiny{QED}}}$, their structure can be studied by classical equations (in which $\alpha_{\mbox{\tiny{QED}}}$ factors out, yielding results for the monopole size and monopole structure, that are independent of $\alpha_{\mbox{\tiny{QED}}}$). Moreover, both monopole-monopole scattering, and the scattering of an electron off a monopole are processes with well-defined, non-vanishing limits for $\alpha_{\mbox{\tiny{QED}}} \to 0$. Processes like the production of monopole-antimonopole pairs via electron-positron annihilation are not captured at any order in weak-coupling expansions of QED. As discussed above, all of this bears a close qualitative resemblance to the description of mesons and baryons in the $1/N$ expansion of QCD.

An interesting prediction for the large-$N$ spectrum of baryons of different spin $J$ was first suggested, in connection to the Skyrme model (see ref.~\cite{Zahed:1986qz} for a review), in ref.~\cite{Adkins:1983ya}, and later discussed also in ref.~\cite{Jenkins:1993zu}: the ``hyperfine splitting'' between the masses of these states should be described in terms of a rotor spectrum:
\begin{equation}
\label{rotor_spectrum}
M = c_1 N + c_2 \frac{J(J+1)}{N}+ O(1/N^2).
\end{equation}

Other important large-$N$ implications for baryon phenomenology, that were discovered during the 1980's, include the determination of the group structure for baryon multiplets~\cite{Gervais:1983wq, Jain:1984gp} and its relation to the $\SU(3)$ chiral model and to current algebra Lagrangians and to the quark model~\cite{Witten:1983tw, Guadagnini:1983uv, Balachandran:1983dj, Mazur:1984yf, Manohar:1984ys, Dhar:1985gh, Gupta:1987dc}. Large-$N$ baryons can also be interpreted as chiral solitons of the Nambu-Jona-Lasinio model~\cite{Bijnens:1995ww, Alkofer:1994ph}.

Later, it was further realized that, combining the large-$N$ counting rules, that characterize the 't~Hooft limit, with unitarity conditions, it is possible to derive quantitative predictions for several baryonic observables, and to organize the large-$N$ expansions in the baryon sector in a systematic way~\cite{Dashen:1993jt, Jenkins:1993zu, Gervais:1983wq, Carone:1993dz, Luty:1993fu, Gervais:1984rc}---see also refs.~\cite{Manohar:1998xv, Jenkins:1998wy} for reviews. This possibility arises from the fact that a contracted $\SU(2 n_f)$ spin-flavor symmetry for baryons emerges in the large-$N$ limit. 

A simple way to see how these predictions can be derived is the following. At leading order, the coupling of a meson to a baryon is obtained by insertion of a fermion bilinear onto one of the valence quark lines in a baryon---which can be done in $N$ different ways. Since, as discussed above, fermion bilinears create mesons with amplitude $O(N^{1/2})$, it follows that the effective baryon-meson coupling must be $O(N^{1/2})$. The process of baryon-meson scattering, on the other hand, can be obtained by insertion of two fermion bilinears onto the valence quark lines of the baryon: if they are inserted on the same line, then the process can occur in $N$ different ways; alternatively, there are $O(N^2)$ possibilities to insert them on different lines, but then, in order to transfer energy from the incoming meson to the outcoming one, a gluon has to be exchanged between the two baryon quark lines, and this involves a $g^2$ factor, which reduces the overall amplitude of the process by one power of $N$. Finally, taking into account that each baryon-meson coupling is $O(N^{1/2})$, one obtains that the total amplitude for baryon-meson scattering is $O(1)$.

From a kinematic point of view, however, in the scattering of a meson (with mass $O(1)$) off a baryon (whose mass is proportional to $N$), the baryon can be considered as static, and the process can be interpreted as the absorption of the incoming meson, followed by the emission of the scattered meson. Since the couplings of the baryon to both the incoming and the outcoming meson are $O(N^{1/2})$, the total scattering amplitude for this process should be proportional to $N$. This, however, would violate unitarity, and would be incompatible with the fact that the baryon-meson scattering amplitude is $O(1)$. The apparent paradox can be resolved, if there are cancellations among the baryon-meson scattering diagrams, and this leads to a set of large-$N$ consistency conditions for baryon-meson couplings~\cite{Gervais:1983wq, Gervais:1984rc, Dashen:1993as}. 

The process of pion-nucleon scattering at low energies in the chiral limit provides a concrete example: the absorption of a pion by a static nucleon, $\mathcal{N} + \pi \to \mathcal{N}^\prime$, can be described by the formula:
\begin{equation}
cN\frac{\partial_i \pi}{f_\pi} (X^{ia})_{\mathcal{N}\mathcal{N}^\prime} \;,\qquad \mbox{with:}\;\;(X^{ia})_{BB^\prime}=\frac{\langle B^\prime |(\overline{\psi}\gamma^i\gamma_5 \tau^a \psi)|B \rangle}{cN},
\end{equation}
where an overall $N$ factor (together with an arbitrary, $N$-independent, normalization constant $c$) has been extracted from the isovector baryon axial current, so that $X^{ia}$ has a finite limit for $N \to \infty$. At large $N$, the leading contributions to the baryon-pion scattering come from the difference of two tree-level diagrams (corresponding to the fact that the nucleon can either first get into its excited state $\mathcal{N}^\prime$ by absorption of the incoming pion, and then decay to the ground state with emission of the outgoing pion, or \emph{vice versa}). Requiring the total scattering amplitude to be $O(1)$, leads to the constraint:
\begin{equation}
\label{consistency_condition}
N[X^{ia},X^{jb}] \le O(1).
\end{equation}
By expanding the $X^{ia}$ operator in powers of $1/N$:
\begin{equation}
X^{ia}=X^{ia}_0+\frac{1}{N}X^{ia}_1+\frac{1}{N^2}X^{ia}_2+\dots
\end{equation}
eq.~(\ref{consistency_condition}) yields a sequence of constraints among the different terms in the expansion; in particular, at the first order in $1/N$, it implies that:
\begin{equation}
\label{consistency_condition_at_leading_order}
[X^{ia}_0,X^{jb}_0]=0.
\end{equation}
Eq.~(\ref{consistency_condition_at_leading_order}) can only be satisfied if, besides the nucleon, there exist other intermediate states, with the same mass. Since a pion in the $p$-wave state coupled to a nucleon can also produce a baryon with spin and isospin both equal to $3/2$ (i.e. a $\Delta$ baryon), the large-$N$ consistency conditions impose a constraint relating $g_{\pi\mathcal{N}\mathcal{N}}$ to $g_{\pi\mathcal{N}\Delta}$. In turn, the same argument can be applied to the $\Delta + \pi \to \Delta + \pi$ scattering process, implying the existence (at least in the large-$N$ limit) of a baryon with the same mass as $\mathcal{N}$ and $\Delta$ and with quantum numbers $ J=I=5/2 $, and fixing a constraint among the respective couplings, and so on. The conclusion is that the consistency condition eq.~(\ref{consistency_condition_at_leading_order}) implies the existence of an infinite tower of degenerate baryons with $ J=I=k+1/2 $, for any $k \in \N$, and fixes their couplings to the pion (up to the overall normalization factor $c$).

Eq.~(\ref{consistency_condition_at_leading_order}) can be interpreted as a subalgebra of a \emph{contracted} $su(4)$ (or $su(2n_f)$, in the case of $n_f$ light quark flavors) spin-flavor algebra: the $X^{ia}_0$ operators can be seen as the large-$N$ limit of the $su(4)$ generators that mix spin and flavor (rescaled by $N$). Note that the emergence of a spin-flavor symmetry, mixing internal and Lorentz degrees of freedom, is possible because in the large-$N$ limit the baryon field is infinitely heavy, and hence static. The irreducible representations of the contracted spin-flavor algebra can be classified using standard techniques for induced representations~\cite{Dashen:1994qi}.

More generally, consistency conditions of the form of eq.~(\ref{consistency_condition}) can be used to systematically organize the $ 1/N $ expansion using group-theoretical methods and operator identities, and to fix constraints on the subleading corrections. In particular, for $n_f=2$, one finds that the first corrections to the mass of low-spin baryons, as well as to the baryon axial couplings, are $ O(N^{-2}) $ relative to the respective leading terms. This framework can be extended to $n_f=3$ light flavors as well, and allows one to derive predictive theoretical expectations for a number of observables, including axial couplings and form factors~\cite{Dai:1995zg, FloresMendieta:1998ii, FloresMendieta:2006ei, FloresMendieta:2012dn}, masses~\cite{Jenkins:1993zu, Bedaque:1995wb, Jenkins:1995td, Cordon:2012nm, Cordon:2012xz}, magnetic moments~\cite{Dashen:1993as, Dai:1995zg, Jenkins:1994md, Luty:1994ub, Lebed:2004fj, Jenkins:2011dr}, masses of excited states~\cite{Carlson:1998vx, Schat:2001xr, Carlson:2000zr, Goity:2002pu,  Cohen:2003tb, Cohen:2003nn, Goity:2003ab, Matagne:2004pm, Matagne:2006zf}, their couplings and decays~\cite{Carlson:2000zr, Carone:1994tu, Goity:1996hk, Pirjol:1997nh, Pirjol:1997bp, Pirjol:1997sr, Carlson:1998si, Cohen:2003jk, Cohen:2003fv, Goity:2004ss}, as well as those of baryons containing a heavy quark~\cite{Jenkins:1993af, Jenkins:1993ta, Jenkins:1996de, Chow:2000qv, Chow:2000fi, AzizaBaccouche:2001jc, Jenkins:2007dm}. In this approach it is also possible to study a large-$N$ generalization~\cite{Jenkins:1995gc} of heavy-baryon chiral perturbation theory~\cite{Jenkins:1990jv, Jenkins:1991es}, the nucleon-nucleon potential and scattering~\cite{Kaplan:1995yg, Kaplan:1996rk, Banerjee:2001js, Belitsky:2002ni, Beane:2002ab, Riska:2002vn, Cohen:2002im, Cohen:2002qn, Kaskulov:2004kr, Cohen:2011hq} (see also ref.~\cite{Bonanno:2011yr}, for work on related topics), and various quantities related to the baryonic structure\footnote{For a different type of approach to hadronic structure functions using the large-$N$ limit, see refs.~\cite{Ji:1998zb, Braun:2000av, Kovchegov:1999ua, Kovchegov:1999yj} and references therein.}~\cite{Buchmann:2000wf, Buchmann:2002mm}. Further phenomenological implications for baryons in the large-$N$ limit are discussed in refs.~\cite{Goity:2004ss, Cherman:2012eg, Cohen:2004qt, Cherman:2009fh, Cohen:2006up, Cohen:2011cw, Buisseret:2011aa} and in the review~\cite{Lutz:2001yb}.

Other articles on related topics include refs.~\cite{Cohen:2003yi, Cohen:2003mc, Cohen:2003nk}, which used large-$N$ consistency conditions to show a flaw in the theoretical prediction of the $ \Theta^+ $ pentaquark state~\cite{Diakonov:1997mm} (various other exotic baryonic states were studied in refs.~\cite{Pobylitsa:2003ju, Jenkins:2004tm, Jenkins:2004vb}), and refs.~\cite{Semay:2007cv, Semay:2007ff}, in which a relation between the large-$N$ approach to baryons and the quark model was pointed out. Finally, large-$N$ QCD expansions based on the contracted spin-flavor symmetry also show the connection between the large-$N$ limits of the Skyrme model~\cite{Adkins:1983ya, Guadagnini:1983uv, Bardakci:1983ev, Kaplan:1989fc} and of the quark model~\cite{Karl:1984cz, Cvetic:1985qn}: this formalism enables one to derive model-independent predictions, purely based on group-theoretical relations.

\subsection{Topological aspects in large-$N$ QCD}
\label{subsec:topology}

Various types of gauge field configurations characterized by non-trivial topological properties are believed to play a prominent r\^ole in determining the vacuum structure in QCD and in other confining gauge theories: these include center vortices~\cite{'tHooft:1977hy, Yoneya:1978dt, Vinciarelli:1978kp, Cornwall:1979hz}, Abelian monopoles~\cite{'tHooft:1981ht}, instantons~\cite{'tHooft:1976up, Polyakov:1975rs, Belavin:1975fg, Shuryak:1981ff, Diakonov:1983hh, Schafer:1996wv} and bions~\cite{Unsal:2007jx, Poppitz:2009uq, Poppitz:2011wy, Poppitz:2012sw, Argyres:2012ka}. In general, the analytical understanding of the way these objects affect the dynamics, however, is limited to some form of semi-classical approximation.

The topological objects, for which the large-$N$ limit has the most interesting implications, are instantons: they are solutions of the Euclidean equations of motion with finite action, which can be interpreted as ``tunneling'' events between the topologically inequivalent classical vacua of the theory, classified by a Pontryagin index taking values in the third homotopy group of the gauge group: $\pi_3(\SU(N)) = \Z$~\cite{Bryan:1993hz}. Their relation to the large-$N$ limit of QCD has been investigated since the late 1970's~\cite{Witten:1978bc, Witten:1979vv, Veneziano:1979ec, Teper:1979tq}, and is still raising continued interest~\cite{Thacker_Xiong}.

As mentioned in subsec.~\ref{subsec:mesons_and_glueballs}, in QCD with $n_f$ massless quarks, the global Abelian axial symmetry is anomalous, due to the non-invariance of the fermion measure;\footnote{An alternative, equivalent way to see the origin of the $\U_A(1)$ anomaly, is from a perturbative analysis, which shows that the anomaly arises from the ultraviolet divergences of a one-loop triangle diagram of virtual quarks.} correspondingly, the conservation of the $\U_A(1)$ current is violated:
\begin{equation}
\label{axial_current_anomaly}
\partial_\alpha j_A^\alpha(x) = - 2n_f q(x),
\end{equation}
by a term proportional to the topological charge density $q(x)$:
\begin{equation}
\label{topological_charge_density}
q(x) = \frac{g^2}{16\pi^2} \Tr \left( \tilde{F}_{\mu\nu}(x) F^{\mu\nu}(x) \right), \;\;\; \mbox{with:}\;\; \tilde{F}_{\mu\nu}(x)=\frac{1}{2} \epsilon_{\mu\nu\rho\sigma}F^{\rho\sigma}(x).
\end{equation}
Note that $q(x)$ can be written as the divergence of the Chern current $K^\alpha(x)$:
\begin{equation}
\label{Chern_current}
q(x) = \partial_\mu K^\mu(x), \;\;\; \mbox{where:}\;\; K^\mu(x) = \frac{g^2}{4\pi^2} \epsilon^{\mu\nu\rho\sigma} A^a_\nu \left( \partial_\rho A^a_\sigma - \frac{g}{3} f^{abc} A^b_\rho A^c_\sigma \right),
\end{equation}
which is proportional to the Hodge dual of the Chern-Simons three-form. Although $q(x)$ is given by the divergence of a current, its spacetime integral (the ``topological charge'' $Q$) can take non-vanishing, integer values in the presence of instantons. In general, the value of $Q$ for a given gauge field configuration is equal to the index of the Dirac operator, i.e. to the difference between the number of left- and right-handed zero-modes of the Dirac operator~\cite{Atiyah:1963zz, Atiyah:1968mp, Atiyah:1968rj, Atiyah:1967ih, Atiyah:1970ws, Atiyah:1971rm}.

In the 't~Hooft limit, the right-hand-side of eq.~(\ref{axial_current_anomaly}), being proportional to $g^2$, is---at least na\"{\i}vely---suppressed like $O(1/N)$: as a consequence, $m^2_{\eta^\prime}$ should vanish like $O(1/N)$ for $N \to \infty$. Thus, in the 't~Hooft limit one expects the meson spectrum to include $n_f^2$---rather than $(n_f^2-1)$---massless (or light) Nambu-Goldstone bosons.

By studying the Ward-Takahashi identities for $j_A^\alpha(x) + 2 n_f K^\alpha(x)$, it is possible to derive the Witten-Veneziano formula~\cite{Witten:1979vv, Veneziano:1979ec}:
\begin{equation}
\label{Witten_Veneziano_formula}
m^2_{\eta}+m^2_{\eta^\prime} = 2 m^2_K + \frac{4 n_f \topsusc}{f_\pi^2}
\end{equation}
(see also ref.~\cite{Shore:2007yn} for a discussion), relating the pseudoscalar meson masses, the pion decay constant, and the topological susceptibility $\topsusc$, which is defined by:
\begin{equation}
\label{topological_susceptibility}
\topsusc = -i \int \dd^4 x \langle 0 | \mathcal{T} \{ q(x) q(0) \} | 0 \rangle,
\end{equation}
where $\mathcal{T}$ denotes time-ordering.

Combining the Witten-Veneziano formula eq.~(\ref{Witten_Veneziano_formula}) with the expectation that the square of the $\eta^\prime$ mass vanishes proportionally to $1/N$ in the 't~Hooft limit (and with the fact that $f_\pi$ scales like $N^{1/2}$ for $N \to \infty$), it follows that the topological susceptibility should have a non-vanishing, $O(1)$ limit at large $N$. This is at odds with traditional pictures of the QCD vacuum as a semiclassical gas of dilute instantons, according to which the probability of tunneling events between topologically different vacua of the theory (i.e. the ``statistical weight'' of instantons) should scale proportionally to $\exp(-8\pi/g^2)$; in particular, the number density $d$ of instantons of (small) size $\rho$ is expected to be proportional to~\cite{Witten:1978bc, Schafer:1996wv}:
\begin{equation}
\label{small_size_instanton_density}
d(\rho) \propto \exp \left[ - \frac{8 \pi^2 N}{\lambda(\rho)} \right],
\end{equation}
i.e. to be exponentially suppressed with $N$ in the 't~Hooft limit. A na\"{\i}ve dilute instanton picture, however, is invalidated by infrared divergences; furthermore, one should note that the exponential suppression which holds for instantons at the cutoff scale, may not be valid for instantons of finite size, see, e.g., refs.~\cite{Shuryak:1993ee, Schafer:2002af, Schafer:2004gy} for a discussion. 

The non-trivial interplay between $1/N$ suppression of hadronic mass differences, the anomaly equation~(\ref{axial_current_anomaly}), and topologically non-trivial non-Abelian gauge fields (in particular instantons) has also been discussed in ref.~\cite{Novikov:1981xi}. Further implications for the so-called proton spin problem~\cite{Ashman:1987hv} were discussed in refs.~\cite{Shore:1991dv, Narison:1998aq, Shore:2006mm}.

It is interesting to consider the issues related to the topological aspects of large-$N$ QCD in the context of a generalization of the theory, including a topological $\theta$-term (see, e.g., ref.~\cite{Ohta:1981ai} for a discussion): in the Standard Model, there exists no fundamental symmetry principle forbidding a term proportional to the topological charge density $q(x)$ in the QCD Lagrangian:
\begin{equation}
\label{QCD_Lagrangian_with_theta_term}
\mathcal{L}_\theta = -\frac{1}{2} \Tr \left( F_{\alpha\beta}F^{\alpha\beta} \right) + \sum_{f=1}^{n_f} \overline{\psi}_f \left( i \gamma^\alpha D_\alpha - m_f \right) \psi_f - \theta q(x)
\end{equation}
(so that, in particular, $\topsusc$ can be expressed in terms of the second derivative of the corresponding effective action, with respect to $\theta$).

Since the topological charge density is a pseudoscalar, it explicitly breaks the discrete parity ($P$) and time-reversal ($T$) symmetries. For massless quarks, a $\theta$-term in the QCD Lagrangian can be reabsorbed via a redefinition of the fermion fields, by a global $\U_A(1)$ rotation. For massive quarks, however, such rotation would make the quark masses complex, leading to phenomenological consequences: these include, in particular, a non-vanishing electric dipole moment for the neutron, proportional to $\theta$. Experimental results on this quantity~\cite{Baker:2006ts, Hewett:2012ns} pose very stringent bounds on the magnitude of the $\theta$ term, indicating that---if non-vanishing at all---it must be extremely (``unnaturally'') small: $|\theta| \le 10^{-10}$. The smallness of $|\theta|$ is usually called the ``strong $CP$ problem''~\cite{Peccei:2006as} of the Standard Model (see also ref.~\cite{Vicari:2008jw} for a discussion).

If the QCD Lagrangian includes a $\theta$-term, as in eq.~(\ref{QCD_Lagrangian_with_theta_term}), then the quantization of $Q$ in integer units implies that the vacuum energy density of the theory is periodic in $\theta$, with period $2\pi$. In addition, the large-$N$ counting rules discussed in subsec.~\ref{subsec:planar_limit} imply that the vacuum energy density should also be $O(N^2)$ at large $N$. Finally, the form of eq.~(\ref{QCD_Lagrangian_with_theta_term}) shows that, in the 't~Hooft limit, the natural scaling variable is $\theta/N$; note that, in particular, this implies that the dependence on $\theta$ in quantities like, e.g., the string tension or the mass gap, should be suppressed like $1/N^2$ in the 't~Hooft limit. A possible way to reconcile these observations consists in assuming that the ground state energy density $\mathcal{E}$ is given by the minimum of a multi-branched function of $\theta/N$~\cite{Witten:1980sp}:
\begin{equation}
\label{vacuum_energy}
\mathcal{E}(\theta) = N^2 \min_{0 \le k < N} f\left( \frac{\theta + 2\pi k}{N} \right).
\end{equation}
The potential barrier between states corresponding to different $k$ values is expected to be proportional to $N$, so that tunneling effects should be suppressed in the large-$N$ limit~\cite{Witten:1998uka, Shifman:1998if, Tytgat:1999yx, Gabadadze:2002ff}: at large $N$, different candidate ground states become stable, but not degenerate. In addition, for $N \gg 1$ the energy densities associated with the lowest few $k$-values at $\theta=0$ can be approximately expressed as:
\begin{equation}
\label{k_vacuum_energy}
\mathcal{E}(0)+2\pi^2 k^2 \topsusc,
\end{equation}
hence the gap with respect to the actual ground state energy is independent of $N$.  Since the relevant scaling parameter of the theory in the large-$N$ limit is expected to be $\theta/N$, it follows that the dependence on $\theta$ in quantities like, e.g., the energy density per unit area of domain walls separating different vacua, or the mass gap, should be suppressed like $1/N^2$ in the 't~Hooft limit. 

Finally, note that, for any $N$, the theory is not only $CP$-invariant for $\theta=0$, but also for $\theta = \pi$. In the latter case, if two degenerate, $CP$-violating vacua exist, $CP$ symmetry may get spontaneously broken~\cite{Dashen:1970et} (a phenomenon that cannot occur at $\theta=0$, if the assumptions of the Vafa-Witten theorem hold~\cite{Vafa:1984xg}): the interesting phenomenological implications of this possibility have been discussed in a number of works, including refs.~\cite{Witten:1980sp, DiVecchia:1980ve, Tytgat:1999yx, Creutz:1995wf, Evans:1996eq, Smilga:1998dh, Lenaghan:2001ur, Akemann:2001ir, Kalloniatis:2004fb}.

\subsection{Large-$N$ theories at finite temperature and/or density}
\label{subsec:phase_diagram}

The focus of this subsection is on QCD at finite temperature and/or net baryon number density\footnote{Since the baryon number is a conserved charge in QCD, it is possible to associate with it a corresponding chemical potential $\mu$, whose value determines the \emph{net} density of quarks over antiquarks in a thermodynamic system.} in thermodynamic equilibrium~\cite{Rischke:2003mt} (which is relevant for the description of heavy-ion collision experiments~\cite{Heinz:2000bk, Muller:2006ee, BraunMunzinger:2007zz}), and on the implications for its properties that can be derived in the large-$N$ limit---a problem which has been studied since the 1980's~\cite{Thorn:1980iv, Gocksch:1982en, Greensite:1982be, Pisarski:1983db, McLerran:1985uh, Toublan:2005rq}.

First of all, if the large-$N$ limit of QCD is confining at zero and low temperatures and densities, then the spectrum of physical states consists of color-singlet states: mesons, glueballs, and baryons. Their number is $O(1)$ for $N \to \infty$, and the masses of mesons and glueballs are also $O(1)$, while baryon masses are $O(N)$. Furthermore, meson-meson interactions scale as $O(1/N)$, while meson-glueball and glueball-glueball interactions are even more suppressed, $O(1/N^2)$. As a consequence of these properties, the equilibrium thermodynamic properties of large-$N$ QCD in the confining phase are expected to scale like $O(1)$, and to be described in terms of a gas of non-interacting mesons and glueballs. Thus, in the thermodynamic limit, the pressure $p$ at a finite temperature $T$ can be written as:
\begin{equation}
\label{confined_pressure}
p = - \sum_s \frac {n_s T}{2\pi^2}\,
\int_0^\infty \dd q~q^2 \ln \left( 1-e^{-\sqrt{M_s^2+q^2}/T} \right) = \sum_s \frac{n_s M_s^4}{2\pi^2} \sum_{k=1}^\infty \left( \frac{T}{k M_s} \right)^2 K_2 \left( k\frac{M_s}{T} \right),
\end{equation}
where the summation is done over meson and glueball states, with the index $s$ labeling a species of particles with mass $M_s$, spin $J_s$, isospin $I_s$ and $n_s=(2I_s+1)(2J_s+1)$ physical degrees of freedom; $K_\nu(z)$ denotes the modified Bessel function of the second kind of index $\nu$. Note that the sum in eq.~(\ref{confined_pressure}) does not include baryon states: since their masses are $O(N)$, their contribution to equilibrium thermodynamic quantities is exponentially suppressed as $e^{-N}$ in the large-$N$ limit.

On the other hand, due to asymptotic freedom~\cite{Gross:1973id, Politzer:1973fx}, in the high-temperature limit $T \to \infty$ the thermodynamics of non-Abelian gauge theories is expected to be described in terms of a gas of free, massless quarks and gluons. Since the number of gluon degrees of freedom is $O(N^2)$, while for quarks it is $O(n_fN)$ (i.e. proportional to $N$ in the 't~Hooft limit), in this limit the pressure scales like $O(N^2)$:
\begin{equation}
\label{SB_limit}
p = \frac{\pi^2 T^4}{180} \left[ 4 (N^2-1) + 7 N n_f \right]
\end{equation}
(see also ref.~\cite{Dumitru:2001xa} for a discussion on the counting of degrees of freedom at finite temperature). Approximating the thermodynamics of a deconfined non-Abelian gauge theory in terms of a relativistic gas of free quarks and gluons is accurate at high enough temperatures, where the physical running coupling becomes negligible. At lower temperatures (but still in the deconfined phase), the physical coupling is finite~\cite{Laine:2005ai}, so that corrections to the Stefan-Boltzmann limit can be expressed in terms of thermal weak-coupling expansions~\cite{Shuryak:1977ut, Chin:1978gj, Kapusta:1979fh, Toimela:1982hv, Arnold:1994ps, Arnold:1994eb, Zhai:1995ac, Braaten:1995jr, Kajantie:2002wa, Kraemmer:2003gd, Blaizot:2003iq, Blaizot:2003tw, Andersen:2009tc, Andersen:2011sf}. It is important to point out that, even at high temperature, and for any number of colors, the deconfined plasma retains some aspects of non-perturbative nature: in particular, the low-frequency modes of the plasma are strongly coupled at \emph{all} temperatures~\cite{Linde:1980ts, Gross:1980br}.\footnote{The non-perturbative aspects of the deconfined plasma are closely related to the so-called ``Linde problem''~\cite{Linde:1980ts}, affecting perturbative expansions in thermal gauge theories, namely to the breakdown of the familiar correspondence between expansions in powers of the coupling, and expansions in the number of loops. This effect, which is due to the existence of infrared divergences in finite-temperature gauge theory (see, e.g., ref.~\cite{Kapusta_book}), is responsible for the non-trivial mathematical structure of perturbative expansions in thermal QCD, which include, for example, terms proportional to odd powers and to logarithms of the coupling, as well as terms whose coefficients receive contributions from infinitely many Feynman diagrams, of arbitrarily complicated topology. A pedagogical introduction to these topics can be found in ref.~\cite{Arnold:2007pg}.} Via dimensional reduction~\cite{Ginsparg:1980ef, Appelquist:1981vg}, the long-wavelength degrees of freedom of the deconfined phase of an $\SU(N)$ gauge theory at high temperature can be mapped to a \emph{confining} effective $\SU(N)$ theory in three dimensions, possibly coupled to an adjoint scalar field~\cite{ Braaten:1995jr, Farakos:1994xh, Braaten:1995cm, Kajantie:1995dw}; the latter effective theory is known as ``electrostatic QCD'' (EQCD), and captures the physics at ``soft'' scales $O(gT)$, while the former goes under the name of ``magnetostatic QCD'' (MQCD), and describes phenomena at ``ultrasoft'' scales, parametrically $O(g^2T)$.

The deconfined phase at high temperature is expected to be separated from the low-temperature, confining phase by a deconfinement transition (or cross-over), taking place at a finite temperature $T_c$~\cite{Cabibbo:1975ig, Polyakov:1978vu, Susskind:1978ua}, which for the 't~Hooft limit of QCD was first discussed in ref.~\cite{Thorn:1980iv}. In $\SU(N)$ Yang-Mills theory, the standard interpretation of the deconfinement transition is in terms of spontaneous breakdown of the exact global $\Z_N$ symmetry associated with the center of the gauge group: in the Euclidean formulation of thermal field theory, bosonic fields obey periodic boundary conditions along the compact Euclidean time direction, and the Yang-Mills dynamics is invariant if gauge fields satisfy periodic boundary conditions up to a global transformation in the center of the gauge group.\footnote{An alternative view on the r\^ole of center symmetry and on the physical relevance of different center sectors in continuum Minkowski spacetime, however, was discussed in ref.~\cite{Smilga:1993vb}.} The corresponding order parameter is the Wilson line along the compactified Euclidean time direction (or Polyakov loop)~\cite{Polyakov:1978vu, Weiss:1980rj, McLerran:1981pb}:
\begin{equation}
\label{Polyakov_loop}
L(\vec{x})= \frac{1}{N} \tr \left[ \mathcal{P} \exp \left( i g \int_0^{1/T} \dd \tau A_\tau(\vec{x},\tau) \right) \right]
\end{equation}
(where $\mathcal{P}$ denotes path ordering). In the thermodynamic (i.e., infinite-volume) limit, the thermal Wilson line defined in eq.~(\ref{Polyakov_loop}) has a vanishing expectation value in the confined phase at low temperature, while it acquires a non-zero expectation value at high temperature.\footnote{Strictly speaking, this holds for a proper definition of the Polyakov loop, which requires renormalization~\cite{Dotsenko:1979wb}.} It can be interpreted in terms of the propagator of an infinitely massive external color probe source, implying that the free energy associated to an isolated color source is infinite in the confined phase, and is finite in the deconfined phase.\footnote{Some expectations about the Polyakov loop temperature dependence in the large-$N$ limit have been recently discussed in refs.~\cite{Andreev:2009zk, Noronha:2009ud, Noronha:2010hb}, using arguments related to the gauge/string duality.} In general, the Polyakov loop free energy depends on the representation of the color source: in ref.~\cite{Damgaard:1987wh}, a theoretical prediction for this dependence was worked out analytically, using an effective lattice theory of Polyakov lines at strong coupling~\cite{Polonyi:1982wz, Bartholomew:1983jv, Green:1983sd, Gross:1984wb, Damgaard:1985bx}, finding that the Polyakov loop free energy is proportional to the eigenvalue of the quadratic Casimir of the representation. This effective model also leads to the prediction that, when approaching the deconfinement temperature from above, the Polyakov loop in the large-$N$ theory tends to the value $1/2$~\cite{Damgaard:1986mx}.

Although in QCD with $N=3$ colors the center symmetry is explicitly broken by the quarks, in the 't~Hooft limit the theory becomes quenched; as a consequence, the finite-temperature deconfinement in large-$N$ QCD is expected to be a genuine transition~\cite{Thorn:1980iv}.\footnote{The existence of a finite-temperature deconfinement transition can also be argued on the basis of an exponential growth of the density of hadronic states $\tilde{\rho} (m)$ as a function of their mass $m$~\cite{Hagedorn:1965st} (see also refs.~\cite{Pisarski:1983db, Buisseret:2011fq}).} In particular, arguments based on the large mismatch between the number of degrees of freedom in the two phases suggest that the large-$N$ transition should be of first order (see, e.g., ref.~\cite{Pepe:2005sz} and references therein), with a latent heat $O(N^2)$---but the possibility of a second-order transition has also been discussed in the literature~\cite{Pisarski:1997yh}. Very recently, ref.~\cite{Poppitz:2012nz} presented an analytical study of the finite-temperature deconfinement phase transition (and of the $\theta$-dependence of the critical temperature) in pure Yang-Mills theory, for all simple gauge groups. Studying the interplay of perturbative and non-perturbative effects (due to monopole-instantons and bions), the authors of ref.~\cite{Poppitz:2012nz} found that the transition is of first order for all $\SU(N\ge3)$ (and $\Gtwo$) gauge theories (while the same approach predicts it to be of second order for $\SU(2)$ gauge group~\cite{Poppitz:2012sw}), and that~$T_c$~decreases when $\theta$ increases (see also ref.~\cite{Unsal:2012zj}).

Building on the approach pioneered in ref.~\cite{Sundborg:1999ue}, an analytical study of the deconfining transition in large-$N$ gauge theories on compact spatial manifolds was presented in refs.~\cite{Aharony:2003sx, Aharony:2005bq}: these works reduced the Yang-Mills partition function to an integral over an appropriate matrix model, which could be tackled with analytical techniques in the large-$N$ limit. The results showed the existence of a high-temperature phase, with a free energy $O(N^2)$, separated from the confining phase by one or two phase transitions. Analytical studies based on a similar approach include refs.~\cite{Schnitzer:2004qt, Dumitru:2004gd, Myers:2012tz}, and ref.~\cite{Hollowood:2006cq} (which considers the case of orientifold gauge theories). Other interesting implications of the large-$N$ limit for the thermal properties of non-Abelian gauge theories can be found in phenomenological models~\cite{Meisinger:2001cq, Pisarski:2000eq, Dumitru:2003hp, Hidaka:2008dr, Hidaka:2009hs, Dumitru:2010mj, Kashiwa:2012wa, Dumitru:2012fw, Pisarski:2012bj}, in quasi-particle models~\cite{Peshier:1995ty, Buisseret:2010mb}, and in computations in the weak-coupling limit, when the theory features various types of matter content~\cite{Myers:2009df}. On the lattice, in addition to the effective models for Polyakov loops mentioned above, there are analytical studies based on approximate renormalization group transformations~\cite{Billo:1994ss} and on strong-coupling expansions~\cite{Langelage:2009jb, Langelage:2010yn}---see also ref.~\cite{Svetitsky:1985ye} for a review of early works.

Besides deconfinement, another important phenomenon taking place in QCD at finite temperature is the restoration of chiral symmetry~\cite{Shuryak:1981fz} (see also ref.~\cite{Tomboulis:1984dd} for a proof in $\SU(2)$ gauge theory on the lattice)---which, depending on the dynamics of the model, can occur at the same temperature as deconfinement or at a different (in particular: higher) temperature~\cite{Neri:1983ic, Mocsy:2003qw, Hatta:2003ga, Cohen:2004cd}. As discussed in subsection~\ref{subsec:mesons_and_glueballs}, in QCD with $n_f$ massless quarks, the $\SU_A(n_f)$ symmetry is spontaneously broken by the existence of a non-vanishing quark condensate, while the $\U_A(1)$ symmetry is explicitly broken by an anomaly---which, however, is suppressed in the 't~Hooft limit. At high temperatures, both these symmetries are expected to get restored~\cite{Shuryak:1993ee} and, as discussed in refs.~\cite{Kharzeev:1998kz, Kharzeev:2000na}, the restoration of the $\U_A(1)$ symmetry may trigger the appearance of parity- and $CP$-violating effects. The qualitative features of the finite-temperature phase diagram of $\SU(N)$ gauge theories (for $N$ generic) with $n_f$ massless quark flavors have been recently discussed in ref.~\cite{Tuominen:2012qu}. In the large-$N$ limit, one can get more quantitative insight, e.g., by calculating the dependence of the topological susceptibility on the temperature semiclassically, in terms of instantons~\cite{Affleck:1979gy, Affleck:1980ij}, or by considering some phenomenological models~\cite{Heinz:2011xq}. 

Asymptotic freedom implies that QCD should also deconfine under conditions of large net quark density~\cite{Collins:1974ky}, where interesting novel phases may be realized~\cite{Alford:1998mk, Son:1998uk, Berges:1998rc, Pisarski:1999bf, Pisarski:1999tv, Schafer:1999pb, Rapp:1999qa, Rajagopal:2000wf}. In particular, since lattice calculations show that at zero density and finite temperature the deconfinement transition for $N=3$ colors and physical quark masses is actually a crossover~\cite{Aoki:2006br, Aoki:2009sc, Bazavov:2010sb, Bazavov:2011nk}, while at zero (or low) temperatures and finite density several model calculations indicate that it is a first-order transition~\cite{Berges:1998rc, Asakawa:1989bq, Barducci:1989wi, Barducci:1989eu, Barducci:1993bh, Halasz:1998qr, Scavenius:2000qd, Hatta:2002sj}, one expects the first-order transition line separating the hadronic phase from the deconfined phase to end at a second-order critical point~\cite{Stephanov:2004wx}. The implications of the 't~Hooft limit for QCD-like theories at finite temperature and finite quark chemical potential $\mu$~have been reviewed in ref.~\cite{McLerran:2007qj},\footnote{For a similar discussion in the Veneziano limit, see refs.~\cite{Ohnishi:2006hs, Hidaka:2008yy}; for a discussion of possible phase transitions depending on the $n_f/N$ ratio, see ref.~\cite{Torrieri:2011dg} and references therein.} in which (under the assumption of \emph{fermionic} baryons---i.e. for odd values of $N$) it was also suggested that, at temperatures below the deconfinement transition, baryons could form a dense phase of ``quarkyonic matter'', with pressure $O(N)$ and with interactions between baryons near the Fermi surface (see also refs.~\cite{Bonanno:2011yr, McLerran:2008ua, Zhitnitsky:2008ha, Glozman:2008fk, Kojo:2009ha, Hidaka:2010ph, Torrieri:2010gz, Adam:2010ds, Kojo:2010fe, Lottini:2011zp, Kojo:2012hf}, for discussions on related topics). At large $N$ (with $\lambda$ and $n_f$ fixed) the deconfinement temperature is expected to be independent of $\mu$: this can already be seen from the leading-order perturbative expression for the trace of the gluon self-energy tensor at vanishing momentum (which is gauge-invariant), which yields the square of the Debye mass~\cite{Freedman:1976ub, Toimela:1984xy, Ipp:2006ij}:
\begin{equation}
\label{gluon_self-energy_tensor}
\left. \Pi^\alpha_{\phantom{\alpha}\alpha}(q^2)\right|_{q^\mu=0} = \lambda \left[ \frac{T^2}{3} + \frac{n_f}{2N} \left( \frac{T^2}{3} + \frac{\mu^2}{\pi^2} \right) \right] = \frac{\lambda T^2}{3} + O(1/N),
\end{equation}
and holds at all orders in perturbation theory. In the confined phase at finite values of $\mu$, the contribution to the free energy due to baryons is exponentially suppressed, because their mass is $O(N)$. When the quark chemical potential exceeds the critical value corresponding to the mass of a constituent quark $m_q=M_B/N$, which is an $O(1)$ quantity, a Fermi sea of baryons is created; if $\mu$ is just slightly larger than $m_q$, the baryons (including, possibly, resonances) form a dilute ideal gas, with pressure $O(1/N)$. For larger values of $\mu$, however, their interactions become non-negligible, and yield an $O(N)$ contribution to the thermodynamics of the system. By going to much larger values of $\mu$ (in particular: $\mu \gg \LambdaQCD$), the free energy can be reliably computed perturbatively, in terms of quarks; however, the degrees of freedom in a region of width of order $\LambdaQCD$ near the Fermi surface are baryonic, i.e. non-perturbative. In addition, in such quarkyonic phase the chiral restoration transition is expected to take place at values of $\mu$ comparable with $m_q$, implying that, for a larger chemical potential, quarkyonic matter should be confined but chirally symmetric, with a parity-doubled hadron spectrum.\footnote{The arguments discussed in refs.~\cite{Casher:1979vw, Banks:1979yr} may not apply in the presence of a Fermi sea. In addition, at finite temperature the implications related to anomalies cannot be immediately translated into a constraint relating the deconfinement and chiral restoration temperatures~\cite{Itoyama:1982up}.} This problem was studied using Schwinger-Dyson equation methods in refs.~\cite{Glozman:2007tv, Guo:2009ma}. The implications of the 't~Hooft limit for a possible color-superconducting state of matter at large values of the chemical potential were discussed in refs.~\cite{Deryagin:1992rw, Shuster:1999tn}. There are also analytical studies of the large-$N$ limit of QCD at finite chemical potential and at weak coupling, when the theory is defined on a compact space~\cite{Hands:2010zp, Hollowood:2011ep, Hollowood:2012nr}, or in the strong-coupling regime on the lattice~\cite{Hollowood:2012nr, Christensen:2012km}.

Various properties of the QCD phase diagram (including, in particular, the coincidence or splitting of the chiral and deconfinement transitions~\cite{Mizher:2010zb}, $\rho$-meson condensation~\cite{Chernodub:2010qx}, the possible existence of second-order critical endpoints~\cite{Agasian:2008tb}, or of the chiral magnetic effect~\cite{Kharzeev:2009fn}) may be influenced by the effects of strong electromagnetic fields: it has been estimated that the magnetic field strength realized in heavy-ion collisions at LHC energies could be as large as $15 m_\pi^2$~\cite{Skokov:2009qp}. The properties of the QCD plasma under strong magnetic fields in the 't~Hooft limit have been recently discussed in refs.~\cite{Fraga:2012ev, Fraga:2012rr}, in which it was found that the magnetic fields decrease the deconfinement temperature---at least if the quarks exhibit paramagnetic behavior.

Finally, there exists a very large number of articles discussing the properties of QCD-like theories at finite temperature and/or density, using holographic methods, which implicitly rely on the large-$N$ limit. Like for the holographic studies of mesons mentioned at the end of subsec.~\ref{subsec:mesons_and_glueballs}, it would not be possible to cover all the relevant literature in the present manuscript. Hence, we recommend some recent review articles and lecture notes on this topic~\cite{Son:2007vk, Mateos:2007ay, Peeters:2007ab, Shuryak:2008eq, Gubser:2009md, Rangamani:2009xk, CasalderreySolana:2011us}, and limit our present discussion to a brief summary of the r\^ole of the large-$N$ limit in the gauge-string correspondence, which is reviewed in subsec.~\ref{subsec:gauge_string}.

\subsection{The r\^ole of the large-$N$ limit in the gauge-string correspondence---A brief summary}
\label{subsec:gauge_string}

An expansion analogous to eq.~(\ref{large_N_amplitude_expansion}) also arises in string theory---with the string coupling $g_s$ replacing $1/N$ as the expansion parameter of the topological series. Were the ``planar'' and ``non-planar'' diagrams contributing to eq.~(\ref{large_N_amplitude_expansion}) to be interpreted as defined in real space (rather than in the internal index space of the theory), then one could imagine that, while the perturbative expansion is organized as a series of diagrams which, at any finite order, look like ``fishnets'', in a fully non-perturbative formulation of the theory the holes may close, and the surfaces of the Feynman diagrams may become \emph{real} surfaces, or the world-sheets spanned by strings. This observation led to speculations that string theory might provide a reformulation of non-Abelian gauge theory since the 1970's---a possibility which, in fact, was already discussed in 't~Hooft's original work~\cite{'tHooft:1973jz}.

More recently, this idea resurfaced again in the context of the holographic correspondence, i.e. in the conjectured duality between gauge theories and string theories in higher-dimensional, curved spacetimes~\cite{Maldacena:1997re, Gubser:1998bc, Witten:1998qj} (see refs.~\cite{Aharony:1999ti, Petersen:1999zh, Klebanov:2000me, D'Hoker:2002aw} for an introduction to the topic).

To give an idea of how the gauge/gravity duality arises, and of the r\^ole of the large-$N$ limit in it, it is helpful to consider the most studied example of this correspondence, which relates $ \mathcal{N}=4 $ $ \U(N) $ Yang-Mills theory in four spacetime dimensions to type IIB superstring theory in ten spacetime dimensions. The former theory is the maximally supersymmetric non-Abelian gauge theory in four dimensions, with four supercharges~\cite{Brink:1976bc, Gliozzi:1976qd}. Its field content includes the gauge field $A_\mu$ (which is a singlet under the global $\SU(4)$ $\mathcal{R}$-symmetry), four Weyl fermions in the fundamental representation of the $\mathcal{R}$-symmetry, and six real scalars in the two-index antisymmetric representation of $\SU(4)$: all of these fields transform in the adjoint representation of the gauge group. In addition, the theory is invariant under conformal transformations: this holds at all orders in perturbation theory~\cite{Mandelstam:1982cb, Howe:1983sr}, and also non-perturbatively~\cite{Seiberg:1988ur}. As a consequence, the theory is scale-invariant, and its coupling does not get renormalized. Type IIB superstring theory is a chiral supersymmetric string theory in ten dimensions with 32 supercharges~\cite{Green:1981yb, Schwarz:1983qr}. Its spectrum includes D3-branes~\cite{Polchinski:1995mt, Johnson:2000ch}: they are 3+1-dimensional hyperplanes, on which open strings end, but they can also be interpreted as topological solutions of the supergravity limit of type IIB theory. In particular, a stack of $ N $ coincident D3-branes is a supergravity solution corresponding to the metric:
\begin{equation}
\label{brane_metric}
\dd s^2 = \frac{1}{\sqrt{ 1 + \frac{R^4}{r^4} } } \left( -\dd t^2 + \dd \mathbf{x}^2 \right) + \sqrt{ 1 + \frac{R^4}{r^4} } \left( \dd r^2 + r^2 \dd \Omega_5^2 \right),
\end{equation}
where $t$ is the time coordinate, the $\mathbf{x}^i$'s are the spatial coordinates on the brane, while $r$ denotes the transverse distance from the branes, and $R$ is related to the string coupling ($g_s$) and length ($l_s$) via: $R^4 = 4 \pi g_s N l_s^4 $. Note that $r \to 0$ is a horizon. At very large distance from the branes, $R \ll r$, the metric in eq.~(\ref{brane_metric}) reduces to a 9+1-dimensional Minkowski spacetime. In the opposite limit, i.e. for $r \ll R$, introducing the variable $z=R^2/r$, the geometry of the spacetime reduces to:
\begin{equation}
\label{AdS5_S5_metric}
\dd s^2 = \frac{R^2}{z^2} \left( -\dd t^2 + \dd \mathbf{x}^2 + \dd z^2 \right) + R^2 \dd \Omega_5^2 ,
\end{equation}
which is the product of a five-dimensional anti-de~Sitter (AdS) spacetime\footnote{The anti-de~Sitter spacetime is a maximally symmetric Lorentz manifold with constant negative scalar curvature, which is a vacuum solution to Einstein equations with a negative cosmological constant.} times a five-dimensional sphere, $AdS_5 \times S^5$. The D3-branes are charged under an antisymmetric 4-form tensor field, and the flux of the corresponding self-dual field strength through the $S^5$ sphere is equal to $N$, i.e. it counts the number of D3-branes.\footnote{Note the analogy with Gau{\ss}'s law of electromagnetism: in four spacetime dimensions, pointlike electric charges (``D0-branes'') are sources of the electromagnetic potential (a 1-form). The flux of the electromagnetic field strength (a 2-form) through a $S^2$ sphere counts the number of charges in its interior.} At this point, the gauge/string correspondence arises as a form of open/closed string duality: it stems from the observation~\cite{Maldacena:1997re} that the low-energy dynamics of this system can be described both in terms of a Dirac-Born-Infeld low-energy effective action for open strings (which, when expanded in derivatives, reduces to the action of $\mathcal{N}=4$ super-Yang-Mills theory), and in terms of gravitational excitations, i.e. closed strings propagating in the bulk of the spacetime. In particular, the description in terms of open strings is most natural when $g_s N$ (which describes the strength of the coupling of $N$ D3-branes to gravity) is small, so that the spacetime is almost flat. On the contrary, when $g_s N$ is large, the spacetime is strongly curved, and the low-energy dynamics of the theory can be described as a supergravity theory in anti-de~Sitter spacetime.

The parameters of the two theories---namely, the number of color charges and the 't~Hooft coupling in the gauge theory, and the string length $l_s$ and coupling, and the spacetime radius $R$ on the string side---are related by:
\begin{equation}
\label{couplings}
\lambda = \frac{R^4}{l_s^4}, \qquad \qquad \frac{\lambda}{N} = 4 \pi g_s.
\end{equation}
As a consequence of the second relation in eq.~(\ref{couplings}), the large-$N$ limit of the gauge theory corresponds to the limit in which loop effects on the string side become negligible, i.e. to a classical string limit. If, furthermore, the 't~Hooft coupling of the gauge theory is large, then the string theory reduces to its classical gravity limit.

The gauge/string correspondence discussed above relates the symmetries in the two theories in a non-trivial way:
\begin{itemize}
\item the $\mathcal{R}$-symmetry in the gauge theory is $\SU(4)$, which is isomorphic to the $\SO(6)$ symmetry of the five-dimensional sphere $S^5$;
\item the conformal invariance group of the gauge theory is isomorphic to $\SO(2,4)$, which is the symmetry group of $AdS_5$.
\end{itemize}
In particular, the latter isomorphism is related to the physical interpretation of the ``radial'' coordinate of the $AdS_5$ spacetime, which corresponds to the energy scale of the dual gauge theory (see, e.g., refs.~\cite{Balasubramanian:1998sn, Balasubramanian:1998de, Danielsson:1998wt} for a discussion).

\begin{figure}[-t]
\centerline{\includegraphics[width=0.8\textwidth]{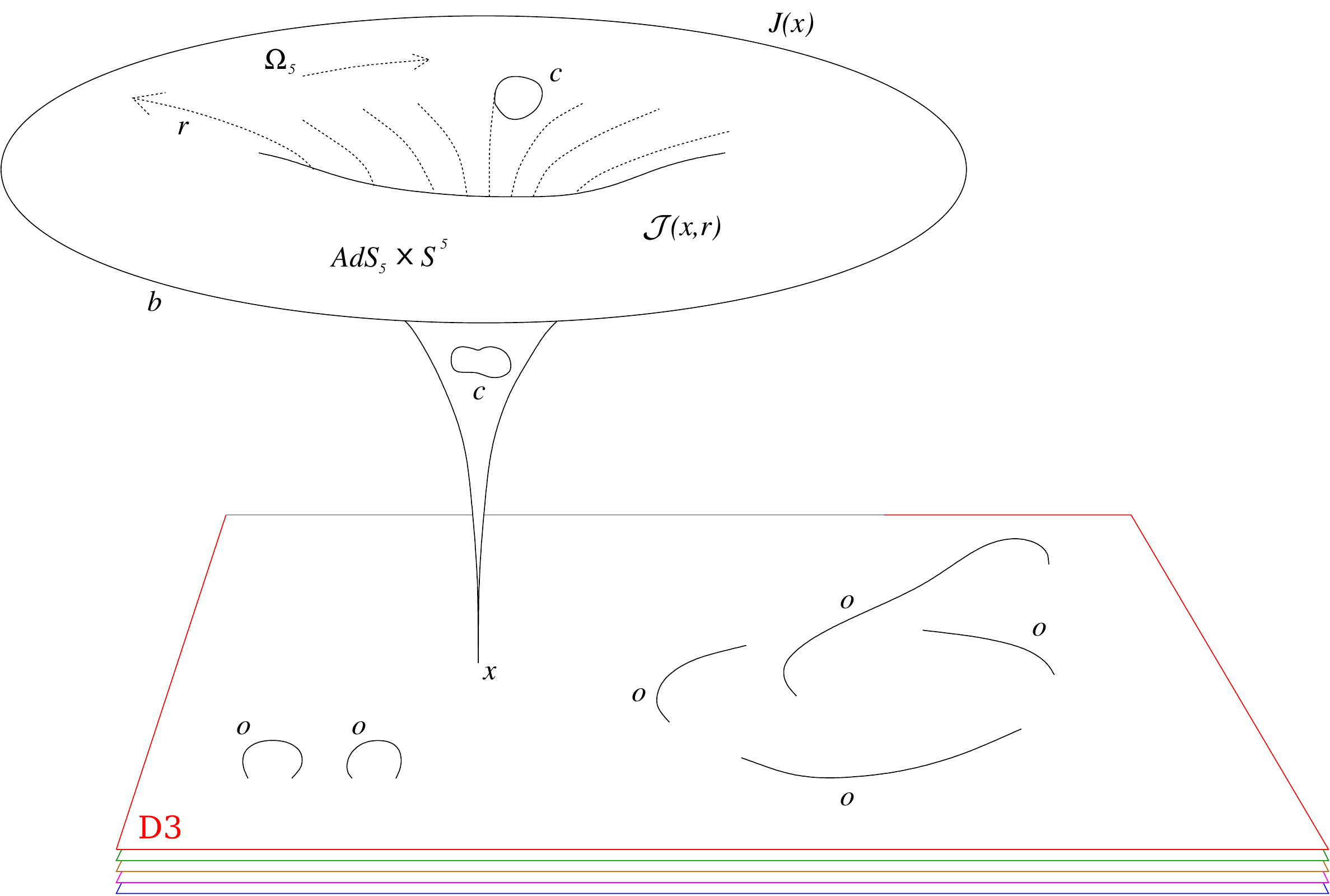}}
\caption{A simplified cartoon of the gauge/string correspondence: the figure shows a set of $N$ coincident D3-branes (D3, at the bottom of the figure), supporting excitations described by open strings ($o$), which describe gauge interactions. The D3-branes are heavy objects, so they curve the six extra dimensions of the spacetime in which they are defined. For the sake of clarity, the cartoon only displays a sketch of the extra dimensions through a given point $x$ on the branes. Starting from $x$, the metric in eq.~(\ref{brane_metric}) shows that at small distance $r$ from the branes ($r \ll R$) the geometry of the ten-dimensional spacetime is approximately that of a five-dimensional anti-de~Sitter spacetime ($AdS_5$), times a five-dimensional sphere ($S^5$), whose coordinates are here symbolically denoted as $\Omega_5$ (while for $R \ll r$ it tends to a ten-dimensional Minkowski spacetime). The gravitational excitations in this space are described by closed strings ($c$). The boundary ($b$) of the $AdS_5$ spacetime is obtained for $r \to \infty$, and is conformally equivalent to four-dimensional Minkowski spacetime. The holographic correspondence expressed by eq.~(\ref{AdS_CFT_mapping}) states that the string partition function with a bulk field $\mathcal{J}$, which reduces to a field $J$ on the boundary---up to a scaling factor expressed by eq.~(\ref{boundary_condition})---, is equivalent to the generating functional of the gauge theory defined in four-dimensional Minkowski spacetime, with a source $J$ coupled to a gauge-invariant operator $\mathcal{O}$, as in eq.~(\ref{source_term}).\label{fig:gramophone}}
\end{figure}

The correspondence can be formulated in a mathematically more precise way by relating the generating functional of connected Green's functions in the gauge theory to the minimum of the supergravity action, with appropriate boundary conditions, via a \emph{field-operator
map}~\cite{Gubser:1998bc, Witten:1998qj}. This means that a deformation of the field theory (defined in a flat spacetime of dimension $D$) via a source term of the form:
\begin{equation}
\label{source_term}
\int \dd^D x \mathcal{O}(x) J(x)
\end{equation}
(where $\mathcal{O}$ is a local, gauge-invariant operator) can be mapped to (the supergravity limit of) a dual string theory with a bulk field $ \mathcal{J} $ which reduces to $ J $ (up to a scaling factor dependent on the conformal dimension) on the boundary\footnote{Here, the word ``boundary'' should be interpreted in the sense of a \emph{conformal}---rather than \emph{topological}---boundary.} of a $(D+1)$-dimensional AdS space (see fig.~\ref{fig:gramophone} for a sketch):
\begin{equation}
\label{AdS_CFT_mapping}
\left\langle \mathcal{T}~\exp \int \dd ^D x \mathcal{O}(x) J(x) \right\rangle = 
\exp\left\{ -S_{\mbox{\tiny{sugra}}} \left[ \mathcal{J} (x,r) \right] \right\},
\end{equation}
where $S_{\mbox{\tiny{sugra}}}$ denotes the on-shell supergravity action, and:
\begin{equation}
\label{boundary_condition}
\lim_{r \to \infty} \mathcal{J} (x,r) = r^{\Delta-D} J(x).
\end{equation}
For this reason, the gauge/string duality is also called ``holographic correspondence'', as it relates the description of dynamics within a volume of space to information encoded on its boundary~\cite{'tHooft:1993gx, Stephens:1993an, Susskind:1994vu}---see also ref.~\cite{Bousso:2002ju} for a review, and ref.~\cite{Susskind:1998dq} for a detailed discussion of the interplay between infrared and ultraviolet effects in the bulk and boundary theories. By taking functional derivatives with respect to the source, eq.~(\ref{AdS_CFT_mapping}) opens up the possibility to compute correlators of composite operators in the field theory by mapping them to integrals on the AdS space~\cite{Freedman:1998tz}.

The construction outlined above can be extended to a finite-temperature setup, via a Wick rotation, and compactification of the resulting Euclidean time direction~\cite{Witten:1998zw}. In this case, the asymptotic boundary geometry is $S^3 \times S^1$, and, in addition to the (Wick-rotated version of the) AdS metric, there exists another supergravity solution, which describes an AdS-Schwarzschild black hole:
\begin{equation}
\label{AdS_Schwarzschild}
\dd s^2 = \frac{r^2}{R^2} \left[ f(r) \dd \tau^2 + \dd \mathbf{x}^2 \right] + \frac{R^2}{r^2} \left[ \frac{1}{f(r)}\dd r^2 + r^2 \dd \Omega_5^2 \right], \qquad f(r) = 1 - \frac{r_H^4}{r^4},
\end{equation}
where $\tau$ denotes the Euclidean time. The partition function of the theory is then dominated by the solution with the smallest Euclidean action. Note that the ``blackness function'' $f(r)$ is monotonic, and tends to $1$ for $r\to \infty$, while it vanishes for $r=r_H$, which corresponds to the location of the black hole horizon. The Hawking temperature of the horizon is $T= r_H/(\pi R^2)$, and corresponds to the temperature of the dual gauge theory.\footnote{Note that a finite temperature breaks explicitly the supersymmetry and conformal invariance of the $\mathcal{N}=4$ super-Yang-Mills theory, making the latter---at least qualitatively---``more similar'' to QCD.} Renowned results that have been obtained for the large-$N$ $\mathcal{N}=4$ super-Yang-Mills theory at finite temperature, using the holographic correspondence, include the value of the entropy density at strong coupling (identified with the Bekenstein-Hawking entropy density)~\cite{Gubser:1996de,Gubser:1998nz}:
\begin{equation}
\label{s_over_s0}
\frac{s}{s_0} = \frac{3}{4} +\frac{45}{32} \zeta(3) (2 \lambda)^{-3/2} + \dots
\end{equation}
(where $s_0=\pi^2 N^2 T^3/2$ is the entropy density of the free $\mathcal{N}=4$ plasma) and the ratio between the shear viscosity $\eta$ and the entropy density~\cite{Policastro:2001yc}:
\begin{equation}
\label{one_over_fourpi}
\frac{\eta}{s}=\frac{1}{4\pi}.
\end{equation}

Although the $\mathcal{N}=4$ theory only contains fields in the adjoint representation of the gauge group, it is possible to study fundamental matter in the probe approximation. On the string side, this can be done by the inclusion of $n_f$ D7-branes~\cite{Karch:2000gx, Karch:2002sh}---called ``flavor branes'', and representing the different quark flavors---which extend along the same spatial directions as, and which can be separated from, the D3-branes: then the theory has a reduced supersymmetry, and a spectrum including open strings stretching from a D7- to a D3-brane (which are interpreted as massive ``quarks''), and open strings with both ends on flavor branes (the ``mesons'' of the theory)---see also ref.~\cite{Erdmenger:2007cm} and references therein for an extensive discussion.

The gauge/gravity correspondence has also been extended to gauge theories qualitatively more similar to QCD---including non-supersymmetric theories with a linearly rising confining potential---in the so-called ``top-down approach''~\cite{Witten:1998zw}: a partial list of relevant references includes~\cite{Polchinski:2000uf, Maldacena:2000yy, Klebanov:2000hb, Aharony:2002up, Kruczenski:2003be, Kruczenski:2003uq, Sakai:2004cn, Sakai:2005yt, Nunez:2010sf}. On the other hand, a different, more phenomenology-oriented, ``bottom-up'' approach has been pursued in a series of works~\cite{Polchinski:2001tt, Polchinski:2002jw, Karch:2002xe, Son:2003et, Brodsky:2003px, deTeramond:2005su, DaRold:2005zs, Erlich:2005qh, Hirn:2005nr, Karch:2006pv, Csaki:2006ji, Csaki:2008dt} trying to reproduce the main features of QCD theories by constructing an appropriate five-dimensional gravitational background for the dual model.

In principle, these gauge/string models allow one to get non-perturbative information about the strongly coupled regime of a field theory, by studying an appropriate limit of the dual string model. While both the top-down and the bottom-up approaches have led to some qualitative and quantitative success, it is fair to say that a completely realistic holographic dual of QCD still remains missing. The main shortcomings of holographic models are inherent in the classical supergravity approximation (which is expected to capture the large-$N$ and strong coupling limits of the dual gauge theory), and include, for example, the lack of a satisfactory description of asymptotic freedom, and certain mismatches between the parametric dependence of typical dynamically generated mass scales characterizing confining gauge theories, like, e.g., the mass gap and the (square root of the) force between color sources at asymptotically large distances~\cite{Gubser:2009md}. In fact, as nicely summarized in ref.~\cite{Reece:2011zz}, the main difficulty appears to be the inclusion of non-negligible $\alpha^\prime$ string corrections, which account for the finiteness of the physical coupling in the dual gauge theory. As eq.~(\ref{couplings}) shows, at fixed 't~Hooft coupling, the string coupling constant is inversely proportional to the number of colors of the gauge theory, so that quantum corrections on the string side of the correspondence are suppressed if $N$ is large. Furthermore, if in this limit the 't~Hooft coupling is large, then the string length $l_s$ becomes negligible with respect to the radius of the space on which the theory is defined, and hence string theory essentially reduces to classical gravity (in an appropriate higher-dimensional spacetime). Since this limit is particularly interesting, because it can be studied analytically or semi-analytically, the correspondence is often called ``gauge/gravity'' duality.\footnote{Note that, strictly speaking, this is actually a misnomer~\cite{Mateos:2011bs}, since the correspondence (in its ``strong'' form) is expected to hold at all values of the parameters of the string theory.} In addition to the vast literature investigating the gauge/gravity duality with analytical methods, it is worth mentioning that, especially during the last few years, there has been a tremendous boost of studies of related problems using numerical tools: for an overview, see, e.g., ref.~\cite{novelnum12} and the links to the online talks therein.

Finally, the large-$N$ limit plays a crucial r\^ole also in the integrability of the $\mathcal{N}=4$ supersymmetric Yang-Mills theory: the planar sector of this theory can be ``solved'' (meaning that scaling dimensions of local operators can be expressed  analytically as a function of the coupling constant) in terms of a system of algebraic equations, which can be derived as a particular limit of the integral equations obtained from the thermodynamic Bethe \emph{Ansatz}~\cite{Minahan:2002ve, Bena:2003wd, Beisert:2003jj, Beisert:2003tq}---see the review~\cite{Beisert:2010jr} and references therein for a discussion.

\section{$\SU(N)$ gauge theories on the lattice}
\label{sec:lattice}
For energies above 1 GeV, QCD is perturbative. Hence, a controlled expansion in
$\alpha_s = g^2/(4 \pi)$ can be used to compute physical quantities. This programme
is successful at explaining experimental data obtained at particle
collider facilities, and is indeed one of the key theoretical
ingredients for the analysis of collision events observed at the
LHC. However, as the energy is lowered below 1 GeV, $\alpha_s$ rapidly
becomes of order one, and the perturbative expansion looses its
predictive power. Hence, in order to derive from first principles
phenomena that are typical of the non-perturbative domain of QCD, like
chiral symmetry breaking and confinement, a different approach is
needed. In the previous section, we have discussed the large-$N$
expansion as a possible systematic expansion to study non-perturbative
QCD and the gauge-gravity dual as a way to compute observables in the zeroth order in
that expansion (corresponding to the $N \to \infty$ limit of the theory). In this
section, we shall discuss a complementary approach, based on numerical
simulations of the theory discretized on a spacetime
lattice using Monte Carlo techniques. Once the continuum limit is
taken, numerical results for QCD obtained in the lattice framework can
be compared to observations. Thanks to recent 
technical advances in the field of lattice gauge theory, not only do
we have now a reasonable numerical proof that QCD is the correct
theory of the strong interactions, but we can also trust the
predictive power of lattice QCD calculations. The progress in the field has
encouraged further development targeting a more theoretical understanding
of gauge theories. Among possible directions, numerical studies of the
't~Hooft large-$N$ limit have generated a large volume of quantitative
results that on the one hand significantly advances our understanding
of the theory and on the other hand can be used to inform analytical approaches. In this
section, we lay the foundations for understanding Monte Carlo results
for $\SU(N)$ gauge theories in the 't~Hooft large-$N$ limit. 
While this part brings only limited benefit to the
reader already familiar with numerical calculations on a spacetime
lattice, it could provide a useful reference for those who are
interested in using lattice results to inform or inspire analytical calculations.
The exposition is pedagogical and aimed at underlining the main conceptual
steps and the technicalities needed for a critical understanding of
the numerical results and fixing the notation for later
chapters rather than at providing detailed derivations of all the basic
lattice results needed later on. The reader interested in this latter
aspect is referred to the excellent textbooks on lattice gauge
theories~\cite{Creutz:1984mg, Rothe:2005nw, Smit:2002ug, Montvay:1994cy, DeGrand:2006zz, Gattringer:2010zz}. 
The works reported in the early original literature on the subject are by
now considered  classic papers in the field. Although we have
mentioned the main original references, there are many more to which
we can not and we are not making justice. Once again, we refer the
reader to the textbooks, which contain a better 
account of the early original contributions to the 
field.

This section is organized as follows. In
subsec.~\ref{sect:lattice:pi} we discuss the discretization of the
free scalar field and some general aspects of the recovery of
the continuum limit. How to construct lattice variables for the gauge
fields so that gauge invariance is respected in a discretized
spacetime will be the subject of subsec.~\ref{sect:lattice:discrete},
which will also deal with subtleties connected with the
discretization of fermionic fields. Numerical calculations will be
discussed in subsec.~\ref{sect:lattice:mc}, while in
subsec.~\ref{sec:lattice:recovery} we shall present how continuum
large-$N$ physics is extracted from numerical
simulations. For the sake of definiteness, unless otherwise stated, in
this section we only deal with four-dimensional theories.
\subsection{The free scalar field}
\label{sect:lattice:pi}
\subsubsection{Path integral approach}
\label{sect:lattice:pi:phi}
The lattice approach is based on the path integral
quantization. For the sake of definiteness, let us consider a free
scalar theory, described by the Lagrangian 
\beq
{\cal L}(\phi(x)) = \frac{1}{2} \partial^{\mu} \phi(x) \partial_{\mu} \phi(x) -
\frac{1}{2} m^2 \phi(x)^2 ,
\eeq
where $\phi$ is a spin zero scalar field of mass $m$. The easiest way
to quantize this theory is to use the canonical approach. With $x =
(x_0,\vec{x})$, with $x_0$ being the temporal component and $\vec{x}$
the spatial component of the quadrivector $x$, we define the momentum
$\Pi(x)$ as 
\beq
\Pi(x) = \frac{\partial {\cal L}(\phi(x))}{\partial_0 \phi(x)} 
\eeq
and we impose the usual equal-time ($x_0 = y_0$) commutation relations
\beq
\nonumber
\left[ \phi(x), \Pi(y) \right] &=& i \delta^3(\vec{x} -
\vec{y}) , \\
\nonumber
\left[ \phi(x), \phi(y) \right] &=& \left[ \Pi(x), \Pi(y) \right] = 0
.
\eeq

This approach naturally leads to the Fock space, in which the base
states are multi-particle states labelled by the momenta of each
single particle.

If we now consider an interaction that can be written
as\footnote{We assume that $V(\phi)$ is bounded
  from below.} $V(\phi) = \alpha f(\phi)$, with $f$ e.g. a polynomial of degree
$k$, for small $\alpha$ we can still start from the canonical quantization 
and treat the effect of the interaction perturbatively in $\alpha$. 
It is then possible to compute systematically the
$n$-point correlation function for arbitrary $n$
\beq
{\mathcal C}_n\left(\phi(x_i) ,  \ \dots , \  \phi(x_n) \right) = \langle \phi(x_1)
  \ \dots \phi(x_n) \rangle
\eeq
as a power series in $\alpha$. Since all the observables can be
expressed using the ${\mathcal C}_n$, all physical processes can be accessed in
this way. The basic physical assumption underlying the perturbative
expansion is that multiparticle states of well-defined momentum are
still a good approximation of the eigenstates of the interacting
theory, with corrections that can be accounted for by a systematic expansion in
powers of $\alpha$. However, in a quantum field theory $\alpha$ is not
constant, but depends on the momentum. Hence, in order for the
perturbative calculation to be valid, $\alpha$ needs to remain small
for the relevant range of energies. 

An alternative approach for computing the correlation functions is the
path integral. In this formulation, they are given by
\beq
{\mathcal C}_n\left(\phi(x_i) ,  \ \dots , \  \phi(x_n) \right) = \frac{\int
 \left( {\cal D} \phi(x) \right) \phi(x_1)  \ \dots \phi(x_n) e^{i S\left(\phi\right)}
}
{\int \left( {\cal D} \phi(x) \right) e^{i S\left(\phi\right)}} ,
\eeq
where $\phi$ has to be interpreted as a classical field and the
integral has to be performed over all possible classical field
configurations
(this is indicated by the expression $\left( {\cal D} \phi(x)
\right)$). The integrand is weighted by the factor $e^{i S(\phi)}$,
where
\beq
S(\phi) = \int {\mbox d}^4 x \ {\cal L}(\phi) 
\eeq
is the action evaluated over a field configuration $\phi(x)$. The
denominator 
\beq
\label{eq:pi:scalar}
Z = \int \left( {\cal D} \phi(x) \right) e^{i S\left(\phi\right)} ,
\eeq
needed to normalize the correlation functions, is what is referred to
as the {\em path integral}. The advantage of the path integral
formulation is that it can be used also for the interacting theory,
irrespectively of the value of the coupling. 
\subsubsection{The scalar field on the lattice}
\label{sect:lattice:pi:discrete}
At this stage, the path integral~(\ref{eq:pi:scalar}) is still a
formal expression: in order to be able to use it, we need to give a
prescription on how to perform the integration. To this purpose, we
first perform a Wick rotation, which consists in the change of
variable $\tau =i x_0$. In the new variables, up to an overall minus
sign, the metric is Euclidean. For this reason, the space of the vectors $x_E =
(i x_0, \vec{x})$ is called {\em Euclidean space}. In Euclidean space,
we write the Lagrangian as
\beq
{\cal L}_E(\phi) = \frac{1}{2} \partial_E^{\mu} \phi(x_E) \partial_{\mu,E} \phi(x_E) +
\frac{1}{2} m^2 \phi(x_E)^2 , 
\eeq
where the metric is the identity (and in fact we could disregard the
convention of lower and upper indices), and the action as
\beq
S_E(\phi) =   \int {\mbox d}^4 x_E \ {\cal L}_E(\phi)  .
\eeq
Then, the path integral becomes
\beq
Z_E =  \int \left( {\cal D} \phi(x_E) \right) e^{- S_E\left(\phi\right)} .
\eeq
Over eq.~(\ref{eq:pi:scalar}), $Z_E$ has the advantage of having the
damping factor $e^{- S_E\left(\phi\right)} $ replacing the oscillating
factor $e^{i S\left(\phi\right)} $ in the integrand, which improves
the convergence of the integral. Real-time correlation functions can
be obtained from Euclidean-time correlation functions by analytic
continuation. From now on, we will work in Euclidean space, and for
convenience we will drop the subscript $E$ from all expressions. 
For instance, we will write the Euclidean path integral as
\beq
Z=  \int \left( {\cal D} \phi(x) \right) e^{- S\left(\phi\right)} ,
\eeq
where it is understood that $\phi$ is defined in Euclidean space and
$S$ is the Euclidean action.

We still need to give an operational implementation of the measure 
$\left( {\cal D} \phi(x) \right)$. To this purpose, we can consider a
spacetime grid of spacing $a$ (the {\em lattice}) and define $\phi$
only on sites of the grid, i.e. on points $x$ such that $x = (n_0 a,
n_1 a, n_2 a, n_3 a)$, with the $n_i$ all integer. If we also impose
that $0 \le n_i \le N_i -1$, i.e. that the lattice size in the $i$-th
direction is $a N_i$, then the path integral measure becomes a
multidimensional integral. In the remainder of this section, we focus
on the discretization of the theory, while in the next section we
shall see how the continuum theory can be recovered from its
discretized version. 

For convenience, we rescale the field $\phi$ by $a$, obtaining the 
dimensionless combination $\varphi(i) = a \phi(a i)$, with $i = (n_0,n_1,n_2,n_3) =
(n_0,\vec{n})$, in terms of which we formulate the free field scalar
theory on the lattice. To this purpose, we replace the integral $\int
\mbox{d}^4 x$ with $a^4 \sum_i$, where $i$ runs over all lattice
points, and the derivative of $\phi$ with the finite difference
\beq
\partial_{\mu} \phi(i) \to \frac{\phi(i + \hat{\mu}) - \phi(i)}{a} ,
\eeq
where $\hat{\mu}$ is the unit vector in the direction $\mu$. 
The lattice action becomes
\beq
\label{eq:latt_phi_act} 
S = \frac{1}{2} \sum_i \left[ - \frac{1}{2} \left( \sum_{\mu = 0}^{3}\varphi(i +
    \hat{\mu})\varphi(i) +  \varphi(i) \varphi(i - \hat{\mu} ) \right) + \left(\hat{m}^2 +
    8 \right) \right] ,
\eeq
where $\hat{m} = a m$, and the discretized path integral reads
\beq
\label{eq:latt_phi_pi}
Z=  \int \left( \prod_i \mbox{d} \varphi(i) \right) e^{- S\left(\varphi\right)} ,
\eeq
with $S$ given in~(\ref{eq:latt_phi_act}) and the product running over
all lattice points. 
$n$-point Green functions are easily computed in momentum space, where
it has to be considered that the momenta are cut-off at $p_{max} =
\pi/a$ (in crystallography, this corresponds to the first Brillouin
zone). For instance, for the two-point function (the lattice
propagator), we have:
\beq
\langle  \varphi(l) \varphi(m) \rangle = 
\int_{- \pi}^{\phantom{-}\pi} \frac{d^4  \hat{p}}{\left( 2 \pi \right)^4} 
\frac{e^ {i \hat{p}\cdot(l-m)}}{4
  \sum_{\mu} \sin^2\left( \hat{p_{\mu}}/2\right) +
  \hat{m}^2} .
\eeq
\subsubsection{Continuum limit}
\label{sect:lattice:pi:continuum}
The lattice path integral~(\ref{eq:latt_phi_pi}) is formally identical
to the partition function of a statistical system with $N_1 N_2 N_3
N_4$ degrees of freedom and Hamiltonian  $S$. The problem of recovering the infinite volume
limit of the original system is then mapped into the problem of
performing the thermodynamic limit of the associated statistical
system. This mapping allows typical concepts of statistical mechanics
to be carried over to lattice field theory. Although we do not pursue
the analogy further, important progress has been achieved exploiting
this correspondence.

Once the infinite volume limit has been performed, the continuum limit
can be recovered by taking the lattice spacing $a$ to zero. At the
classical level, this implies that the discretized action should
reproduce the continuum action in the limit $a \to 0$. This request is
easily fulfilled, since the lattice action has been constructed as a
simple discretization of the continuum action. For the quantum theory,
we need to systematically compute all lattice $n$-point functions and
show that they converge to the corresponding continuum functions when $a \to
0$. Since a quantum field theory is uniquely specified by its
$n$-point functions, this would suffice to prove that the lattice field
theory reproduces the wanted field theory in the continuum limit. The
above procedure can look tautological. However, there is no guarantee
that the lattice theory obtained by na\"{\i}ve discretization of the path
integral describe the wanted field theory in the continuum limit. 

Since for the free theory the issue is trivial, let us consider the
case of the interacting theory. Once the theory is discretized on a
lattice, observables depend on the dimensionless couplings
$(\hat{m},\alpha,\dots)$, which in turn are functions of the lattice
spacing. If we now take the lowest mass of the physical spectrum, $M$,
if a continuum limit exist, we must have
\beq
\lim_{a \to 0} \hat{M}/a = M , 
\eeq 
where $\hat{M}$ is the dimensionless mass determined on the lattice,
and $a$ reinstates the dimensions. The previous equation implies
\beq
\lim_{a \to 0} 1/\hat{M} = \infty ,
\eeq 
where $1/\hat{M} = \xi$ can be interpreted as the correlation length
(in dimensionless units) of the
statistical system associated to the regularized quantum field
theory. Hence, in the language of Statistical Mechanics, reaching the
continuum limit means finding the values of the couplings for which
the system is critical. Physically, this means that the correlation
length of the system is much larger than the lattice spacing $a$,
which implies that the system looses memory of the discretization. 
In general, whether values of the coupling exist such that 
the system is critical is a dynamical problem. If it happens, it might not be immediate
to identify the corresponding quantum field theory in the infrared,
since this is determined by the value of the $n$-point functions at
the critical point. For $\SU(N)$ gauge theories, thanks to asymptotic
freedom, the system is critical at the ultraviolet fixed point, where the theory is perturbative. Hence,
the lattice provides a way to regularize the theory in the ultraviolet, and removing the
ultraviolet cut-off $a$ corresponds to a well-defined renormalization
procedure, in which an infrared scale (e.g. the string tension in the
pure gauge case or the mass of the $\rho$ meson in the presence of
dynamical quarks) is fixed, and all spectral quantities are determined in terms of this scale.\footnote{An alternative way to set the lattice scale is 
based on Sommer's scale $r_0$~\cite{Sommer:1993ce}, which
is defined as the distance at which the
force $F$ between a pair of external, infinitely heavy, fundamental color sources (i.e., a static quark-antiquark pair) satisfies: $r_0^2 F(r_0)=1.65$. Comparison
with phenomenological potential models shows that this length scale
corresponds to approximately $r_0=0.5$~fm. For the lattice regularization of $\SU(3)$ Yang-Mills with the Wilson gauge action, a 
high-precision determination of the lattice spacing $a$ in units
of $r_0$, as a function of the lattice parameter $\beta$, was reported in ref.~\cite{Necco:2001xg}.} It is important to emphasize that different
ways to set the lattice scale (in terms of different physical observables) 
may lead to slightly different results at finite values of the 
lattice spacing, because different observables can be affected by 
different lattice artifacts. However, these differences disappear 
when results are extrapolated to the continuum limit $a \to 0$. 
In fact, once one has proved that the lattice discretized version of
QCD has the right continuum limit, one could invert the logic and
define QCD starting from the construction on the lattice. This would
give a more rigorous way to specify the quantum field theory than
canonical quantization, which has the problem of defining the theory
starting from the perturbative Fock vacuum, which is very far from the
real QCD vacuum. By the same token, using the lattice we can
rigorously define the large-$N$ limit {\em ~\`a la }~'t~Hooft of
$\SU(N)$ gauge theories as the theory defined by all the $n-$point
correlators when this limit is taken. In this approach, the lattice
could be used to prove the existence of this limit, even in the
non-perturbative regime.

In passing by, we notice that different prescriptions could have been used to construct
the lattice action from the continuum one. For instance, we could
request that the discrete derivative be defined as
\beq
\label{eq:scalar:forder}
\partial_{\mu} \phi(i) \to \frac{\phi(i + \hat{\mu}) - \phi(i -
  \hat{\mu})}{2 a}  .
\eeq
The corresponding action would have differed from~(\ref{eq:latt_phi_act}) 
by corrections that go to zero in the continuum limit. This ambiguity
in defining the lattice action can be exploited to improve at the
quantum level the convergence of the lattice theory to the continuum
one as $a \to 0$. 

\subsection{Discretization of $\SU(N)$ gauge theories}
\label{sect:lattice:discrete}
\subsubsection{Gauge fields}
\label{sect:lattice:discrete:gauge}
A good regularization of a quantum theory respects the crucial
properties of the original theory, with the others recovered when the
ultraviolet cut-off is removed. On the lattice, the property that we
absolutely need to preserve is gauge invariance. This can be
accomplished through parallel transports.  Consider for instance
continuum scalar electrodynamics (in Euclidean space). If the (complex) scalar field is coupled to an
Abelian gauge field $A_{\mu}(x)$, the continuum derivative is replaced by the
covariant derivative:
\beq
\partial_{\mu} \to D_{\mu} = \partial_{\mu} + i g_0 A_{\mu}(x) ,
\eeq
where $g_0$ is the gauge coupling. Under gauge transformations defined
by the function $\Lambda(x)$
\beq
A_{\mu}(x) \to A_{\mu}(x) - \partial_{\mu}  \Lambda(x) , \qquad
\phi(x) \to e^{- i g_0 \Lambda(x)} \phi(x) ,
\eeq
so that the Lagrangian
\beq
\label{eq:scalar_qed}
{\cal L}(\phi, A_{\mu}) = D_{\mu} \phi(x) D_{\mu}
  \phi^{\ast}(x) + m^2 \phi(x) \phi^{\ast}(x) + \frac{1}{4}
  F_{\mu \nu}(x) F_{\mu \nu}(x) \ , \ F_{\mu \nu}(x)
  = \partial_{\mu} A_{\nu} - \partial_{\nu} A_{\mu}
\eeq
is invariant. It is immediate to see that a na\"{\i}ve discretization of
this Lagrangian will not preserve gauge invariance, the problem being
that the finite difference mixes fields defined on different lattice
points. This can be remedied as follows. We introduce the parallel transport along
the link joining the sites $i$ and $i + \hat{\mu}$ as
\beq
U_{\mu}(i) = e^{i g_0 \int_x^{x + a \hat{\mu}} A_{\mu}(s) \dd s} ,
\eeq
where $x = a i$. On the lattice, we make the replacement
\beq
\label{eq:scalar:covforder}
D_{\mu} \phi(i) \to \frac{U_{\mu}(x) \phi(i + \hat{\mu}) -  \phi(i)}{a}  .
\eeq
Gauge transformations act as usual on $\phi$:
\beq
\phi(j) \to e^{- i \lambda(j)} \phi(j); \qquad \phi^{\ast}(j)
\to e^{i \lambda(j)} \phi^{\ast}(j) .
\eeq
For the link variables, we have
\beq
U_{\mu}(j) \to e^{- i \lambda(j)}  U_{\mu}(j) e^{ i \lambda(j +
  \hat{\mu})} .
\eeq
Using the fact that $U_{-\mu}(i)$ = $U_{\mu}^{\ast}(i - \hat{\mu})$, it is easy to
see that the terms involving $\phi$ in eq.~(\ref{eq:scalar_qed}) are
invariant under discretized gauge transformations when using the
prescription~(\ref{eq:scalar:covforder}) for the covariant
derivative. The part of the Lagrangian involving the field tensor
$F_{\mu \nu}$ can also be expressed in terms of the link
variables. The simplest possibility is given by the Wilson action~\cite{Wilson:1974sk}
\beq
\label{eq:wilsonact_u1}
S = \beta \sum_{j, \mu > \nu} \left( 1 - \frac{1}{2} \left( U_{\mu
      \nu}(i) + U_{\mu \nu}^\ast (i) \right) \right) 
\eeq
where $\beta = 1/g_0^2$ and 
\beq 
U_{\mu \nu}(j) = U_{\mu}(j)  U_{\nu}(j+\hat{\mu}) U^{\ast}_{\mu}(j+\hat{\nu}) U^{\ast}_{\nu}(j) 
\eeq
is the parallel transport of the gauge field around an elementary
lattice square ({\em plaquette}). It is worth noticing that $S$
defined in eq.~(\ref{eq:wilsonact_u1}) is manifestly gauge
invariant. Expanding the exponentials defining the links at the
leading order in $a$, we get 
\beq
S \simeq \frac{a^4}{4}  \sum_{j, \mu,\nu} F_{\mu \nu}(j) F_{\mu
  \nu}(j) ,
\eeq
which is the na\"{\i}ve discretization of the gauge field action.

The generalization of the above discussion to $\SU(N)$ is
immediate. Now, the link variables $U_{\mu}(j)$ are defined as the
path-ordered exponential
\beq
U_{\mu}(j) = \mathcal{P} e^{i g_0 \int_x^{x + a \hat{\mu}} A_{\mu}(s) \dd s} ,
\eeq
where now $U_{\mu}(j)$ is a matrix in the $\SU(N)$ group, and the plaquette is
the path-ordered product of links around the lattice plaquette:
\beq 
U_{\mu \nu}(j) = U_{\mu}(j)  U_{\nu}(j+\hat{\mu})
U^{\dag}_{\mu}(j+\hat{\nu}) U^{\dag}_{\nu}(j) \, 
\eeq
where as before  $U_{-\mu}(j) =  U^{\dag}_{j - \hat{\mu}}(j)$. Under gauge
transformation implemented by the $\SU(N)$-valued function $\Omega(x)$,
the $U_{\mu}(j)$ transforms as 
\beq
U_{\mu}(j)\to \Omega(j) U_{\mu}(j) \Omega^{\dag}(j + \hat{\mu})
,
\eeq
which is the lattice version of the continuum gauge transformation
\beq
A_{\mu}(x) \to \Omega(x) A_{\mu}(x) \Omega^{\dag}(x) - \frac{i}{g_0}
\Omega(x) \partial_{\mu} \Omega^{\dag}(x) .
\eeq
In terms of $U_{\mu \nu}(j)$, the Wilson action reads 
\beq
\label{eq:wilsonact_ym}
S = \beta \sum_{j, \mu > \nu} \left( 1 - \frac{1}{N} \real \Tr
  U_{\mu \nu}(j) \right) .
\eeq
where $\real \Tr$ indicates the real part of the trace and $\beta
= 2 N /g_0^2$. Gauge invariance is guaranteed by the trace. The path
integral reads
\beq
Z = \int \left( \prod_{j,\mu} \dd U_{\mu}(j) \right) e^{- S} ,
\eeq
where each factor $\dd U_\mu(j)$ in the measure is the Haar measure
of $\SU(N)$ associated to the link $U_\mu(j)$. For our purposes,
the most important property of the Haar measure is that it is uniform
in the group. The action~(\ref{eq:wilsonact_ym}) yields the na\"{\i}ve
discretization of the $\SU(N)$ Yang-Mills theory at the leading order
in $a$ as $a \to 0$, as it should. The lattice theory can also be
shown to reproduce the correct continuum limit when the lattice
spacing goes to zero. As we will see in more detail later, this
is a non-trivial consequence of asymptotic freedom. 
\subsubsection{Fermions}
\label{sect:lattice:discrete:fermions}
Lattice discretization of fermion fields is not immediate. To see the
origin of the problem and the possible solutions, we start by
following closely the strategy we have used for the bosonic
field. Given a continuum field $\psi(x)$, we define the
(dimensionless) discretized fermion field $\hat{\psi}(i)$ as
\beq
\label{eq:latferm}
\hat{\psi}(i) = a^{3/2} \psi(i a)
\eeq
and its discretized derivative as
\beq
\label{eq:latfermder}
\hat{\partial}_{\mu} \hat{\psi}(i) =  \frac{\hat{\psi}(i + \hat{\mu})
  - \hat{\psi}(i - \hat{\mu})}{2} .
\eeq
Note that, differently from the bosonic field case, we have used a
symmetric definition for the derivative. By substituting these
equations in the Euclidean Dirac action, we get the lattice fermionic action
\beq
\label{eq:latfermlag}
S_f = \sum_{i,j,\alpha,\beta} \hat{\overline{\psi}}_{\alpha} (i) 
\left[ \frac{1}{2} \sum_{\mu} \left( \gamma^E _{\mu} \right)_{\alpha \beta} 
\left( \delta_{i+\hat{\mu},j}- \delta_{i-\hat{\mu},j} \right) +
\hat{m} \delta_{i j } \delta_{\alpha \beta} \right]
\hat{\psi}_{\beta}(j) ,
\eeq
where $\hat{m} = a m$, with $m$ the fermion mass. In this equation,
Dirac indices are expressed with Greek letters, while Latin letters
indicate lattice points. The $\gamma_E$ are the Euclidean $\gamma$
matrices, satisfying the anticommutation relations
\beq
\{ \gamma_{\mu}^E, \gamma_{\nu}^E\} = 2 \delta_{\mu \nu} .
\eeq
In terms of the Minkowskian $\gamma$ matrices, the Euclidean $\gamma$
matrices are given by
\beq
\gamma^E_0 = \gamma_0 , \qquad \gamma_i^E = - i \gamma_i ,\qquad i=1,2,3
.
\eeq
It is convenient to define the Dirac operator $D$ as
\beq
D_{\alpha \beta}(i j) = \left[ \frac{1}{2} \sum_{\mu} \left( \gamma^E _{\mu} \right)_{\alpha \beta} 
\left( \delta_{i+\hat{\mu},j}- \delta_{i-\hat{\mu},j} \right) +
\hat{m} \delta_{i j } \delta_{\alpha \beta} \right] ,
\eeq
so that the fermion action can be written in a more compact form as
\beq
S_f = \sum_{i,j,\alpha,\beta} \hat{\overline{\psi}}_{\alpha} (i)
D_{\alpha \beta}(i j) \hat{\psi}_{\beta}(j) .
\eeq
Fermionic correlation functions can be expressed in terms of the
inverse Dirac operator. For instance, for the correlator, we have
\beq
\langle \hat{\psi}_{\alpha} (i) \hat{\overline{\psi}}_{\beta} (j)
  \rangle = D^{-1}_{\alpha \beta}(i j) .
\eeq
Using the momentum representation, we can rewrite this expression as
\beq
\label{eq:propnaive}
\langle \hat{\psi}_{\alpha} (i) \hat{\overline{\psi}}_{\beta} (j)
  \rangle = 
\int _{- \pi}^{\phantom{-}\pi} \frac{d^4  \hat{p}}{\left( 2 \pi \right)^4} 
\frac{- i \sum_{\mu} \gamma_{\mu}^E \sin \left( \hat{p}_{\mu} \right) + \hat{m} }{
  \sum_{\mu} \sin^2\left( \hat{p_{\mu}}\right) +
  \hat{m}^2} e^ {i \hat{p}\cdot(i-j)}  ,
\eeq
where $\hat{p} = a p$ and $p$ is the continuum momentum. This
expression reproduces the continuum Euclidean propagator of a free
fermion when $a \to 0$, as it should. However, the same continuum form is
obtained also when at least one of the components of the momentum
$p_{\mu} = \pi/a$. Hence, each corner of the Brillouin zone
contributes equally to the propagator, which means that the
so-called na\"{\i}ve discretization, given by 
eq.~(\ref{eq:latferm}), eq.~(\ref{eq:latfermder}) and 
eq.~(\ref{eq:latfermlag}), 
gives rise to 16 fermion flavors in the continuum limit. This is the
problem of fermion doubling. A well-known no-go theorem due to Nielsen
and Ninomiya~\cite{Nielsen:1980rz, Nielsen:1981xu} implies that 
doubling is an unavoidable consequence if one
requires a discretized fermion theory that preserves chiral symmetry
and is ultralocal (i.e. the action only involves couplings between
fields in a localized region of space). Hence, in order to avoid
doubling, one has to relax the request of chirality or the request of
ultralocality. The Wilson solution to the problem of doubling was to
relax the request of chirality~\cite{Wilson:1975id}. In 
his approach, an irrelevant operator
in the limit $a \to 0$ provides the doublers with an infinite mass in
the continuum limit.\footnote{A variant of Wilson fermions, including a ``chirally twisted'' mass term, was proposed in refs.~\cite{Frezzotti:1999vv, Frezzotti:2000nk, Frezzotti:2001ea}.} A different approach was proposed by Kogut and
Susskind~\cite{Kogut:1974ag, Banks:1975gq, Susskind:1976jm}, who were able to 
reduce the number of doublers from $2^D$ 
(in $D$ spacetime dimensions, assuming $D$ to be even) to
$2^{D/2}$ by spreading the four components of the Dirac spinor on the
corners of the Brillouin zone (hence the name of ``staggered fermions''
for this discretization). More recently, a series of fermion discretizations, 
that avoid the doubling problem, by satisfying a \emph{modified} 
form of chiral symmetry on the lattice~\cite{Ginsparg:1981bj, Luscher:1998pqa} (which 
goes over to the usual continuum chiral symmetry for $a \to 0$), have 
been proposed: these include the domain wall 
formulation~\cite{Kaplan:1992bt, Shamir:1993zy, Shamir:1993bi, Furman:1994ky}, the overlap formulation~\cite{Neuberger:1997fp, Neuberger:1998wv, Narayanan:1992wx, Narayanan:1993sk, Narayanan:1993ss, Narayanan:1994gw}, and the
fixed-point formulation~\cite{Bietenholz:1995cy, Bietenholz:1995nk, Hasenfratz:1998ri}.\footnote{A ``chirally improved'' variant of Wilson fermions was proposed in ref.~\cite{Gattringer:2000js}.} For a general discussion 
about lattice fermions and chiral symmetry,
see refs.~\cite{Bietenholz:1998ut, Luscher:2000hn, Jansen:2008vs}.
We stress that the physics in the continuum limit is independent of the
lattice discretization used. However, a particular formulation could
be more suited for a particular problem. In lattice calculations of
$\SU(N)$ gauge theories with fermions, mostly the Wilson formulation
has been used. The main motivations for this choice are the following:
\begin{itemize}
\item simulations using Wilson fermions are much faster than
  simulations using non-ultralocal fermions;
\item unlike in the staggered fermion case, a generic number of flavors can
  be simulated in the Wilson approach;
\item chiral symmetry can be recovered by tuning the bare quark mass
  to a critical value.\footnote{Although this value has to be found as
    a part of the simulation, this does not create particular problems
    in practical applications.}
\end{itemize}
The Wilson discretization starts from a modification of the Dirac operator
with an additive term that goes like the Laplacian of $\hat{\psi}$ with strength
controlled by a parameter $r$, conventionally taken to be equal to
1. Explicitly, this term reads 
\beq
\Delta S_f = - \frac{r}{2} \sum_i \hat{\overline{\psi}}(i) 
\sum_{\mu} \left( \hat{\psi}(i+\hat{\mu}) + \hat{\psi}(i-\hat{\mu})
  - 2 \hat{\psi}(i) \right) ,
\eeq
This yields the following Wilson Dirac operator
\beq
  D^W_{\alpha \beta}(ij) = 
  - \frac{1}{2} \sum_{\mu} \left[\left( r - \gamma_{\mu}^E \right)_{\alpha
      \beta} \delta_{i+\hat{\mu},j} + \left(r +
      \gamma_{\mu}^E\right)_{\alpha \beta} \delta_{i-\hat{\mu},j}
  \right] + \left(\hat{m}+ 4r \right) \delta_{ij}
  \delta_{\alpha \beta} .
  \eeq
The resulting two-point function is
\beq
\langle \hat{\psi}_{\alpha(i)} \hat{\overline{\psi}}_{\beta} (j)
  \rangle = 
\int_{- \pi}^{\phantom{-}\pi} \frac{d^4  \hat{p}}{\left( 2 \pi \right)^4} 
\frac{- i \sum_{\mu} \gamma_{\mu}^E \sin \left( \hat{p}_{\mu} \right) + \hat{m}(p) }{
  \sum_{\mu} \sin^2\left( \hat{p_{\mu}}\right) +
  \left(\hat{m}(p)^2 \right)} e^ {i \hat{p}\cdot(i-j)} ,
\eeq
which is similar in form to the na\"{\i}ve propagator, the difference being
that now the mass depends on the momentum:
\beq
m(p) = m + \frac{2r}{a} \sum_{\mu} \sin^2 \left(
  \frac{a p_{\mu}}{2} 
\right) 
\eeq
(note that we have reinstated dimensionful units in this expression). 
When $a p_{\mu} \to 0$ for all $\mu$, $m(p) \simeq m$, and the
expected continuum result is recovered. Conversely, at the other
corners of the Brillouin zone, $m(p)$ diverges. Hence, in the
continuum limit the unphysical doublers get an infinite mass, 
decoupling from the action.
\subsubsection{Gauge theories with fermionic matter}
\label{sect:lattice:discrete:full}
We conclude this brief introduction to lattice gauge theories with the
case of fermionic matter coupled to gauge fields. For simplicity, we
consider the case in which there are $n_f$ fermion flavors having the
same mass $\hat{m}$. In the presence of gauge interactions, for Wilson
fermions the Dirac operator is given by 
\beq
  D^W_{\alpha \beta}(ij) = 
  - \frac{1}{2} \sum_{\mu} \left[\left( r - \gamma_{\mu}^E \right)_{\alpha
      \beta} U_{\mu}(i) \delta_{i+\hat{\mu},j} + \left(r +
      \gamma_{\mu}^E\right)_{\alpha \beta} U_{\mu}^{\dag}(j) \delta_{i-\hat{\mu},j} 
  \right] + \left(\hat{m}+ 4r \right) \delta_{ij}
  \delta_{\alpha \beta} .
  \eeq
For the sake of clarity, wherever the context is unambiguous, from now
on we suppress the Dirac indices and the spacetime coordinates in
$D^W$ and in the fermion fields. In the most straightforward
formulation, the path integral of the theory is given by
\beq
\label{eq:pifull}
Z = \int \left( \prod_{j,\mu} \dd U_{\mu}(j) \right)  \left(
  \prod_{j,\alpha, l} \dd \hat{\psi}_{\alpha}^l (j) \right)
  \left( \prod_{j,\alpha, l} \dd \hat{\overline{\psi}^l}_{\alpha} (j)
\right)  e^{- S - \hat{\overline{\psi}^l} D_W \hat{\psi}^l} .
\eeq
where $l$ is the flavor index, running from 1 to $n_f$, and $S$ is the
gauge action~(\ref{eq:wilsonact_ym}). Performing explicitly the
integration over the fermionic variables gives
\beq
\label{eq:pifulldet}
Z = \int \left( \prod_{j,\mu} \dd U_{\mu}(j) \right) 
\left( \rm{det} \ D_W \right)^{n_f} e^{- S}  ,
\eeq
in which the determinant of $D_W$ to the power of $n_f$ appears.

Among fermionic observables, we shall consider only zero-momentum
correlation functions of isovector meson operators, which take the form
\beq
\label{eq:meson_corr_def}
{\mathcal C}_{\Gamma \Gamma^{\prime}}(t) = \sum_{\vec{x}} \left\langle \left( \hat{\overline{\psi}^l}
    (t, \vec{x}) \Gamma \hat{\psi}^k(t, \vec{x})
  \right)^{\dag}  \left( \hat{\overline{\psi}^l}
    (0, \vec{0}) \Gamma^{\prime}\hat{\psi}^k(0,\vec{0})
  \right) \right\rangle ,
\eeq
where $\Gamma$ and $\Gamma^{\prime}$ are two Euclidean Dirac matrices,
$(0,\vec{0})$ is the conventional origin of the
lattice and $l \ne k$. Integrating the previous expression over the fermion fields
yields
\beq
{\mathcal C}_{\Gamma \Gamma^{\prime}}(t)=  - \sum_{\vec{x}} 
 \left\langle \mathrm{tr} \left( \Gamma^{\dag} (D^W)^{-1}(x,0)
     \Gamma^{\prime} (D^W)^{-1}(0,x) \right) \right\rangle ,
\eeq
with $0 \equiv (0,\vec{0})$ and $x \equiv (t,\vec{x})$ and $\mathrm{tr}$
being the trace over Dirac indices. More explicitly, in the path
integral formulation this expression reads
\beq
\label{eq:picorr}
{\mathcal C}_{\Gamma \Gamma^{\prime}}(t) = - \frac{1}{Z} \int \left( \prod_{j,\mu} \dd U_{\mu}(j) \right) 
\left( \rm{det} \ D_W \right)^{n_f} \mathrm{tr} \left( \Gamma^{\dag} (D^W)^{-1}(x,0)
     \Gamma^{\prime} (D^W)^{-1}(0,x) \right)  e^{- S}  .
\eeq
\subsubsection{Quenched approximation}
\label{sect:lattice:fermions:quenched}
Fermionic observables like ${\mathcal C}_{\Gamma \Gamma^{\prime}}$ require the
evaluation of the fermionic determinant. This determinant can be
expanded in fermionic loops. As an expansion parameter, one can use
for instance the fermion mass. This gives rise to the so-called
hopping parameter expansion. The leading order in this expansion
simply consists in neglecting all fermionic loops, which means setting
$\mathrm{det} \ D^W = 1$. This infinite fermion mass limit defines the
{\em quenched approximation}. In practical terms, working in the
quenched approximation means neglecting the back-reaction
of fermions on the gauge fields. Note in addition that the quenched
theory is non-unitary. 

In lattice QCD, in order to obtain an accurate numerical result it is
crucial to evaluate the full fermionic determinant. However, in
large-$N$ QCD, at fixed quark mass fermion loops become less and less 
relevant. In fact, the quenched large-$N$ limit coincides with the
large-$N$ limit in the theory with dynamical fermions. Hence, if we
are interested only in the large-$N$ result, the calculation can be
performed in the quenched theory (note that, obviously, this does not
hold for the evaluation of finite-$N$ corrections). 

A phenomenon that has been observed in lattice QCD is the delayed
onset of unquenching effects: the quenched calculation proves to work
even in a regime in which one would expect significant contributions
from fermion loops~\cite{Aoki:1999yr}. This feature can be seen as another indication
that the physical strong interaction is close to its large-$N$ limit.
\subsection{Monte Carlo calculations}
\label{sect:lattice:mc}
In a $\SU(N)$ gauge theory, the evaluation of the path integral on a
lattice of size $\mathcal{S}$ (in lattice units) involves the evaluation of $d =
4(N^2 - 1)\mathcal{S}$ integrals. Such a large number of integrals 
are impractical to be 
performed using grid methods. A stochastic approach, based on the
observation that the path integral measure is reminiscent of the
Boltzmann measure in statistical mechanics---and, hence, at fixed bare
parameters only a subset of possible values of the variables will
dominate the integral---is preferable. Moreover, grid methods are affected by a
systematic error that is $O(1/\mathcal{S}^{s/d})$, where $s$ is a number
that depends on the approximation used by the given grid method (for
instance, $s = 4$ for the popular Simpson method). Hence, when 
$d$ is large, the error becomes unavoidably of order one.  Monte Carlo methods
provide the wanted stochastic approach: in a Monte Carlo calculation,
field configurations are generated according to the path integral
measure, which means that configurations recur according to the weight
of their contribution to the path integral. Thanks to this property, the vacuum
expectation value of an observable can be computed as a simple average
over the configurations generated during the Monte Carlo simulation. Moreover, 
the error (which in this case is statistical and not systematic) can be kept
under control, since it scales as $1/\sqrt{N_K}$, where $N_K$ is the
number of generated configurations.

The theory behind Monte Carlo calculations is based on Markovian
processes. Consider a system that evolves through a sequence of
states over discrete time. We indicate a generic state as $C_m$ and
the ensemble of all states (or configurations) as $\{ C_m \}$. The
evolution is determined by a probability $P_{n m}$ to transition from
$C_n$ to $C_m$ at any given time. The dynamics is said to be Markovian
if the configuration realized at time $t$ only depends on the 
configuration realized at time $t - 1$, and not on the configurations
realized at previous times. Under some technical assumptions that we do
not specify here, one can prove that an asymptotic probability
distribution characterizes a Markovian dynamics. This is called the
equilibrium distribution. Monte Carlo
algorithms are recipes to construct Markovian dynamics that have the
path integral measure as the equilibrium distribution. In general,
several different Markovian dynamics can generate 
the same equilibrium distribution. For 
some Markovian dynamics that satisfies the detailed balance relation
\beq
P_{n m}  e^{- S(C_n)} = P_{m n} e^{- S(C_m)} ,
\eeq
where $S(C)$ is the action evaluated on the configuration $C$, the
equilibrium distribution $\rho_m$ is given by
\beq
\rho_m = \frac{e^{- S(C_m)}}{\sum_n e^{- S(C_n)}} ,
\eeq
in which it is easy to recognize the Boltzmann equilibrium 
distribution. 

The problem of generating an ensemble of configurations dominating the
path integral and approximating the observables with controlled
precision becomes then the problem of defining an 
appropriate Markovian dynamics for
our theory. Once we have done this, we can start from any arbitrary
state and let the system evolve. After discarding a sufficient number
of configurations at the beginning of the chain, the remaining ones
would be distributed with the right statistics.\footnote{A correct
  identification of the transient requires an analysis of the
  numerical stability of the observables against the length of this
  cut.} 

An algorithm can be characterized by the efficiency with which it
explores the configuration space. A good measure of the efficiency is
the correlation time. In general, for a given observable ${\mathcal O}$, 
\beq
\langle {\mathcal O}(t) {\mathcal O}(t  + \tau) \rangle \propto e^{- t /\tau_{\mathcal O}},
\eeq
where the correlation time $\tau_{\mathcal O}$ depends on both ${\mathcal O}$ and the chosen
Markovian dynamics.  At fixed observable, $\tau_{\mathcal O}$ provides a measure
of the efficiency of the algorithm: the smaller $\tau_{\mathcal O}$, the faster
the configuration space is explored. Topological observables are
notoriously more difficult to decorrelate than local observables. An
algorithm that decorrelates fast topological observables can be
considered efficient in general terms. 

For $\SU(2)$ pure gauge theory, an
efficient algorithm is the one proposed by Creutz~\cite{Creutz:1980zw} 
and later refined by Kennedy and 
Pendleton~\cite{Kennedy:1985nu}. This algorithm belongs to the wider
class of the heath-bath algorithms, i.e. of those algorithms in which
detailed balance is obtained by generating the new configuration
according to its Boltzmann weight. To further decrease the correlation
time and increase ergodicity, it is possible to supplement the heath-bath with an
overrelaxation step, in which the link variables are changed in such a
way that the action remains constant~\cite{Adler:1981sn, Brown:1987rra}.

For $\SU(N)$, a Kennedy-Pendleton or overrelaxation update can be
performed by a sequence of updates on different $\SU(2)$ 
subgroups. The idea of updating the 
$\SU(2)$ subgroups was proposed by Cabibbo and
Marinari~\cite{Cabibbo:1982zn}. Although the Kennedy-Pendleton 
algorithm with a Cabibbo-Marinari cycle works well for $\SU(3)$, at 
very large $N$ cycling over the $\SU(2)$
subgroups may become inefficient. To overcome this 
potential issue, it has been proposed to
supplement the Cabibbo-Marinari update with an overrelaxation step
over the whole $\SU(N)$ group~\cite{Kiskis:2003rd, deForcrand:2005xr}.

After a sequence of thermalized updates, the expectation value of an
observable can be computed as the average of the values of the
observable evaluated over the various configurations of the Markov
chain. For instance, if ${\mathcal O}(i)$ is the value of the observable ${\mathcal O}$ at
the $i$-th configuration in a Markov chain of length $N_K$, we define
\beq
\overline{{\mathcal O}}_{N_K} = \frac{1}{N_K} \sum_{i=1}^{N_K} {\mathcal O}(i) .
\eeq
$\langle {\mathcal O} \rangle$, the vacuum expectation value of ${\mathcal O}$, is found as
\beq
\langle {\mathcal O} \rangle = \lim_{N_K \to \infty} \overline{{\mathcal O}}_{N_K}  .
\eeq
$\overline{{\mathcal O}}_{N_K}$ is a controlled estimate for $\langle {\mathcal O} \rangle$,
in the sense that
\beq
\left( \frac{ \Delta \langle {\mathcal O} \rangle} {\langle {\mathcal O} \rangle}\right)^2 = \frac{ \langle {\mathcal O}
  \rangle^2  -\overline{{\mathcal O}}_{N_K}^2} {\langle {\mathcal O}
  \rangle ^2} \propto N_K^{-1} . 
\eeq
Therefore, in a numerical simulation, $\langle {\mathcal O} \rangle$ can in
principle be obtained with the desired accuracy, by tuning the length of
the Markov chain. The statistical uncertainty $\Delta {\mathcal O}$ can also be
quantified. However, due to the correlation of the configurations,
simple Gaussian statistics is not applicable in a straightforward
way. Nevertheless, Gaussian statistics can be applied after the values
of the observable have been averaged over bins of size equal to or larger than 
$\tau_{\mathcal O}$. Practically, this means that the number of independent
estimates of ${\mathcal O}$ is not $N_K$, as one would na\"{\i}vely think, but
rather $N_K/\tau_{\mathcal O}$. 

Another issue that needs to be taken into account in order to provide
a reliable estimator for $\langle {\mathcal O} \rangle$ is the bias. Technically,
one says that an estimate is biased if it does not agree with the
analytical value when the latter is computable. While any bias will
disappear for $N_K \to \infty$, a bias could appear for na\"{\i}ve
estimates at finite $N_K$. There are standard methods to remove biases,
the most popular ones being jack-knife and bootstrap. Although the
details of those methods will not be discussed any further (we refer to ref.~\cite{jun1995jackknife} for a pedagogical introduction), it is
important to be aware that extracting a numerical value and the
corresponding error for an observable measured in numerical
simulations using Monte Carlo methods requires a careful analysis, the
reason being that the data are correlated. This also applies to fits
of Monte Carlo data~\cite{Michael:1993yj}. Modern lattice
calculations, including those discussed in this review, use robust
procedures to estimate errors on observables and fitting
parameters. The particular procedure used in each calculation is
generally discussed in the corresponding original publication, to
which we refer for details.  

\subsection{Lattice simulations of $\SU(N)$ gauge theories}
\label{sec:lattice:recovery}
Monte Carlo calculations in lattice gauge theory aim at computing 
numerically values of physical observables. Typical quantities that
are computed are connected correlation functions of two operators at
zero momentum:
\beq
{\mathcal C}(\tau) =\frac{1}{N_t} \sum_{t=0}^{N_t - 1} \left( \langle {\mathcal O}_1^\dag(t) {\mathcal
  O}_2(t + \tau)  \rangle -  \langle {\mathcal O}_1^{\dag} (t) \rangle \langle
{\mathcal O}_2(t + \tau) \rangle \right) ,
\eeq
where $N_t$ is the number of sites in the Euclidean-time direction of the lattice, and
the operator is averaged over all the other three spatial coordinates,
to project onto zero-momentum states. We have assumed 
that the Euclidean-time direction is compact, and that 
periodic (antiperiodic) boundary conditions along it are imposed for 
bosonic (respectively: fermionic) fields. 
For asymptotically large $N_t$,
\beq
{\mathcal C}(\tau) \propto e^{- m \tau} ,
\eeq
where $m$ is the mass of the lowest-lying state that connects
${\mathcal O}_1|0\rangle$ to ${\mathcal O}_2|0 \rangle$. If
${\mathcal O}_i$ are traces of closed
loop operators carrying well-defined $J^{PC}$ quantum numbers
(note that in this context $J$ refers to the dihedral group, to which the group of
spatial rotations is reduced by the lattice structure), ${\mathcal C}(\tau)$ will identify
glueball states. If the ${\mathcal O}_i$ operator is a fermion 
bilinear with well-defined
$J^{PC}$, then ${\mathcal C}$ will be saturated by mesonic states. Finally, 
if the ${\mathcal O}_i$ are Polyakov loops wrapping around 
a compact spatial direction 
(with periodic boundary conditions), then the propagating states will
be torelons, which will have a mass that is proportional to the string
tension. Note that, in practice, extracting a signal
with sufficient accuracy for a meaningful determination of the
mass can be challenging. This problem can be overcome,
 by building various operators with
the same quantum numbers that can be used in a variational approach
(see e.g. ref.~\cite{Lucini:2004eq}). The variational calculation
also allows one to extract masses of excitations in the given channels.

Another important remark is that the
quality of the numerical data does not decrease with $N$: simple
large-$N$ counting arguments show that the ratio noise over signal is
constant in $N$ for pure gluonic correlators, while improves as
$1/\sqrt{N}$ for fermionic correlators.

Thermodynamic properties can be studied in a finite temperature setup,
in which the lattice has $N_s^3 \times N_t$ sites ($N_s$ is the number
of sites in each spatial direction, while $N_t$ is the number of sites
in the temporal direction), with $N_s \gg N_t$,
and, again, periodic (antiperiodic) boundary conditions for bosonic
(fermionic) fields along the compact 
direction, whose size is related to the temperature $T$ by
$T = 1/(a N_t)$. The deconfinement temperature is identified by
looking at the peak of the susceptibility of the order parameter, which
for a deconfining transition in the Yang-Mills theory is the Polyakov
loop.\footnote{Strictly speaking, the Polyakov loop is a well-defined
order parameter only on an infinite lattice, because there can not be
any phase transition in a system with finitely many degrees of
freedom. For this reason, one normally considers the \emph{modulus} of the
average Polyakov loop in a configuration as a (pseudo-)order parameter, and
identifies the location of the critical point by monitoring its
susceptibility, when the system parameters are varied.} 
The scaling of the position (in $\beta$) of this susceptibility
as $N_s \to \infty$ allows to identify the critical value of $\beta$
(which is a function of $N_t$ and of the order of the phase
transition). The framework that enables us to perform those studies is
the theory of finite size scaling.

Numerical studies of a lattice gauge theory with Monte Carlo techniques
involve the following steps:
\begin{enumerate}
\item on a lattice of fixed size $\mathcal{S}$ and at a 
  fixed value of the lattice
  coupling $\beta$ (and of the hopping parameter $\kappa$, in 
  the presence of dynamical fermions), 
  evaluate numerically the vacuum expectation values of
  operators corresponding to physical observables;
\item at the same lattice couplings, perform numerical simulations on
  lattices of larger volume, in order to extrapolate to the
  thermodynamic limit;
\item repeat these two steps at various couplings in order to
  determine the value of the observables in the continuum limit.
\end{enumerate}
In the programme, there are two extrapolation processes. Both of them
involve fitting the Monte Carlo data to analytical behaviors whose
leading order in the correcting parameters is known. For instance,
on a lattice of linear size $L$, in the chiral limit the mass of the
pseudoscalar meson $m_{\pi}$ receives finite size corrections that, at the leading
order, are proportional to $e^{ - m_{\pi} L }$~\cite{Colangelo:2003hf}. Fitting the lattice data
on various sizes with this \emph{Ansatz} in the region in which the
asymptotic behavior is reached (which has to be determined as a part
of the simulation) allows one to extract $m_{\pi}$ in the infinite
volume limit.

Asymptotic freedom dictates that the continuum limit of the
theory is reached for 
$\beta \to \infty$. In fact, at the lowest order, perturbation
theory\footnote{For $\SU(N)$ gauge theories, in the lattice scheme the
  Symanzik $\beta$-function, which determines the variation of $a$ as a 
  function of the coupling $g_0$, is known to three
  loops~\cite{Alles:1996cy, Alles:1998is}.} predicts that
\beq
a = \frac{1}{\LambdaLat} e^{ - \frac{12 \pi^2} {11 N^2} \beta} 
\eeq
(up to subleading corrections), where $\LambdaLat$ denotes the 
dynamically generated mass scale in the
lattice regularization scheme. This formula implies that the lattice
spacing goes (exponentially) to 
zero when $\beta \to \infty$. The variation of an
observable with $a$ is predicted by the Callan-Symanzik equation.  
The existence of a well-defined continuum limit implies that for two
observables of the same mass dimension ${\mathcal O}_1$ and ${\mathcal
  O}_2$ 
\beq
\lim_{\beta \to \infty} \frac{\hat{\mathcal O}_1}{\hat{\mathcal O}_2} =
\frac{{\mathcal O}_1}{{\mathcal O}_2},
\eeq
where $\hat{\mathcal O}_i = a^{- d_i} {\mathcal O}_i$ and $d_i$ is the mass
dimension of ${\mathcal O}_i$. At the leading non-trivial order in $a$, near the continuum
limit\footnote{We are assuming that boundary conditions do not
  introduce corrections that are proportional to $a$, which is not
  always the case, but it is true in most simulations---including, in 
particular, those that use (anti-)periodic boundary conditions.}
\beq
\label{eq:scaling}
\frac{\hat{\mathcal O}_1}{\hat{\mathcal O}_2} = 
\frac{{\mathcal O}_1}{{\mathcal O}_2}  + O( a^2
M^2) ,
\eeq
where $M^2$ denotes an observable with mass dimension
$2$~\cite{Symanzik:1983dc, Symanzik:1983gh}.
This expression implies that, asymptotically, lattice corrections are
quadratic in the lattice spacing. Values of $\beta$ for which
observables fulfill eq.~(\ref{eq:scaling}) are said to be in the
scaling region. 

We stress again that, in order for the extrapolations described above
to be meaningful, the system must be in the correct regime. For
instance, in a small volume, deconfinement might arise and as a result
the numerical data extracted in this phase are not simply related with
their infinite volume limit. Similarly, the limit corresponding to strong (bare)
lattice gauge coupling ($\beta \to 0$) is heavily dominated by discretization
artifacts, whose properties (like, e.g., the existence of
confinement~\cite{Wilson:1974sk} and the finiteness of the mass
gap~\cite{Munster:1981es}, which were proven analytically in the early
days of lattice QCD) are not necessarily relevant for the continuum
theory.\footnote{Indeed, in the limit of strong bare lattice gauge
  coupling, even compact $\U(1)$ lattice gauge theory is
  confining~\cite{Banks:1977cc}---while it is in a Coulomb phase at
  weak 
  coupling~\cite{Guth:1979gz, Frohlich:1982gf, Seiler:1982pw, Bellissard:1979ms}.} Nevertheless, strong coupling expansions are
a useful theoretical tool in lattice gauge theory, and are
characterized by a finite convergence 
radius~\cite{Osterwalder:1977pc, Seiler:1982pw}. For 
the Wilson discretization of $\SU(N)$ Yang-Mills theories in four spacetime
dimensions, it is known that, for $N \ge 5$, the range of (weak)
couplings, which is analytically connected to the continuum limit, is
separated from the strong-coupling regime by a strong, first-order
bulk transition---which is signalled by a discontinuity in the
expectation value of the plaquette, and which is \emph{not} related to
any symmetry breaking pattern---, while for $N<5$ a crossover connects
the strong- and weak-coupling regimes~\cite{Lucini:2001ej}. In order
to probe the region of couplings analytically connected to the
continuum limit, the simulations have to be performed at values of
bare lattice 't~Hooft coupling $\lambda_0$ smaller than the critical
value corresponding to the bulk transition leading to the
strong-coupling regime---which, for $N\to \infty$, has been 
numerically estimated 
to be at $0.3596(2)$~\cite{Campostrini:1998zd}.\footnote{It is
interesting to note that this value is close to some 
estimates, worked out with (truncated) analytical expansions
based on the large-$N$ limit, which were already obtained 
in the early 1980's~\cite{Green:1980bs, Green:1981mx}.}

$\SU(N)$ Yang-Mills theory dynamically generates a scale. The whole
physical spectrum can be expressed in terms of this scale. Hence, in
order to meaningfully compare theories at different $N$, a scale needs
to be fixed. This could be for instance the string tension $\sigma$ 
(i.e. the asymptotic slope of the confining potential at large distances) 
or the deconfinement temperature $T_c$. Once the choice has been made,
large-$N$ arguments predict the scaling with $N$ of all other 
quantities relevant for the infrared dynamics. In particular, 
if a well-defined large-$N$ theory exists,
all spectral quantities should have a finite large-$N$ limit. From
perturbation theory, we expect that the leading finite-$N$ corrections
are of order $1/N^2$ for the gauge theory, and of order $1/N$ for the
theory with dynamical fermions. Taking further the perturbative
argument, one would expect that pure gauge observables can be
expressed in a power series in $1/N^2$, while in the presence of
dynamical fermions the power series is in $1/N$. In this language, the
proximity of $\SU(3)$ to $\SU(\infty)$ means that the series converges for
$N = 3$. Moreover, the large-$N$ approach is useful to describe QCD if
for a comprehensive set of observables a reasonable approximation (to
the order of few percents) can be obtained by retaining only few
leading corrections, with the quality of the approximation
systematically improving when higher-order corrections are added. Note 
that, in the large-$N$ limit, the quantity of reference, 
that one uses for comparing the results in theories with a
different number of colors $N$, does not play
any r\^ole: different quantities may be affected by finite-$N$ corrections
with different coefficients, but each of them tends to a 
well-defined value in the $N \to \infty$ limit. 

We conclude this section with a brief discussion about the question,
whether the continuum and the large-$N$ limits commute. As 
pointed out in ref.~\cite{Neuberger:2002bk}, in 
general the interchange of these two limits may be  
non-trivial~\cite{Kessler:1985ge, Neuberger:1989kd}---especially 
if there exists a set of degrees of 
freedom, whose number does not grow with $N$, but which 
nevertheless have a strong effect on the dynamics at 
the cut-off scale at any finite $N$. However, as discussed in 
refs.~\cite{Neuberger:2002bk, 'tHooft:2002yn}, for the gauge
theories that we are presently interested in, 
one can safely assume that the 
continuum and the large-$N$ limits commute. In other 
words, if we want to study
the theory at infinite $N$, we can either take first the continuum
limit at fixed $N$ and then the large-$N$ limit, or take first the
large-$N$ limit at fixed cut-off $a$ and then the continuum limit. In
the latter approach, the lattice spacing is kept fixed across the
various $N$, by simulating the various theories for $\beta$ such that
the value of a physical quantity (e.g. the string tension or the
critical temperature) has a predefined value in units of 
the lattice spacing $a$. While
performing the large-$N$ limit at fixed lattice spacing should be seen
as an intermediate step towards getting continuum large-$N$ physics,
this approach can prove convenient in calculations that are
particularly demanding from the computational point of
view. In addition to various examples in the continuum limit, in
sec.~\ref{sec:results}, we will discuss some results for which the
large-$N$ limit has been taken at a fixed lattice spacing in the
scaling region.

\section{Factorization, loop equations, and large-$N$ equivalences}
\label{sec:from_factorization_to_orbifold}

Besides the phenomenological implications and the connections with string theory discussed in sec.~\ref{sec:large_N_limit}, large-$N$ field theories and statistical models exhibit many further interesting mathematical properties: the expectation values of products of physical operators factorize, up to $ O(1/N) $ corrections (see subsection~\ref{subsec:factorization} below), which suggests an analogy with the classical limit of a quantum theory (subsec.~\ref{subsec:classical_limit}), and indicates the suppression of fluctuations for $N \to \infty$. This led to conjecture that the large-$N$ dynamics may be determined by a \emph{master field} (subsec.~\ref{subsec:master_field}). The factorization properties also imply that one can formulate a closed set of equations for the expectation values of gauge invariant operators, which are presented in subsec.~\ref{subsec:loop_equations}. For the lattice formulation of the Yang-Mills theory, Eguchi and Kawai discovered that these equations reveal the volume independence of the theory in the large-$N$ limit, so that, in four spacetime dimensions, the theory can be reduced to a single-site model of only four matrices (subsec.~\ref{subsec:EK_model}), provided that center symmetry remains unbroken. Since the latter condition is not satisfied at weak couplings (which are relevant to take the continuum limit), different variants of the original model have been proposed in the literature: these include the quenched  (subsec.~\ref{subsec:quenched_EK_model}) and the twisted (subsec.~\ref{subsec:twisted_EK_model}) versions of the Eguchi-Kawai model, its generalization with dynamical adjoint fermions (subsec.~\ref{subsec:adjoint_EK_model}) or with double-trace deformations of the Yang-Mills action (subsec.~\ref{subsec:deformed_EK_model}). In parallel to these theoretical developments (and to the related numerical studies), a complementary approach has been pursued in a series of works (discussed in subsec.~\ref{subsec:partial_EK_reduction}) exploiting the \emph{partial} volume reduction of the original Eguchi-Kawai model, down to the minimal lattice volume, in which center symmetry is unbroken. Finally, in subsec.~\ref{subsec:orbifold_projections} we discuss how large-$N$ volume independence and certain correspondences between different large-$N$ theories can be interpreted in terms of \emph{orbifold equivalences}.

\subsection{Factorization}
\label{subsec:factorization}

Many of the mathematical simplifications characteristic of large-$N$ field theories and statistical models are immediately made manifest by the combinatorics of large-$N$ counting rules. In particular, in the computation of expectation values of products of (appropriate) gauge-invariant operators\footnote{Such operators include, in particular, local gauge-invariant purely gluonic operators, fermionic bilinear operators, and Wilson loop operators. Examples of operators which, on the contrary, do not satisfy eq.~(\ref{factorization}) have also been pointed out in the literature~\cite{Haan:1981ks, Green:1980bg}.} $ \mathcal{O}_i $ in large-$N$ gauge theories, the latter imply that the leading (in $N$) contributions come from disconnected terms, because they are associated with the maximum number of color traces, and, hence, with the largest number of independent color indices to be summed over:
\begin{equation}
\label{factorization}
\left\langle \mathcal{O}_1 \mathcal{O}_2 \right\rangle = \left\langle \mathcal{O}_1 \right\rangle \left\langle \mathcal{O}_2 \right\rangle + O(1/N).
\end{equation}
Eq.~(\ref{factorization}) shows that the large-$N$ limit bears analogies with two different limits in quantum field theory. On the one hand, it can be interpreted as a thermodynamic limit~\cite{Haan:1981ks}: for a system characterized by a finite correlation length, eq.~(\ref{factorization}) is analogous to the cluster decomposition property of statistical, volume averages in a large volume $V$, which holds up to $ O(1/V) $ corrections. An interesting facet of this analogy is that both $ N $ and $ V $ are related to the number of degrees of freedom of the system (so that the integration measure for the generating functional, or for the partition function of the system, depends on $N$, or on $V$).

\subsection{A classical-mechanics description for the large-$N$ limit}
\label{subsec:classical_limit}

A different interpretation of eq.~(\ref{factorization}) is motivated by the  analogy with the classical limit of a quantum field theory~\cite{Yaffe:1981vf}. As it is well-known, a classical system arises explicitly in the $ \hbar \to 0 $ limit of a quantum theory, when one studies the behavior of a basis of coherent states. The latter form an overcomplete basis, which allows one to write all operators solely in terms of their \emph{diagonal elements}, and have vanishing overlaps in the $ \hbar \to 0 $ limit. Together, these two properties lead to the factorization of expectation values of products of operators, and the deterministic nature of the limit becomes clear from the fact that, for $\hbar \to 0$, coherent states are characterized by simultaneously vanishing uncertainties in conjugate variables. In this limit, the classical phase space can be defined as the manifold of the coordinates labeling different coherent states, the quantum Hamiltonian (or, more precisely: the set of its diagonal matrix elements---also called its \emph{symbol}---in the coherent state basis) can be mapped to its classical counterpart (which is minimized by the solutions of the classical equations of motion), and the classical Lagrange function is then obtained by Legendre transform.

This construction can be repeated for the large-$N$ limit (more precisely: for the $ 1/N \to 0 $ limit) of a family of statistical systems or of quantum field theories,\footnote{Following ref.~\cite{Yaffe:1981vf}, we refer to a \emph{family} of statistical systems or of quantum field theories, rather than to just a statistical system or a theory with one parameter, in order to remark that also the very structure of the theory (including, in particular, its number of degrees of freedom) can depend on the parameter.} by generalizing the familiar Heisenberg group to an appropriate coherence group~\cite{Klauder_Skagerstam}, generated by suitable ``coordinates'' and ``momenta''. Choosing a coherence group that, for each value of $N$, acts irreducibly on the Hilbert space $ \mathscr{H}_N $ of the corresponding theory, and a reference state $\left| 0 \right\rangle_N$, one can construct generalized coherent states $ \left| \alpha \right\rangle_N $ by acting with elements of the coherence group on $\left| 0 \right\rangle_N$. The irreducibility condition allows one to express any bounded operator as a linear combination of elements of the coherence group~\cite{Bratteli:1979tw}. For gauge theories, the coherence group is generated by Wilson loops, possibly decorated by a (chromo-)electric field insertion.

Assuming that the correspondence between linear operators in $ \mathscr{H}_N $ and their symbols on coherent states is injective, one can consider those operators $ \mathcal{O}_N $ whose elements in the coherent state basis (appropriately normalized) have a smooth large-$N$ limit:
\begin{equation}
\label{classical_operator_definition}
\exists \lim_{N \to \infty} \frac{ \left\langle \alpha | \mathcal{O} | \beta \right\rangle_N }{\left\langle \alpha | \beta \right\rangle_N}
\end{equation}
and introduce an equivalence relation among coherent states, $\left| \alpha \right\rangle \sim \left| \beta \right\rangle$, defined by the requirement that, for all operators satisfying eq.~(\ref{classical_operator_definition}), one has:
\begin{equation}
\label{classically_equivalent_states}
\lim_{N\to\infty} \left( \left\langle \alpha | \mathcal{O} | \alpha \right\rangle_N - \left\langle \beta | \mathcal{O} | \beta \right\rangle_N \right) = 0.
\end{equation}
If representatives of distinct equivalence classes defined by this relation have exponentially suppressed overlaps for $N \to \infty$, and if the $ \mathcal{H}/N $ operator satisfies eq.~(\ref{classical_operator_definition}), then one can prove that, for $N \to \infty$, the original theory reduces to a classical mechanics theory defined on the coadjoint orbit of the coherence group, and $ \mathcal{H}/N $ tends to the corresponding classical Hamiltonian~\cite{Yaffe:1981vf}.

We conclude this subsection with an important observation: while the discussion presented above shows that it is possible to construct a suitable mapping of a large-$N$ quantum field theory to a classical-mechanics system, it is important to remark that, in general, this does not imply that fundamental fields of the large-$N$ quantum theory are described by the classical equations of motion derived from their Lagrangian density. As discussed, e.g., in ref.~\cite{Makeenko:1999hq}, the reason is that the generating functional of the system features an exponent, that receives $ O(N^2) $ contributions not only from the action, but also from the measure, due to the integration over $ O(N^2) $ gauge field degrees of freedom. For a discussion of somewhat related concepts, see also ref.~\cite{Bochicchio:1998fn}.

\subsection{Spacetime independence and the master field}
\label{subsec:master_field}

The construction outlined above can be carried out to find an explicit large-$N$ solution for certain classes of models (e.g., vector models~\cite{Coleman:1974jh} and single-matrix models~\cite{Brezin:1977sv, Marchesini:1979yq}), but not for the most interesting case of gauge theories\footnote{In this context, the difficulty of gauge theories is related to the \emph{representation} of their fundamental fields---see, e.g., ref.~\cite{McGreevy:2009xe} for a discussion.}---with the exception of one-plaquette lattice models~\cite{Jevicki:1980zq, Wadia:1980cp}. While, at first blush, this latter case may seem completely unphysical, it turns out that the factorization properties expressed by eq.~(\ref{factorization}) also have potential implications for the (lack of) spacetime dependence of the large-$N$ theory. A first, intuitive discussion of this feature was already expounded in ref.~\cite{Witten:1979pi}: for $\mathcal{O}_1=\mathcal{O}_2$, eq.~(\ref{factorization}) shows that quantum fluctuations are suppressed in the large-$N$ limit, hence it is reasonable to expect that, for $N \to \infty$, the path integral is dominated by a unique gauge configuration (or, more precisely, gauge orbit). The latter was interpreted as a classical field by Witten~\cite{Witten:1979pi}, and later dubbed ``master field'' by Coleman~\cite{Coleman:1980nk}. 

Explicit results based on the master field approach were derived for some simple models in refs.~\cite{Levine:1980mr, Halpern:1980ar}, while an algebraic equation for the master field was proposed in ref.~\cite{Greensite:1982mf}, in the form of a quenched Langevin equation. Poincar\'e invariance of vev's then implies that the master field should also be Poincar\'e invariant---possibly up to gauge transformations, meaning that there exists at least a gauge in which it is Poincar\'e invariant. While, strictly speaking, the interpretation of the master field as a classical, $c$-valued field is not correct~\cite{Haan:1981ks}, a more rigorous treatment can be formulated in terms of non-commutative probability theory (see refs.~\cite{Voiculescu:1985, Voiculescu:1991, Voiculescu:1992, Singer:1994zz, Accardi:1994gd, www.uni-math.gwdg.de/mitch/free.pdf} and references therein), as illustrated in refs.~\cite{Douglas:1994kw, Douglas:1994zu, Gopakumar:1994iq}.

\subsection{Loop equations}
\label{subsec:loop_equations}

As discussed above, spacetime independence of the master field is related to the large-$N$ factorization of expectation values of products of gauge-invariant operators, expressed by eq.~(\ref{factorization}). Another important consequence of the same equation is that it implies that, for $N \to \infty$, one can derive a \emph{closed} set of Schwinger-Dyson equations for expectation values of physical operators, such as traces of Wilson loops along a contour $C$ in a gauge theory~\cite{Makeenko:1979pb},
\begin{equation}
\label{Wilson_loop}
\frac{1}{N} \tr W(C)= \frac{1}{N} \tr \left[ \mathcal{P} \exp \left( i g \oint_C A^a_\mu(x) \dd x^\mu \right)\right]
\end{equation}
(where $ \mathcal{P} $ denotes path-ordering), which allow one to reformulate the theory in a gauge-invariant way~\cite{Makeenko:1980vm}. Such geometric reformulation of gauge theories in terms of ``loop calculus''~\cite{Migdal:1984gj} maps the familiar objects appearing in gauge theories, like (non-)Abelian phase factors, covariant derivatives and field strength to loop functionals, path derivatives and area derivative, respectively---see ref.~\cite{Makeenko:1999hq} for a pedagogical introduction to the subject. Loop calculus deals with the Hilbert space of square-integrable functions which describe closed loops $ C $ (up to reparametrizations), and functionals thereof, $\mathcal{F}(C)$. For the latter, the operations of area and path derivative can be defined,\footnote{We restrict our attention to functionals of the Stokes' type, for which the area derivative is well-defined, i.e., independent of the shape of the infinitesimal loop.} by considering the variations obtained by deforming $ C $ through the addition of an infinitesimal loop or, respectively, a backtracking infinitesimal path at one of its points. 

Taking the area derivative (sometimes also called ``keyboard derivative'') of a Wilson loop at a point corresponds to inserting the field strength at that point~\cite{Mandelstam:1968hz}:
\begin{equation}
\label{Mandelstam_formula}
\frac{\delta}{\delta \sigma_{\alpha\beta}(x)} 
\frac{1}{N} \tr W(C)= \frac{i}{N} \tr \left[ F^{\alpha\beta}(x) W(C) \right].
\end{equation}
Acting with the path derivative on a Wilson loop decorated by the insertion of a local operator $ \mathcal{O} $ at a ``marked'' point $x$ has the effect of replacing $\mathcal{O}$ with its covariant derivative (evaluated at $x$):
\begin{equation}
\label{path_derivative}
\frac{\partial}{\partial x^\mu} 
\frac{1}{N} \tr \left[ \mathcal{O}(x) W(C) \right] = 
\frac{1}{N} \tr \left[ \left( D_\mu \mathcal{O}(x) \right) W(C) \right]
\end{equation}
(on the other hand, for functionals of the Stokes' type, the path derivative is identically vanishing at ``regular'', i.e. non-marked, points along a loop).

The loop calculus outlined above can be used to derive the equations describing the  invariance of the following functional integral:
\begin{equation}
\label{variation}
\int \mathcal{D}A \frac{1}{N} \tr \left( T^a W(C) \right) \exp \left( i \int \dd t~\dd^3x~\mathcal{L} \right) 
\end{equation}
(in which the base point of the loop $ C $ is taken to be $ x $, i.e. $ T^a $ is inserted at the point $x$) under (functional) variation of the components of the gauge field, $A^b_\mu(x)$: the ``source'' terms obtained when the variation is applied to $ T^a W(C) $ must cancel against the ``equation-of-motion'' terms obtained applying it to the exponential. Upon contraction of the color indices according to the Fierz identities of the color algebra, this results in the equation~\cite{Makeenko:1979pb}:
\begin{equation}
\label{loop_equation_generic_N}
\partial_\mu \frac{\delta }{\delta \sigma_{\mu\nu}(x)} \left\langle \frac{1}{N} \tr W(C) \right\rangle = \lambda \oint \dd y^\nu \delta^{(D)} (x-y) \left[ \left\langle \frac{1}{N^2} \tr W(C_{x,y}) \tr W(C_{y,x}) \right\rangle - \frac{1}{N^2} \left\langle \frac{1}{N} \tr W(C) \right\rangle \right],
\end{equation}
where $ C_{x,y} $ and $ C_{y,x} $ denote two complementary portions\footnote{Note that, if $ C $ is a self-intersecting loop, the $\delta$-distribution does not necessarily force $ C_{x,y} $ or $ C_{y,x} $ to be trivial.} of the loop $ C $. 

Eq.~(\ref{loop_equation_generic_N}) holds for every value of $N$, however it is not a \emph{closed} equation in loop space, because, generally, the product of traces appearing on the right-hand side is not a linear combination of single-trace loop operators. In the large-$N$ limit, however, factorization allows one to write such term as the product of the expectation values of two single-trace operators (and to discard the last term on the right-hand side of eq.~(\ref{loop_equation_generic_N}) as subleading), leading to~\cite{Makeenko:1979pb}:
\begin{equation}
\label{Migdal_Makeenko_equation}
\partial_\mu \frac{\delta }{\delta \sigma_{\mu\nu}(x)} \left\langle \frac{1}{N} \tr W(C) \right\rangle = \lambda \oint \dd y^\nu \delta^{(D)} (x-y) \left\langle \frac{1}{N} \tr W(C_{x,y})  \right\rangle \left\langle \frac{1}{N}  \tr W(C_{y,x}) \right\rangle + O(1/N^2),
\end{equation}
which is a closed equation in loop space.\footnote{Strictly speaking, eq.~(\ref{Migdal_Makeenko_equation}) should be formulated taking an appropriate, gauge-invariant loop renormalization procedure into account; this does not pose any particular problem, as it is possible to prove that the loop renormalization can be expressed through a purely multiplicative factor: this holds both for smooth~\cite{Dotsenko:1979wb} and for self-intersecting loops~\cite{Brandt:1981kf}.} 

Eq.~(\ref{Migdal_Makeenko_equation}) is a geometric equation\footnote{As discussed in ref.~\cite{Makeenko:1999hq}, the geometric aspects of eq.~(\ref{Migdal_Makeenko_equation}) can be exhibited by expressing it in terms of the L\'evy operator~\cite{Feller:2005zz}.} with deep implications.\footnote{Note, however, that solving eq.~(\ref{Migdal_Makeenko_equation}) for all loops would not provide the \emph{complete} data about the theory: in particular, physical quantities such as the mass spectrum and scattering amplitudes are encoded in \emph{connected} correlation functions of gauge-invariant operators, which vanish if exact factorization holds, and hence cannot be captured by eq.~(\ref{Migdal_Makeenko_equation}).} In particular, the construction of an iterative solution, starting from the expansion of the path-ordered Wilson loop $ W(C) $ in a series of cyclically ordered Green's functions, reproduces the set of planar diagrams which give the non-vanishing contributions to $W(C)$ in the large-$N$ limit. 

At the non-perturbative level, an explicit solution of eq.~(\ref{Migdal_Makeenko_equation}) is not known; in ref.~\cite{Makeenko:1980wr}, it was shown that a confining area-law \emph{Ansatz} for asymptotically large loops is consistent with eq.~(\ref{Migdal_Makeenko_equation}), but the corresponding string tension could not be computed. Interestingly, the Nambu-Goto bosonic string action~\cite{Nambu:1974zg, Goto:1971ce} (which describes confining Wilson loops in terms of fluctuating surfaces, with an action proportional to their area---see the discussion in subsec.~\ref{subsec:4D_results}), which has been a candidate effective string model for the infrared QCD dynamics since the 1970's~\cite{Nambu:1978bd, Polyakov:1979gp, Gervais:1978mp}, is \emph{not} consistent with eq.~(\ref{Migdal_Makeenko_equation}). A \emph{formal} string solution to eq.~(\ref{Migdal_Makeenko_equation}) was discussed in refs.~\cite{Migdal:1981np,Migdal:1981ht}: it takes the form of a fermionic string model, describing the dynamics of point-like Majorana spinor fields (``elves'') on the surface bounded by the loop; the effective string tension was found to be related to the bare elf mass by a scaling law with critical exponent $12/11$, but a complete solution of the theory was not found.

\subsection{The Eguchi-Kawai model}
\label{subsec:EK_model}

One can also formulate equations, analogous to eq.~(\ref{Migdal_Makeenko_equation}), on the lattice~\cite{Forster:1979wf, Weingarten:1979zj, Eguchi:1979nk}: this led Eguchi and Kawai (EK) to discover the surprising property of volume reduction~\cite{Eguchi:1982nm}, which, \emph{if} the necessary conditions are satisfied, provides a concrete realization of spacetime independence at large $ N $~already alluded to by the idea of a translationally invariant master field. In short, the statement of EK reduction is that, if:
\begin{enumerate}
\item factorization of the vev's of physical operators, eq.~(\ref{factorization}), holds, and
\item the global $ \Z_N $ center symmetry of the system (which, for the Euclidean time direction, has already been discussed in subsection~\ref{subsec:phase_diagram} in the context of finite-temperature gauge theories) along each of the four Euclidean directions is not spontaneously broken,
\end{enumerate} 
then the Schwinger-Dyson equations satisfied by vev's of (topologically trivial)
Wilson loops in the large-$N$~$ \SU(N) $~Yang-Mills theory on the lattice are independent of the physical hypervolume of the system. As a consequence---assuming that these equations have a unique solution---also the physical observables of the theory are independent of the system hypervolume. Note that the second condition mentioned above arises from the fact that, due to gauge invariance, vev's of open Wilson lines of finite length are vanishing in the large-volume theory, while, in the reduced-volume theory, if the linear size of the system equals the line length, they vanish only if the global $ \Z_N^4 $ center symmetry (which tends to $ \U(1)^4 $ in the $ N \to \infty $ limit) is unbroken.

As a consequence, if the conditions above are realized, one could study the large-$N$ lattice theory by reducing it down to a single-site lattice model, i.e. to a model of only four matrices $ U_\mu $ (which describe the parallel transporters along the four Euclidean spacetime directions---see section~\ref{sec:lattice}). The importance of this observation is not purely academic: if the two conditions stated above are satisfied, then volume independence of the large-$N$ theory could open up the possibility of studying its non-perturbative dynamics analytically, by reducing the equations of the quantum field theory in an infinite spacetime to the equation of the theory reduced to a single point, i.e. to the Schr\"odinger equation of ordinary quantum mechanics~\cite{Mithat_talk_at_Lattice_2011}.

\subsection{The quenched momentum prescription and the quenched Eguchi-Kawai model}
\label{subsec:quenched_EK_model}

Even though a theoretically appealing idea, it was almost immediately realized that large-$N$ volume reduction for pure Yang-Mills theory along the lines of the original EK proposal could not work.\footnote{This was already acknowledged in a note added to the original paper~\cite{Eguchi:1982nm}, mentioning the result of the work by Bhanot, Heller and Neuberger~\cite{Bhanot:1982sh}---who found evidence for the spontaneous breakdown of center symmetry in the regime of couplings connected to the continuum limit, i.e. in the perturbative regime---and similar findings by Wilson and by Peskin.} The reason is the failure of the second necessary requirement listed above: by reducing the lattice sizes, center symmetry gets spontaneously broken (at least in the range of couplings relevant for approaching the continuum limit). This can already be seen at the perturbative level: denoting the eigenvalues of the $ U_\mu $ matrices in the EK model as $ \exp(i\theta_\mu^a) $ (with $ 1 \le a \le N$), the one-loop effective potential experienced by the $ \theta_\mu^a $ phases turns out to be attractive for all spacetime dimensions $D>2$, leading to spontaneous breakdown of center symmetry~\cite{Bhanot:1982sh, Kazakov:1982gh}. The Monte~Carlo simulations reported in ref.~\cite{Okawa:1982ic} confirmed this also non-perturbatively. 

A first, possible solution to preserve center symmetry was proposed in ref.~\cite{Bhanot:1982sh}: the quenched EK model.\footnote{Related approaches were also discussed in refs.~\cite{Migdal:1982gi, Chen:1982uz, Das:1982ux}, while a Hamiltonian version of the quenched EK model was discussed in refs.~\cite{Neuberger:1982ne, Levine:1982fa}.} The quenched EK model is based on the idea of computing expectation values in the reduced model at ``frozen'' values of the $\theta_\mu^a$'s, and then performing an average over such values according to a suitable probability distribution for the $\theta_\mu^a$'s, which is chosen in such a way, that center symmetry is explicitly enforced. 

The way volume independence could arise in the EK model, the way it fails due to the spontaneous breakdown of center symmetry, and the way this problem is avoided (at least at the perturbative level) in the quenched EK model can be understood via  a suitable mapping of the degrees of freedom associated with different four-momenta in the original planar theory in a large volume to the matrix entries of the one-site model~\cite{Heller:1982gg, Gross:1982at, Parisi:1982gp, Parisi:1982tj, Parisi:1982nm}. This is based on the following, elementary observation: for a generic theory (either one with global or with gauge invariance under a $ \U(N) $ symmetry group) with fields in a two-index representation of the group, like, e.g., the adjoint representation, the fundamental lines in a generic Feynman diagram expressed in double-line notation are closed and non-intersecting. Hence, momentum conservation at each vertex of the diagram is automatically satisfied, if one associates a generic four-momentum $ p_\mu^a $ to each fundamental line (with index $a$), and takes $ p_\mu^{ab}=p_\mu^a-p_\mu^b $ to be the momentum associated with the propagator obtained from the (oppositely oriented) lines of indices $ a $ and $b$. In view of this observation, one can prove that, by replacing each occurrence of $ i \partial_\mu $ in the original action of the continuum theory with the adjoint action of the matrix:
\begin{equation}
\label{P_mu}
P_\mu = \mbox{diag} \left( p_\mu^1,  p_\mu^2, \dots ,  p_\mu^N \right),
\end{equation}
(or, equivalently, by replacing the finite translation operator $ \exp (a\partial_\mu ) $ with $ \exp (i a [P_\mu, \cdot ] ) $ in the lattice theory), one obtains propagators which are equivalent to those of the original theory~\cite{Gross:1982at, Parisi:1982gp}. This prescription, which goes under the name of ``quenched momentum prescription''~\cite{Gross:1982at}, removes the spacetime dependence of the original matrix field from the theory, turning the action of the original theory into a function of a constant matrix and of the diagonal matrices $P_\mu$, so that, for example, the vev of a generic observable $ \mathcal{O} $ could be obtained by computing, first, its value at fixed momenta $P_\mu$, and then integrating over the distribution of the $P_\mu$'s. The theory obtained from the quenching momentum prescription has the same planar limit as the original one, at least at any finite order in perturbation theory~\cite{Gross:1982at}.

Note that, by identifying each $ i \partial_\mu $ with the corresponding $ P_\mu $ matrix defined in eq.~(\ref{P_mu}), applying the quenched momentum prescription corresponds to representing the group of spacetime translations in terms of the degrees of freedom of the $ \U(1)^N $ diagonal subgroup of the internal $ \U(N) $ symmetry group (this holds for a global as well as for a gauge symmetry).

For gauge theory, restricting our attention to pure Yang-Mills theory,\footnote{Generalizations of the quenched EK model to include quarks (in the Veneziano limit) were discussed in refs.~\cite{Levine:1982uz, Kazakov:1982zr, Klinkhamer:1983na}.} the quenched momentum prescription reads:
\begin{equation}
A_\mu(x)= \exp (i P_\nu x^\nu) A_\mu \exp (-i P_\nu x^\nu), \qquad iD_\mu \to P_\mu + A_\mu;
\end{equation}
when applied to the lattice formulation of the theory, using $D_\mu=\exp(i a P_\mu )$, this yields:
\begin{equation}
\label{S_tilde}
\tilde{S}(U_\mu,P_\mu) = \frac{2N}{\lambda} \sum_{1 \le \alpha < \beta \le 4} \mbox{Re} \tr \left[ U_\alpha D_\alpha U_\beta D_\beta (U_\alpha D_\alpha)^\dagger (U_\beta D_\beta)^\dagger \right],
\end{equation}
which (by a change of variables: $U_\alpha \to U_\alpha D_\alpha^\dagger$) is equivalent to the Eguchi-Kawai model~\cite{Eguchi:1982nm}:
\begin{equation}
\label{EK}
\tilde{S}(U_\mu,P_\mu) = \frac{2N}{\lambda} \sum_{1 \le \alpha < \beta \le 4} \mbox{Re} \tr \left( U_\alpha U_\beta U_\alpha^\dagger U_\beta^\dagger \right).
\end{equation}
Note that the classical solutions minimizing $ \tilde{S} $ are arbitrary diagonal unitary (or special unitary, in the case of $ \SU(N) $ gauge group) matrices.

The partition function and expectation values of the original lattice model are obtained by computing the corresponding quantities in the reduced model, described by eq.~(\ref{EK}), at fixed $P_\mu$'s, and then averaging over the distribution of the $D_\mu$'s. In taking the latter step, it is convenient to express the Haar integration measure over each $ D_\mu $ in terms of the eigenvalues $\exp(i a p_\mu^k)$, which yields a Vandermonde determinant $\Delta(D_\mu)$:
\begin{equation}
\dd D_\mu = \prod_{k=1}^N \frac{\dd p_\mu^k}{2\pi}\prod_{1 \le b < c \le N} \sin^2 \left( \frac{p_\mu^b-p_\mu^c}{2} a \right) = \prod_{k=1}^N \frac{\dd p_\mu^k}{2\pi} \Delta(D_\mu)
\end{equation}
(where, in the case of a special unitary gauge theory, the phases of the eigenvalues of each $ D_\mu $~matrix are constrained to sum up to an integer multiple of $2\pi$). The dynamics of the EK model in the weak-coupling regime can then be investigated by studying the effect of quantum fluctuations around a classical solution in a leading-order perturbative computation. This requires an appropriate gauge-fixing (e.g. to Feynman gauge) and the corresponding Faddeev-Popov determinant, so that the integration measure for the $D_\mu$'s~(in $ D $ spacetime dimensions) turns out to be proportional to~\cite{Bhanot:1982sh, Kazakov:1982gh}:
\begin{equation}
\label{EK_measure}
\prod_{\mu=1}^D \prod_{k=1}^N \frac{\dd p_\mu^k}{2\pi} \exp \left\{ -(D-2) \sum_{1 \le b < c \le N} \ln \left[ \sum_{\nu=1}^D \sin^2 \left( \frac{p_\nu^b-p_\nu^c}{2} a \right)\right] \right\}.
\end{equation}
As anticipated above, eq.~(\ref{EK_measure}) shows that a uniform distribution of the $p_\mu^a$'s is a stationary point of the argument of the exponential, but one which corresponds to a minimum of the statistical weight for $D>2$. As a consequence, quantum fluctuations lead to the collapse of the $p_\mu^a$'s, and, hence, to spontaneous breaking of center symmetry, which, in turn, invalidates the correspondence of the reduced model to the large-volume theory.

A possible way to prevent spontaneous symmetry breaking (at least perturbatively), consists in modifying the integration measure for the $U_\mu$'s,\footnote{An alternative possibility consists in modifying the action~\cite{Chen:1982uz}.} so that the dependence on the $P_\mu$'s cannot be eliminated. In ref.~\cite{Gross:1982at} it was proposed to do this by requiring the eigenvalues of the lattice covariant derivative $ U_\mu D_\mu $ to be equal to those of $D_\mu$, for each $\mu$, i.e. by imposing the constraints:
\begin{equation}
\label{QEK_constraint}
U_\mu D_\mu = V_\mu D_\mu V_\mu^\dagger,
\end{equation}
which are explicitly gauge-invariant.\footnote{Alternative approaches, based on non-gauge-invariant constraints, were discussed in refs.~\cite{Das:1982ux, Parisi:1982tj}.} This leads to the integration measure:
\begin{equation}
\prod_\mu \dd U_\mu \dd V_\mu \Delta(D_\mu) \delta(U_\mu D_\mu - V_\mu D_\mu V_\mu^\dagger),
\end{equation}
which includes integration over the degrees of freedom of the unitary matrix $ V_\mu $ that maps $ D_\mu $ to $ U_\mu D_\mu $ by a similarity transformation. Shifting $U_\mu \to U_\mu D_\mu^\dagger$, the integration over the $U_\mu$'s can be immediately performed, leading to the reduced action:
\begin{equation}
\label{reduced_action_quenched_EK}
\tilde{S} = \frac{2N}{\lambda} \sum_{1 \le \alpha < \beta \le 4} \mbox{Re} \tr \left[ V_\alpha D_\alpha V_\alpha^\dagger V_\beta D_\beta V_\beta^\dagger V_\alpha D_\alpha^\dagger V_\alpha^\dagger V_\beta D_\beta^\dagger V_\beta^\dagger \right],
\end{equation}
and, finally, one is left with an integration over the $p_\mu^a$'s with a uniform distribution, yielding the quenched EK model introduced in ref.~\cite{Bhanot:1982sh}. Thus, at the perturbative level, the effect of the gauge-invariant constraints in eq.~(\ref{QEK_constraint}) is to replace the non-uniform weight for the $p_\mu^a$'s of the original EK model, eq.~(\ref{EK_measure}), which was the cause of spontaneous breaking of the center symmetry, with a uniform distribution for the $p_\mu^a$'s, which explicitly enforces center symmetry.

Although early numerical tests found evidence confirming the validity of volume reduction in the quenched EK model also at the non-perturbative level~\cite{Bhanot:1982cm, Okawa:1982nn, Bhanot:1982bh, Neuberger:1983xc, Carlson:1983kb, Lewis:1984xq}, these conclusions have been recently disproven by the high-precision study reported in refs.~\cite{Bringoltz:2008av, Bringoltz:2008ek}, which also discuss analytical arguments for this failure. The subtle dynamical mechanism responsible for the breakdown of volume reduction in the quenched EK model can be understood by a careful inspection of eq.~(\ref{QEK_constraint}), which plays a pivotal r\^ole in the model, and that, when expressed for the redefined $U_\mu$'s appearing in eq.~(\ref{reduced_action_quenched_EK}), takes the form of a polar decomposition of the four $ U_\mu $ matrices:
\begin{equation}
\label{polar_decomposition}
U_\mu = V_\mu D_\mu V_\mu^\dagger.
\end{equation}
The crucial observation is that eq.~(\ref{polar_decomposition}) fixes the eigenvalues of the $ U_\mu $ matrices \emph{only up to permutations}~\cite{Neuberger:1982ne, Parsons:1983dm}. In general, configurations of the reduced model, in which the eigenvalues of the $U_\mu$'s differ only by permutations, yield different values of the reduced action defined in eq.~(\ref{reduced_action_quenched_EK}). Hence, if dynamical fluctuations of the $V_\mu$'s can produce tunneling events between such configurations, the quenched system will choose the energetically most favored configuration(s), introducing non-trivial correlations between the $ U_\mu $ components along different directions (``momentum locking''), and resulting in non-uniform sampling of the gluon momenta. In fact, such phenomenon does occur: at fixed coupling, there exist configurations mediating eigenvalue permutations, whose energy cost is not enhanced in the $ N \to \infty $ limit. The factorization condition, which is necessary for volume reduction to hold, is generally violated in the presence of correlations between the $U_\mu$'s in different directions. This can be seen by considering, for example, Wilson lines of the form: $M_{\alpha\beta}=(1/N)\Tr (U_\alpha U_\beta)$, with $ \alpha \neq \beta$, whose value on a typical momentum-locked configuration is a complex number with modulus $ O(1) $ and with a generic, momentum-dependent phase. Averaging over the momentum values, $ \left\langle M_{\alpha\beta} \right\rangle $ will generally vanish, whereas $\left\langle M_{\alpha\beta} M_{\alpha\beta}^\dagger \right\rangle$, in which the momentum-dependent phase drops out, remains finite.

\subsection{The twisted Eguchi-Kawai model}
\label{subsec:twisted_EK_model}

A different variant of the EK model was proposed by Gonz\'alez-Arroyo and Okawa: the twisted EK model~\cite{GonzalezArroyo:1982ub, GonzalezArroyo:1982hz}. It is based on the observation that the perturbative behavior of the EK model can be altered, in a way that avoids the spontaneous breaking of center symmetry, by modifying the boundary conditions with appropriate twist factors $ z_{\alpha \beta} $ in the center of the group~\cite{'tHooft:1979uj, Groeneveld:1980tt}. 

Similarly to what happens for the quenched EK model, the twisted EK model can be written in a volume-independent form. This is done by representing the group of spacetime translations of the original theory in terms of an Abelian subgroup of $\SU(N)$. The key observation is that, for a theory with fields in a representation
of the group which is blind to the action of center transformations (i.e. a representation with zero $N$-ality), such as the adjoint representation, it is possible to represent translations in a $D$-dimensional spacetime\footnote{Here, we assume $ D $ to be even.} in terms of $ D $ traceless $ \SU(N) $ matrices, satisfying the 't~Hooft-Weyl algebra~\cite{'tHooft:1979uj}:
\begin{equation}
\label{tHooft_algebra}
\Gamma_\beta \Gamma_\alpha = z_{\alpha\beta} \Gamma_\alpha \Gamma_\beta = \exp( 2 \pi i n_{\alpha\beta}/N ) \Gamma_\alpha \Gamma_\beta, \qquad \alpha, \beta \in \{ 1, 2, \dots D \}
\end{equation}
(where $ n_{\alpha\beta} $ is an antisymmetric $ D \times D $ matrix, whose entries are integers modulo $N$), i.e., commuting only up to an element of the center of the group (see also, e.g., ref.~\cite{Eguchi:1982ta}). Starting from the lattice formulation of $ \SU(N) $ Yang-Mills theory, the twisted EK model can be obtained by defining the following products of strings of $ \Gamma_\alpha $ matrices at each site $x$:
\begin{equation}
\label{Vs}
V(x)=\prod_{\mu} \Gamma_\mu^{x_\mu},
\end{equation}
by replacing the $U_\mu(x)$ link variables with:
\begin{equation}
\label{TEK_link_replacement}
U_\mu(x) \to V(x) U_\mu V^\dagger(x)
\end{equation}
and finally by doing a change of variables: $U_\mu \to U_\mu \Gamma_\mu^\dagger$. This leads to the following action~\cite{GonzalezArroyo:1982ub, GonzalezArroyo:1982hz}:
\begin{equation}
\label{twisted_EK_action}
S_{\mbox{\tiny{TEK}}} = - \frac{N}{\lambda} \sum_{1 \le \alpha < \beta \le 4} [ z_{\alpha \beta} \tr( U_\alpha U_\beta  U_\alpha^\dagger U_\beta^\dagger) + \mbox{H.c.}].
\end{equation}
For suitable choices of the $ z_{\alpha\beta} $ twist factors, it is possible to prove that, at the perturbative level, the model described by eq.~(\ref{twisted_EK_action}) satisfies the same Schwinger-Dyson equations as the theory defined in an infinite volume in the large-$N$ limit. In particular, since the classical solution is given by $U_\alpha=\Gamma_\alpha$,~in the weak-coupling limit an open Wilson line from the origin to a generic point $x$ fluctuates around $ V(x) $ (up to a $ \Z_N $ factor, which depends on the shape of the line). One can prove that any two $ \SU(N) $ matrices satisfying eq.~(\ref{tHooft_algebra}) with a twist factor different from $ 1 $ are traceless. Using this fact, it is possible to show that a sufficient condition for the traces of open Wilson lines in the reduced model to vanish is that $V (x) $ commutes with all of the $\Gamma_\alpha$'s. Since, for the theory in a lattice of finite volume, the only ``open'' Wilson lines whose trace can have a non-vanishing expectation value are those winding around the lattice (like, e.g., Polyakov loops), the latter condition reduces to:
\begin{equation}
\label{twist_tensor_condition}
x_\alpha n_{\alpha\beta}/N \in \Z.
\end{equation}
In four spacetime dimensions, a simple solution for the case when $ N $ is a perfect square ($N=L^2$, $L\in\N_0$) is given by the symmetric twist: $ n_{\alpha\beta}=L $ for all $\alpha<\beta$. This also reveals that, in general, finite-$N$ corrections in the twisted EK model can be interpreted as finite-volume corrections, with the four-dimensional volume scaling like $L^4=N^2$.

If the conditions that ensure the equivalence between the twisted EK model and the theory defined on a lattice of finite (large) volume are satisfied, expectation values in the twisted EK model are trivially related to those in the theory with trivial boundary conditions (i.e., no twists): for example, vev's of Wilson loops of area $A$ in the $ (\alpha,\beta) $ plane get simply multiplied by a $ z_{\alpha,\beta}^A $ factor.

As mentioned above, for the simplest symmetric twist, the classical solutions of the twisted EK model are of the form:
\begin{equation}
U_\alpha=\Gamma_\alpha,
\end{equation}
where the $ \Gamma_\alpha $'s are a set of $ D $ traceless, $ N \times N $ special unitary matrices satisfying eq.~(\ref{tHooft_algebra}); such configurations correspond to the absolute minimum that the action in eq.~(\ref{twisted_EK_action}) can take, and are hence called ``twist eating'' configurations~\cite{Groeneveld:1980tt}. As discussed in ref.~\cite{'tHooft:1981sz}, in two spacetime dimensions the $ \Gamma_\alpha $~matrices can be identified with the ``shift'':
\begin{equation}
S_N = \left(
\begin{array}{ccccc}
0 & 1 & 0 & 0 & \dots \\
0 & 0 & 1 & 0 & \dots \\
0 & 0 & 0 & 1 & \dots \\
\dots & \dots & \dots & \dots & \dots \\
1 & 0 & 0 & 0 & \dots 
\end{array}
\right)
\end{equation}
and ``clock'':
\begin{equation}
C_N = \left(
\begin{array}{ccccc}
1 & 0 & 0 & 0 & \dots \\
0 & \omega & 0 & 0 & \dots \\
0 & 0 & \omega^2 & 0 & \dots \\
0 & 0 & 0 & \omega^3 & \dots \\
\dots & \dots & \dots & \dots & \dots \\
0 & 0 & \dots & 0 & \omega^{N-1}
\end{array}
\right), \qquad \omega=\exp(2\pi i /N),
\end{equation}
matrices, which satisfy the little 't~Hooft algebra:
\begin{equation}
\label{little_t_Hooft_algebra}
S_N C_N = \exp(2\pi i /N) C_N S_N.
\end{equation}
In four dimensions, analogous matrices can be obtained from tensor products of these matrices (taking $ N $ to be a perfect square, $ N = L^2 $, with $L \in \N_0$, and replacing $ N $ with $ L $ in the definitions of the shift and clock matrices):\footnote{Strictly speaking, this construction yields the twist eaters in a representation in which the twist tensor is skew-diagonal: $n = iL\sigma_2 \otimes \ide_{2}$. However, $ n $ can then be brought to the standard form by an $ \SL(4,\Z) $ transformation.}
\begin{equation}
\label{twist_eaters_tensor_construction}
\Gamma_\alpha = C_L^{\delta_{\alpha,1}} S_L^{\delta_{\alpha,2}} \otimes C_L^{\delta_{\alpha,3}} S_L^{\delta_{\alpha,4}}.
\end{equation}
A systematic classification of the solutions was given by van~Baal in ref.~\cite{vanBaal:1983eq} (see also ref.~\cite{GonzalezArroyo:1997uj}), showing the existence of non-trivial configurations which could survive in the large-$N$ limit.

The distribution of eigenvalues of the $ U_\mu $ matrices in the (unphysical) strong-coupling limit is uniform over (four copies of) the $ \U(1) $ circle, hence center symmetry is not broken in this limit. In the opposite limit, the configuration corresponding to the classical solution is invariant under $\Z_L \subset \Z_N$. As a consequence, the expectation values of the four Polyakov loops in the twisted EK model are vanishing both at strong coupling and in the classical limit. To understand how the continuum limit is approached, a leading-order weak-coupling expansion around the twist-eating solution was first discussed in ref.~\cite{GonzalezArroyo:1982hz}: it revealed that, while the propagators coincide with those of the lattice theory, the vertices are generally modified by momentum-dependent phase factors, and the overall phase of a given diagram is related to the leg ordering. In the large-$N$ limit, this leads to strongly oscillating factors, which suppress all non-planar diagrams~\cite{GonzalezArroyo:1982hz, Aldazabal:1983ec}. In a continuum formulation~\cite{GonzalezArroyo:1983ac}, the presence of momentum-dependent vertices, with relative suppression of diagrams which differ by non-cyclic permutations of the legs, can be interpreted in the context of field theories defined in non-commutative spaces~\cite{Douglas:2001ba, Szabo:2001kg, Landi:1997sh} (see also ref.~\cite{Filk:1996dm}), which are relevant for certain low-energy limits of M- and string theory~\cite{Seiberg:1999vs, Aoki:1999vr, Connes:1997cr, Douglas:1997fm}. The twisted EK model provided a way to regularize theories defined in such spaces and to study them non-perturbatively~\cite{Ambjorn:1999ts, Ambjorn:2000cs, Ambjorn:2000nb, Bietenholz:2002ch}.\footnote{Alternative regularizations for field theories defined in non-commutative spaces are reviewed in refs.~\cite{Panero:2006bx, Panero:2006cs}.} In fact, this formulation can also be carried out at finite values of $N$; in contrast to the quenched EK model (which only captures the dynamics associated with planar graphs) the twisted EK model is well-defined order by order in an expansion in powers of $1/N$.

Non-perturbative Monte~Carlo studies of the twisted EK model have been carried out since the 1980's; for Yang-Mills theories in four spacetime dimensions,\footnote{The twisted EK model has been studied numerically also in two~\cite{Bietenholz:2002ch, Fabricius:1983au, Bietenholz:2002vj} and in six dimensions~\cite{Nishimura:1996pe}.} early works include refs.~\cite{GonzalezArroyo:1982hz, GonzalezArroyo:1983pw, Fabricius:1984un, Gocksch:1983jj, Das:1984xs, Fabricius:1984ac, Das:1984xz, Das:1984jh, Migdal:1983xp}. These articles found numerical evidence that the model, with the simple symmetric twist described above, correctly describes the physics of the large-$N$ theory in a large volume, and reported results for quantities like $T_c/\Lambda_{\mbox{\tiny{L}}}$,~the deconfinement temperature in units of the lattice $\Lambda$-parameter. Recently, however, some accurate numerical studies disproved these claims~\cite{Teper:2006sp, Azeyanagi:2007su, Bietenholz:2006cz}, showing that, although the center symmetry in the twisted EK model is preserved both in the weak- and strong-coupling limits, it does get spontaneously broken at intermediate couplings, and the width of the range of couplings in which this occurs grows with $N$, making the investigation of the correct continuum physics at large $ N $ challenging---if possible at all. The spontaneous breakdown of center symmetry at intermediate couplings appears to be due to the system getting stuck in metastable phases, characterized by the fields fluctuating around center-symmetry breaking configurations (with action $O(N)$ above the twist-eating configurations, and separated from the latter by effective potential barriers of order $O(N^2)$), that become arbitrarily long-lived for $N \to \infty$. 

\begin{figure}[-t]
\centerline{\includegraphics[width=0.7\textwidth]{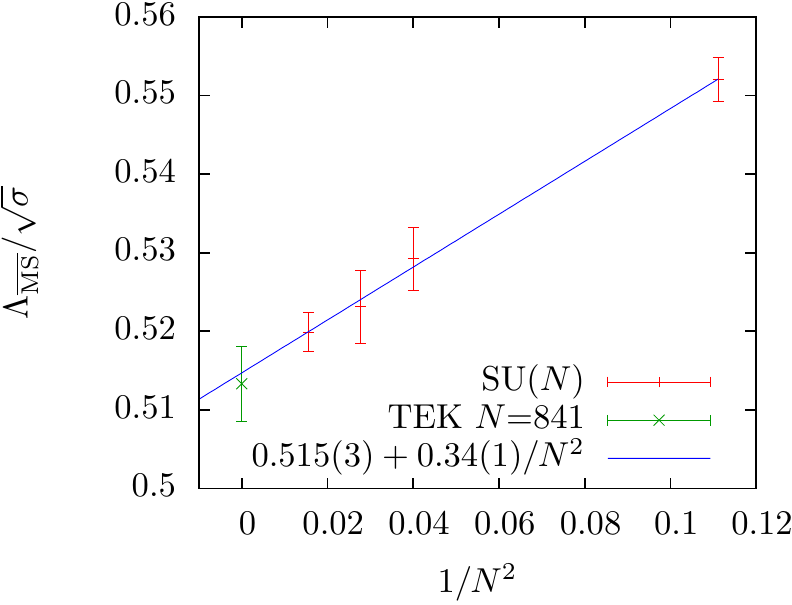}}
\caption{A comparison of the results of simulations of the twisted EK model performed in ref.~\cite{GonzalezArroyo:2012fx} using the twist formulation discussed in ref.~\cite{GonzalezArroyo:2010ss} (green symbol), with those from lattice simulations performed in a large volume (in the conventional formulation of the theory, without exploiting the volume-independence properties), taken from ref.~\cite{Allton:2008ty} (red symbols), and their extrapolation as a function of the number of colors (blue line). The quantity which is plotted is the ratio (extrapolated to the continuum limit) between the $ \LambdaQCD $ parameter in the \MSbar~scheme, and the square root of the string tension, i.e., of the asymptotic force between static color sources at large separations.\label{fig:GonzalezArroyo}}
\end{figure}

A possible solution to this problem was recently proposed in ref.~\cite{GonzalezArroyo:2010ss}: the idea is to lift the action of the center-symmetry breaking configurations $ O(N^2) $ above the twist-eating ones, by changing the twist to: $ n_{\alpha\beta}=kL $ for all $ \alpha<\beta $, with $ k=O(L) $ (in particular: $k>L/4$) and co-prime with $N$. Non-perturbative investigations of this new formulation of the twisted EK model were initiated in ref.~\cite{GonzalezArroyo:2010ss}, and encouraging new results have been recently reported in ref.~\cite{GonzalezArroyo:2012fx}. An example is shown in fig.~\ref{fig:GonzalezArroyo}, taken from ref.~\cite{GonzalezArroyo:2012fx}, in which the value of the string tension (i.e., of the asymptotic slope of the confining potential at large distances) extracted from simulations in this new version of the twisted EK model at $N=841$, and suitably extrapolated to the continuum, is compared with those reported in ref.~\cite{Allton:2008ty}, from a conventional lattice study in a large volume: the result obtained from the twisted EK model simulation is consistent with the extrapolation of the latter.

However, it is worth mentioning that the numerical study of ref.~\cite{Azeyanagi:2007su}, which also discussed the case of a twist tensor $n=O(N)$, concluded that the critical value of 't~Hooft coupling, below which center symmetry is not spontaneously broken, tends to zero like $ 1/N $ when $ N $ is increased. Similar conclusions were obtained in another numerical study~\cite{Bietenholz:2006cz}, that used a twist tensor with $k=(L+1)/2$. To completely clarify the viability of the twisted EK model with a non-minimal twist to investigate the behavior of large-$N$ Yang-Mills theory, further numerical studies are needed. 

Other recent numerical studies related to the twisted EK model include those discussed in ref.~\cite{Perez:2012fz} (which considered Yang-Mills theory in three spacetime dimensions in a finite volume and in the presence of a chromomagnetic flux realized with a twist, studying the interplay between the large-$N$ limit and the physical size of the system) and in ref.~\cite{GonzalezArroyo:2012st} (which investigated volume reduction in the twisted model including fermionic fields in the adjoint representation of the gauge group---see subsec.~\ref{subsec:adjoint_EK_model} for a discussion).

\subsection{Volume reduction with adjoint fermions}
\label{subsec:adjoint_EK_model}

A different method to preserve center symmetry in the EK model, which has been proposed in the literature, consists in modifying the theory by adding one or more flavors of dynamical massless Majorana fermions in the adjoint representation of the gauge group, and with periodic boundary conditions~\cite{Kovtun:2007py}.\footnote{This idea is related to circle compactifications and supersymmetry-preserving deformations in supersymmetric gauge theories~\cite{Poppitz:2011wy, Seiberg:1996nz, Aharony:1997bx}.} When the theory is compactified on a small spatial torus of length $R$, a perturbative calculation of the effective potential $V_{\mbox{\tiny{eff}}}$ for the phases of the eigenvalues $ \exp (i \theta_a) $ of the Polyakov line along that direction shows that the effect of fermions can (over-)compensate the symmetry-breaking terms~coming from gluons~\cite{Gross:1980br, Barbon:2005zj, Unsal:2006pj}:
\begin{equation}
\label{effective_potential_adj_fermions}
V_{\mbox{\tiny{eff}}}(\theta) = \frac{2(n_f -1)}{\pi^2 R^4} \sum_{n=1}^{\infty} \frac{1}{n^4} \left| \sum_{a=1}^N \exp( i n \theta_a) \right|^2 = \frac{\pi^2 (n_f-1)}{45 R^4} \sum_{a=1}^N \sum_{b=1}^N \left\{ 1 - \frac{15}{8\pi^4} \left[ \pi^2 - f^2(\theta_a - \theta_b) \right]^2 \right\},
\end{equation}
where $f(x)=(x$~mod~$2\pi)-\pi$. In the na\"{\i}ve EK model ($n_f=0$), the minimum of the potential is obtained when the term between the square brackets (which is positive semi-definite, because $ |f(x)|\le \pi $ for any $x$) vanishes: this corresponds to the case when all eigenvalues collapse to the same value, inducing spontaneous center-symmetry breaking. By contrast, in the $ n_f=1 $ case (which corresponds to $ \mathcal{N}=1 $ supersymmetric Yang-Mills theory) perturbatively one finds a flat one-loop effective potential, but the flatness is actually lifted by non-perturbative effects due to center-stabilizing bions, or ``neutral bions''~\cite{Poppitz:2011wy, Poppitz:2012sw,  Argyres:2012ka, Argyres:2012vv}, which lead to a repulsive potential for Wilson-line eigenvalues, and stabilize center-symmetric configurations; finally, for $ 1 < n_f < 6 $ the inclusion of adjoint fermions in the theory has the effect of favoring vacua in which center symmetry is preserved\footnote{For $n_f \ge 6$, the theory is no longer asymptotically free.} (see also ref.~\cite{Meisinger:2001fi} for a related computation). Analogous formul\ae~hold in the case when two or more directions are compactified to small tori. It is worth emphasizing that studying QCD-like theories including fermionic matter with periodic boundary conditions along a compact, spatial direction is interesting on its own, as it may reveal transitions of \emph{quantum} (rather than \emph{thermal}) nature---and the non-trivial parametric dependence of the corresponding scales on $N$~\cite{Unsal:2008eg}.

An important aspect of gauge theories coupled to adjoint fermions with periodic boundary conditions along a short spatial direction is that they give some \emph{analytical} control over phenomena such as confinement and chiral symmetry breaking---see ref.~\cite{Unsal:2007vu} for an explicit example in the theory with $\SU(2)$ gauge group.

While the inclusion of adjoint fermions in a QCD model may appear artificial, it should be noted that, in the large-$N$ limit, the orientifold equivalence relates QCD with adjoint fermions to QCD with Dirac fermions in a two-index, symmetric or anti-symmetric representation~\cite{Armoni:2003gp, Armoni:2003fb, Armoni:2004uu, Armoni:2004ub}. Given that in real-world QCD the quark fields are in the fundamental representation of the $ \SU(3) $ gauge group, which is equivalent (up to charge conjugation) to the two-index antisymmetric representation, the latter theory can be regarded as a natural generalization of QCD to the large-$N$ limit. Incidentally, we mention that a quenched lattice study of the orientifold equivalence (in a large volume) was reported in ref.~\cite{Lucini:2010kj}: it was found that the masses of vector mesons corresponding to quarks in different representations tend to compatible values in the chiral and large-$N$ limits, in agreement with the theoretical expectations. A similar result holds for the chiral condensate in the quenched theory~\cite{Armoni:2008nq}.

\begin{figure}[-t!]
\centerline{\includegraphics[height=0.5\textheight]{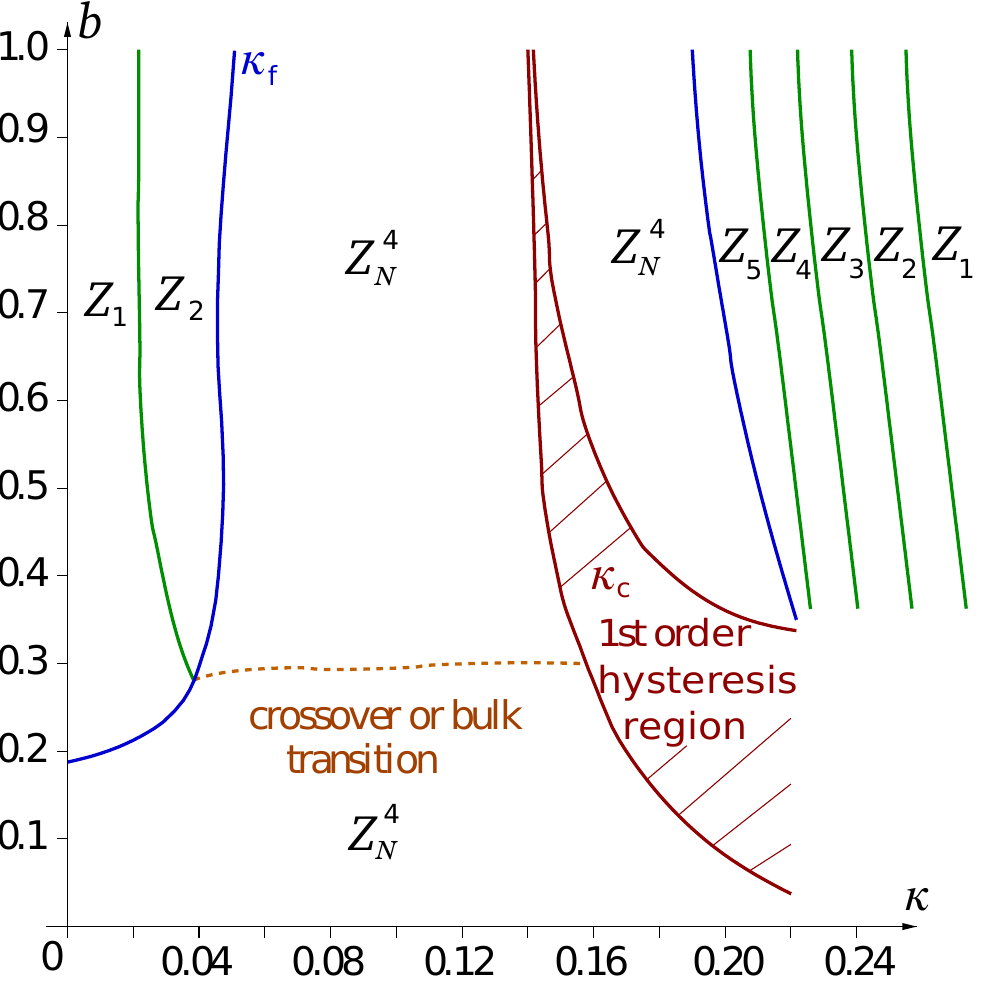}}
\caption{Lattice simulations of large-$N$ gauge theories with $ n_f=1 $ and $ n_f=2 $ flavors of dynamical adjoint fermions in a small volume indicate that center symmetry is preserved in a wide region of quark masses, whose width remains finite in the continuum limit. The figure, taken from ref.~\cite{Bringoltz:2011by}, is a sketch of the symmetry realizations in different regions of the space of the simulation parameters. In particular, the regions in which center symmetry is preserved in all the four directions are denoted by $Z_N^4$. The quantity on the vertical axis ($b$) is the inverse of the bare lattice 't~Hooft coupling, so that (by virtue of asymptotic freedom) the continuum limit is obtained by taking $b \to \infty$---while the region corresponding to small values of $ b $ is the strong-coupling region of the lattice theory. The quantity on the horizontal axis ($\kappa$) is a simulation parameter related to the fermion mass: in particular, the latter is infinite for $ \kappa=0 $ (corresponding to the pure Yang-Mills theory), while it decreases when $ \kappa $ is increased towards the dashed region, reaching zero at the critical value $ \kappa_c $. The figure, based on information obtained from simulations with $ n_f=2 $ dynamical quarks in the adjoint representation of the gauge group in a small volume, shows that, in the presence of dynamical adjoint fermions, the spontaneous breakdown of center symmetry is avoided, for a rather wide range of quark masses, in a region that extends to the continuum limit.\label{fig:funnel}}
\end{figure}

Following the proposal of ref.~\cite{Kovtun:2007py}, various works investigated the EK model with adjoint fermions and variants thereof, both analytically and numerically~\cite{Unsal:2007fb, Hollowood:2009sy, Bedaque:2009md, Cossu:2009sq,  Bringoltz:2009mi, Bringoltz:2009kb, Bringoltz:2009fj, Poppitz:2009fm, Hietanen:2009ex, Hietanen:2010fx, Azeyanagi:2010ne, Catterall:2010gx, Bringoltz:2011by, Armoni:2011dw, Lee_parallel, Okawa_parallel}. The results of the most recent non-perturbative studies via lattice simulations are encouraging, and indicate that EK volume reduction with adjoint fermions works as expected, both with $ n_f=1 $ and $ n_f=2 $ Dirac flavors: as an example, ref.~\cite{Bringoltz:2011by} reported numerical evidence that center symmetry in the model with two flavors is preserved, for a finite (and rather large---see the discussion in ref.~\cite{Azeyanagi:2010ne}) range of bare quark masses, over an interval of couplings which appears to extend all the way to the continuum limit (see the sketch in fig.~\ref{fig:funnel}, from ref.~~\cite{Bringoltz:2011by}). For the case with one adjoint quark flavor, a similar study was presented in ref.~\cite{Bringoltz:2009kb}. 

After establishing that volume reduction in the EK model with adjoint fermions does work, the natural next step in this agenda of lattice studies consists in investigating how cost-effective these simulations are.

\subsection{Volume reduction in Yang-Mills theories with double-trace deformations}
\label{subsec:deformed_EK_model}

A related variant of the EK model was proposed in ref.~\cite{Unsal:2008ch} (see also ref.~\cite{Shifman:2008ja}): it leads to a one-loop effective potential of the type appearing in eq.~(\ref{effective_potential_adj_fermions}) by deforming the ordinary lattice Yang-Mills action adding products of traces of Polyakov loops,\footnote{Related ideas have been considered in studies of the phase diagram of Yang-Mills theory at finite temperature~\cite{Myers:2007vc, Simic:2010sv, Ogilvie:2012fe, Ogilvie:2012is}.} with suitably chosen (in particular: sufficiently large) positive coefficients, which suppress the weight of center-symmetry breaking configurations in the path integral. One advantage of this approach is that it allows one to reduce the extent of an arbitrary number of sizes of the system. For example, in order to preserve center symmetry when only one direction is compactified, the deformation term may be of the form:
\begin{equation}
\label{deformation_term}
\sum_{\vec{x}} \sum_{n=1}^{\lfloor N/2 \rfloor} a_n |\mbox{tr}(L^n(\vec{x}))|^2
\end{equation}
(where the summation is done over the points of a hyperplane orthogonal to the compactified direction), while more complicated deformations, which also include products of loops along different directions, can be added, when two or more directions are compactified to small sizes. The deformation terms also modify the expectation values of the observables, however it is expected that they only do so by $ O(1/N) $ corrections, which become negligible in the large-$N$ limit.

Note that the deformation terms appearing in eq.~(\ref{deformation_term}) are (1) non-local and (2) non-linear functions of the $ U_\mu(x) $ fields on the lattice, and that (3) their number grows $ O(N) $ (this latter requirement is necessary, in order to prevent partial breakdown of the $ \Z_N $ symmetry down to a $ \Z_k $ subgroup, if $ k $ is a divisor of $N$). Even though these three features make the numerical simulation of the deformed model considerably more demanding than that of ordinary Yang-Mills theory, it should be noted that dedicated numerical algorithms for the simulation of this model are already available~\cite{Vairinhos:2010ha}, and Monte Carlo investigations are presently in progress~\cite{volume_reduction_paper}.

\subsection{Partial volume reduction}
\label{subsec:partial_EK_reduction}

Finally, a different approach to EK volume reduction has been proposed by Kiskis,  Narayanan and Neuberger in ref.~\cite{Narayanan:2003fc}, and successfully carried out in a series of works, which investigated the large-$N$ limit of various observables, in two, three and four spacetime dimensions~\cite{Kiskis:2003rd, Narayanan:2003fc, Narayanan:2004cp, Narayanan:2006rf, Narayanan:2007dv, Narayanan:2007ug, Narayanan:2006sd, Kiskis:2005hf, Kiskis:2008ah, Kiskis:2009xj, Kiskis:2009rf} (see also refs.~\cite{Narayanan:2007fb, Narayanan:2005gm} for reviews). The idea underlying this approach consists in reducing the lattice size down to the minimum value, for which center symmetry is preserved: this is possible because the critical size $ l_c $ at which the symmetry gets spontaneously broken (the ``inverse deconfinement temperature'') is a physical quantity with a finite, non-vanishing limit for $N \to \infty$. For this reason, in this approach one only achieves a partial volume reduction, and the number of lattice variables that have to be simulated still grows like $ (1/a)^D $ when the lattice spacing $ a $ is reduced towards zero to approach the continuum limit.\footnote{Note that, volume independence for lattices of linear sizes larger than $ l_c $ is equivalent to temperature independence of physical operators in the confining phase~\cite{Gocksch:1982en}.} An advantage of this approach, however, is that it does not rely on any ``trick'' (whose validity may possibly fail at the non-perturbative level---perhaps through some non-trivial mechanism); moreover, its conceptual simplicity makes it straightforward to study numerically. 

\begin{figure}[-t]
\centerline{\includegraphics[width=0.8\textwidth]{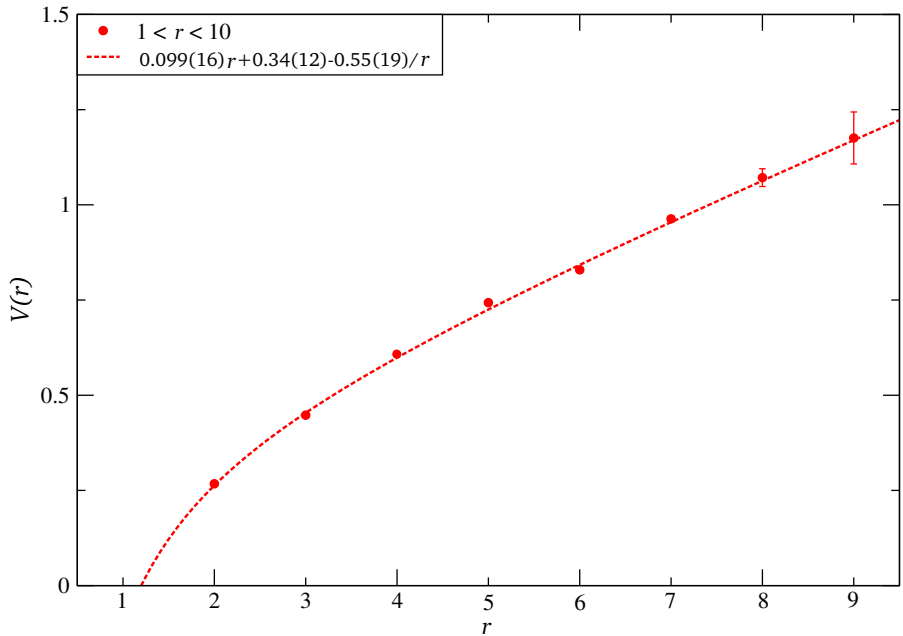}}
\caption{Computation of the confining potential $V$, as a function of the distance $r$ between static color sources, using partial volume reduction. The figure shows the results obtained in ref.~\cite{Kiskis:2009rf} for $ V $ (red circles), up to distances of nine lattice spacings, i.e. one and a half the linear size ($L$) of the lattice used in those simulations. A three-parameter fit (dashed red curve) to a Cornell-type potential is also displayed.\label{fig:Kiskis}}
\end{figure}

In particular, following this approach, in ref.~\cite{Narayanan:2003fc} it was shown that, in three spacetime dimensions, the distributions of eigenvalues for Wilson and Polyakov loops agree when extracted from simulations on a large volume lattice and on a partially reduced lattice. The analysis of the four-dimensional case was carried out in subsequent work, reported in ref.~\cite{Kiskis:2003rd}, which studied the details of the phase diagram of the lattice theory, as a function of the lattice size in physical units: it was found that there exists a cascade of transitions, corresponding to the breaking of center symmetry in one, then two, three, and eventually in all four directions. This analysis was refined in ref.~\cite{Narayanan:2006rf}, by addressing the loop renormalization~\cite{Dotsenko:1979wb} with a suitable smearing procedure, to get a well-defined continuum limit. In parallel with these works, the chiral symmetry realizations were studied in refs.~\cite{Narayanan:2004cp, Narayanan:2006sd}; the results confirmed the validity of partial volume reduction at large $N$: for finite lattices of linear extent larger than $l_c$, chiral symmetry gets spontaneously broken, and a chiral condensate, independent of the lattice volume, appears. Finally, a series of works studied the confining potential and the string tension in the partial volume reduction approach~\cite{Kiskis:2008ah, Kiskis:2009xj, Kiskis:2009rf}. In particular, this was done for the four-dimensional case in ref.~\cite{Kiskis:2009rf}, from which the results shown in fig.~\ref{fig:Kiskis} are taken: the plot shows the behavior of the confining potential $ V $ as a function of the distance $ r $ (the numerical values for both quantities are shown in the appropriate units of the lattice spacing $a$), as extracted from simulations with $ N=47 $ at inverse 't~Hooft coupling $b=0.348$, on a lattice of linear size $ L=6a $ in all directions (which is small, but sufficiently large to preserve center symmetry). The potential, which was computed up to a distance of nine lattice spacings (i.e., one and a half the linear size of the lattice), shows the behavior that is expected in a large volume.

\subsection{Volume independence and large-$N$ equivalences from orbifold projections}
\label{subsec:orbifold_projections}

Having discussed how the EK equivalence between theories defined in different spacetime volumes arises at large $ N $ (and how it can fail), we conclude this section presenting its interpretation in the broader context of a general class of equivalences among large-$N$ theories. Various types of correspondences between theories, which are different at every finite value of $N$, but become coincident in the large-$N$ limit, were already discovered during the early 1980's~\cite{Makeenko:1981bb, Lovelace:1982hz, Samuel:1982bd}, but it is only during the last decade, that systematic methods to construct such equivalences (and to precisely identify the conditions for their validity) have been worked out. 

One of the first papers pointing out the interpretation of EK volume reduction in large-$N$ lattice gauge theories in terms of these equivalences was ref.~\cite{Neuberger:2002bk}. The discussion that we present here mostly follows the developments of this subject through a series of articles by Kovtun, Poppitz, \"Unsal and Yaffe~\cite{Kovtun:2007py, Unsal:2006pj, Poppitz:2009fm, Unsal:2008ch, Kovtun:2003hr, Kovtun:2004bz, Kovtun:2005kh,  Unsal:2010qh, Poppitz:2010bt}. For historical reasons related to their origin in the context of string theory~\cite{Douglas:1996sw, Kachru:1998ys, Bershadsky:1998cb, Strassler:2001fs},\footnote{In string theory, equivalences of this type arise when considering the low-energy dynamics of D-branes defined in spacetimes which are orbifolds, i.e. manifolds admitting points whose neighborhood is locally diffeomorphic to a quotient of a real vector space by a finite group.} these equivalences are called ``orbifold'' equivalences.

In this context, an orbifold equivalence is defined as a mapping between a ``parent'' theory, and a ``daughter'' theory, which is obtained from the former by removing (``projecting out'') the degrees of freedom which are not invariant under the action of a discrete global symmetry group.\footnote{The degrees of freedom which are not invariant under the discrete symmetry group used for the projection are said to form the ``twisted sector'' of the theory. Note, however, that here the adjective ``twisted'' has no direct connection with our discussion of the twisted EK model.} More precisely, in the large-$N$ limit these equivalences relate vacuum expectation values and correlation functions of a subset of gauge-invariant observables in the parent theory and in the daughter theory (possibly up to a rescaling of couplings and volume factors), provided that the global symmetry of the parent theory used to construct the orbifold projection is not spontaneously broken, and that possible global symmetries of the daughter theory, which interchange its gauge group factors, are not spontaneously broken either. 

At the perturbative level, this equivalence is based on the observation that planar graphs of the parent and daughter theories coincide at all orders, and that generic correlation functions of gauge-invariant operators (from which one can extract physical observables, including the mass spectrum) obey the same set of closed equations. To extend the validity of the correspondence to the non-perturbative domain, the authors of ref.~\cite{Kovtun:2003hr} focused on $ \U(N) $ gauge theories (possibly coupled to $ n_s $ species of massive scalar fields and $ n_f $ species of fermion fields, both of which in the adjoint representation of the gauge group) regularized on a lattice, and on the Migdal-Makeenko-like equations for gauge-invariant operators (which can be written as traces of closed Wilson loops\footnote{As discussed in ref.~\cite{Kovtun:2003hr}, operators involving adjoint scalar or fermion fields can also be expressed through closed Wilson loops (without clumsy insertions of matter field decorations along loops made only of gauge link variables), by generalizing the lattice to include one further dimension for each matter species.}) and their correlators. Consider first the pure Yang-Mills case ($n_s=n_f=0$): assuming that $ N $ is not prime (so that $N=k n$, with both $ k $ and $ n $ integers larger than $1$), one can introduce an orbifold projection under a $ H=\Z_k $ subgroup of the global $ \U(N) $ group of coordinate-independent gauge transformations, by requiring that a generic $ \Phi $ field variable of the parent theory be invariant under:
\begin{equation}
\label{pure_gauge_orbifold}
\Phi \to \gamma \Phi \gamma^{-1},
\end{equation}
where:
\begin{equation}
\label{orbifold_gamma_definition}
\gamma = \Omega \otimes \ide_n, \qquad \Omega = \diag ( 1, \omega, \omega^2 , \dots, \omega^{k-1} ), \qquad \omega = \exp ( 2\pi i / k ).
\end{equation}
The construction can be easily generalized to an orbifold under a $ \Z_k^d $~subgroup (if $N=k^d n$), by introducing $ d $ matrices $ \gamma_\alpha $, and requiring invariance of $ \Phi $ under the transformations in eq.~(\ref{pure_gauge_orbifold}), for each $ \gamma_\alpha $ separately. Matter fields can also be included in a straightforward way: for non-zero $ n_s $ and/or $n_f$, the global symmetry of the parent theory is enlarged to $\U(N) \otimes \U(n_s) \otimes \U(n_f)$, and each matter field can have a generic charge $ r_\alpha $ under the $ \gamma_\alpha $ projection, so that the orbifolding condition reads:
\begin{equation}
\label{orbifold}
\Phi = \exp \left[ 2 \pi i r_\alpha(\Phi)/k \right] \gamma_\alpha \Phi \gamma_\alpha^{-1}.
\end{equation}
Under this projection, the gauge degrees of freedom of the parent theory are mapped to block-diagonal matrices, with $ k^d $ blocks of $ \U(n) $ matrices, and the gauge invariance group of the daughter theory is $ \U(n)^{k^d} $. The action for the daughter theory can then be written in the form of a standard Wilson lattice action (possibly coupled to the matter field terms), with a sum over the ``sites'' of a $ \Z_k^d $ lattice, and with a trivial rescaling of couplings and volume factors. 

This mapping can be extended to all gauge-invariants observables of the parent theory, which are ``neutral'' under the orbifold projection, and, finally, one can derive the loop equations for the daughter theory. If the $ \Z_k^d $ symmetry in the daughter theory is not spontaneously broken, these equations are equivalent to those in the parent theory, provided that the latter does not break the discrete symmetry used in the orbifold projection. For $ N \to \infty $, these equations become a closed set of loop equations, and, in the strong coupling/large mass limit, they can be solved by an iterative algorithm, which generates the large-$N$ lattice strong-coupling expansion for the corresponding observables.

Note that the construction of an orbifold duality, that we outlined above, provides an explicit example of how the spacetime-dependent degrees of freedom of the daughter theory can be obtained from a projection of internal degrees of freedom in the parent theory.

The analysis of ref.~\cite{Kovtun:2003hr} was later extended beyond the strong-coupling domain in ref.~\cite{Kovtun:2004bz}, which used the large-$N$ coherent-state methods described in ref.~\cite{Yaffe:1981vf} (and reviewed in subsec.~\ref{subsec:classical_limit}) to demonstrate that the symmetry conditions in the parent and daughter theories are both necessary and sufficient for (i.e., completely characterize) the validity of large-$N$~orbifold dualities.

Using these tools, the validity of the orientifold equivalence, relating the planar limits of QCD with an adjoint Majorana fermion (i.e., $ \mathcal{N}=1 $ supersymmetric QCD) and of QCD with a Dirac fermion in the two-index antisymmetric representation~\cite{Armoni:2003gp, Armoni:2003fb, Armoni:2004uu, Armoni:2004ub}, was discussed in ref.~\cite{Unsal:2006pj}. In particular, it was shown that the equivalence between these two theories can be interpreted in terms of a ``daughter-daughter'' orbifold equivalence: both theories can be obtained from a common parent theory, namely $ \SO(2N) $ $ \mathcal{N}=1 $ supersymmetric QCD, by applying orbifold projections based on two different $ \Z_2 $ groups. Projecting out the degrees of freedom which are not invariant under $ J=i \sigma_2 \otimes \ide_N $ yields the $ \mathcal{N}=1 $ supersymmetric $ \U(N) $ theory as a daughter, whereas a projection using a ``graded'' variant of the same operator, $J (-1)^F$, leads to the $ \U(N) $ theory coupled to a  fermion in the two-index antisymmetric representation of the gauge group. Furthermore, it was also pointed out that the validity or invalidity of the planar orientifold equivalence (in the neutral sectors of both daughter theories), being intimately related to realization of the discrete symmetries used in the projections, depends crucially on the dynamics, and, as an example, it was shown that the orientifold planar equivalence may fail, when the theories are compactified on a small\footnote{Here, ``small'' means ``small with respect to the inverse of the characteristic dynamical scale at which the theory becomes strongly interacting'', i.e. small as compared to $1/\LambdaQCD$.} spatial torus: this regime can be reliably studied with weak-coupling expansion methods, which show that charge conjugation gets spontaneously broken---and, as a consequence, the orientifold equivalence fails.

As mentioned above, large-$N$ EK volume independence can also be interpreted in terms of an orbifold equivalence: as discussed in detail in ref.~\cite{Kovtun:2007py}, in this case the parent theory (defined, say, on a $D$-dimensional hypertorus of linear size $L$) can be mapped to the daughter (defined on a smaller hypertorus of size $ L^\prime $, with $ L $ an integer multiple of $L^\prime$, i.e. $L = k L^\prime$), by projecting out the degrees of freedom that are not invariant under the $ \Z_k^D $ translations by integer multiples of $ L^\prime $ in each direction---or, equivalently, removing the Fourier components that are not quantized in units of $ 2\pi/L^\prime $. This leads to a correspondence between theories defined in two different volumes, and, for a parent theory on a lattice of spacing $ a $, yields the EK model as the daughter theory, if $ L^\prime=a $. Conversely, it is also possible to map a theory defined in a smaller volume to one in a larger volume: this can be done by associating some of the internal degrees of freedom of the parent theory to the spacetime degrees of freedom of the daughter. For example, the $ \U(N) $~EK model with $ N=k^d n $ can be mapped to ordinary $ \U(n) $ Yang-Mills theory on a $ k^d $ lattice, following the construction that we outlined in the discussion of orbifold projections earlier in this subsection, which amounts to imposing a set of constraints of the form:
\begin{equation}
\label{volume_enlarging_projection}
U_\alpha = \exp ( 2 \pi i \delta_{\alpha \beta} / k ) \gamma_\beta U_\alpha \gamma_\beta^{-1},
\end{equation}
with the $\gamma_\alpha$'s defined analogously to eq.~(\ref{orbifold_gamma_definition}). The action (as well as the other neutral-sector observables) of the EK model is then mapped to the action of the lattice gauge theory in the enlarged volume (up to a trivial rescaling of the couplings), and the two theories are equivalent, as long as center symmetry is not spontaneously broken; furthermore, the inclusion of matter fields in various representations is also possible (see ref.~\cite{Kovtun:2007py} for details).

An interesting application of combined orbifold equivalences consists in the possibility of studying the large-$N$ limit of QCD with fermions in a two-index symmetric or antisymmetric representation in a large volume, by first mapping this theory to QCD with adjoint fermions via the orientifold equivalence (which holds in a large volume), and then studying the latter theory in a small volume (possibly reducing it to a matrix model, or to a single-site lattice model), using the fact that, for adjoint QCD, the orbifold equivalence relating the theory in different volumes hold all the way down to infinitesimally small system sizes.

Orbifold equivalences relating large-$N$~theories defined in systems of different sizes also have interesting physical implications for the phase diagram of these theories: in particular, volume-reducing or volume-expanding projections, that change the size of a Euclidean gauge system only along one direction, increase or reduce the temperature of the system by a factor $k$. If center symmetry is not spontaneously broken (i.e. if the system is in its confining phase), then the correspondence of physical observables in the neutral sectors of the parent and daughter theory implies that, at large $N$, such observables are temperature-independent in the confining phase.\footnote{Related topics have been discussed in the recent paper~\cite{Cherman:2012gn}, in which the temperature independence of a three-dimensional Yang-Mills theory with adjoint fermions was used to map it to a two-dimensional gauge theory, which was then studied using non-Abelian bosonization methods.} Another potentially interesting application of orbifold equivalences for the study of the phase diagram of QCD has been proposed in ref.~\cite{Cherman:2010jj} and discussed in refs.~\cite{Hanada:2011ju, Cherman:2011mh, Hanada:2012nj, Armoni:2012jw, Hanada:2012es, Bursa:2012ab, Hanada:2012aj}, and stems from the observation that the large-$N$ equivalence of theories based on orthogonal and unitary gauge groups may allow one to get information about the latter in the finite baryon density regime (in which lattice simulations are hindered by a severe computational sign problem~\cite{Philipsen:2010gj, deForcrand:2010ys, Philipsen:2012nu}), by performing numerical simulations of the former, for which the sign problem is absent. In addition, these equivalences also allow one to analytically derive interesting implications for the critical point in the QCD phase diagram, and for the order of the chiral symmetry restoration transition for massless quarks---which is of first (second) order when it occurs at a critical temperature equal to (larger than) that of the deconfinement transition~\cite{Hidaka:2011jj}. Recently, the problem of baryons in the context of orbifold equivalences has also been discussed in ref.~\cite{Blake:2012dp}.

Finally, one further important application of orbifold constructions is in the context of lattice supersymmetry~\cite{Kaplan:2002wv, Cohen:2003xe, Cohen:2003qw, Kaplan:2005ta, Endres:2006ic, Damgaard:2007be, Damgaard:2007xi, Damgaard:2007eh}---for further details, we refer the reader to the review~\cite{Catterall:2009it} and to the references therein.

\section{Large-$N$ results from lattice simulations}
\label{sec:results}

In this section, we review the main large-$N$ results for physical observables from lattice simulations. Unless otherwise stated, these are obtained from simulations on large lattices (i.e., without exploiting the large-$N$ volume-reduction property). First, in subsection~\ref{subsec:4D_results}, we discuss the numerical results relevant for four spacetime dimensions; then, we briefly review the results that have been obtained for large-$N$ theories in three spacetime dimensions in subsec.~\ref{subsec:3D_results}. Finally, we conclude this section with a summary of results from studies in two spacetime dimensions in subsec.~\ref{subsec:2D_results}.

\subsection{Results in four spacetime dimensions}
\label{subsec:4D_results}

\subsubsection{Is the large-$N$ limit of QCD a confining theory?}
\label{subsubsec:4D_confining}

As mentioned in subsection~\ref{subsec:mesons_and_glueballs}, many phenomenological implications derived from the large-$N$ counting rules are based on the assumption that QCD be a confining theory in the large-$N$ limit. Since confinement is a non-perturbative phenomenon, the validity of this assumption should be assessed with non-perturbative methods. 

The first studies addressing this issue were reported in refs.~\cite{Teper:1998kw, Teper:1997tq}, comparing the $\SU(2)$, $\SU(3)$ and $\SU(4)$ Yang-Mills theories. A possible way to assess whether these theories are confining consists in computing the two-point correlation function of zero-transverse-momentum, purely gluonic, spatial string operators\footnote{The formation of string-like objects in gauge theories is an idea that dates back to almost forty years ago~\cite{Nielsen:1973cs}.} winding around the lattice (torelons) along a periodic direction of size $L$, and studying its behavior as a function of the torelon separation $\tau$: at large $\tau$, the correlator was found to decay exponentially, $\propto \exp [ - m(L) \tau ]$, where $m(L)$ represents the mass of the lightest torelon. For long torelons, $m(L)$ tends to become linear in $L$ (indicating that the torelon is characterized by a non-vanishing linear energy density, or string tension, $\sigma$), so the mass of an infinitely long torelon diverges, implying confinement.\footnote{Note that, as discussed, e.g., in ref.~\cite{Greensite:2003bk}, this criterion for establishing the confining behavior of a theory is conceptually well-defined, and free from the potential ambiguities between confinement and screening, that may hinder alternative definitions of confinement. In addition, it also provides a viable operational way to study the phenomenon in lattice simulations.} The results from refs.~\cite{Teper:1998kw, Teper:1997tq} showed that confinement persists in all the theories studied.

\subsubsection{Running of the coupling}
\label{subsubsec:4D_coupling}

Another important problem, that was investigated in the first numerical studies of large-$N$ gauge theories, is the relation between the scale and the gauge coupling. 
A key statement in large-$N$ arguments is that the $N \to \infty$ limit exists if one keeps $g^2 N$ fixed. Equivalently, one might say that, perturbatively, the dynamically generated scale $ \LambdaQCD $ is the same for all $N$, at fixed $g^2 N$. To define non-perturbatively the large-$N$ limit, it is useful to look at the dependence of $ \LambdaQCD $ (or of a related, dimensionful, dynamically generated scale, such as the string tension $\sigma$) on $g^2 N$ in the non-perturbative regime. By determining non-perturbatively how the lattice spacing $a$ varies, as a function of the bare lattice gauge coupling, ref.~\cite{Lucini:2001ej} showed that the square of the bare lattice coupling scales roughly like $1/N$ (\emph{'t~Hooft scaling}), as expected from perturbative arguments. 

\begin{figure}[-t]
\centerline{\includegraphics[width=0.8\textwidth]{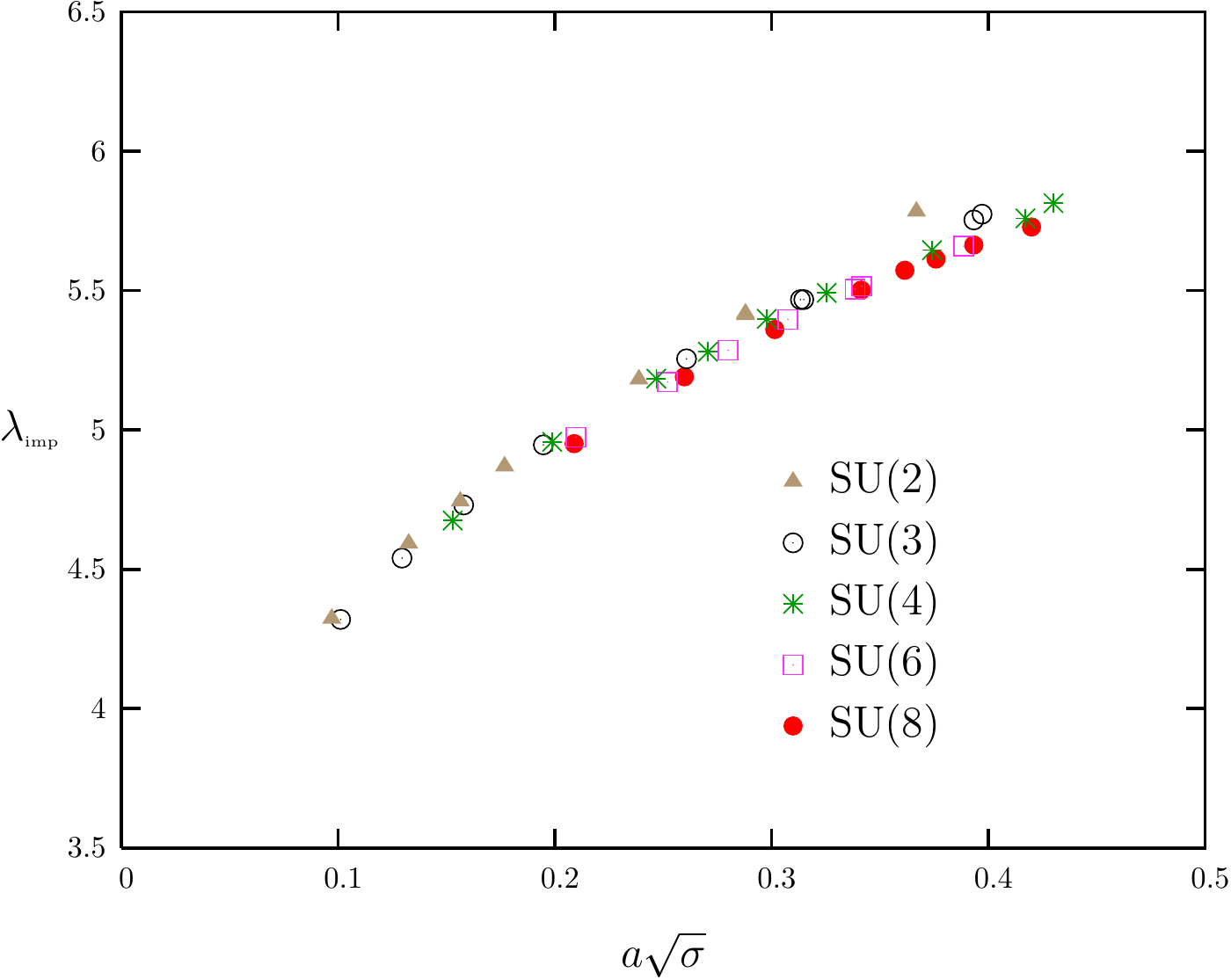}}
\caption{Relation between the mean-field improved 't~Hooft coupling and the lattice spacing (in units of $1/\sqrt{\sigma}$), as determined non-perturbatively in ref.~\cite{Allton:2008ty}, in Yang-Mills theories with a different number of colors, from $2$ to $8$.\label{fig:Trivini}}
\end{figure}

In fact, even stronger evidence for 't~Hooft scaling can be obtained, by using a ``mean-field improved'' definition of the bare coupling~\cite{Parisi:1980pe, Lepage:1992xa}, obtained by dividing the square of the bare coupling by the average plaquette (normalized to $1$ in the weak-coupling limit). This improved definition of the lattice coupling comes from the observation that, expanding the relation between lattice fields and continuum fields as:
\begin{equation}
U_\mu(x) = \exp \left[ i a g_0 A_\mu (x) \right] = \ide +  i a g_0 A_\mu (x) - \frac{1}{2} a^2 g_0^2  A^2_\mu (x) + \dots
\end{equation}
reveals that, in addition to the terms reproducing the continuum Yang-Mills action, the lattice action also includes infinitely many vertices with multiple gluons, which are pure artifacts of the lattice discretization. At the classical level, these terms are irrelevant in the continuum limit, because they are multiplied by higher powers of the lattice spacing $a$. At the quantum level, however, the tadpole diagrams arising from the contraction of fields in these terms turn out to be proportional to (a power of) the lattice cut-off $\pi/a$, so that, in fact, in the continuum limit the artifacts are only suppressed by powers of $g_0$. To alleviate this problem, one can introduce a tadpole- or mean-field-improved coupling, which is obtained by integrating out the high-frequency modes of the gauge fields, and reabsorbing the resulting constant into a redefinition of the coupling~\cite{Parisi:1980pe, Lepage:1992xa}. In particular, a convenient definition of the rescaling factor for the lattice coupling $\beta$ is given by the average plaquette,\footnote{An alternative definition of the rescaling factor is based on the average value of the link variables in Landau gauge. However, the computation of the average plaquette is numerically much easier.} so that the corresponding improved coupling can be interpreted as the effective coupling experienced by the dynamical variables in a uniform background field. The scaling of the mean-field improved bare lattice coupling with $N$ (at a fixed physical scale) was studied in refs.~\cite{Lucini:2001ej,DelDebbio:2001sj} and later in more detail in ref.~\cite{Allton:2008ty}, from which the plot shown in fig.~\ref{fig:Trivini} is taken: the figure shows the mean-field improved 't~Hooft coupling as a function of the lattice spacing, determined non-perturbatively in units of the inverse of the string tension square root: the collapse of results obtained from different $\SU(N)$ Yang-Mills theories (for $N$ in the range from $2$ to $8$) confirms that, indeed, a physically meaningful large-$N$ limit is the one at fixed 't~Hooft coupling---also at the non-perturbative level.

While the bare (possibly improved) lattice gauge coupling can be interpreted as a physical coupling at distance scales of the order of the lattice spacing, it is also interesting to investigate how the coupling runs at lower energies---in particular, because, by definition, lattice gauge theory describes continuum physics for phenomena whose characteristic energy scales are well below the intrinsic scale set by the inverse of the lattice spacing. A possible scheme choice is provided by the Schr\"odinger functional (SF)~\cite{Luscher:1991wu, Luscher:1992an, Sint:1993un}: in a nutshell, the idea is to extract the running coupling $\bar{g}(L)$ at a given momentum scale proportional to $1/L$, by studying the effective action of the gauge system with fixed boundary conditions at the opposite ends of a direction of size $L$ (and periodic boundary conditions along the other directions, which are taken to be of sufficiently large extent). Denoting the field configurations on the two boundaries as $\mathcal{C}_i$ and $\mathcal{C}_f$, the SF is defined as the probability amplitude for evolution from the state $\mathcal{C}_i$ to the state $\mathcal{C}_f$. In the Euclidean setting, this can be written as:
\begin{equation}
\mathcal{Z}\left[ \mathcal{C}_i, \mathcal{C}_f \right] = \int_{\mathcal{C}_i, \mathcal{C}_f} D U \exp(-S) = \exp(-\Gamma)
\end{equation}
where the path integral is done at fixed boundary conditions, and $\Gamma=\Gamma\left[ \mathcal{C}_i, \mathcal{C}_f \right]$ denotes the corresponding effective action. For $\mathcal{C}_i$ and $\mathcal{C}_f$, one takes the lattice fields to satisfy inhomogeneous Dirichlet boundary conditions, depending on a dimensionless real parameter $\eta$, so that a physical running coupling at the length scale $L$ can be obtained, by comparing the derivative of the classical action and of the full effective action, with respect to the $\eta$ parameter.\footnote{This allows one to trade the explicit computation of the effective action, for the computation of expectation values of the operators obtained by deriving the lattice action with respect to $\eta$.} 

\begin{figure}[-t]
\centerline{\includegraphics[width=0.49\textwidth]{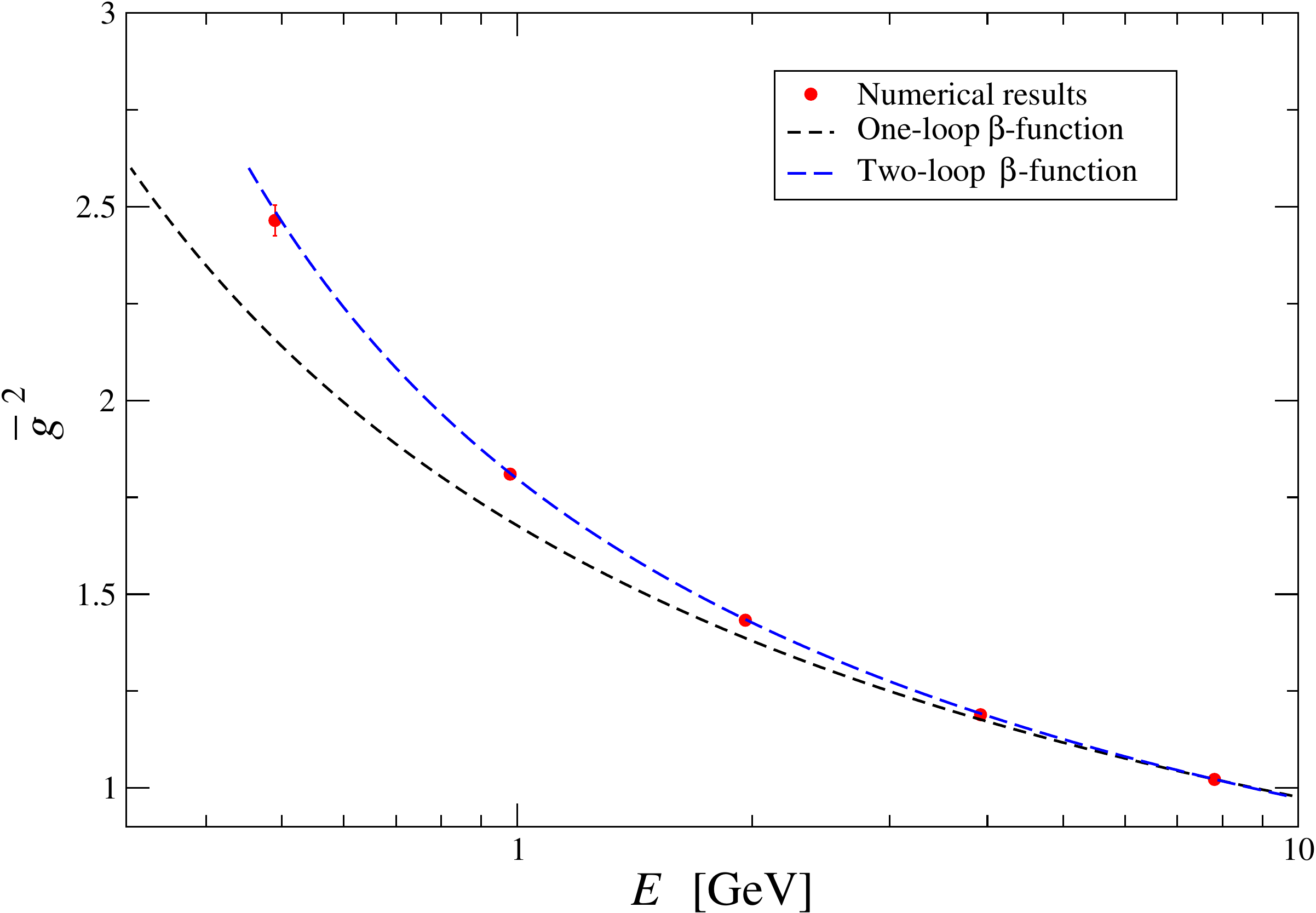} \hfill \includegraphics[width=0.48\textwidth]{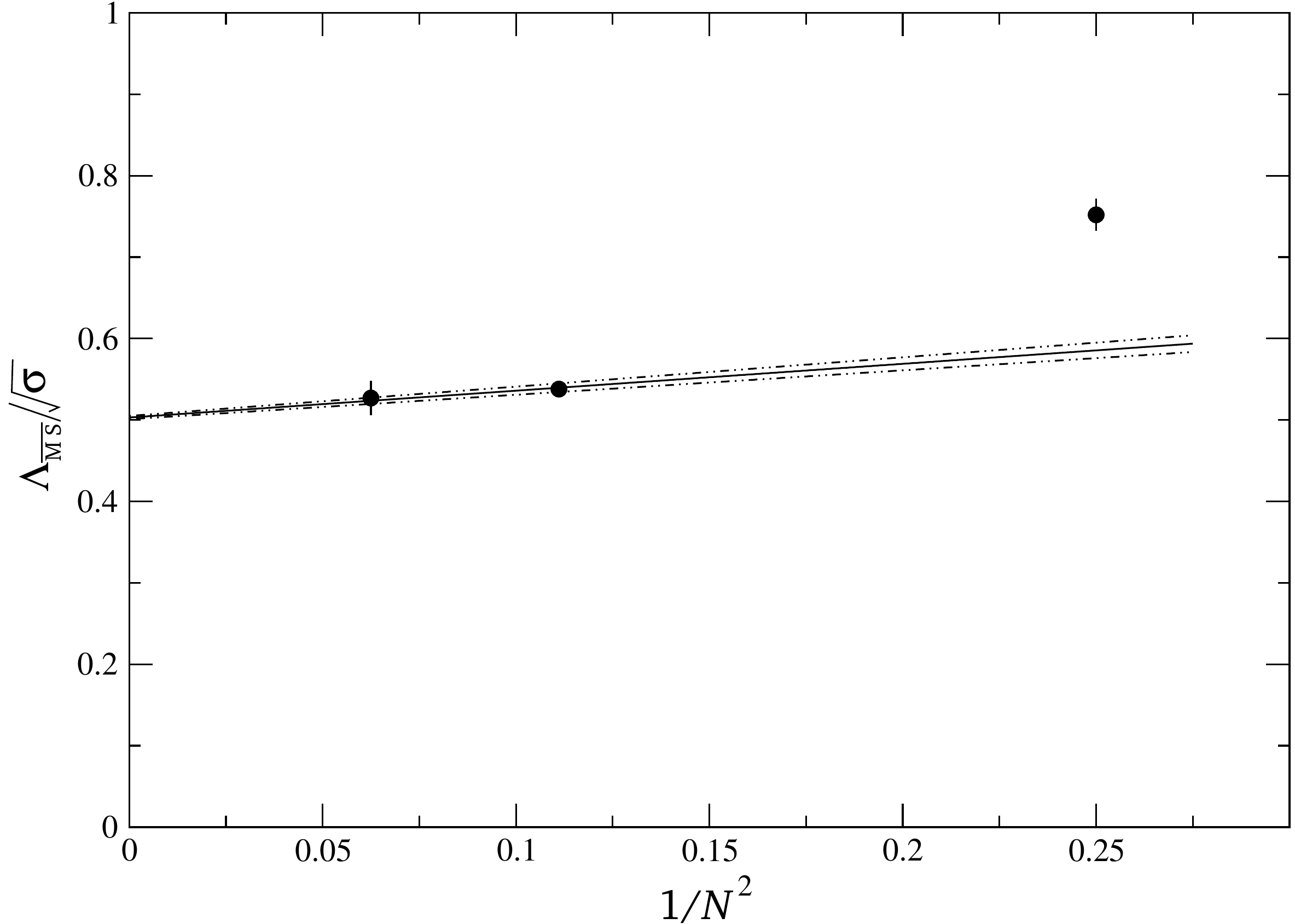}}
\caption{Left panel: The running coupling in the SF scheme for $\SU(4)$ Yang-Mills theory, as determined in ref.~\cite{Lucini:2008vi}, assuming $\sqrt{\sigma}=420$~MeV. The black and blue dashed curves are the one- and two-loop perturbative predictions. Right panel: Dependence of the $\LambdaQCD/\sqrt{\sigma}$ ratio (after conversion to the \MSbar \ scheme) on the number of colors, as obtained from computations of the physical coupling in the SF scheme in $\SU(2)$~\cite{Luscher:1992zx}, $\SU(3)$~\cite{Luscher:1993gh} and $\SU(4)$ Yang-Mills theory~\cite{Lucini:2008vi}. The plot, taken from ref.~\cite{Lucini:2008vi}, also shows the comparison with the large-$N$ extrapolation (solid straight line) obtained from the study of the mean-field improved lattice coupling in ref.~\cite{Allton:2008ty}; the region bounded by the dash-dotted lines corresponds to $68.3\%$ confidence level.\label{fig:Moraitis}}
\end{figure}

In ref.~\cite{Lucini:2008vi}, this approach was used to compute the running coupling in the SF scheme in $\SU(4)$ Yang-Mills theory (and to discuss a large-$N$ extrapolation, by comparison with analogous results for the theories with two~\cite{Luscher:1992zx} and three colors~\cite{Luscher:1993gh}): the simulation results showed that the running coupling is in very good agreement with the two-loop perturbative $\beta$-function down to momentum scales of the order of a few hundreds MeV, and that the ratio of the $\LambdaQCD$ scale (in the modified minimal subtraction scheme) over the square root of the string tension has a smooth dependence on $1/N^2$, with the value in the $\SU(3)$ theory already very close to the extrapolated large-$N$ limit (see fig.~\ref{fig:Moraitis}).

\begin{figure}[-t]
\centerline{\includegraphics[width=0.8\textwidth]{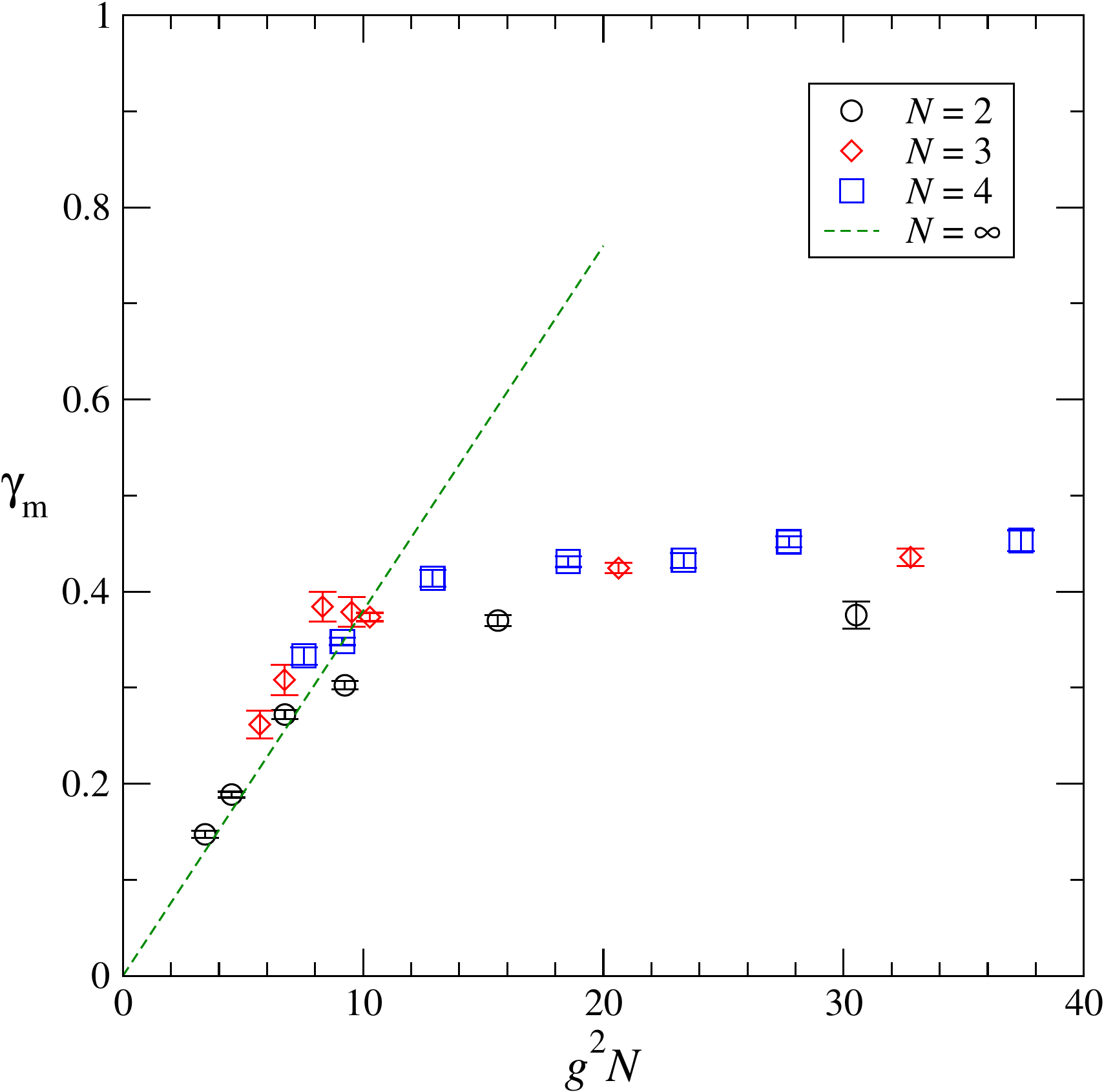}}
\caption{Mass anomalous dimension in $\SU(N)$ gauge theory with $N=2$, $3$ and $4$ colors and two flavors of dynamical fermions in the two-index symmetric representation, as a function of the renormalized 't~Hooft coupling in the SF scheme, from ref.~\cite{DeGrand:2012qa}. The dashed green line denotes the leading-order perturbative prediction in the $N \to \infty$ limit.\label{fig:Svetitsky}}
\end{figure}

Very recently, a similar type of comparison has been performed for $\SU(N)$ theories with $N=2$, $3$ and $4$ colors and with two flavors of \emph{dynamical} fermions in the two-index symmetric representation of the gauge group~\cite{DeGrand:2012qa}. The motivation for this study comes from the suggestion~\cite{Sannino:2004qp, Dietrich:2005jn, Dietrich:2006cm} (see also ref.~\cite{Sannino:2009za} and references therein) that these theories are among the potentially interesting candidates for walking technicolor models of dynamical electro-weak symmetry breaking, which recently have been the subject of many lattice studies (see, e.g., refs.~\cite{Fleming:2008gy, DeGrand:2010ba, DelDebbio:2010zz, Rummukainen:2011xv, Giedt_Lattice2012} for reviews). For this reason, the authors of ref.~\cite{DeGrand:2012qa} focused on the investigation of the $\beta$-function, and on the behavior of the mass anomalous dimension $\gamma_m$ (which relates the pseudoscalar renormalization $Z_{\mbox{\tiny{PS}}}$ to the length scale $L$ via: $Z_{\mbox{\tiny{PS}}}(L) \propto L^{-\gamma_m}$) as a function of the coupling, finding remarkable similarities between the theories with two, three and four colors, as shown in fig.~\ref{fig:Svetitsky}.

Preliminary results of another study of the mass anomalous dimension in large-$N$ QCD (with two flavors of adjoint fermions) have recently been reported in ref.~\cite{Keegan:2012xq}.

\subsubsection{Confining flux tubes as strings}
\label{subsubsec:4D_confining_flux_tubes}

\begin{figure}[-t]
\centerline{\includegraphics[width=0.6\textwidth]{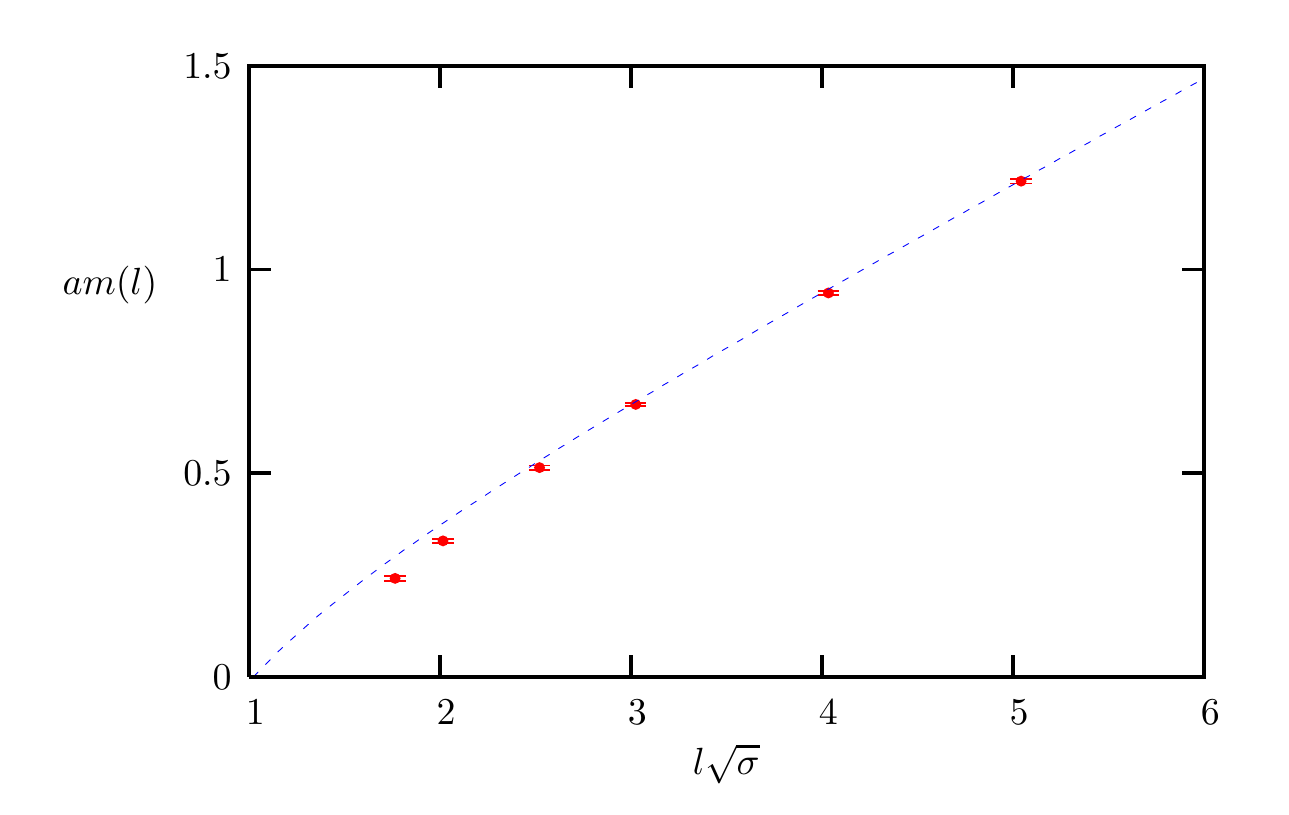} \hfill \includegraphics[height=0.3\textheight]{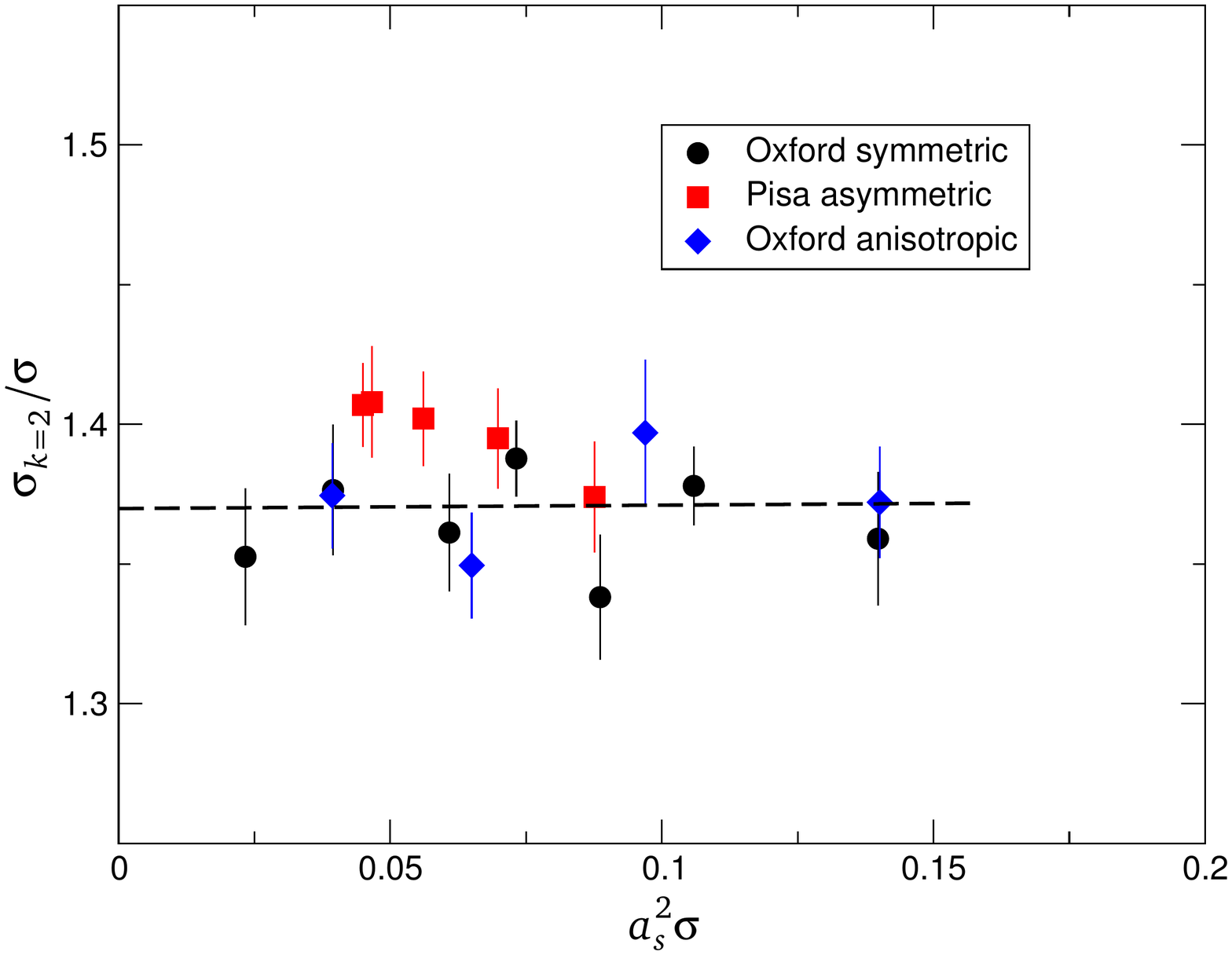}}
\caption{Left panel: Torelon mass, in lattice units, as a function of the length of the loop (in units of the string tension). Note that the dashed line is not a best fit to the data, but has been obtained by extracting the string tension at the point for which $l \sqrt{\sigma} = 5$ and inserting this value in eq.~(\protect\ref{torelon_mass}). The figure is taken from ref.~\cite{Teper:2009uf}. Right panel: Ratio of the tension of the string with $N$-ality $k=2$ over the fundamental string tension. The Pisa group data are taken from ref.~\cite{DelDebbio:2001sj}, while Oxford data are taken from ref.~\cite{Lucini:2004my}. The long dashed line is a continuum extrapolation of the Oxford data, obtained on symmetric lattices (see ref.~\cite{Lucini:2004my} for technical details).\label{fig:loop_mass}}
\end{figure}

Coming back to the problem of the heavy quark-antiquark potential in large-$N$ pure Yang-Mills theory, the first lattice studies~\cite{Teper:1998kw, Lucini:2001ej, Teper:1997tq, Meyer:2004hv} also showed that, besides the term linear in $L$, the masses of long torelons also include a correction proportional to the inverse of their length, with a coefficient, which can be evaluated analytically, known as the L\"uscher term~\cite{Luscher:1980fr, Luscher:1980ac} (see also ref.~\cite{deForcrand:1984cz}):
\begin{equation}
\label{torelon_mass}
m(L)= \sigma L - \frac{\pi}{3L} + \dots
\end{equation}
where the ellipsis denotes subleading terms, suppressed by higher powers of $1/L$. The left panel of fig.~\ref{fig:loop_mass} shows a lattice calculation of torelon masses in $\SU(6)$ gauge theory. The \emph{Ansatz} in eq.~(\ref{torelon_mass}) describes the numerical data down to lengths of the order of $3/\sqrt{\sigma}$.

As discussed, for example, in refs.~\cite{Kuti:2005xg, Mykkanen:2012dv}, the L\"uscher term can be interpreted as a Casimir effect (namely: as a quantum, finite-size effect), which arises from \emph{massless} fluctuations of the flux tube along the transverse directions,\footnote{These massless fluctuations can be interpreted as the Nambu-Goldstone modes associated with the spontaneous breakdown of Lorentz-Poincar\'e symmetry, due to the formation of the flux tube.} and signals that the low-energy excitations of the flux tube can be described in terms of a vibrating, bosonic string.\footnote{As discussed in refs.~\cite{Pisarski:1982cn, Olesen:1985ej}, it is interesting to note that an effective string model for the gauge theory also suggests the existence of a deconfinement transition at finite temperature.} For an overview of recent studies of this subject, see, e.g., the slides of the presentations at the ``Confining Flux Tubes and Strings'' workshop held in ECT$^{\star}$, Trento, Italy, in 2010~\cite{ECTstar_workshop}.

While the existence of string-like excitations is a generic feature of all confining gauge theories~\cite{Bali:1994de, Luscher:2002qv, Juge:2002br, Koma:2003gi, Panero:2004zq, Panero:2005iu, Greensite:2006sm, Bonati:2011nt}, the possibility that a particularly simple effective string model for the low-energy dynamics of QCD could become exact at large $N$ is suggested by theoretical arguments~\cite{Polchinski:1992vg}, and has been studied in a number of lattice works---see, e.g., refs.~\cite{Teper:1998kw, Lucini:2001ej, Teper:1997tq, Meyer:2004hv, Mykkanen:2012dv, Lucini:2001nv, Lohmayer:2012ue} and references therein.

The lattice analysis of the confining potential between static probe sources in $\SU(N)$ Yang-Mills theories has also been carried out for color sources in higher representations, and for excited states. In particular, a series of works~\cite{DelDebbio:2001sj, Meyer:2004hv, Lucini:2004my, Lucini:2000qp, DelDebbio:2001kz, DelDebbio:2002yp} investigated the behavior of the potential between static sources in representations of various $N$-alities (i.e., with different transformation properties with respect to the $\Z_N$ center of the gauge group). Part of the motivation to investigate the confining potential for sources in representations of different $N$-ality stems from the observation that, in supersymmetric $\SU(N)$ gauge theories, the associated string tensions $\sigma_k$ (for stable strings) are related to each other by the sine-formula~\cite{Douglas:1995nw, Hanany:1997hr, Witten:1997sc, Herzog:2001fq, Armoni:2003nz}:
\begin{equation}
\label{sine_formula}
\frac{\sigma_k }{ \sigma_1} = \frac{ \sin(k\pi/N) }{\sin(\pi/N)},
\end{equation}
where $\sigma_1$ denotes the fundamental string tension. This relation can be compared with the one that can be obtained from Casimir scaling~\cite{Ambjorn:1984mb}, which, for the totally antisymmetric irreducible representation of $N$-ality $k$, implies:
\begin{equation}
\label{Casimir_k_strings}
\frac{\sigma_k }{ \sigma_1} = \frac{ k(N-k) }{ N-1 }.
\end{equation}
In the large-$N$ limit at fixed $k$, both eq.~(\ref{sine_formula}) and eq.~(\ref{Casimir_k_strings}) are consistent with the expectation from factorization: $\sigma_k / \sigma_1 \to k$. In particular, the large-$N$ behavior of $\sigma_k/\sigma_1$ has been discussed in refs.~\cite{Armoni:2006ri,Greensite:2011gg}. Lattice results show that at finite $N$ the energy of $k$-strings turns out to be lower than $k$ times the energy of a fundamental string, and, in particular, appears to favor sine scaling~\cite{DelDebbio:2001sj, Lucini:2000qp, DelDebbio:2001kz} or to lie in between the predictions of eq.~(\ref{sine_formula}) and eq.~(\ref{Casimir_k_strings})~\cite{Meyer:2004hv, Lucini:2004my}. A summary of the situation is presented in the right panel of fig.~\ref{fig:loop_mass}: the different conclusion is not due to the numerical data, which are compatible, but to the extrapolation to the continuum limit. To resolve the apparent discrepancy, more simulations closer to the continuum limit need to be performed. On the other hand, for different irreducible representations of the same $N$-ality, the lattice results are broadly consistent with Casimir scaling, for $\SU(3)$ Yang-Mills theory~\cite{DelDebbio:2003tk} as well as for $N>3$~\cite{Lucini:2004my}.

Recently, a lot of work has been devoted to the study of excited string states in large-$N$ Yang-Mills theories (see ref.~\cite{Teper:2009uf} and references therein for a thorough discussion of the topic): this is important for understanding the nature of the effective action describing the low-energy dynamics of confining flux tubes. A large number of lattice studies (recently reviewed by Mykk\"anen in ref.~\cite{Mykkanen:2012dv}) show that the main features of the dependence of the torelon mass on its length, or, equivalently, of the ground-state quark-antiquark potential on the distance between the color sources, are consistent with the hypothesis that the flux tube fluctuates like a bosonic string described by the Nambu-Goto action~\cite{Nambu:1974zg, Goto:1971ce}, which is simply proportional to the area of the string world-sheet.\footnote{It is well-known that (unless the number of spacetime dimensions is $D=26$) the Nambu-Goto action is not a consistent string action at the quantum level, due, for example, to the Weyl anomaly, and to the existence of tachyonic states~\cite{Lovelace:1970sj, Goddard:1973qh}. Nevertheless, these problems are not relevant for the case of an effective, low-energy model, as discussed in the present context.} Due to the simplicity of the string action, the Nambu-Goto spectrum can be computed exactly: for an open string of length $L$~with fixed ends, it reads~\cite{Arvis:1983fp, Alvarez:1981kc}: 
\begin{equation}
\label{open_string_spectrum}
E_n(L) = \sigma L \sqrt{ 1 + \frac{2 \pi}{ \sigma L^2} \left( n - \frac{D-2}{24} \right)  } \;\; ,  \;\;\;\; n \in \N
\end{equation}
(where $D$ denotes the number of spacetime dimensions, and $\sigma$ is the string tension). Expanding the lowest-lying energy around the $\sigma L^2 \to \infty$ limit yields (besides the classical, linear term $\sigma L$) the L\"uscher term for the open string:
\begin{equation} 
\label{Cornell_potential}
E_0(L) = \sigma L - \frac{\pi}{24L}(D-2) + O( L^{-3} ).
\end{equation}
Similarly, for a closed string, the Nambu-Goto spectrum reads (see, e.g., ref.~\cite{Billo:2006zg}): 
\begin{equation}
\label{closed_string_spectrum}
E_{n_l, n_r}(L) = \sigma L \sqrt{ 1 + \frac{4 \pi }{\sigma L^2} \left( n_l + n_r - \frac{D-2}{12} \right)  + \frac{4 \pi^2 (n_l - n_r)^2 }{\sigma^2 L^4}}
\end{equation}
(where $n_l$ and $n_r$ are non-negative integers denoting the number of left- and right-moving modes, respectively), which, when expanded around the long-string limit in four spacetime dimensions, reproduces eq.~(\ref{torelon_mass}) for the ground state.

During the last decade, various theoretical works tried to determine the form of the subleading (in an expansion in terms of the inverse of the string length) terms which the effective string action describing a confining flux tube should include, using arguments related to open-closed string duality~\cite{Luscher:2004ib} and to Lorentz-Poincar\'e symmetry and its non-linear realization~\cite{Meyer:2006qx, Aharony:2009gg, Aharony:2010cx, Aharony:2010db, Aharony:2011gb, Billo:2012da, Gomis:2012ki, Gliozzi:2012cx} (a different approach was proposed by Polchinski and Strominger in ref.~\cite{Polchinski:1991ax}---see refs.~\cite{Drummond:2004yp, Aharony:2011ga, Dubovsky:2012sh} for a discussion of the relation between the two approaches). For a string of length $L$ in four spacetime dimensions, this shows that the string energy levels deviate from the Nambu-Goto spectrum at order $1/L^4$ for an open string~\cite{Aharony:2010cx}, whereas for a closed string the deviations occur at order $1/L^5$ for excited states, but only at order $1/L^7$ for the ground state~\cite{Aharony:2010cx}.

\begin{figure}[-t]
\centerline{\includegraphics[width=0.8\textwidth]{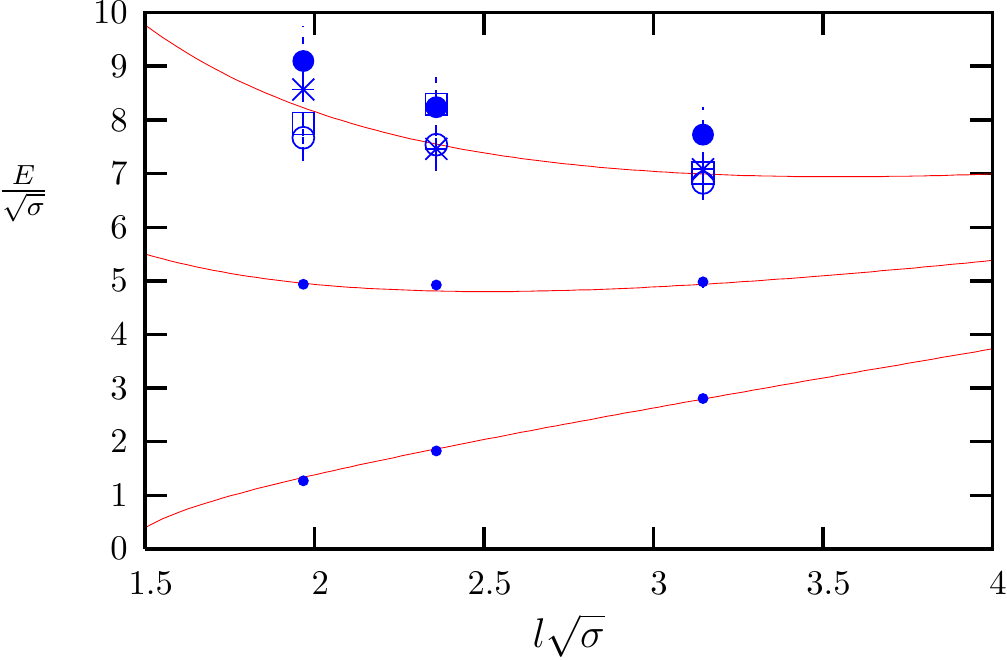}}
\caption{Closed string spectrum in $\SU(5)$ Yang-Mills theory, taken from ref.~\cite{Athenodorou:2010cs}, in comparison with the predictions for the Nambu-Goto  model.\label{fig:4D_strings}}
\end{figure}

From the numerical point of view, the spectrum of closed strings in four spacetime dimensions was studied in ref.~\cite{Athenodorou:2010cs}, using a sophisticated variational technique involving about 700 operators, in $\SU(3)$, $\SU(5)$ and $\SU(6)$ Yang-Mills theories. The results showed general agreement with the Nambu-Goto spectrum, even down to surprisingly short values of $L$, of the order of the inverse square root of the string tension---although some discrepancies were also reported. An example of these results is shown in fig.~\ref{fig:4D_strings}. As discussed in ref.~\cite{Aharony:2010db}, the precision of these very challenging lattice computations is still insufficient to clearly distinguish deviations from the Nambu-Goto spectrum in the range of $L$ values within the convergence domain of the expansion in inverse powers of $L$.

\subsubsection{Hadronic spectrum}
\label{subsubsec:4D_spectrum}

Having ascertained that non-Abelian gauge theories are confining in the 't~Hooft limit, i.e., that a linearly rising potential develops between a static quark and antiquark, the next problem to be addressed non-perturbatively, via lattice simulations, is determining the values of the hadron masses at large $N$. In the literature, there exist computations of the masses of glueballs~\cite{Teper:1998kw, Lucini:2001ej, Lucini:2004my, Meyer:2004jc, Lucini:2010nv}, of mesons~\cite{DelDebbio:2007wk, Bali:2008an, Bali:2013kia, Hietanen:2009tu} and baryons~\cite{DeGrand:2012hd}---all in the quenched approximation.

\begin{figure}[-t]
\centerline{\includegraphics[width=0.5\textwidth]{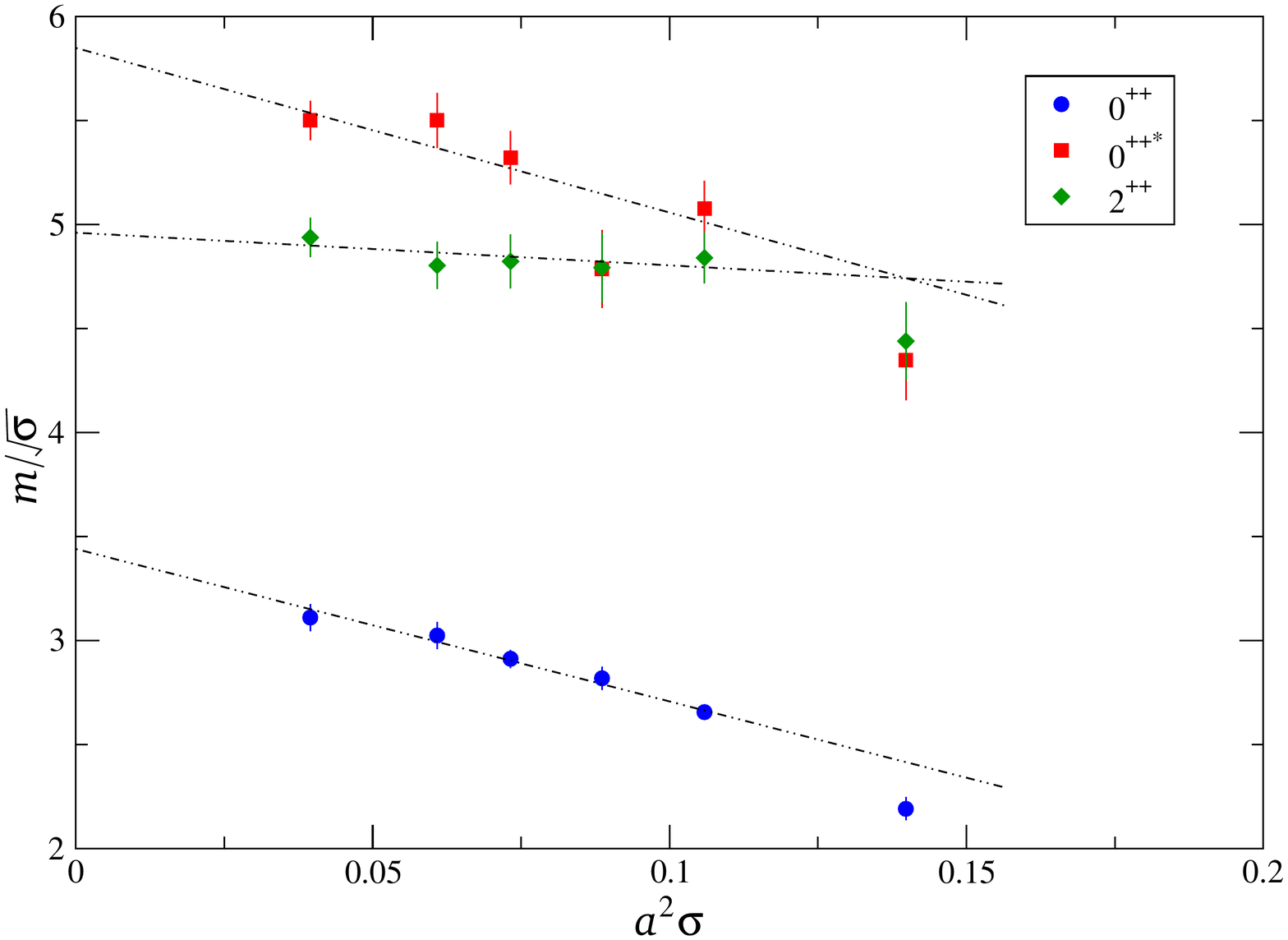} \hfill \includegraphics[width=0.5\textwidth]{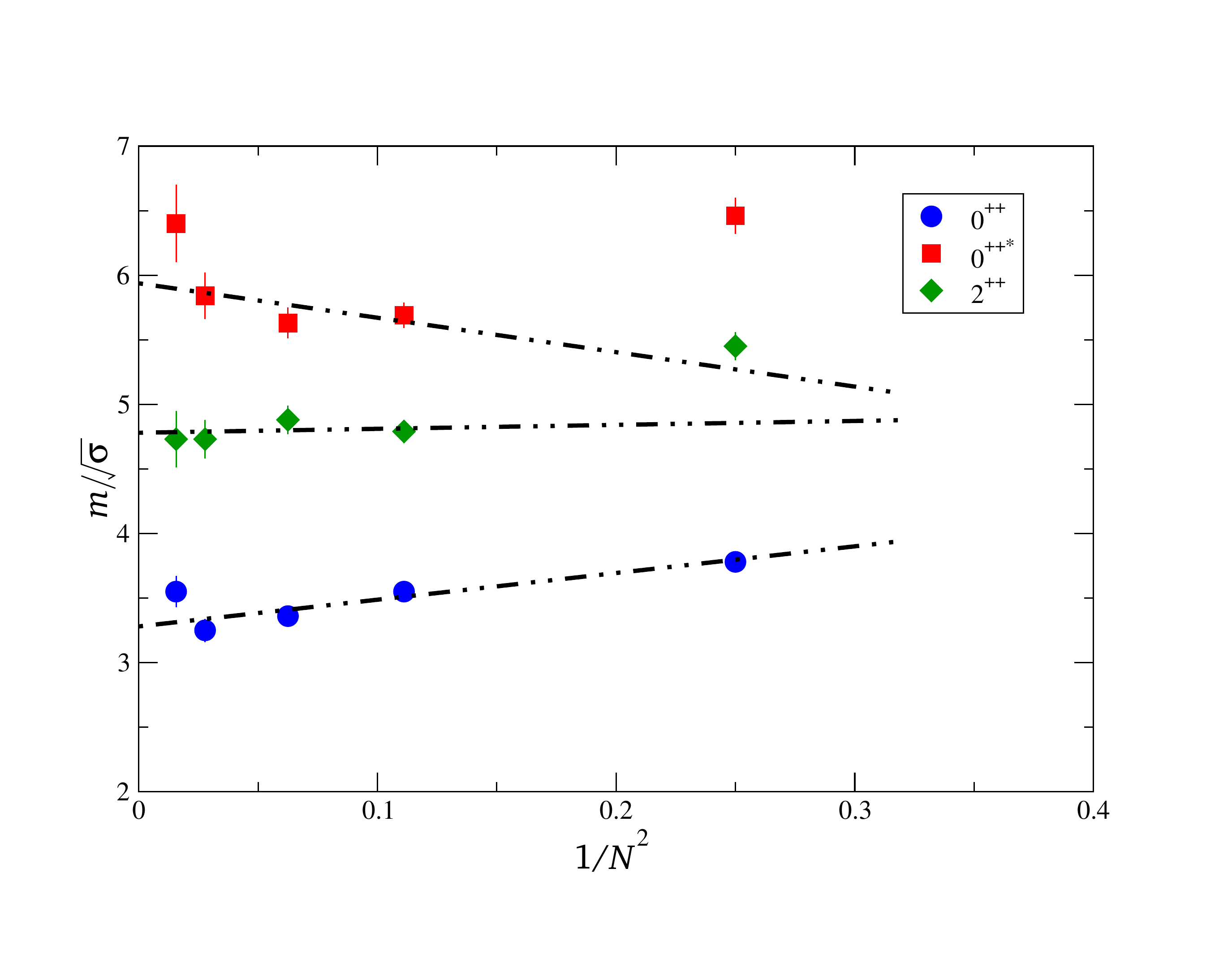}}
\caption{Left panel: Extrapolation to the continuum limit of the lowest-lying glueball spectrum in $\SU(4)$ gauge theory. Right panel: Extrapolation of the lowest-lying glueball spectrum for $N \to \infty$, using only the leading finite-$N$ correction. Both figures are plots of results published in ref.~\cite{Lucini:2004my}.\label{fig:4D_LTW}}
\end{figure}

In general, the dependence of glueball masses on $N$ turns out to be very smooth: typically, for each physical state, the lattice results obtained in Yang-Mills theories at different values of $N$ can be nicely fitted to a constant plus a term linear in $1/N^2$ (as expected theoretically); within the precision of the numerical results, this holds all the way down to $N=3$---or even $N=2$, in some cases. Results in the continuum limit for the lowest-lying states were presented in refs.~\cite{Lucini:2001ej,Lucini:2004my}. In particular, the calculation of ref.~\cite{Lucini:2004my} leads to the following results:
\begin{eqnarray}
 & & \frac{m_{0^{++}}}{\sqrt{\sigma}} = 3.28(8) + \frac{2.1(1.1)}{N^2} \ , \\
 & & \frac{m_{0^{++\star}}}  {\sqrt{\sigma}} = 5.93(17) - \frac{2.7(2.0)}{N^2} \ , \\
 & & \frac{m_{2^{++}}}{\sqrt{\sigma}} = 4.78(14) + \frac{0.3(1.7)}{N^2} \ .
\end{eqnarray}
The quality of the data and the extrapolation to the continuum limit is shown in fig.~\ref{fig:4D_LTW}.

\begin{figure}[-t]
\centerline{\includegraphics[width=0.8\textwidth]{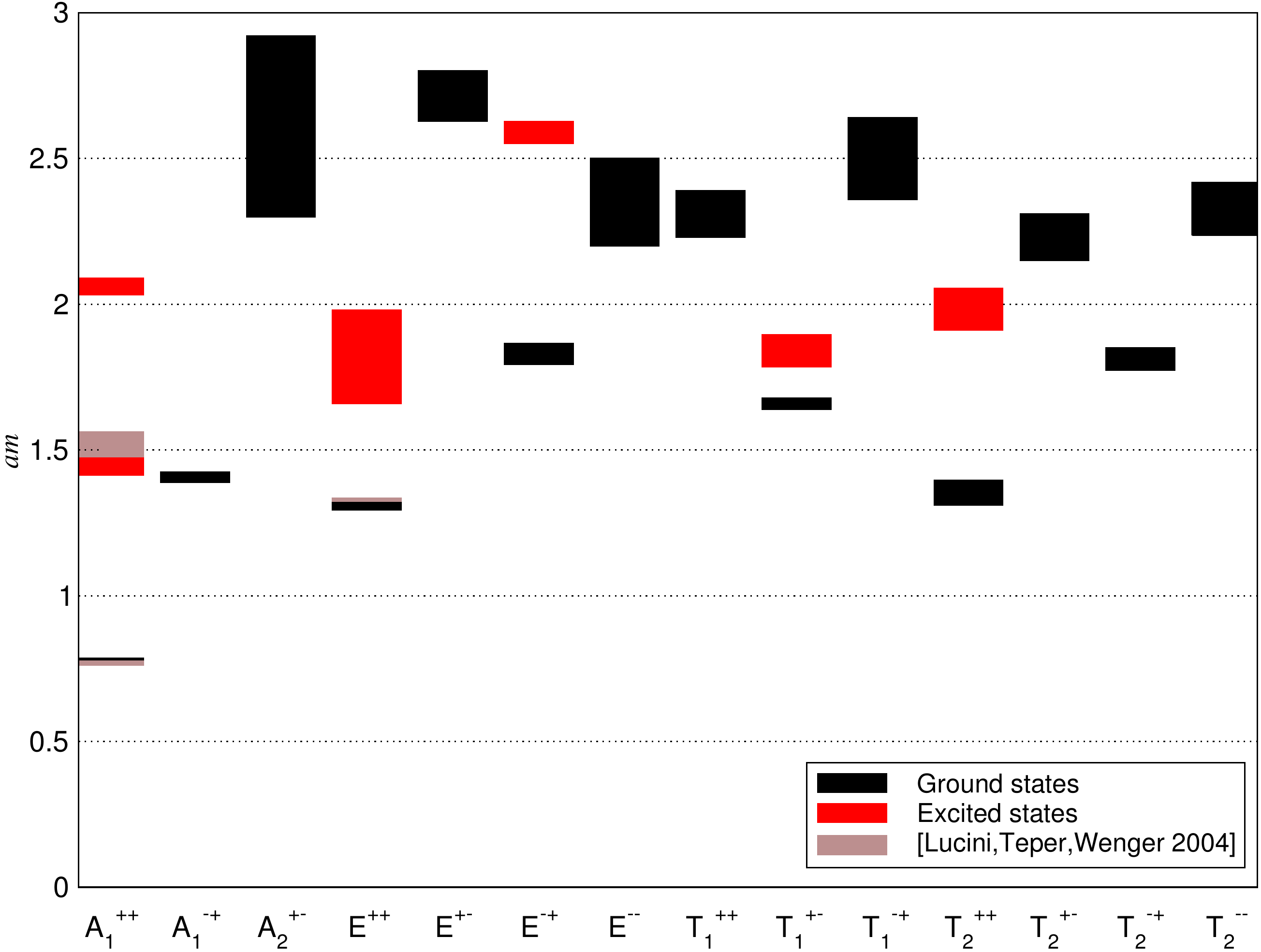}}
\caption{The spectrum of glueballs in the large-$N$ limit of $\SU(N)$ Yang-Mills theory, as determined in ref.~\cite{Lucini:2010nv}. The plot shows the results (at a fixed, finite value of the lattice spacing) for the ground state and some of the excitations, in various channels corresponding to different irreducible representations of the cubic group, and different values of the parity and charge-conjugation quantum numbers. The figure also shows a comparison with the results obtained in ref.~\cite{Lucini:2004my} for the ground state and for the first excited $J^{PC}=0^{++}$ glueball, and for the $2^{++}$ ground state.\label{fig:4D_glueballs}}
\end{figure}

The most recent computation of the glueball spectrum was reported in ref.~\cite{Lucini:2010nv}, and is summarized in fig.~\ref{fig:4D_glueballs}, showing the masses of the ground state (and of some excited states) in the different channels associated with the irreducible representations of the cubic group\footnote{The computation is done on a hypercubic lattice, at finite lattice spacing.} and distinguishing states that are ``even'' and ``odd'' under parity and/or charge conjugation. For comparison, the figure also shows the mass of the ground state and of the first radial excitation in the scalar channel, and the ground state in the tensor channel, taken from an earlier work~\cite{Lucini:2004my}. In addition, a lattice study of the relation between mass and spin of the various glueballs in $\SU(3)$ and $\SU(8)$ Yang-Mills theories has also been carried out, in ref.~\cite{Meyer:2004jc}.

\begin{figure}[-t]
\centerline{\includegraphics[width=0.8\textwidth]{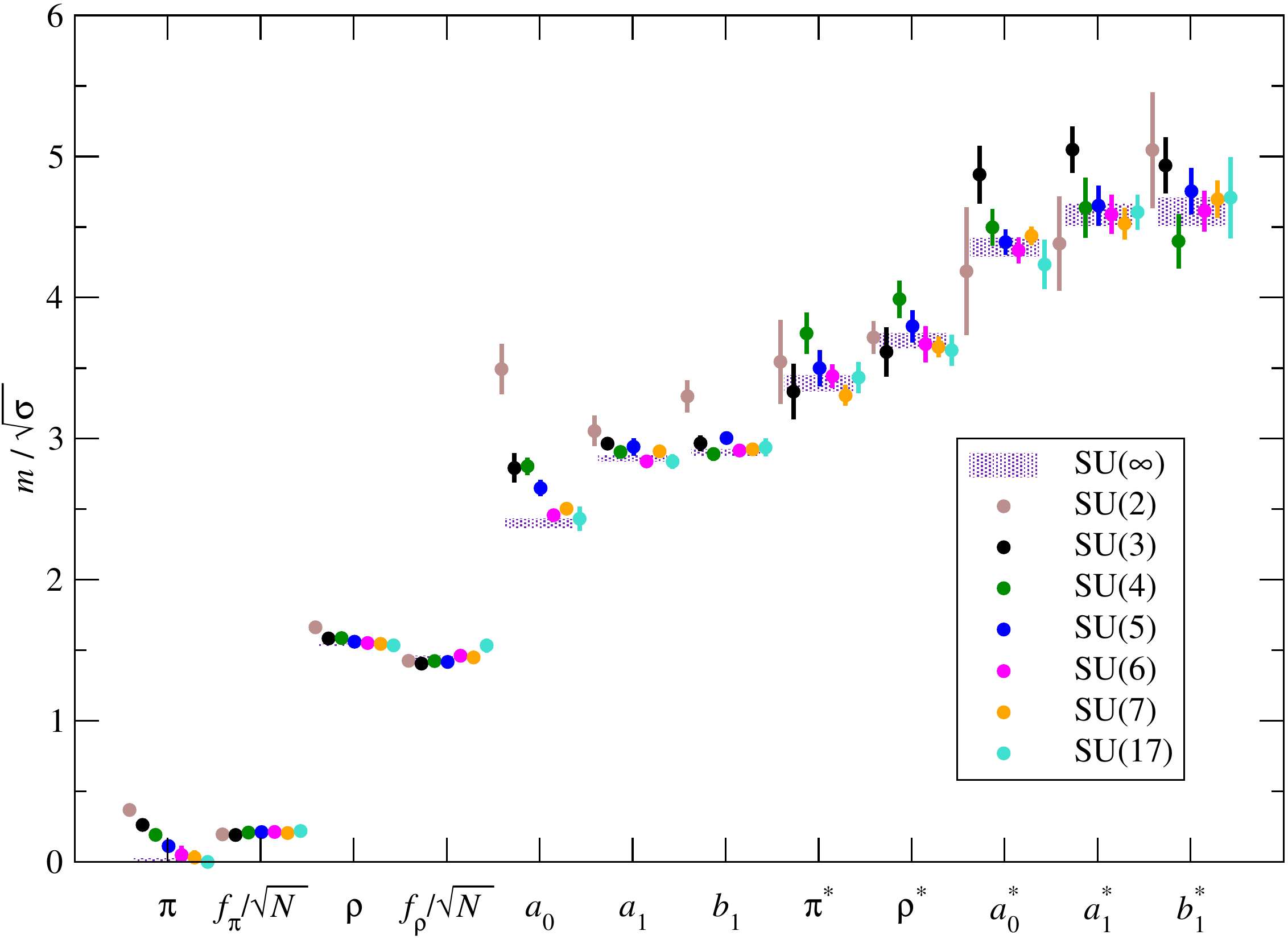}}
\caption{Results from a quenched computation of the mesonic spectrum at large $N$~\cite{Bali:2013kia}: the figure shows the masses (in units of the square root of the string tension) of various meson states, and the decay constants of the pion and the $\rho$ meson (divided by $\sqrt{N}$), as obtained from simulations for different numbers of colors, in the chiral limit. The values extrapolated to $N \to \infty$ are shown by the horizontal bands.\label{fig:meson_spectrum}}
\end{figure}

Similarly, a smooth dependence on $N$ has also been observed for the meson spectrum~\cite{DelDebbio:2007wk, Bali:2008an, Bali:2013kia}: this is shown in fig.~\ref{fig:meson_spectrum}, reporting the results from ref.~\cite{Bali:2013kia}, in which the masses of various meson states in the chiral limit are shown---together with the pion and $\rho$ decay constants (rescaled by $\sqrt{N}$). Symbols of different colors refer to different values of $N$, and the band denotes the extrapolation to the 't~Hooft limit.

These results are consistent with the studies previously presented in refs.~\cite{DelDebbio:2007wk, Bali:2008an}, which confirmed that the values of the pion and $\rho$ masses in the large-$N$ limit are close to those in the real world: for example, combining original results with those of~\cite{DelDebbio:2007wk},  ref.~\cite{Bali:2008an} reported:
\begin{equation}
\lim_{N \to \infty} \frac{m_\rho}{\sqrt{\sigma}} = 1.79(5)
\end{equation}
as an estimate for the ratio in the continuum limit. This result is consistent with the value (approximately equal to $1.75$) obtained from the experimentally measured mass of the $\rho$ meson, and from a reasonable phenomenological estimate\footnote{The value of the string tension $ \sigma $ can be estimated from the analysis of Regge trajectories in experimentally observed meson states, and from studies of charmonium and bottomonium spectra~\cite{Eichten:1976jk, Quigg:1979vr, Eichten:1979ms}, which suggest values in the range between $ (400~\mbox{MeV})^2 $ and $(450~\mbox{MeV})^2$---see refs.~\cite{Bali:2000gf, Bali:1993zj} for a discussion.} of $\sigma$. Furthermore, the dependence of the $\rho$ mass on the pion mass appears to be approximately quantitatively consistent with the holographic computations discussed in ref.~\cite{Erdmenger:2007cm}.

We note that, by contrast, ref.~\cite{Hietanen:2009tu} reported very different results, indicating that the mass of the $\rho$ meson in the large-$N$ limit could be about twice as large as in the real world. A possible explanation of the discrepancy between the results of ref.~\cite{Hietanen:2009tu} and those of refs.~\cite{DelDebbio:2007wk, Bali:2008an} in terms of an unexpectedly large finite-$N$ correction (the simulations presented in refs.~\cite{DelDebbio:2007wk, Bali:2008an} were limited to $N \le 6$, whereas those in ref.~\cite{Hietanen:2009tu} were done for $N=17$ and $19$) has now been ruled out by the results of the most recent study~\cite{Bali:2013kia}, which also included simulations for $N$ up to $17$, and found consistency with refs.~\cite{DelDebbio:2007wk, Bali:2008an}. Perhaps, a more likely interpretation of the difference between the results in ref.~\cite{Hietanen:2009tu} and those in the other three works~\cite{DelDebbio:2007wk, Bali:2008an, Bali:2013kia} could be the one already suggested in ref.~\cite{Teper:2009uf}, which noted that the computation presented in ref.~\cite{Hietanen:2009tu} was based on the evaluation of quark propagators in momentum space, and was carried out for only a few small momenta, so it may have suffered from contamination from some excited state. 

\begin{figure}[-t]
\centerline{\includegraphics[width=0.8\textwidth]{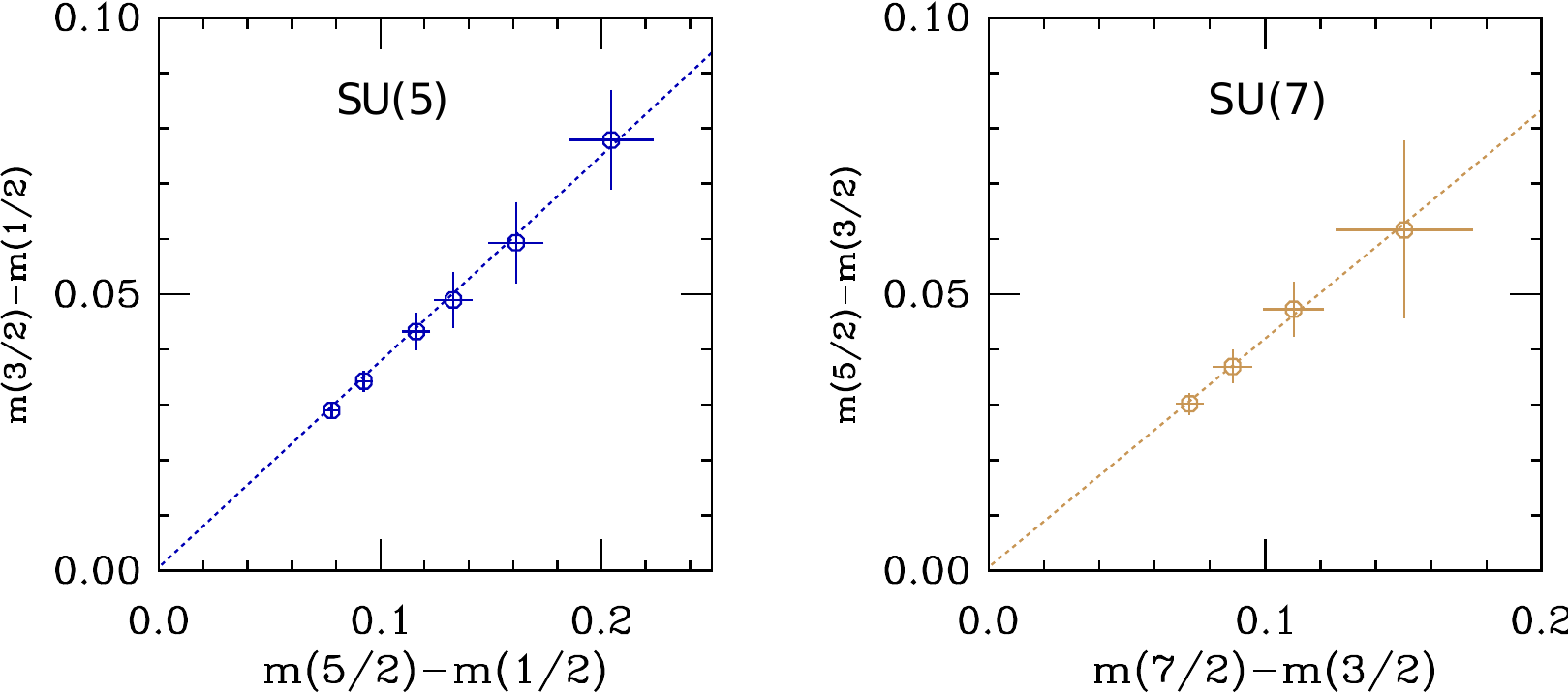}}
\caption{A recent quenched computation~\cite{DeGrand:2012hd} of the baryon spectrum at large $N$ reported evidence for a rotor-type spectrum in the masses of different states as a function of their spin, as predicted in refs.~\cite{Adkins:1983ya, Jenkins:1993zu}. The plot, taken from ref.~\cite{DeGrand:2012hd}, shows differences between the masses of states of various spin, plotted against each other, in QCD with $N=5$ (left panel) and $N=7$ colors (right panel).\label{fig:baryon_spectrum}}
\end{figure}

The recent article~\cite{DeGrand:2012hd} presented a quenched computation of the baryonic spectrum for large values of $N$, up to $7$; this study was restricted to odd values of $N$---so that the baryons are fermions, like in the real world. Once again, the results confirmed the theoretical expectations for the dependence of physical quantities on the number of colors (including, in particular, baryon masses approximately linearly increasing with $N$), up to numerically modest relative finite-$N$ corrections. In addition, this study also revealed that states of larger spin are heavier, and, more precisely, the dependence of the mass of a baryon on its spin is consistent with a rotor spectrum, eq.~(\ref{rotor_spectrum}), as predicted in refs.~\cite{Adkins:1983ya, Jenkins:1993zu}. This can be seen in fig.~\ref{fig:baryon_spectrum}, showing the linear relation between various mass splittings for baryon states of different spin, on each other. Another interesting comparison of lattice baryon spectroscopy and large-$N$ predictions was reported in ref.~\cite{Jenkins:2009wv}: this work analyzed a set of configurations in $N=3$~QCD (including dynamical fermions)~\cite{WalkerLoud:2008bp}, and investigated how the baryon mass splitting predicted in a $1/N$-expansion~\cite{Jenkins:1995td} compares with lattice results. Using different values of the quark mass to vary the flavor-symmetry breaking, the authors of ref.~\cite{Jenkins:2009wv} were able to show that the results from lattice simulations of QCD with $N=3$~colors satisfy the expected $1/N$-flavor scaling laws. A related work was presented in ref.~\cite{WalkerLoud:2011ab}.

\subsubsection{Topological properties at large $N$}
\label{subsubsec:4D_topology}

\begin{figure}[-t]
\centerline{\includegraphics[width=0.5\textwidth]{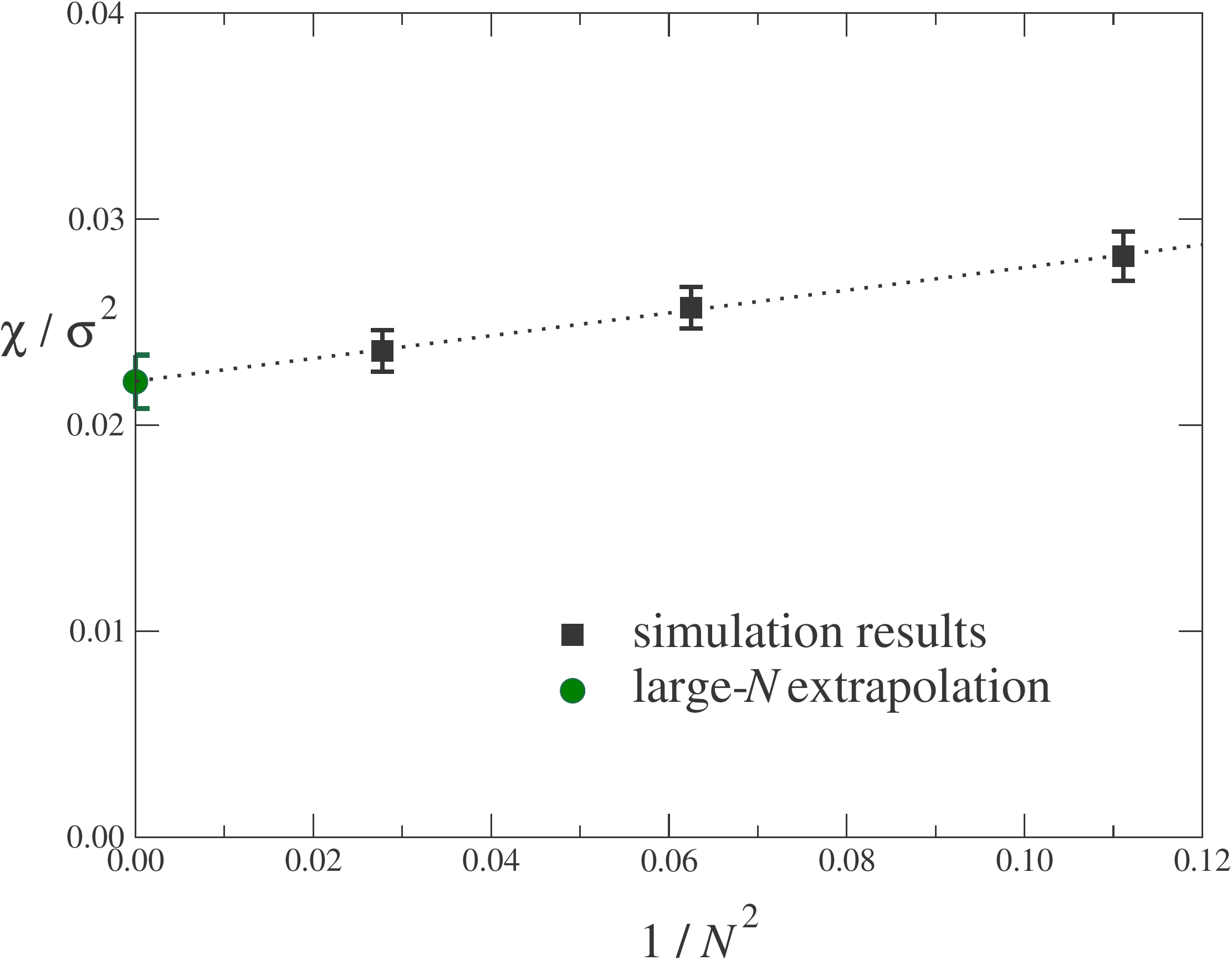} \hfill \includegraphics[height=0.5\textheight]{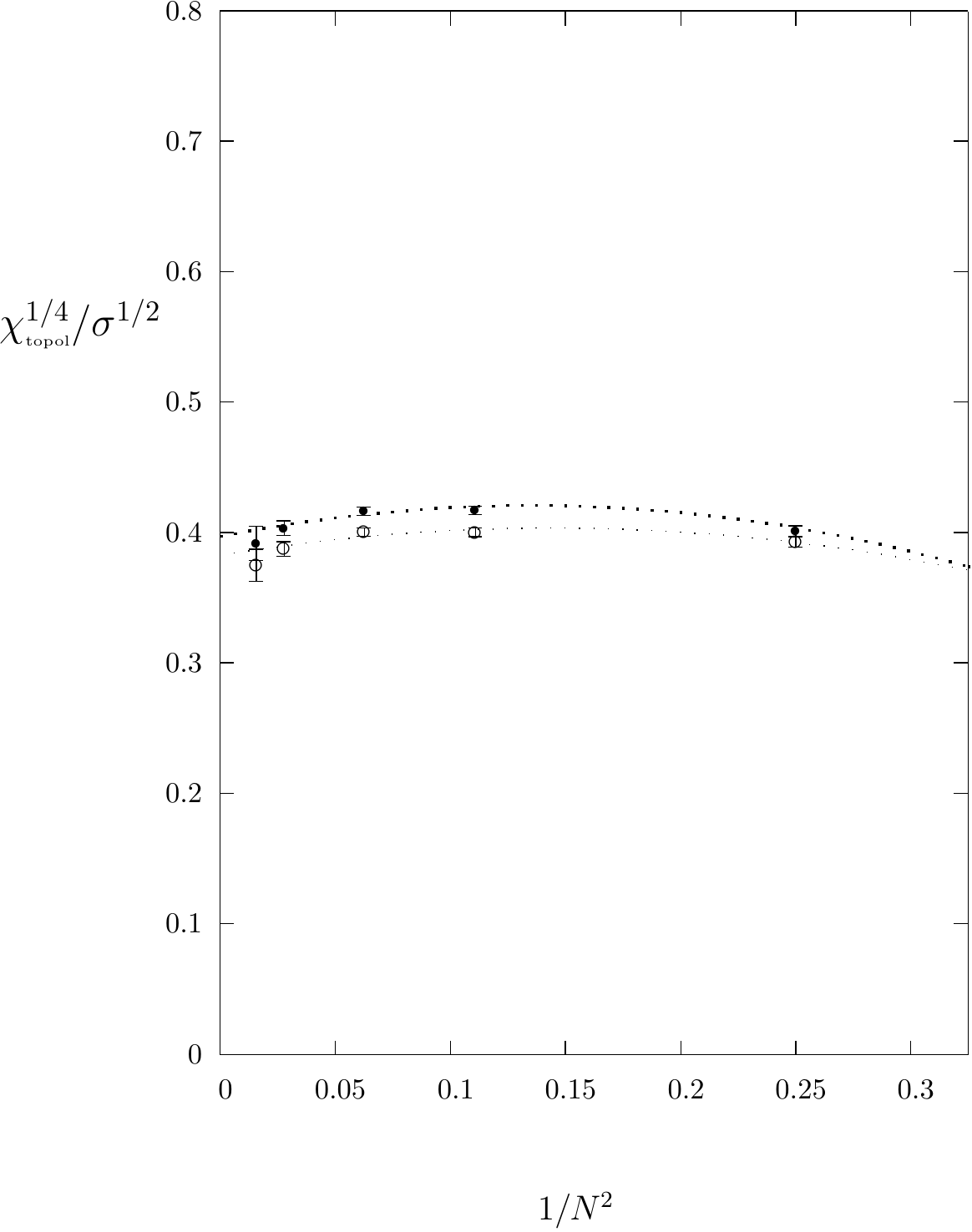}}
\caption{Left panel: Results for the topological susceptibility (in units of the square of the string tension), from the simulations performed in ref.~\cite{DelDebbio:2002xa}, and their extrapolation to the large-$N$ limit. Right panel: The fourth root of the topological susceptibility (in units of the square root of the string tension) in $\SU(2)$, $\SU(3)$, $\SU(4)$, $\SU(6)$ and $\SU(8)$ Yang-Mills theory at zero temperature, as determined in ref.~\cite{Lucini:2004yh} using a real- (open symbols) or integer-valued (filled symbols) definition of the lattice topological charge operator. The corresponding quadratic fits, as a function of $1/N^2$, are also shown.\label{fig:Luigi_and_fig1_heplat0401028}}
\end{figure}

The topological properties in large-$N$ Yang-Mills theories have been studied numerically in various works~\cite{Lucini:2001ej, Lucini:2001rc, Cundy:2002hv, DelDebbio:2002xa, Lucini:2004yh}, and are summarized in the review~\cite{Vicari:2008jw}. In particular, ref.~\cite{Lucini:2001ej} considered $\SU(N)$ Yang-Mills theory with $N=2$, $3$, $4$ and $5$ colors, and found that, in agreement with the theoretical expectation~\cite{Witten:1978bc}, the number density $d$ of instantons in the Yang-Mills vacuum is exponentially suppressed when $N$ increases---see eq.~(\ref{small_size_instanton_density})---and that the density of instantons of small size $\rho$ at fixed $N$ scales compatibly with:
\begin{equation}
\label{small_size_instanton_density_at_fixed_N}
d(\rho) \propto \rho^{\frac{11}{3}N-5},
\end{equation}
as one can predict using perturbation theory (which should become reliable for $\rho \to 0$). In addition, ref.~\cite{Lucini:2001ej} also found that the topological susceptibility tends to a non-vanishing value for $N \to \infty$:
\begin{equation}
\label{topsusc_Biagio_Mike}
\frac{ \topsusc^{1/4} }{ \sigma^{1/2} } = 0.376(20) + \frac{0.43(10)}{N^2}.
\end{equation}
Fig.~\ref{fig:Luigi_and_fig1_heplat0401028} shows the results from two subsequent, similar studies~\cite{DelDebbio:2002xa, Lucini:2004yh}, which found results consistent with each other and with ref.~\cite{Lucini:2001ej}. In particular, ref.~\cite{DelDebbio:2002xa} reported simulations for $N=3$, $4$ and $6$ colors, yielding: $\topsusc/\sigma^2 = 0.0221(14) + 0.055(18)/N^2$, which corresponds to:
\begin{equation}
\label{topsusc_Luigi_Haris_Ettore}
\frac{ \topsusc^{1/4} }{ \sigma^{1/2} } =  0.386(6) + \frac{0.24(8)}{N^2},
\end{equation}
while the authors of ref.~\cite{Lucini:2004yh} studied the theories with $N=2$, $3$, $4$, $6$ and $8$ colors, obtaining:
\begin{equation}
\label{topsusc_Biagio_Mike_Urs}
\frac{ \topsusc^{1/4} }{ \sigma^{1/2} } = 0.382(7) + \frac{0.30(13)}{N^2} - \frac{1.02(42)}{N^4}, \qquad \mbox{with: $\chi^2$/d.o.f.}=1.3,
\end{equation}
from the real values for the topological charge computed using a twisted plaquette operator (open symbols in fig.~\ref{fig:Luigi_and_fig1_heplat0401028})---and very similar results with a definition of the topological charge truncated to integer values (filled symbols in fig.~\ref{fig:Luigi_and_fig1_heplat0401028}). To summarize, these studies indicate that the large-$N$ limit of the topological susceptibility is non-vanishing, and around $(170~\mbox{MeV})^4$, rather close to its value in the $\SU(3)$ theory (which is about $(180~\mbox{MeV})^4$).

The relation between the topological charge of the gauge fields (evaluated on ``smoothed'' gluon field configurations) and the chirality of the quark fields, as probed by the low-lying eigenmodes of the overlap Dirac operator~\cite{Neuberger:1997fp}, was investigated in ref.~\cite{Cundy:2002hv}, which found that the numerical results for these two quantities become consistent when $N$ gets large, and that the low-lying eigenmodes of the Dirac operator tend to lose their chirality properties and to become less and less localized for $N \to \infty$, suggesting that instantons do not survive and are not responsible for chiral symmetry breaking in this limit.

As discussed in subsection~\ref{subsec:topology}, the topological susceptibility is closely related to the dependence of non-Abelian gauge theory on the $\theta$-term, whose lattice investigation in the large-$N$ limit has been carried out in various works (including, e.g., refs.~\cite{DelDebbio:2002xa, DelDebbio:2006df}), and is reviewed in ref.~\cite{Vicari:2008jw}---see also refs.~\cite{D'Elia:2003gr, Giusti:2007tu, Panagopoulos:2011rb} and references therein for analogous studies in the $\SU(3)$ theory. It should be pointed out that the formulation of the theory with a topological $\theta$-term on a Euclidean lattice suffers from a sign problem which prevents its direct numerical simulation, due to the complex nature of the action. A possible way to  circumvent this problem consists in studying the Taylor expansions of the physical quantities of interest around $\theta=0$. In particular, following this strategy, the authors of ref.~\cite{DelDebbio:2006df} studied the $\theta$-dependence of the string tension and of the mass gap in $\SU(3)$, $\SU(4)$ and $\SU(6)$ Yang-Mills theory, finding results suggesting that such dependence vanishes in the 't~Hooft limit, in agreement with the theoretical expectation that the relevant parameter at large $N$ should be $\theta/N$.

\subsubsection{Large-$N$ QCD at finite temperature}
\label{subsubsec:4D_finite_T}

In addition to these studies at zero temperature, in the literature there exist a number of works about large-$N$ lattice gauge theories at finite temperature~\cite{Lucini:2002ku, Lucini:2003zr, Lucini:2005vg, Bursa:2005yv, Bringoltz:2005rr, Bringoltz:2005xx, deForcrand:2005rg, Panero:2008mg, Panero:2009tv, Datta:2009jn, Datta:2010sq, Mykkanen:2012ri, Lucini:2012wq}. Typically, in these simulations the temperature (which, in Euclidean thermal field theory, is just the inverse of the system size along a compactified direction: $T=1/L$) is varied by changing the value of the lattice spacing $a$, which, in turn, means changing the parameter $\beta=2N/g_0^2=2N^2/\lambda_0$ in front of the Wilson gauge action, i.e. the inverse of the bare gauge coupling.\footnote{A different approach, based on varying the temperature by changing the number of lattice sites along the compactified direction, is also possible~\cite{Umeda:2008bd}, however it has the practical disadvantage that it does not allow one to vary the temperature continuously.} As mentioned above, in order to probe the region of couplings analytically connected to the continuum limit, the numerical simulations have to be run at sufficiently small values of the bare lattice 't~Hooft coupling $\lambda_0$. However, it is remarkable that interesting implications for $\SU(N)$ thermodynamics can also be obtained from lattice strong-coupling techniques~\cite{Langelage:2010yn}.

In four spacetime dimensions, the simulation results show that all $\SU(N)$ Yang-Mills theories have a deconfinement transition at a finite critical temperature, which---when expressed in physical units---is in the range between $250$ and $300~\mbox{MeV}$. The transition turns out to be of second order for the $\SU(2)$ gauge theory~\cite{Engels:1990vr, Fingberg:1992ju, Engels:1994xj}---and its critical exponents are consistent with those of the three-dimensional Ising model~\cite{condmat0012164}, in agreement with the general conjecture that, for continuous transitions, the critical behavior of a gauge theory in $D$ spacetime dimensions is characterized by the critical exponents of statistical spin model in $D-1$ dimensions, with the order parameter taking values in the center of the gauge group~\cite{Svetitsky:1982gs}---, while it is a weak first-order one in the $\SU(3)$ theory~\cite{Boyd:1996bx, Borsanyi:2012ve} and a stronger first-order one for all $N \ge 4$~\cite{Lucini:2002ku, Lucini:2003zr, Lucini:2005vg, Datta:2009jn}. By comparison, note that, as mentioned in subsec.~\ref{subsec:phase_diagram}, lattice simulations of QCD with $n_f=2+1$ flavors of dynamical quarks, for physical values of the quark masses, predict the deconfinement transition to be a crossover,\footnote{In QCD with dynamical quarks of finite mass, there exists no \emph{bona fide} order parameter for the finite-temperature transition, since center symmetry is explicitly broken by the existence of quarks, and chiral symmetry is explicitly broken by their finite mass.} which takes place in a temperature range between $150$ and $170~\mbox{MeV}$~\cite{Aoki:2006br, Aoki:2009sc, Bazavov:2010sb, Bazavov:2011nk},\footnote{Previous studies~\cite{Bernard:2004je, Cheng:2006qk, Cheng:2009zi, Bazavov:2009zn} reported somewhat larger values (up to around $190~\mbox{MeV}$) for the crossover pseudo-critical temperature, which sparked off some heated but constructive debate in the community~\cite{Karsch:2007dt, Fodor:2007sy}. However, it seems that this disagreement can be explained in terms of the uncertainties related to the extrapolation to physical quark masses and to the continuum limit~\cite{Huovinen:2009yb}, and, more recently, one of the two collaborations presented a new study~\cite{Bazavov:2011nk}, based on a highly improved lattice formulation for the quark fields~\cite{Follana:2006rc}, in which they found results compatible with those obtained by the other group~\cite{Aoki:2006br, Aoki:2009sc}.} and appears to be compatible with experimental evidence~\cite{Andronic:2009qf}. Further details about the lattice investigation of the QCD equation of state can be found in refs.~\cite{Philipsen:2012nu, DeTar:2009ef}.

\begin{figure}[-t]
\centerline{\includegraphics[width=0.8\textwidth]{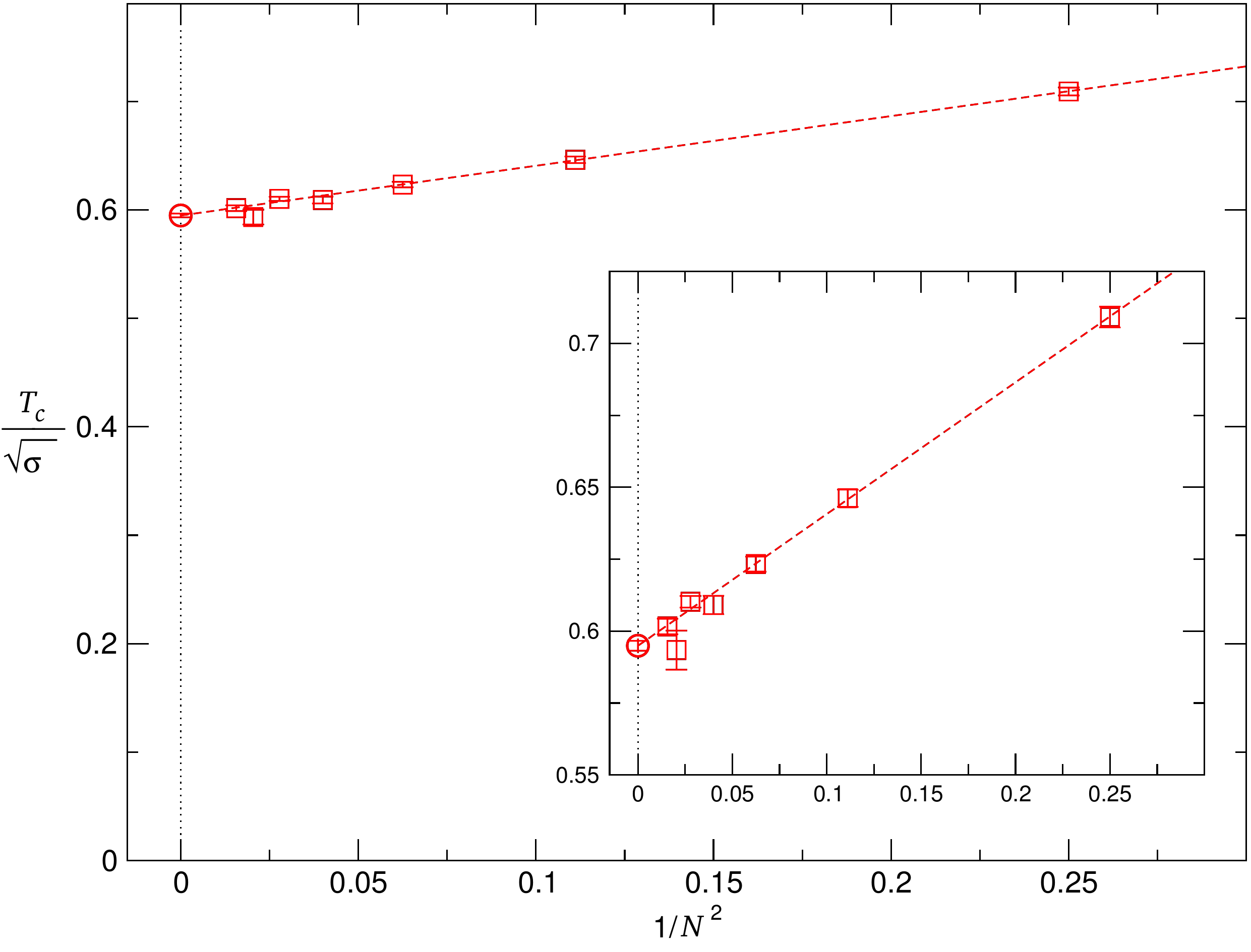}}
\caption{Dependence of the deconfinement critical temperature (in units of the square root of the string tension) on the number of colors in $\SU(N)$ Yang-Mills theory. This figure, taken from ref.~\cite{Lucini:2012wq} also shows the interpolation of the numerical results described by eq.~(\protect\ref{large_N_Tc}). The inset shows a zoomed view of the simulation results.\label{fig:Tc_over_root_sigma}}
\end{figure}

In large-$N$ pure Yang-Mills theory, the deconfinement critical temperature $T_c$~can be determined unambiguously, and the most recent, high-precision results show a mild dependence of this quantity on the number of colors (see fig.~\ref{fig:Tc_over_root_sigma}, taken from ref.~\cite{Lucini:2012wq}), which can be summarized by the relation:
\begin{equation}
\label{large_N_Tc}
\frac{T_c}{\sqrt{\sigma}} = 0.5949(17) + \frac{0.458(18)}{N^2}, \qquad \mbox{with: $\chi^2$/d.o.f.}=1.18,
\end{equation}
obtained in ref.~\cite{Lucini:2012wq} using the results for all values of $N$ from $2$ to $8$.

As mentioned above, the latent heat $L_h$ associated with the deconfinement transition is finite for all $N \ge 3$, and scales proportionally to $N^2$ in the large-$N$ limit~\cite{Lucini:2005vg, Panero:2009tv}; in particular, ref.~\cite{Panero:2009tv} reported:
\begin{equation}
\label{latent_heat}
\lim_{N \to \infty} \frac{ L_h^{1/4} }{ N^{1/2} T_c} = 0.759(19),
\end{equation}
in full agreement with the result obtained in ref.~\cite{Lucini:2005vg} for the same quantity, which reads: $0.766(40) - 0.34(1.60)/N^2$. 

In addition to the finiteness of the latent heat, the first-order nature of the transition for $N \ge 3$ is also related to the finiteness of the surface tension (i.e., energy per unit area) $\gamma_W$~associated with interfaces separating different center domains---which, in real-world QCD, might be of experimental interest~\cite{Asakawa:2012yv} (although the relevance of these objects for physics in Minkowski spacetime has been questioned~\cite{Smilga:1993vb}). In particular, the probability for formation of a ``bubble'' of surface $\mathcal{A}$ separating two different phases near the critical point is proportional to $\exp\left( -\gamma_W \mathcal{A} / T \right)$---so that tunneling events in a finite volume are exponentially suppressed by the finiteness of $\gamma_W$. Note that, from the point of view of numerical simulations, this makes it practically \emph{necessary} to study the deconfinement transition at large $N$ on lattices of relatively small volume, in order to have a sufficient number of tunnelings during the Monte Carlo history. On the other hand, however, the very first-order nature of the transition also makes it \emph{sufficient} to consider small lattices, since it suppresses finite-volume effects (for a more detailed discussion of finite-volume effects in large-$N$ gauge theories at finite temperature, see also ref.~\cite{Panero:2008mg}). In the limit of very high temperatures, the surface tension separating two different domains labelled by Polyakov loops with values $1$ and $\exp \left( 2 \pi i k/N\right)$ in $\SU(N)$ Yang-Mills theory can be computed perturbatively~\cite{Bhattacharya:1990hk, Bhattacharya:1992qb, KorthalsAltes:1993ca, Giovannangeli:2001bh, Giovannangeli:2002uv, Giovannangeli:2004sg}---see also ref.~\cite{Enqvist:1990ae}---with the leading-order result:
\begin{equation}
\label{deconfined_deconfined_interface_tension}
\gamma_W^{0 \to k} = \frac{4k(N-k)\pi^2 T^3}{3\sqrt{3 \lambda}} \left[ 1 + O(\lambda) \right].
\end{equation}
In particular, this shows that the largest of these interface tensions (for $k=\lfloor N/2 \rfloor$) is $O(N^2)$ in the 't~Hooft limit. As discussed in ref.~\cite{Lucini:2003zr}, for temperatures close to $T_c$ it is reasonable to expect that the tunneling from the center sector with Polyakov loop equal to $1$ to the one with $k=\lfloor N/2 \rfloor$ goes through an intermediate tunneling to the center-symmetric phase, with a probability related to the interface tension $\gamma_W^{c \to d}$ between the confining and deconfined phases, so that: $\gamma_W^{c \to d}=\gamma_W^{0 \to k}/2=O(N^2)$. The $N$-dependence of $\gamma_W^{c \to d}$ was studied in ref.~\cite{Lucini:2005vg}, with the result:
\begin{equation}
\label{confined_deconfined_interface_tension}
\frac{\gamma_W^{c \to d}}{N^2 T_c^3} = 0.0138(3) - \frac{0.104(3)}{N^2}, \qquad \mbox{with: $\chi^2$/d.o.f.}=2.7.
\end{equation}

As discussed in ref.~\cite{Giovannangeli:2001bh}, the surface tension $\gamma_W^{0 \to k}$ associated with walls separating different center domains in the deconfined phase is also trivially related to the ``dual'' string tension $\tilde{\sigma}$ characterizing the area-law decay of large 't~Hooft loops~\cite{'tHooft:1977hy} in the deconfined phase: $\tilde{\sigma}T=\gamma_W^{0 \to k}$. The behavior of 't~Hooft loops in large-$N$ Yang-Mills theories at finite temperature was studied numerically in refs.~\cite{Bursa:2005yv, deForcrand:2005rg}, which computed the values of $\tilde{\sigma}$ corresponding to 't~Hooft loops carrying $k$ units of flux (to be denoted by $\tilde{\sigma}_k$), and found clear evidence for nearly perfect Casimir scaling:
\begin{equation}
\label{t_Hooft_loop_tension_Casimir_scaling}
\tilde{\sigma}_k \propto k(N-k).
\end{equation}
While Casimir scaling is predicted by perturbation theory, at least at the lowest orders---see eq.~(\ref{deconfined_deconfined_interface_tension})---, refs.~\cite{Bursa:2005yv, deForcrand:2005rg} found that ratios of dual string tensions are consistent with Casimir scaling not only in the weakly coupled regime, but also down to surprisingly low temperatures, with only very small deviations close to $T_c$.

\begin{figure}[-t]
\centerline{\includegraphics[width=0.48\textwidth]{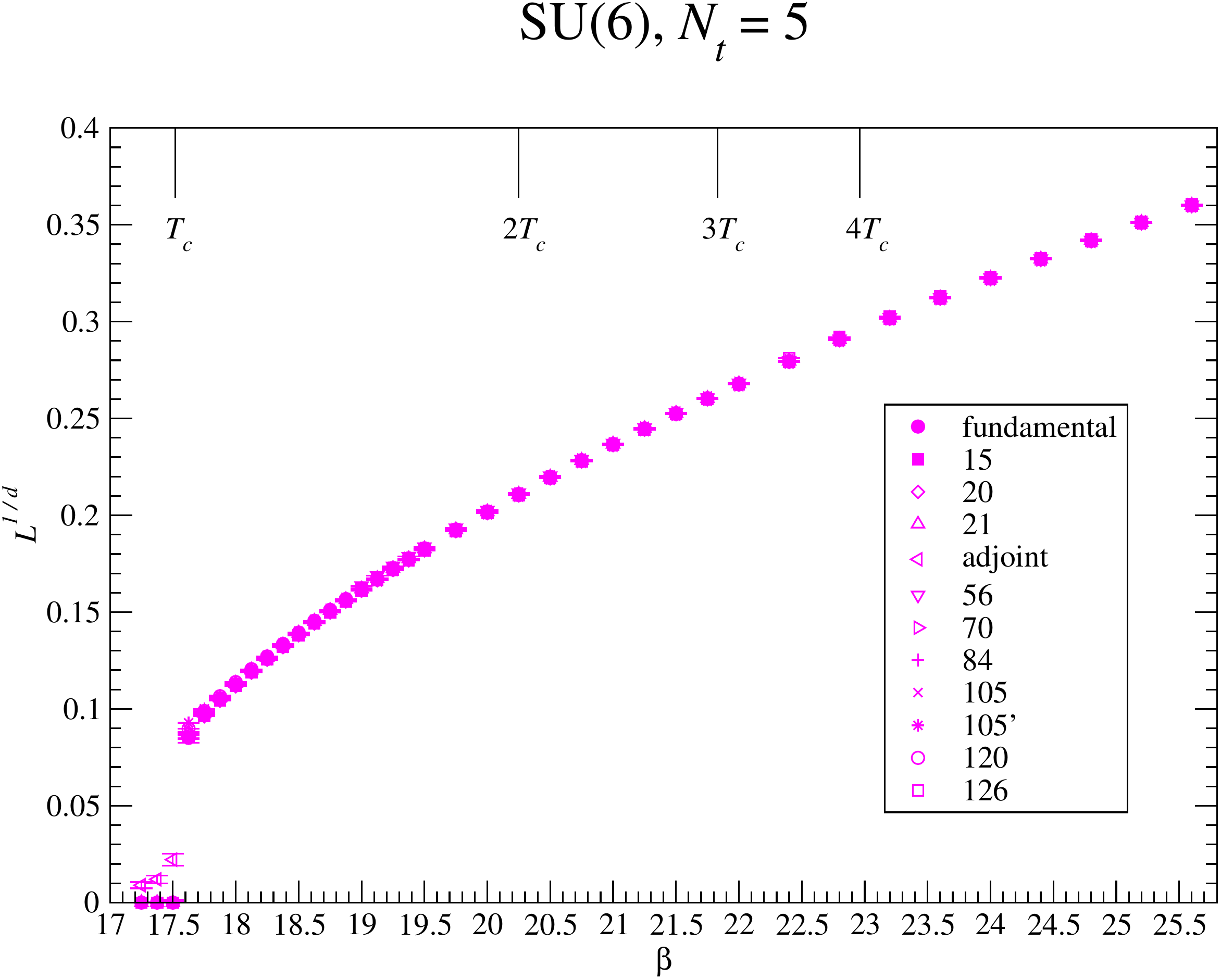} \hfill \includegraphics[width=0.48\textwidth]{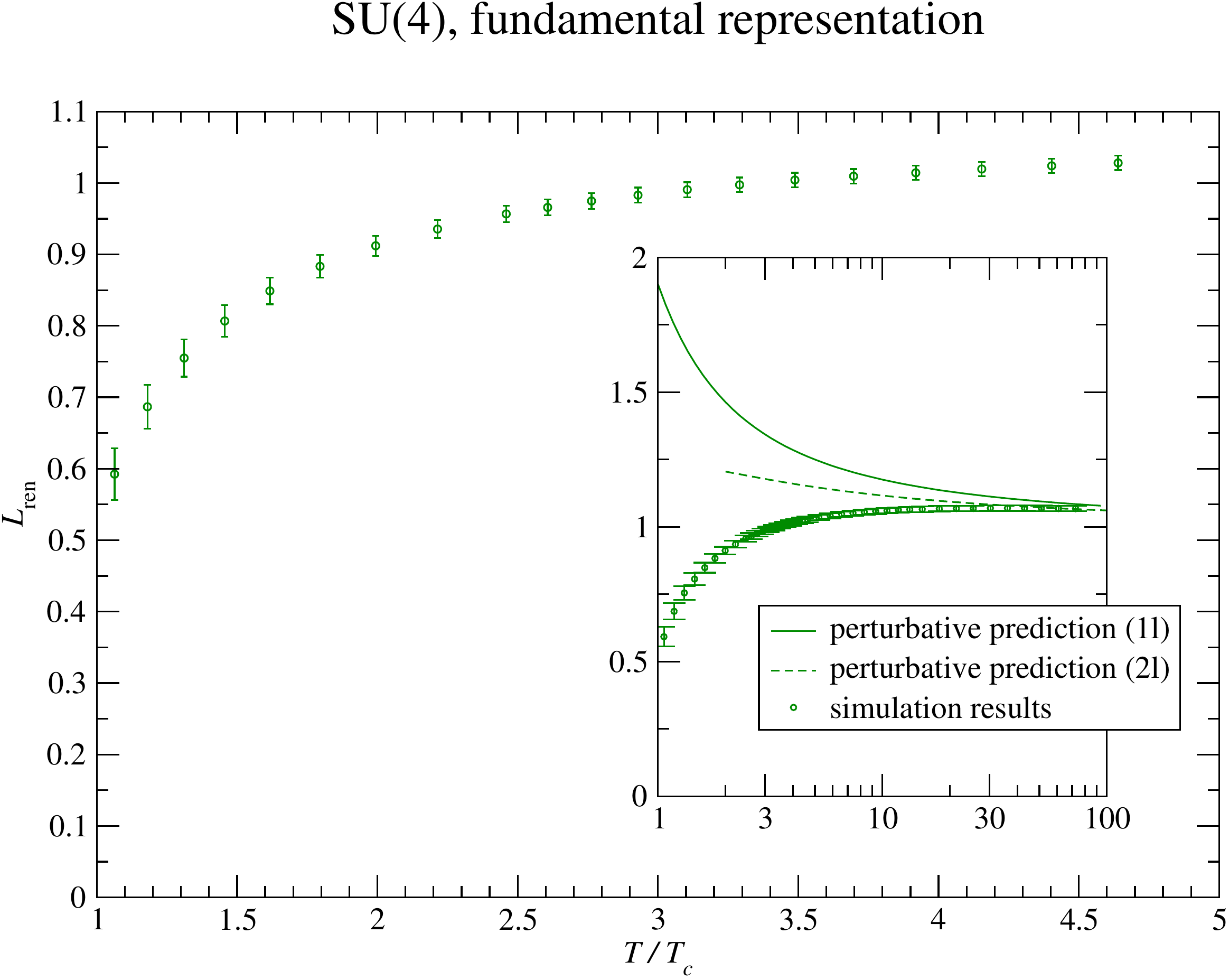}}
\caption{Left panel: The bare Polyakov loops in twelve different irreducible representations of $\SU(6)$ Yang-Mills theory, when their free energies are rescaled dividing them by the quadratic Casimir of the corresponding representation, fall on the same curve; the figure shows the results obtained from simulations on a lattice of spacing $a=1/(5T)$, plotted as a function of the Wilson gauge action parameter $\beta$, with the corresponding values of the temperature, in units of $T_c$, indicated along the upper horizontal axis. Right panel: The renormalized Polyakov loop in the fundamental representation extracted from numerical simulations in the deconfined phase of $\SU(4)$ Yang-Mills theory (open symbols), in comparison with the temperature dependence predicted perturbatively~\cite{Burnier:2009bk, Brambilla:2010xn} at one- (solid line) or two-loop order (dashed line). Note that the Polyakov loop is normalized so that it tends to $1$ in the infinite-temperature limit, and that the leading-order perturbative correction for the renormalized loop is \emph{positive}. Both plots are taken from ref.~\cite{Mykkanen:2012ri}.\label{fig:Polyakov_loops}}
\end{figure}

Interestingly, similar results have also been obtained in a recent study~\cite{Mykkanen:2012ri}, which investigated Polyakov loops in different representations and for different numbers of colors, from $2$ to $6$. This work found that the free energies associated with bare Polyakov loops in different representations (of various $N$-alities) are nearly exactly proportional to the eigenvalue of the corresponding quadratic Casimir (as first predicted in ref.~\cite{Damgaard:1987wh}), even down to temperatures close to $T_c$, while the renormalized Polyakov loops agree with perturbation theory~\cite{Burnier:2009bk, Brambilla:2010xn} at high temperatures, but receive large non-perturbative contributions at low temperatures (see fig.~\ref{fig:Polyakov_loops}). Analogous results had been previously obtained for the $\SU(3)$ theory~\cite{Dumitru:2003hp, Gupta:2007ax}. In addition, ref.~\cite{Mykkanen:2012ri} also found that the value of the renormalized fundamental Polyakov loop for $T \to T_c^{+}$ (defined with a common normalization to $1$ in the weak-coupling limit at high temperatures, and with a consistent choice of renormalization scheme for all groups) appears to be very close to $1/2$, for all the gauge groups investigated. As we mentioned in subsection~\ref{subsec:phase_diagram}, this is the value predicted in the large-$N$ limit~\cite{Damgaard:1986mx} in an effective theory of Polyakov lines; however, it is worth emphasizing that the numerical value of the renormalized Polyakov loop is a scheme-dependent quantity.

The equation of state (i.e., the temperature dependence of the free energy density) in $\SU(N)$ Yang-Mills theory at large $N$ has been independently investigated on the lattice by three different groups in a series of recent studies~\cite{Bringoltz:2005rr, Bringoltz:2005xx, Panero:2008mg, Panero:2009tv, Datta:2009jn, Datta:2010sq}. These works obtained consistent results, yielding a clear physical picture: the main finding is that, in the deconfined phase, the equilibrium thermodynamic quantities (pressure $p$, trace anomaly $\Delta$, energy density $\epsilon$ and entropy density $s$---which are related to each other by elementary thermodynamic identities) \emph{per gluon degree of freedom} have essentially the same dependence on $T/T_c$ in all $\SU(N \ge 3 )$ Yang-Mills theories. This means that, at least as far as it concerns equilibrium properties, the thermodynamic behavior of the ``physical'' theory with $N=3$ colors is the same as that of the theory in the large-$N$ limit. This result gives some confidence that the infinite-$N$ approximation underlying all holographic models of the QCD plasma~\cite{Son:2007vk, Mateos:2007ay, Shuryak:2008eq, Gubser:2009md, Rangamani:2009xk, CasalderreySolana:2011us} should not be a major source of systematic uncertainty. In fact, a quantitative comparison between the lattice results for the large-$N$ equation of state and some holographic models was carried out in ref.~\cite{Panero:2009tv}, considering, in particular, the equation of state constructed from the bottom-up model discussed in refs.~\cite{Gursoy:2007cb, Gursoy:2007er, Gursoy:2008bu, Gursoy:2009jd} (see also refs.~\cite{Andreev:2006vy, Kajantie:2006hv, Alanen:2009ej, Alanen:2009xs, Galow:2009kw, Megias:2010ku, Veschgini:2010ws, Jarvinen:2011qe, Alho:2012mh} for related work) and the relation between the entropy density and the finite 't~Hooft coupling for the $\mathcal{N}=4$ supersymmetric Yang-Mills theory,\footnote{As noted in ref.~\cite{Panero:2009wr}, a similar type of comparison (but with lattice QCD results for $N=3$ and with dynamical quarks) was also carried out in ref.~\cite{Gubser:2006qh}, finding a similar value for the 't~Hooft coupling at which the $\mathcal{N}=4$ theory best ``mimics'' QCD: $\lambda \simeq 5.5$.} as determined in refs.~\cite{Gubser:1996de,Gubser:1998nz}---see eq.~(\ref{s_over_s0}). 

\begin{figure}[-t]
\centerline{\includegraphics[width=0.48\textwidth]{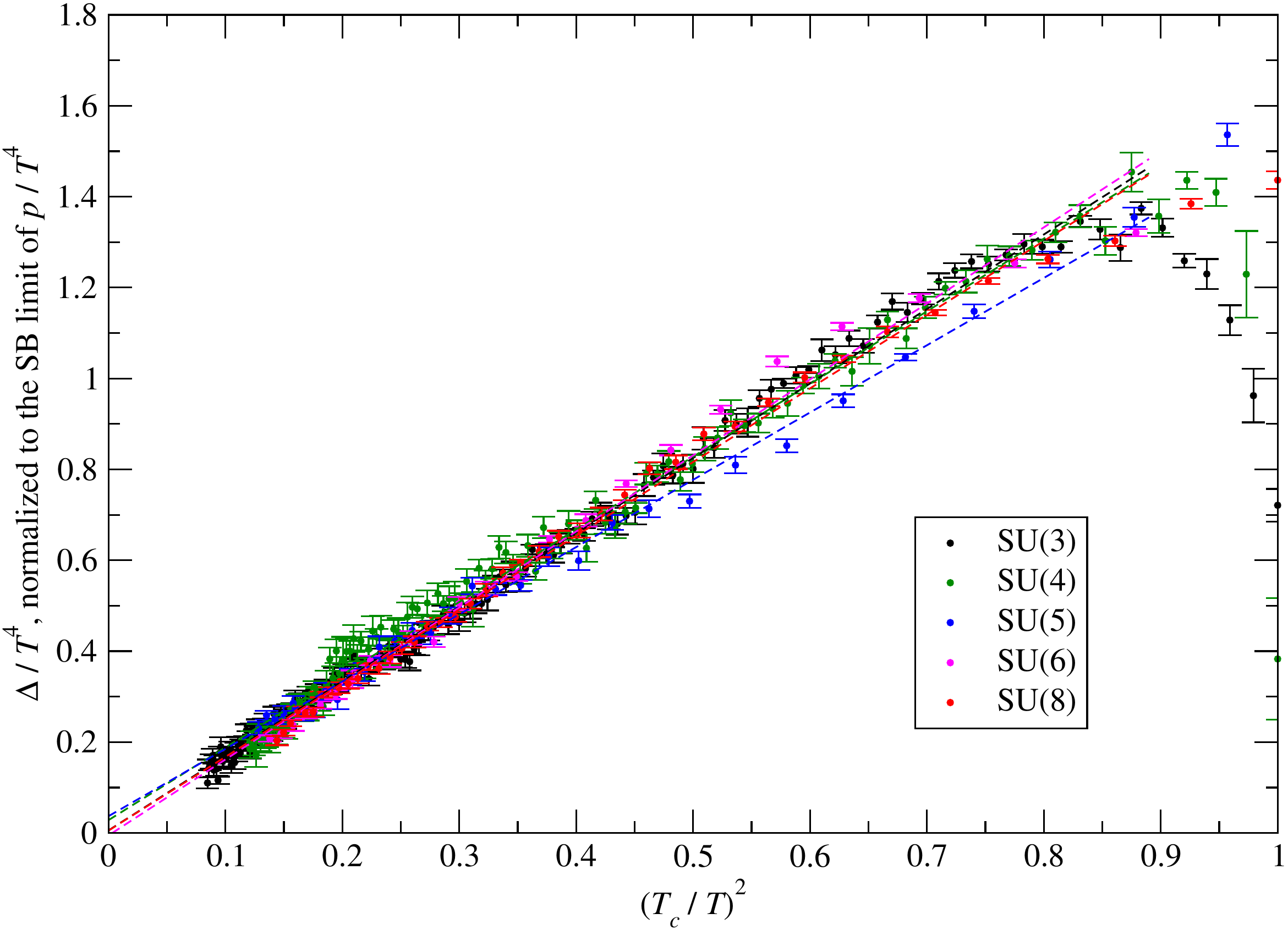} \hfill \includegraphics[width=0.56\textwidth]{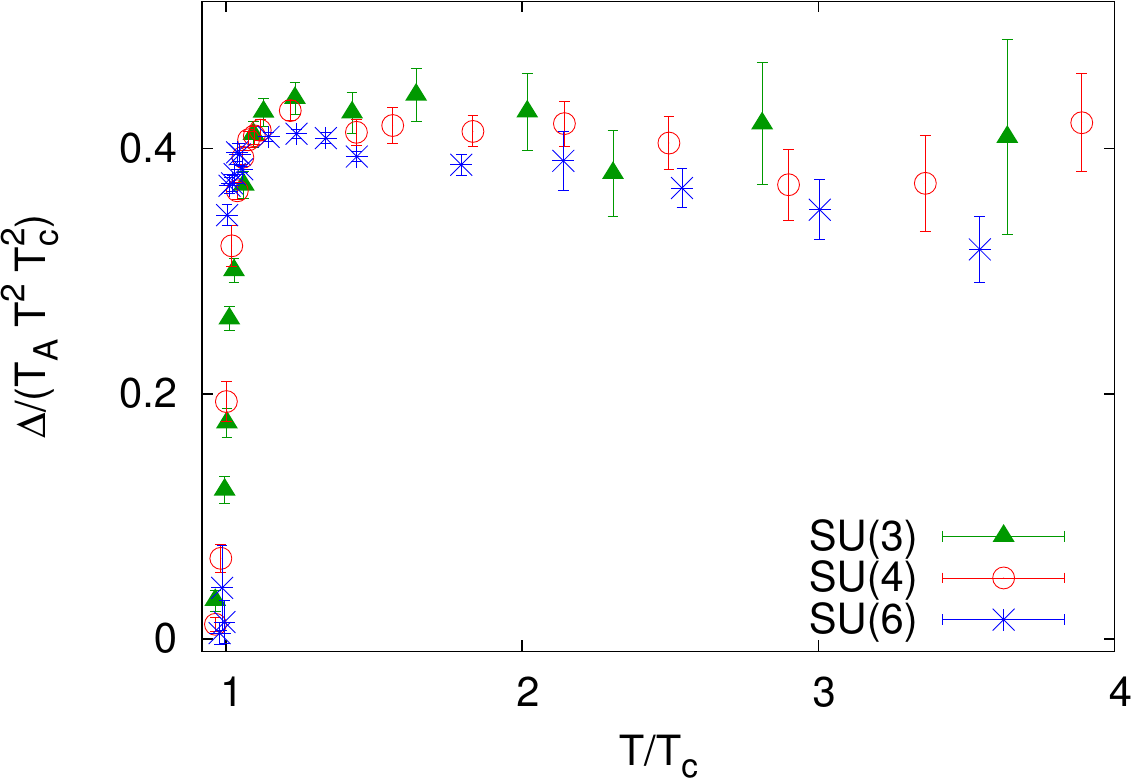}}
\caption{Left panel: At temperatures close to the deconfinement transition, the quadratic dependence of the trace anomaly on $T$ is exhibited by the linear dependence of the dimensionless ratio $\Delta/T^4$ per gluon degree of freedom (with an appropriate overall normalization) on $1/T^2$ (from ref.~\cite{Panero:2009tv}). Right panel: The same effect---and the same consistency of results obtained in theories with a different number of color charges---can also be seen from the approximate temperature-independence of the quantity $\Delta/T^2$ per gluon degree of freedom (from ref.~\cite{Datta:2010sq}).\label{fig:Delta_T2}}
\end{figure}

A further, interesting piece of information obtained in refs.~\cite{Panero:2009tv, Datta:2010sq} concerns the temperature dependence of the trace anomaly $\Delta$. Up to temperatures of a few times $T_c$ (see fig.~\ref{fig:Delta_T2}), this dependence seems to be quadratic---and possibly due to some contribution of non-perturbative origin.\footnote{As discussed in subsec.~\ref{subsec:phase_diagram}, the existence of non-perturbative effects in deconfined gauge theories is not surprising. Such effects can be investigated numerically, via lattice simulations of dimensionally reduced effective theories: this was done, e.g., in refs.~\cite{Hietanen:2004ew, Hietanen:2008tv} for the $\SU(3)$ theory, and in ref.~\cite{Hietanen:2006rc} for the generalization to the large-$N$ limit.} While this characteristic temperature dependence might be accommodated in phenomenological models~\cite{Meisinger:2001cq, Dumitru:2012fw, Megias:2005ve, Pisarski:2006yk, Brau:2009mp, Megias:2009mp, Giacosa:2010vz, Gogokhia:2010dw, Lacroix:2012pt}, its physical origin at the fundamental level remains to be understood. At higher temperatures, the lattice results exhibit a smooth approach to the perturbative predictions. 

Other interesting physical quantities at finite temperature were investigated in ref.~\cite{Lucini:2005vg}: the Debye mass $m_D$, which characterizes the gluon screening in the deconfined phase, and which, at the leading order in a weak-coupling expansion, can be expressed as~\cite{Gross:1980br}:
\begin{equation}
\label{perturbative_Debye_mass}
m_D = \sqrt{\frac{\lambda}{3}} T,
\end{equation}
and the string tension associated to large \emph{spatial} Wilson loops, $\sigma_s$, which, due to the strongly coupled nature of the long-wavelength modes of the plasma (see the discussion in subsec.~\ref{subsec:phase_diagram}), is non-vanishing at any temperature. In particular, in ref.~\cite{Lucini:2005vg} the Debye mass was extracted from the connected correlator of Polyakov loops; the results showed the independence of $m_D$ from $N$ (except close to $T_c$), and also its (approximate) proportionality to the temperature in a temperature range $ 1.5 T_c \lesssim T \lesssim 2.5 T_c$. The linear relation between $m_D$ and $T$ is compatible with the perturbative prediction (assuming that the coupling is only mildly varying in the narrow temperature range considered), although there is no reason to believe that, at those temperatures, the coupling should be sufficiently small to validate a leading-order perturbative result. As for the spatial string tension, the simulations performed in ref.~\cite{Lucini:2005vg} indicated that, when $N$ is large, $\sigma_s$ is approximately constant for $0 \le T < T_c$, whereas it is discontinuous at $T=T_c$ (with $\lim_{T \to T_c^{-}} \sigma_s(T) < \lim_{T \to T_c^{+}} \sigma_s(T)$, by about $15\%$), and increasing with $T$ in the deconfined phase.

The temperature dependence of the topological susceptibility $\topsusc$ at large $N$ was studied in refs.~\cite{Lucini:2004yh, DelDebbio:2004rw}. It was found that $\topsusc$ is nearly temperature-independent (and finite in the 't~Hooft limit) for $0 \le T < T_c$, but it is strongly suppressed (and, likely, vanishing for $N \to \infty$) in the deconfined phase.

Conversely, the authors of the recent work~\cite{D'Elia:2012vv} investigated the dependence of the critical temperature $T_c$ on $\theta$ in the $\SU(3)$ theory, by monitoring the Polyakov loop susceptibility in simulations at imaginary values of $\theta$ (to avoid the sign problem), and then performing an analytical continuation to real $\theta$. They showed that $T_c$ is decreasing when $\theta$ varies from zero to a finite, real value, and, following ref.~\cite{Witten:1998uka}, they obtained:
\begin{equation}
\label{Tc_dependence_on_theta}
\frac{T_c(\theta)}{T_c(0)} = 1 - \frac{\topsusc}{2 L_h} \theta^2 + O(\theta^4) \simeq 1 - \frac{0.253(56)}{N^2} \min_k (\theta + 2 \pi k)^2
\end{equation}
at large $N$, having combined their results with large-$N$ values for the topological susceptibility, for the latent heat and for the critical temperature from refs.~\cite{Vicari:2008jw, Lucini:2004yh, Lucini:2005vg}. The result that, at finite $N$, the critical temperature decreases when $\theta$ takes non-zero values, is consistent with theoretical expectations~\cite{Poppitz:2012nz, Unsal:2012zj} (see subsec.~\ref{subsec:phase_diagram}).

\subsection{Results in three spacetime dimensions}
\label{subsec:3D_results}

Large-$N$ gauge theories in three spacetime dimensions (3D) have been investigated in several lattice studies, with many findings qualitatively similar to those obtained in four spacetime dimensions (4D).

$\SU(N)$ Yang-Mills theories in 3D are characterized by a \emph{dimensionful} gauge coupling: $g^2$ has the dimensions of an energy, so that a natural dimensionless parameter for organizing perturbative expansions for physical processes at the momentum scale $k$ is the $g^2/k$ ratio (see, e.g., refs.~\cite{Farakos:1994xh, Reuter:1993nn} and references therein). They are asymptotically free at high energy, and super-renormalizable; in addition, one can also prove that the Coulomb potential in 3D is characterized by a logarithmic dependence on the distance---a behavior which, in particular, would already be sufficient to imply color confinement. However, perturbation theory is not reliable in the limit of large distances, in which these theories become strongly coupled. The computational tool to study them in the non-perturbative regime is, thus, lattice simulations---in which the $\beta$ parameter appearing as a coefficient in front of the Wilson gauge action is defined as $\beta=2N/(ag_0^2)=2N^2/(a\lambda_0)$, where $\lambda_0$ denotes the bare (lattice) 't~Hooft coupling. Like in 4D, $\SU(N)$ lattice gauge theories in 3D are characterized by a strong-coupling regime at small $\beta$, and by a weak-coupling regime at large $\beta$; in 3D, the two are separated by a crossover, which seems to turn into a third-order phase transition for $N \to \infty$~\cite{Bursa:2005tk}. Recently, a similar study has also been performed for the phase structure of 3D $\SO(2N)$ Yang-Mills theories~\cite{Bursa:2012ab}.

\subsubsection{Is the large-$N$ limit of QCD in 3D a confining theory?}
\label{subsubsec:3D_confining}

The numerical study of large-$N$ Yang-Mills theories in 3D was initiated during the 1990's, and the main results are summarized in refs.~\cite{Teper:1997tp, Teper:1998te, Lucini:2002wg}: in particular, at zero (or low) temperature, all 3D $\SU(N \ge 2)$ gauge theories are confining, with a non-perturbatively generated, asymptotically linear quark-antiquark potential $V(r)$ at large distance $r$. In particular, the string tension $\sigma$ (i.e., the asymptotic slope of the potential at large $r$) is related to the bare 't~Hooft coupling $\lambda_0$ by~\cite{Lucini:2002wg}:
\begin{equation}
\label{sigma_over_lambda_0_3D}
\frac{\sqrt{\sigma}}{\lambda_0} = 0.19755(34) - \frac{0.1200(29)}{N^2}.
\end{equation}

More recently, the numerical determination of the string tension in 3D Yang-Mills theories was pushed to very high precision in ref.~\cite{Bringoltz:2006zg}: the goal was to test the predictions formulated in a series of works by Karabali, Kim and Nair~\cite{Karabali:1995ps, Karabali:1997wk, Karabali:1998yq, Karabali:2000gy}, in which the vacuum wavefunction and the string tension of 3D Yang-Mills theories were computed analytically in a Hamiltonian approach. The results of previous lattice studies~\cite{Teper:1998te, Lucini:2002wg} had shown some discrepancies with respect to these predictions, but could not give a conclusive answer about their validity for $N \to \infty$. On the contrary, the authors of ref.~\cite{Bringoltz:2006zg} presented an accurate extrapolation to the large-$N$ limit, which revealed a discrepancy of six standard deviations with the predictions of ref.~\cite{Karabali:1998yq}. The interesting analytical program discussed in refs.~\cite{Karabali:1995ps, Karabali:1997wk, Karabali:1998yq, Karabali:2000gy} is presently ongoing: a computation of the glueball spectrum was presented in refs.~\cite{Leigh:2005dg, Leigh:2006vg}, while refinements of the string tension predictions have been recently proposed in ref.~\cite{Karabali:2009rg}; however, it is fair to say that its theoretical foundations are still under debate~\cite{Witten_SC_slides}.

\subsubsection{Confining flux tubes in 3D as strings}
\label{subsubsec:3D_confining_flux_tubes}

Inspection of the numerical results for $V(r)$ at intermediate distances reveals that the potential also includes a $1/r$ contribution, with a coefficient that is compatible with the L\"uscher term prediction: this means that, similarly to what happens in 4D, also in 3D the low-energy dynamics of confining flux tubes can be described in terms of (massless) fluctuations of a bosonic string~\cite{Mykkanen:2012dv, Luscher:2002qv, Teper:1998te, Athenodorou:2007du, Bringoltz:2008nd, Athenodorou:2008cj, Bialas:2009pt, Athenodorou:2011rx, Caselle:2011vk}. Again, this is not specific of the large-$N$ limit only, but rather appears as a generic feature of all confining models in 3D: the L\"uscher term (or the broadening of confining flux tubes with their length~\cite{Luscher:1980iy, Armoni:2008sy, Pfeuffer:2008mz, Allais:2008bk, Gliozzi:2010zv, Gliozzi:2010zt, Caselle:2012rp}, which is a related implication of the same effective picture) is also observed in high-precision studies of $\SU(2)$ Yang-Mills theory~\cite{Caselle:2011vk, Ambjorn:1984me, Majumdar:2002mr, Juge:2004xr, Caselle:2004er, Brandt:2009tc, Brandt:2010bw}, of 3D lattice models with local invariance under a discrete gauge group~\cite{Caselle:2011vk, Juge:2004xr, Caselle:1996ii, Caselle:2002rm, Caselle:2002ah, Caselle:2004jq, Caselle:2005xy, Caselle:2005vq, Caselle:2006dv, Giudice:2006hw, Caselle:2007yc, Giudice:2007sk, Rajantie:2012zn} and even in random percolation models (with an appropriate, topological definition of Wilson loop operators)~\cite{Gliozzi:2005ny, Giudice:2009di}. 

\begin{figure}[-t]
\centerline{\includegraphics[height=0.5\textheight]{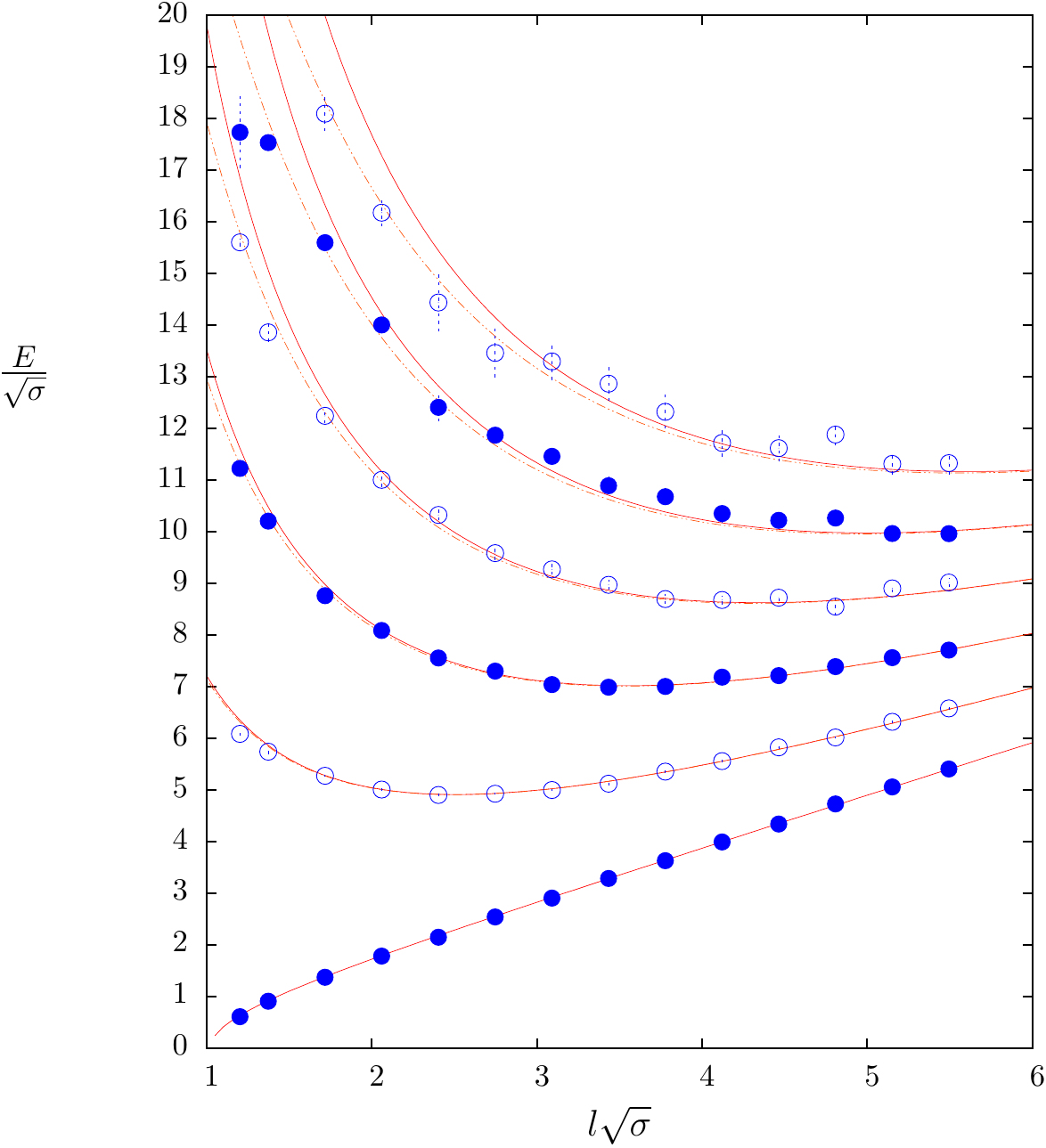}}
\caption{Ground state and excitation spectrum of a closed confining flux tube, winding around a spatial size of the lattice (torelon) in $\SU(6)$ Yang-Mills theory in 3D, from ref.~\cite{Athenodorou:2011rx}. The symbols show the lattice results for the energies of different states, as a function of the torelon length, while the lines denote the corresponding predictions from the Nambu-Goto string model, i.e., eq.~(\protect\ref{closed_string_spectrum}) for $D=3$. In particular, the solid lines refer to the predictions using the continuum definition of the momentum, while the dashed lines are obtained from the lattice momentum (in a free theory)---which differs from the former at scales of the order of the inverse lattice spacing.\label{fig:Athenodorou_et_al_11035854_SU6_3d_string_plot_EgsQall_n6f}}
\end{figure}

In fact, these studies are now reaching a level of numerical precision, at which it becomes possible to investigate subleading (in $1/r$) corrections to the confining potential, and possibly compare them with analytical computations that predict at which order the correct effective string model should deviate from the Nambu-Goto string~\cite{Aharony:2010cx, Aharony:2010db, Aharony:2011gb, Billo:2012da, Gliozzi:2012cx}. As an example, fig.~\ref{fig:Athenodorou_et_al_11035854_SU6_3d_string_plot_EgsQall_n6f}, taken from ref.~\cite{Athenodorou:2011rx}, shows the torelon spectrum obtained from simulations in $\SU(6)$ Yang-Mills theory, in comparison with the closed-string energy levels predicted by the Nambu-Goto model, see eq.~(\ref{closed_string_spectrum}). As shown by this figure, in which both the torelon length and the energies are measured in appropriate units of the relevant energy scale (set by the square root of the string tension), a somewhat surprising feature is that the agreement with the Nambu-Goto model appears to hold down to very short distances, at which \emph{a priori} there is no reason to expect that the approximation of the confining flux tube as a uni-dimensional string should still be valid. In fact, the deviations from the curves predicted by the Nambu-Goto model are mostly accounted for, by replacing the continuum definition of the momentum with its lattice counterpart (the two become significantly different only at scales comparable with the inverse lattice spacing).

\subsubsection{Glueball spectrum in 3D}
\label{subsubsec:3D_spectrum}

The spectrum of 3D Yang-Mills theories at large $N$ (including several excitation states) was studied via lattice simulations in various works~\cite{Meyer:2004jc, Teper:1997tp, Teper:1998te, Lucini:2002wg, Meyer:2003wx} (and compared with a variant of the Isgur-Paton model~\cite{Isgur:1984bm} in ref.~\cite{Johnson:2000qz}): the physical states are hadrons (glueballs), which can be classified according to the irreducible representations of the $\SO(2)$ group in the continuum (or to the corresponding subduced representations restricted to the square group, if the theory is regularized on a cubic Euclidean lattice), as well as a ``mirror'' parity\footnote{Note that, for any even number $d$ of spatial dimensions, the ``usual'' parity transformation, corresponding to inversion of all spatial coordinates, is nothing but a rotation, because $-\ide \in \SO(d)$ for $d$ even.} and charge conjugation. The simulations show that these theories have a finite mass gap (which remains non-vanishing in the large-$N$ limit), and that ratios of different masses are almost independent of $N$---up to small, $O(1/N^2)$ corrections.

A lattice study of correlation functions of gauge-dependent quantities (gluon and ghost propagators) in the Landau gauge in 3D Yang-Mills theories was carried out in ref.~\cite{Maas:2010qw}. This work compared the results obtained in $\SU(N)$ theories with a number of colors ranging from $2$ to $6$, and in $\Gtwo$ Yang-Mills theory, finding essentially no dependence on the rank of the gauge group.\footnote{Analogous results have also been obtained from the comparison of the $\SU(2)$ and $\SU(3)$ theories in four spacetime dimensions~\cite{Bogolubsky:2009dc, Cucchieri:2007rg, Cucchieri:2008fc, Sternbeck:2007ug}.}

\subsubsection{Large-$N$ QCD in 3D at finite temperature}
\label{subsubsec:3D_finite_T}

The properties of 3D $\SU(N)$ \ Yang-Mills theories at finite temperature are qualitatively similar to those found in 4D: there exists a deconfinement transition at a critical temperature $T_c$, at which color-singlet hadronic states give way to a plasma of deconfined particles, and the value of $T_c$ (in units of the square root of the string tension) remains finite in the 't~Hooft limit~\cite{Liddle:2008kk}:
\begin{equation}
\label{Tc_root_sigma_3D}
\frac{T_c}{\sqrt{\sigma}} = 0.9026(23) + \frac{0.880(43)}{N^2}.
\end{equation}
The transition is of second order for ``small'' values of $N$, i.e. for $\SU(2)$~\cite{Christensen:1990vc, Teper:1993gp} and $\SU(3)$~\cite{Christensen:1991rx, Bialas:2012qz} gauge theory, with critical indices respectively in agreement with those of the $\Z_2$ and $\Z_3$ spin models in two dimensions~\cite{Svetitsky:1982gs}. For $\SU(4)$, identifying the order of the transition has been a challenging problem~\cite{deForcrand:2003wa}, but the most recent and accurate studies indicate that it is a first-order one~\cite{Holland:2007ar}, like for $N \ge 5$~\cite{Liddle:2008kk, Holland:2005nd}---in agreement with the intuitive picture of a more and more abrupt transition, due to an increasing imbalance between the free energy in the confined phase (which is $O(1)$) and the same quantity in the deconfined phase,  in which it scales like $O(N^2)$.

\begin{figure}[-t]
\centerline{\includegraphics[height=0.5\textheight]{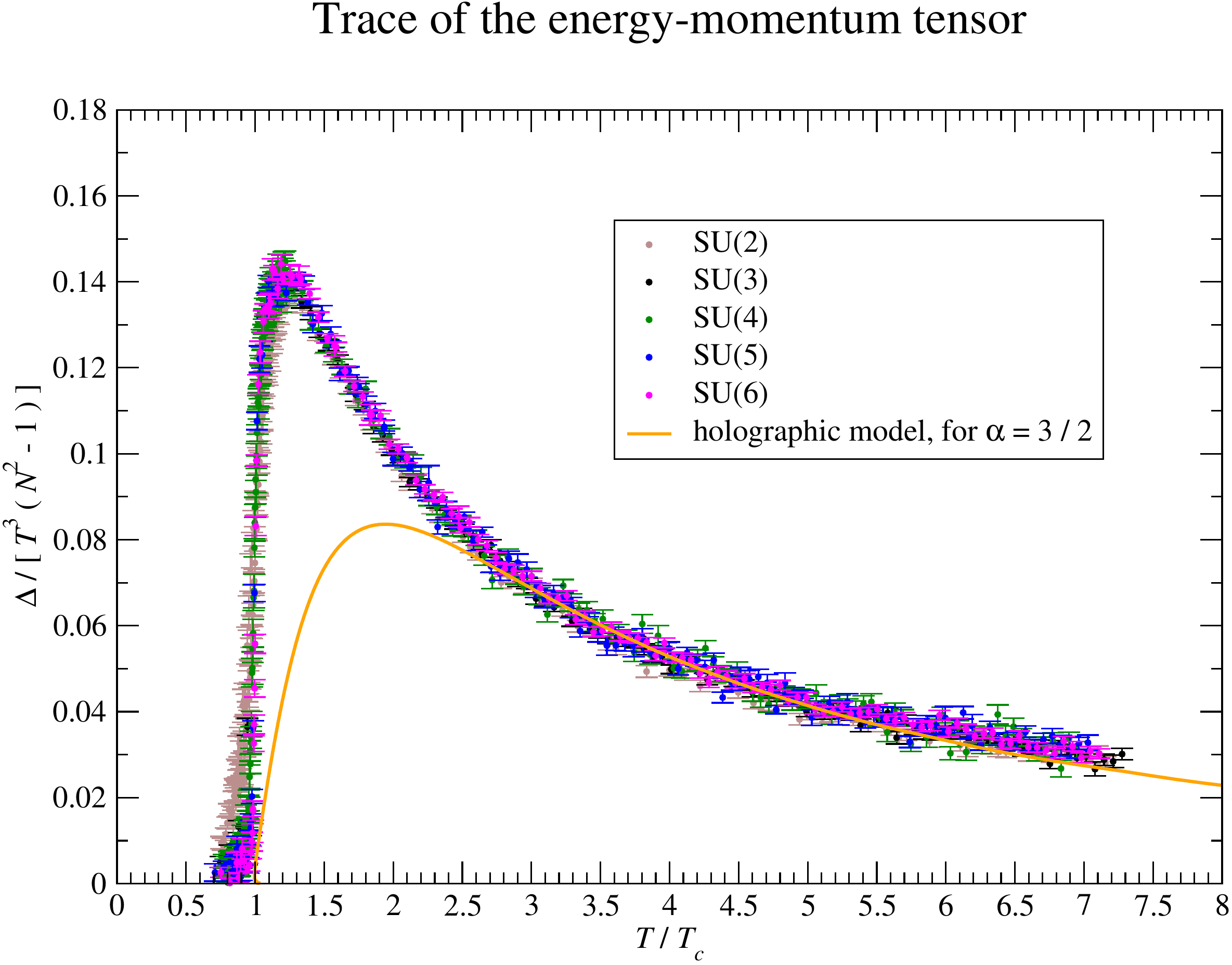}}
\caption{Temperature dependence of the trace of the energy-momentum tensor in units of $T^3$ per gluon from simulations of the deconfined phase of 3D $\SU(N)$ Yang-Mills theories, from ref.~\cite{Caselle:2011mn}. The plot shows the simulation results for $N=2$ (brown symbols), $3$ (black), $4$ (green), $5$ (blue) and $6$ (magenta), and their comparison with the expectation (yellow curve) obtained from a generalization of the holographic model proposed in refs.~\cite{Gursoy:2008bu, Gursoy:2009jd} to 3D Yang-Mills theory.\label{fig:3D_trace_anomaly_deconfined}}
\end{figure}

The large-$N$ lattice computations performed in refs.~\cite{Caselle:2011fy, Caselle:2011mn} have shown that in the confining phase the equation of state can be adequately modelled in terms of a gas of almost non-interacting, massive glueballs, as obtained by summing the known glueball states~\cite{Teper:1998te, Lucini:2002wg}---except very close to $T_c$, where heavier states (whose density can be modelled in terms of a simple bosonic string model, following an approach similar to the one discussed in ref.~\cite{Johnson:2000qz}) appear to become relevant. With the exception of a narrow region of temperatures in the vicinity of $T_c$, the equation of state is essentially independent of $N$. On the contrary, in the deconfined phase the pressure (like the other equilibrium thermodynamics properties) is nearly perfectly proportional to the number of gluon degrees of freedom $(N^2-1)$, although the approach to the asymptotic value given by the Stefan-Boltzmann limit:
\begin{equation}
\label{Stefan_Boltzmann_3D}
\lim_{T \to \infty} \frac{p}{T^3} = (N^2-1) \frac{\zeta(3)}{2\pi}
\end{equation}
(where $\zeta(3)$ denotes Ap\'ery's constant, approximately equal to $1.20205690316\dots$) is rather slow: at temperatures around $7~T_c$, the pressure is still about $15\%$ smaller than the limiting value in eq.~(\ref{Stefan_Boltzmann_3D}). This indicates that, at those temperatures, the deconfined plasma is still far from a gas of non-interacting gluons (for the $\SU(3)$ theory, analogous results have also been obtained in ref.~\cite{Bialas:2008rk}). Fig.~\ref{fig:3D_trace_anomaly_deconfined}, taken from ref.~\cite{Caselle:2011mn}, shows a plot of the trace of the energy-momentum tensor $\Delta$ in units of $T^3$ (which equals the derivative of $p/T^3$ with respect to $\ln T$) per gluon d.o.f., as a function of $T/T_c$: the perfect agreement between the simulation results for different gauge groups shows that, in the deconfined phase, the equilibrium thermodynamic quantities are exactly proportional to $N^2$ in the large-$N$ limit. An interesting feature emerging from these results is that, like in 4D, also in 3D $\Delta$ is proportional to $T^2$ in this range of temperatures. Finally, the solid yellow curve in the figure shows the prediction from a generalization of the holographic model proposed in refs.~\cite{Gursoy:2008bu, Gursoy:2009jd} for 3D Yang-Mills theory.

\subsection{Results in two spacetime dimensions}
\label{subsec:2D_results}

QCD toy models in two spacetime dimensions (2D) are interesting theoretical laboratories, in which, by virtue of a limited number of physical degrees of freedom, various problems can be studied analytically (or semi-analytically) in the large-$N$ limit. As we discussed in sec.~\ref{sec:large_N_limit}, classical examples include the determination of the meson spectrum in the 't~Hooft model~\cite{'tHooft:1974hx}, the discovery of a phase transition separating the weak- and strong-coupling phases of the lattice regularization of the theory with the Wilson gauge action~\cite{Gross:1980he, Wadia:1979vk},\footnote{Note, however, that this phase transition is not physical, and its nature and very existence depend on the lattice discretization that is used~\cite{Lang:1980sz, Lang:1980ws, Menotti:1981ry}---see also ref.~\cite{Jurkiewicz:1982iz} for a discussion.} and the analysis of the spectral density of Wilson loops in the continuum~\cite{Durhuus:1980nb} (which continues to attract attention to this day~\cite{Narayanan:2007dv, Rossi:1994xg, Olesen:2006gt, Olesen:2007rf, Blaizot:2008nc, Neuberger:2008mk, Neuberger:2008ti, Lohmayer:2009aw, Lohmayer:2011nq}).

Recent examples of analytical results in 2D models of QCD at large $N$ include, e.g., those obtained in the study of form factors in the principal chiral model~\cite{Orland:2011rd, Orland:2012sk, Cubero:2012xi}---a theory which has also been studied numerically~\cite{Narayanan:2008he} (see also ref.~\cite{Rossi:1996hs}). 

Finally, another problem that has been recently investigated in the context of 2D large-$N$ QCD, is the one of baryonic matter at finite density~\cite{Bringoltz:2008iu, Bringoltz:2009ym, Galvez:2009rq}. In particular, in ref.~\cite{Bringoltz:2009ym} it was found that, at finite baryon density, the ground state of the system breaks translation invariance, with the formation of a chiral crystal; related topics have been discussed in refs.~\cite{Schon:2000he, Kojo:2011fh}.

\section{Conclusions}
\label{sec:conclusions}

The idea put forward by 't~Hooft in his seminal article~\cite{'tHooft:1973jz} proved to be among the most fruitful ones of the last few decades in theoretical elementary particle physics, and paved the way to a number of developments in different directions. The implications of the large-$N$ limit for non-Abelian gauge theories have been studied in fields ranging from string theory to QCD phenomenology, and in many cases have even generated whole new research areas. Along with this rich variety of developments, however, also came an increasing specialization and fragmentation of the different subfields---a trend presently common to all scientific disciplines (or, more generally, in nearly all branches of human knowledge). In our view, this excessive specialization (together with a tendency towards increasing precarity in the scientific job market) may have harmful consequences for the long-term progress of science, as it naturally limits the possibility of fruitful exchanges of ideas between scholars working in different, albeit closely related, areas. 

This state of affairs gave us our first motivation to write the present review, whose primary purpose consists in offering a broad, unitary, panoramic view of large-$N$ gauge theories. While describing the general aspects of this topic in section~\ref{sec:large_N_limit}, we tried to emphasize the main ideas and the fundamental physical concepts, leaving a technical account of the details out of our discussion. We tried to be as pedagogical as possible, precisely in order to facilitate the comprehension of the various problems, studied by different research communities, as clearly as possible. In doing so, on several occasions we were, unfortunately, bound to be incomplete, or not as accurate as we wished. There are many topics, which we could only touch upon, or mention very briefly: in apologizing for our omissions, we hope that, in this review, the interested readers can find at least a reference to an article or to a review presenting a more detailed discussion of some specific subject---but, at the same time, we also encourage them to read the sections that may look not directly relevant for the topic they are interested in. 

Our second motivation for writing this manuscript was, admittedly, more directly related to our own research work, and consisted in presenting the current status of lattice studies of large-$N$ gauge theories. These studies (both the analytical and the numerical ones) are now reaching a level of maturity, and can give precise answers to many theoretical questions, which are relevant in a large variety of contexts. We hope that this manuscript serve as a vehicle, to inform the large community of researchers interested in topics related to 't~Hooft's limit about the recent progress in the lattice field. We found it important to devote an entire section (sec.~\ref{sec:lattice}) to a pedagogical introduction of the lattice formulation of field theory, in order to facilitate the interpretation of the results reviewed in sec.~\ref{sec:results} also for non-experts. In doing so, we hope that we succeeded in clarifying the main virtues and strengths of lattice computations, and in clearing out some misconceptions or prejudices which are, unfortunately, not so uncommon among non-practitioners. 

The overview of lattice results presented in sec.~\ref{sec:results} shows the convincing numerical results that have been obtained for large-$N$ gauge theories during the last fifteen years, and highlights the progress in areas which, until recently, appeared to be out of the reach of lattice calculations. Highly excited states of confining flux tubes, hadron spectroscopy, thermal observables (to name but a few) are all quantities for which lattice computations can now provide very precise and accurate results---while the first few dynamical simulations at $N>3$ are beginning to appear. On the theoretical side, the last decade has also witnessed very significant developments on the topics related to the time-honored idea of large-$N$ volume independence, for which, in particular, there has been a lot of progress in understanding the transition from the strong- to weak-coupling regimes. At the same time, mathematical tools inspired by orbifold equivalences in string theory led to deep conceptual advances in our understanding of large-$N$ volume independence and large-$N$ equivalences between theories with different field content. As we discussed in sec.~\ref{sec:from_factorization_to_orbifold}, the techniques that have been developed in the process are also finding practical applications for some of the most challenging problems to be studied on the lattice---supersymmetry is one, might QCD at finite density be the next?

To summarize the state of the art in this field, we can say that, while many problems have already been successfully studied, many others are currently still open, and being actively investigated. In view of the advances in lattice QCD computations in the last few years, we foresee further, dramatic progress for large-$N$ studies on the lattice in the near future. For these reasons, we would like to conclude this review encouraging especially the young, ambitious and talented researchers to join this field.

\vskip1.0cm 
\noindent{\bf Acknowledgements}\\
The perspective offered by this review is the result of countless discussions with our collaborators and many colleagues working in this field, whom it is impossible to mention one by one. While most of the credit for the (numerical and analytical) results we have presented goes to them, the responsibility for any error and omission lies exclusively with us. This work is supported by the Academy of Finland, project 1134018, by the Royal Society (grant UF09003) and in part by STFC under grant ST/G000506/1 and by the US National Science Foundation under grant PHY11-25915. M.P. acknowledges the Kavli Institute for Theoretical Physics in Santa Barbara, California, US, for support and hospitality during the ``Novel Numerical Methods for Strongly Coupled Quantum Field Theory and Quantum Gravity'' program (January 17-March 9,~2012)~\cite{novelnum12}, during which part of this work was done.

\newpage



%% References


% \bibliography{<your-bib-database>}
%%%%%%%%%%%%%%%%%%%%%%%%%%%%%%%%%%%%%%%%%%%%%%%%
\addcontentsline{toc}{section}{References}
\bibliographystyle{h-elsevier3}
\bibliography{largenlattice,latt-gen}






% \end{thebibliography}

\end{document}